\documentclass[a4paper,10pt]{article}

\pdfoutput=1

\usepackage{jheppub}

\usepackage{graphicx}         
\usepackage{dcolumn}        
\usepackage{bm}                
\usepackage{amssymb}     
\usepackage{subfigure}
\usepackage{verbatim}
\usepackage{environ}

\usepackage[english]{babel}
\usepackage{xcolor}
\usepackage{graphicx}
\usepackage{amsmath}
\usepackage{dsfont}
\usepackage[makeroom]{cancel}
\usepackage{url}
\usepackage{hyperref}
\usepackage{slashed}
\usepackage{mathabx}
\usepackage{mathtools}

\newcommand{\secn}[1]{Section~\ref{#1}}
\newcommand{\appn}[1]{Appendix~\ref{#1}}
\newcommand{\beq}{\begin{eqnarray}}
\newcommand{\eeq}{\end{eqnarray}}

\newcommand{\as}{\alpha_s}
\newcommand{\eps}{\epsilon}

\newcommand{\sig}{\sigma}
\newcommand{\Gam}{\Gamma}
\newcommand{\gam}{\gamma}

\newcommand{\npo}{{n+1}}
\newcommand{\npt}{{n+2}}
\newcommand{\pn}{\Phi_n}

\newcommand{\mc}{\mathcal}

\newcommand{\Li}{\mbox{Li}}
\newcommand{\dt}{\!\cdot\!}

\newcommand{\one}{\, (\mathbf{1})}
\newcommand{\two}{\, (\mathbf{2})}
\newcommand{\otwo}{\, (\mathbf{12})}

\newcommand{\RV}{(\mathbf{RV})}
\newcommand{\twoRV}{(\mathbf{2+RV})}
\newcommand{\IK}{K}
\newcommand{\otwoRV}{(\mathbf{RV+12})}
\newcommand{\LO}{{\mbox{\tiny{LO}}}}
\newcommand{\NLO}{{\mbox{\tiny{NLO}}}}
\newcommand{\NNLO}{{\mbox{\tiny{NNLO}}}}

\newcommand{\nnb}{\nonumber}

\newcommand{\kb}{\bar{k}}

\newcommand{\kt}{\tilde{k}}

\newcommand{\Si}{{\bf S}_i}

\newcommand{\Cj}{{\bf C}_{ij}}

\newcommand{\bSi}{\overline{\bf S}_i}

\newcommand{\bCj}{\overline{\bf C}_{ij}}

\newcommand{\tk}{\tilde k}

\newcommand{\q}{{q}}

\def\eq#1{Eq.~(\ref{#1})}

\newcommand{\bL}[2]{{\bf L}_{#1}^{#2}}
\newcommand{\bS}[1]{{\bf S}_{#1}}
\newcommand{\bC}[1]{{\bf C}_{#1}}
\newcommand{\bSC}[1]{{\bf SC}_{#1}}

\newcommand{\bbL}[1]{\overline{\bf L}_{#1}}
\newcommand{\bbS}[1]{\overline{\bf S}_{#1}}
\newcommand{\bbC}[1]{\overline{\bf C}_{#1}}
\newcommand{\bbSC}[1]{\overline{\bf SC}_{#1}}

\newcommand{\bbHC}[1]{\overline{\bf HC}_{#1}}
\newcommand{\bbSHC}[1]{\overline{\bf SHC}_{#1}}

\newcommand{\W}[1]{\mc{W}_{#1}}
\newcommand{\bW}[1]{\overline{\mc{W}}_{#1}}

\newcommand{\Z}[1]{\mc{Z}_{#1}}
\newcommand{\bZ}[1]{\bar{\mc{Z}}_{#1}}

\newcommand{\kkl}[1]{\{\bar k\}^{(#1)}}
\newcommand{\kk}[2]{\bar k_{#1}^{(#2)}}
\newcommand{\sk}[2]{\bar s_{#1}^{(#2)}}

\newcommand{\Norm}{\mc{N}_1}

\newcommand{\Bn}{B}
\newcommand{\Vl}{V}
\newcommand{\Rl}{R}
\newcommand{\RR}{RR}

\newcommand{\bB}{\bar{\Bn}}
\newcommand{\bR}{\bar{\Rl}}
\newcommand{\bV}{\bar{\Vl}}

\newcommand{\varsi}{\varsigma}

\newcommand{\euler}{\gamma_{\!_{_E}}}

\newcommand{\zg}{({\rm 0g})}
\newcommand{\og}{({\rm 1g})}
\newcommand{\tg}{({\rm 2g})}

\newcommand{\sS}{{\scriptscriptstyle\rm S}}

\newcommand{\sSS}{{\scriptscriptstyle\rm SS}}
\newcommand{\sSHC}{{\scriptscriptstyle\rm SHC}}
\newcommand{\sHCC}{{\scriptscriptstyle\rm HCC}}
\newcommand{\sHCHC}{{\scriptscriptstyle\rm HCHC}}
\newcommand{\sRV}{{\scriptscriptstyle\rm sRV}}

\newcommand{\hcRV}{{\scriptscriptstyle\rm hcRV}}

\preprint{\\  \rightline{MPP-2022-293, \, TTP23-001, \, P3H-23-002}}

\title{
NNLO subtraction for any massless final state: a complete analytic expression
}

\author[a]{Gloria Bertolotti,}
\author[a]{Lorenzo Magnea,}
\author[b]{Giovanni Pelliccioli,}
\author[b]{Alessandro Ratti,}
\author[c]{Chiara Signorile-Signorile,}
\author[a]{Paolo Torrielli,}
\author[a]{Sandro Uccirati}

\affiliation[a]{Dipartimento di Fisica, Universit\`a di Torino,
and INFN, Sezione di Torino, \\ Via P. Giuria 1, I-10125 Torino,
Italy}
\affiliation[b]{
Max-Planck-Institut f\"ur Physik, F\"ohringer Ring 6, 80805 M\"unchen, Germany}
\affiliation[c]
{Institut f\"ur Theoretische Teilchenphysik, 
Karlsruher Institut f\"ur Technologie (KIT), 
76128 Karlsruhe, Germany, and 
Institut f\"ur Astroteilchenphysik, 
Karlsruher Institut f\"ur Technologie (KIT), 
D-76021 Karlsruhe, Germany}

\emailAdd{gloria.bertolotti@unito.it}
\emailAdd{lorenzo.magnea@unito.it}
\emailAdd{gpellicc@mpp.mpg.de}
\emailAdd{ratti@mpp.mpg.de}
\emailAdd{chiara.signorile-signorile@kit.edu}
\emailAdd{paolo.torrielli@unito.it}
\emailAdd{sandro.uccirati@unito.it}

\abstract{
We use the Local Analytic Sector Subtraction scheme to construct a completely
analytic set of expressions implementing a fully local infrared subtraction at NNLO 
for generic coloured massless final states. The cancellation of all explicit infrared 
poles appearing in the double-virtual contribution, in the real-virtual correction and 
in the integrated local infrared counterterms is explicitly verified, and all finite 
contributions arising from integrated local counterterms are analytically evaluated 
in terms of ordinary polylogarithms up to weight three. The resulting {\it subtraction 
formula} can readily be implemented in any numerical framework containing the 
relevant matrix elements up to NNLO.}

\begin{document}
\maketitle


\section{\vspace{-.5mm}Introduction}
\label{Intro}

\vspace{.5mm}
The coming decades will see a vast increase in the experimental precision
of collider data, as the LHC experiments move into the high-luminosity era. 
At the same time, the complexity of the observables being probed in hadronic 
collisions is likely to increase as well, as more detailed information becomes
available about multi-particle final states. This future evolution on the experimental
side poses a significant challenge for the theory community, which is called upon
to provide increasingly precise predictions for ever more intricate observables.
As a result, a number of innovative theoretical tools for perturbative calculations 
have been developed over the last two decades, and continue to be refined and 
extended (for a recent review, see Ref.~\cite{Huss:2022ful}). Predictions 
at the next-to-next-to-leading order (NNLO) in the strong coupling are rapidly 
becoming standard, even for relatively complex final states (see, for example, 
Ref.~\cite{Czakon:2021mjy,Hartanto:2022qhh,Buonocore:2022pqq,Catani:2022mfv}), 
while the frontier has moved to the third perturbative order in the strong coupling 
(N$^3$LO) for relatively simple processes~\cite{Baglio:2022wzu,Caola:2022ayt}.

A necessary ingredient for the calculation of differential distributions to the 
required accuracy is an efficient and automatic treatment of infrared singularities, 
which must cancel between virtual corrections and the phase-space integrals of 
unresolved final-state radiation, or must be factorised in  a universal manner in 
the case of collisions involving hadrons in the initial state. The theoretical foundations 
of this treatment are well understood (for a recent review, see~\cite{Agarwal:2021ais}): 
the cancellation (or factorisation) is guaranteed by general theorems valid to all 
orders in perturbation theory~\cite{Bloch:1937pw,Kinoshita:1962ur,Lee:1964is,
Grammer:1973db,Collins:1989gx}, and hinges upon the factorisation properties
of virtual corrections to scattering amplitudes~\cite{Sen:1982bt,Collins:1989bt,Magnea:1990zb,
Sterman:1995fz,Catani:1998bh,Sterman:2002qn,Dixon:2008gr,Gardi:2009qi,
Gardi:2009zv,Becher:2009cu,Becher:2009qa,Feige:2014wja} and of real-radiation 
matrix elements~\cite{Catani:1998nv,Catani:1999ss,Kosower:1999xi}. 
The anomalous dimensions required for the infrared factorisation of virtual corrections
are fully known up to three loops~\cite{Almelid:2015jia,Almelid:2017qju}, while the 
real-radiation splitting kernels have been computed at order $\as^2$~\cite{Campbell:1997hg,
Catani:1998nv,Bern:1999ry,Catani:1999ss,Kosower:1999xi,Catani:2000pi}, with 
near-complete information available also at $\as^3$~\cite{DelDuca:1999iql,Badger:2004uk,
Duhr:2013msa,Li:2013lsa,Banerjee:2018ozf,Bruser:2018rad,Dixon:2019lnw,
Catani:2019nqv,DelDuca:2020vst,Catani:2021kcy,DelDuca:2022noh, Czakon:2022dwk,Czakon:2022fqi,Catani:2022hkb}.

Notwithstanding this extensive body of knowledge, the construction of general
and efficient algorithms for infrared subtraction beyond NLO has proved to be
a very difficult task. At NLO, the task of handling infrared singularities was first
approached with {\it phase-space slicing} methods~\cite{Giele:1993dj,Giele:1994xd},
by isolating the phase-space regions where real radiation is singular, introducing
for those regions approximate expressions of the relevant matrix elements, and
integrating analytically up to the slicing parameter. To avoid residual dependence
on the slicing parameter, {\it subtraction} methods~\cite{Frixione:1995ms,
Catani:1996vz,Nagy:2003qn,Bevilacqua:2013iha}, see also \cite{Prisco:2020kyb}, 
were later introduced, which work by defining local counterterms in all regions of phase 
space affected by singularities, subtracting them from the full real-radiation matrix elements, 
and then adding back their exact integrals. Some of these methods have been developed 
in full generality, and versions of the corresponding algorithms are implemented in
a number of multi-purpose NLO event generators~\cite{Campbell:1999ah,
Gleisberg:2007md,Frederix:2008hu,Czakon:2009ss,Hasegawa:2009tx,Frederix:2009yq,
Alioli:2010xd,Platzer:2011bc,Reuter:2016qbi}, providing a solution of the problem at this accuracy.

Beyond NLO, the handling of infrared singularities becomes significantly
more difficult, both conceptually and practically, due to the rapid increase in
the number of overlapping singular regions, to the need for considering 
strongly-ordered infrared limits, and to the mixing between virtual poles
and phase-space singularities. As a consequence, efforts to reach the same
degree of universality and efficiency as was achieved at NLO already span 
almost two decades. Many different approaches have been proposed 
and pursued~\cite{Frixione:2004is,GehrmannDeRidder:2005cm,Currie:2013vh,
Somogyi:2005xz,Somogyi:2006da,Somogyi:2006db,Czakon:2010td,Czakon:2011ve,
Anastasiou:2003gr,Caola:2017dug,Catani:2007vq,Grazzini:2017mhc,Boughezal:2011jf,
Boughezal:2015dva,Boughezal:2015aha,Gaunt:2015pea,
Cacciari:2015jma,Sborlini:2016hat,Herzog:2018ily,Magnea:2018hab,
Magnea:2018ebr,Capatti:2019ypt}, as recently reviewed in 
Ref.~\cite{TorresBobadilla:2020ekr}. Some of the methods proposed belong to the 
{\it slicing} family, or define non-local subtractions, as is the case for Ref.~\cite{Grazzini:2017mhc}, 
while others adopt the {\it local-subtraction} viewpoint
(for example~\cite{Somogyi:2005xz,Czakon:2010td}); they also
range from predominantly numerical methods, as in~\cite{Capatti:2019ypt}, to 
predominantly analytical ones, as for example~\cite{Caola:2017dug};
finally, they have reached varying degrees of practical implementation, culminating 
with the first differential NNLO calculations for $2\to 3$ collider processes with 
at least two QCD particles in the final state at Born level, in Refs.~\cite{Chawdhry:2019bji,
Chawdhry:2021hkp,Czakon:2021mjy,Hartanto:2022qhh,Buonocore:2022pqq,Catani:2022mfv}.

All approaches to infrared subtraction beyond NLO are affected by considerable
computational complexity, either at the level of the analytic integration of counterterms,
or at the level of numerical implementation. Even if the underlying physical mechanism
for the cancellation is essentially simple and well understood, concrete technical 
implementations are intricate, and it is clear that there is room for improvement
in the universality, versatility and efficiency of existing algorithms. With these 
goals in mind, we have developed an approach to infrared {\it subtraction}, which
we call Local Analytic Sector Subtraction~\cite{Magnea:2018hab,Bertolotti:2022ohq}. We attempt
to optimise the structure of the calculation at all stages, while maintaining full 
locality of the counterterms and complete universality for all hadronic final states,  
as well as providing completely analytic expressions for all required counterterms 
and their phase-space integrals, including finite contributions. We believe that the 
completion of this programme will provide an extremely versatile tool: once fully analytic 
expressions are available, the method can in principle be implemented within any 
existing numerical framework, and applications to multi-particle final states will be 
limited only by the available computing power and multi-loop matrix elements (see for instance Ref.~\cite{FebresCordero:2022psq}). In parallel, we are studying more formal
aspects of subtraction, from the point of view of factorisation~\cite{Magnea:2018ebr},
with the hope of further optimising the structure of local counterterms, taking full
advantage of the highly non-trivial structure of infrared factorisation and
exponentiation. In that context, we provided a set of definitions for soft and 
collinear local counterterms which apply to all orders in perturbation theory,
and we are currently studying the necessary organisation of strongly-ordered
infrared configurations~\cite{Magnea:2022twu}.

In the present paper, we complete our subtraction programme for the case of generic massless
coloured final states. All relevant integrals were computed analytically in~\cite{Magnea:2020trj}, 
requiring only standard techniques. In order to achieve this simplicity, we exploited 
as much as possible the existing freedom in the definition of local infrared counterterms. 
Specifically, a crucial element of our approach is the smooth partition of phase space in sectors, 
each of which contains only a minimal set of soft and collinear singularities, along the 
lines of Ref.~\cite{Frixione:1995ms}. The next important ingredient is a flexible family 
of phase-space parametrisations, which can be applied sector by sector, and in fact 
can be varied for each contribution to the local counterterms. This ultimately leads to 
a minimal and simple set of phase-space integrals to be performed. Our final result is 
a completely analytic subtraction formula, which gives the NNLO contribution to the 
differential distribution for any infrared-safe observable built out of massless coloured 
final states (as well as with an arbitrary set of massive or massless colourless final-state 
particles), and requires as input only the relevant matrix elements: the double-virtual 
correction to the Born-level process, the one-loop correction to the single-radiation 
process, and the tree-level expression for the double-real-emission contribution.

We present the architecture of our method in \secn{Frame}, beginning with 
a quick review of our approach at NLO, for massless final states, to introduce 
the relevant notations in a simple context\footnote{Our method provides a complete 
subtraction formalism at NLO, including the case of initial state hadrons, as discussed 
in details in Ref.~\cite{Bertolotti:2022ohq}.}. The following sections give the details 
for the construction of all the ingredients entering the subtracted formula.
\secn{RRsub} discusses the subtracted double-real contribution, which is
integrable over the entire radiative phase space. Explicit expressions for all required 
counterterms are included, as well as a detailed analysis of phase-space mappings.
\secn{RRint} organises the integration procedure for all counterterms associated
with double-real radiation, expressing the required integrals in terms of a small 
set of basic integrals, which were discussed in Ref.~\cite{Magnea:2020trj} and 
are collected here in \appn{app:master integrals}. 
\secn{RVsub} presents the subtracted real-virtual 
correction, providing an explicit expression for the real-virtual counterterm. By combining 
together the real-virtual correction with its local counterterm, and the integrals of 
the single-unresolved and the strongly-ordered counterterms, we build an expression
that is both free of infrared poles and integrable in the radiative phase space.
The integration of the real-virtual counterterm is discussed in \secn{RVint}, and 
again can be organised in terms of simple integrals. 
\secn{VVsub} gives the subtracted double-virtual contribution, which is free of infrared poles. 
Finally, \secn{Explicit} provides an explicit example of cancellation of 
phase-space singularities and
\secn{Persp} summarises our results, putting them in perspective. Several
appendices collect useful formulas and some technical details of the formalism.


\section{Subtraction for massless final states: a framework}
\label{Frame}

We consider a generic process where an electroweak initial state with total momentum 
$q$, $q^2 \equiv s$, produces $n$ massless final-state coloured particles at Born level, 
and we denote with $\mathcal{A}_n (k_i)$, $i = 1, \ldots, n$, the relevant scattering 
amplitude\footnote{The subtraction presented in the following applies with no modifications 
to the case of an arbitrary number of colourless particles accompanying the $n$ coloured 
ones in the final state, so that in general $\sum_i k_i \neq q$. Just for the sake of notational 
simplicity, we will assume $n$ to coincide with the total number of final-state particles and 
the total momentum to be $q$.}. The perturbative expansion of the amplitude reads 
\beq
  {\cal A}_n (k_i) \, = \, {\cal A}_n^{(0)} (k_i) \, + \, {\cal A}_n^{(1)} (k_i) \, + \,
  {\cal A}_n^{(2)} (k_i) \, + \, \ldots \, , 
\label{pertexpA}
\eeq
where $\mathcal{A}_n^{(k)}$ is the $k$-loop correction, and includes the appropriate 
power of the strong coupling constant. For such a process, we consider a generic 
infrared-safe observable $X$, and we write the corresponding differential distribution as
\beq
  \frac{d \sigma}{d X} \, = \, \frac{d \sigma_\LO}{d X} \, + \, 
  \frac{d \sigma_\NLO}{d X} \, + \, 
  \frac{d \sigma_\NNLO}{d X} \, + \, \ldots \, .
\label{pertexpsig}
\eeq
Our task is to express such differential distributions in a manifestly finite form, which is 
free of infrared poles, and integrable over the appropriate phase spaces.
In order to introduce our method and notations, we begin with a brief review of
the NLO calculation for massless final states.


\subsection{Local Analytic Sector Subtraction at NLO}
\label{FrameNLO}

The standard expression for the NLO term in the distribution in \eq{pertexpsig} 
requires combining virtual corrections to the Born term, which contain IR poles in
$\eps \equiv (4 - d)/2$, where $d$ is the number of space-time dimensions,
and the phase-space integral of unresolved radiation, 
which is also singular in $d = 4$. One must then compute the combination
\beq
  \frac{d \sigma_\NLO}{d X} & = & \lim_{d \to 4} 
  \bigg[ \! \int d \Phi_n \, \Vl \, \delta_n (X) + 
  \int d \Phi_\npo \, \Rl \, \delta_\npo (X) \bigg] \, .
\label{pertO}  
\eeq
Here $\delta_m (X) \equiv \delta (X - X_m)$ fixes $X_m$, the expression 
for the observable $X$, computed for an $m$-particle configuration, to its
prescribed value, $d \Phi_m$ denotes the Lorentz-invariant phase-space 
measure for $m$ massless final-state particles, and
\beq
  \Rl \, = \, \left| {\cal A}_\npo^{(0)} \right|^2 \, , \qquad
  \Vl \, = \, 2 \, {\bf Re} \left[ {\cal A}_n^{(0) \dag} \, {\cal A}_n^{(1)} 
  \right] \, ,
\label{pertA2}
\eeq
are the real and the ($\overline{\rm MS}$-renormalised) virtual contributions, respectively.
To rewrite \eq{pertO} in terms of finite quantities we need a sequence of steps.
First, we must define a {\it local counterterm}, denoted here by $K$, 
which is required to reproduce the singular IR behaviour of the real-radiation matrix 
element $\Rl$ {\it locally} in phase space. At the same time, it is expected 
to be simple enough to be analytically integrated in the phase space of the 
unresolved radiation. In order to perform this integration, we need to introduce a 
parametrisation of the radiative phase space $d \Phi_\npo$, which must factorise 
as
\beq
  d \Phi_{\npo} \, = \, \frac{\varsi_\npo}{\varsi_n} \, d \Phi_n \, d \Phi_{{\rm rad}} \, ,
\label{phaspafac}
\eeq
where, as before, $d \Phi_n$ is the phase space for $n$ massless particles, 
while $d \Phi_{{\rm rad}}$ is the measure of integration for the degrees of freedom 
of the unresolved radiation, and we explicitly extracted the ratio of the relevant 
symmetry factors $\varsi_\npo$ and $\varsi_n$. The factorisation theorems for real 
radiation guarantee that the function $K$ will be a combination of products of Born-level 
squared amplitudes (to be integrated over $d \Phi_n$) and infrared kernels, to be 
integrated in $d \Phi_{{\rm rad}}$.  Once a parametrisation yielding \eq{phaspafac} 
is in place, one can compute the {\it integrated counterterm}
\beq
  I \, = \, \int d \Phi_{{\rm rad}} \, K  \, .
\label{intcountNLO}
\eeq
\eq{intcountNLO} will reproduce, by construction, the infrared poles arising 
from the integration of the real-radiation squared matrix element. It is now 
possible to rewrite \eq{pertO} identically as a combination of virtual corrections and 
real contributions that are separately finite, and therefore phase-space integrals 
can be performed numerically when needed. Using $\int d \Phi_\npo K = \int d \Phi_n \, I$, 
we obtain
\beq
\frac{d \sigma_\NLO}{d X} 
\, = \, 
\int d \pn \Vl_{\,\rm sub}(X) \, 
+
\int d \Phi_\npo \, \Rl_{\,\rm sub}(X)
\, ,
\label{subtNLO}
\eeq
with 
\beq
\Vl_{\,\rm sub}(X) 
& = & 
\big( \Vl + I \big) \, \delta_n (X)  
\, ,
\qquad\qquad
\Rl_{\,\rm sub}(X)
\; = \;
\Rl \; \delta_\npo (X) - K \, \delta_n (X)
\, .
\eeq
The subtracted real matrix element $\Rl_{\,\rm sub}(X)$ is free of 
phase-space singularities by construction, while $\Vl_{\,\rm sub}(X)$ is 
finite as $\epsilon \rightarrow 0$ as a consequence of the KLN theorem, 
and both contributions are now suitable for a numerical implementation in 
four space-time dimensions. Notice that the IR safety of the observable 
$X$ is necessary for the cancellation, which requires that $\delta_\npo 
(X)$ turns smoothly into $\delta_n (X)$ in all unresolved limits.

Eqs.~(\ref{pertO})-(\ref{subtNLO}) provide just an outline of NLO subtraction task:
the actual definition of the required local counterterm is in fact not unique, and 
characterises the subtraction scheme. Furthermore, it is necessary to include a 
prescription to perform the phase-space mapping implied by \eq{phaspafac}. 
Within the context of Local Analytic Sector Subtraction at NLO, we proceed as follows.
\begin{itemize}
\item We define {\it projection operators} ${\bf S}_i$  and ${\bf C}_{ij}$ that extract
from the real-radiation squared matrix element $\Rl$ its singular behaviour 
in soft and collinear limits. In practice,
one must pick specific phase-space variables in order to perform the projection: 
one could for example choose a Lorentz frame and define the soft limit in terms 
of the energy of particle $i$ in that frame, and the collinear limit in terms of 
the angle between $i$ and $j$, as  was done in Ref.~\cite{Frixione:1995ms}.
We prefer to use Lorentz-invariant quantities, as discussed in detail in 
Refs.~\cite{Magnea:2018hab} and~\cite{Bertolotti:2022ohq}. Concretely, we 
introduce the variables
\beq
 e_i \, \equiv \, \frac{s_{qi}}{s} \, , \qquad \quad w_{ij} \, \equiv \, \frac{s s_{ij}}{s_{qi} s_{qj}} \, ,
\label{sectprelim}
\eeq 
where $s_{q\ell} = 2 q \cdot k_\ell$. We then define ${\bf S}_i$ as extracting 
the leading power in $e_i$, and ${\bf C}_{ij}= {\bf C}_{ji}$ as extracting the leading power 
in $w_{ij}$. It is not difficult to verify that, with this definition, the two operators
commute when acting on the squared matrix element, ${\bf S}_i \, {\bf C}_{ij} \, \Rl = 
{\bf C}_{ij} \, {\bf S}_i \, \Rl$.
\item We then partition the radiative phase space into {\it sectors}, defined
by introducing a set of {\it sector functions}, ${\cal W}_{ij}$, along the lines of 
Ref.~\cite{Frixione:1995ms}, which constitute a partition of unity, namely a set 
of kinematical weights smoothly dampening all radiative singularities but those 
due to particle $i$ becoming soft, or becoming collinear to a second particle 
$j$. Our sector functions are 
constructed in terms of Lorentz invariants. We indeed define 
\beq
  \sigma_{ij} \, \equiv \, \frac{1}{e_i w_{ij}} \, , \qquad \quad
  {\cal W}_{ij} \, \equiv \, \frac{\sigma_{ij}}{\sum_{k \neq l} \sigma_{kl}} \, ,
\label{sectfunc}
\eeq
satisfying $\sum_{i \neq j} {\cal W}_{ij} = 1$. These sector functions have 
the further defining property that their soft and collinear limits still form a partition of unity.
Indeed, one easily verifies that
\beq
  {\bf S}_i  \sum_{k \neq i} {\cal W}_{ik} \, = \, 1 \, , \qquad 
  {\bf C}_{ij} \Big[ {\cal W}_{ij} + {\cal W}_{ji} \Big] \, = \, 1 \, .
\label{SCsumrules}
\eeq
\eq{SCsumrules} guarantees that, upon summing over sectors, the full soft and 
collinear singularities will be recovered, and sector functions will not explicitly 
appear in counterterms to be integrated. 
\item 
The purpose of introducing sector functions is to minimise the number of singular 
limits of $\Rl\,\W{ij}$, so that we can easily identify a combination which is by construction 
integrable in the radiative phase space. Indeed, in sector $(ij)$
\beq
\label{intij}
\left( 1 - \bS{i} \right)
\left( 1 - \bC{ij} \right)
\,
\Rl \, \W{ij}
\; = \; 
\Rl \, \W{ij}
-
\bL{ij}{\!\one} \; \Rl \, \W{ij}
\; & \to & \;
\mbox{integrable}
\, ,
\eeq
where we introduced $\bL{ij}{\!\one} \equiv \bS{i}+\bC{ij}-\bS{i}\,\bC{ij}$.  We stress 
here that the operators $\bS{i}$ and $\bC{ij}$ are defined to act on all elements that 
lie to their right: therefore, if $\bL{}{}$ denotes a generic singular limit, the relation 
$\bL{}{} \, \Rl \, \W{ij} \equiv (\bL{}{} \, \Rl) \, (\bL{}{} \,\W{ij})$  is understood.
Summing over sectors we get the expression 
\beq
  \sum_i  \sum_{j \neq i} \, {\bf L}^{\one}_{ij} \Rl \, {\cal W}_{ij}  \, \equiv \, 
  \sum_i  \sum_{j \neq i} \Big[ {\bf S}_i + {\bf C}_{ij} \big( 1 - 
  {\bf S}_i \big) \Big] \Rl \, {\cal W}_{ij} \, ,
\label{candcount}
\eeq
which satisfies the requirement of reproducing the singular behaviour of $\Rl$ in all 
soft and collinear regions. \eq{candcount}, however, cannot yet be used directly in 
\eq{subtNLO}, since it does not properly factorise a Born-level squared matrix element 
involving $n$ on-shell particles.
\item For this purpose, we must introduce a set of {\it mappings} of the $(\npo)$-particle 
momenta $\{k\}$ onto the $n$-particle momenta $\{\bar k\}$, which must not affect soft and 
collinear limits at leading power. We adopt the Catani-Seymour mappings~\cite{Catani:1996vz}
\beq
  \hspace{-4mm}
  \bar{k}_i^{(abc)} \, = \, k_i \,,  \,\,\,  i \neq a,b,c \, ;  
  \quad \,
  \bar{k}_b^{(abc)} \, = \, k_a + k_b - \frac{s_{ab}}{s_{ac} + s_{bc}} k_c \, ; \quad
  \bar{k}_c^{(abc)} \, = \, \frac{s_{abc}}{s_{ac} + s_{bc}} k_c \, ,
\label{mapNLO}
\eeq
where $i$ runs from 1 to $n+1$. The mappings above satisfy the on-shell and momentum-
conservation conditions 
\beq
  \big( \bar{k}_j^{(abc)} \big)^2 \, = \, 0 \, , \,\,\,  j \, = \, 1, \ldots , n \, ; \qquad \,\,
  \sum_{j = 1}^n \bar{k}_j^{(abc)} \, = \,  \sum_{i = 1}^\npo k_i \, .
\label{momconsCS}
\eeq
One easily verifies the two sets of momenta coincide when $k_a$ becomes soft, 
and when $k_a$ becomes collinear to $k_b$. 
\item 
Finally, we can turn \eq{candcount} into a {\it local counterterm}, by using the factorised 
expressions for soft and collinear limits of $\Rl$, and evaluating the Born-level squared 
matrix elements with the mapped momenta defined \eq{mapNLO}, sector by sector in 
the radiative phase space. We do this by introducing {\it improved projection operators} $\bSi$ 
and $\bCj$, which are defined at NLO to project on leading-power soft and collinear limits, 
and at the same time apply the selected phase-space mappings. For NLO massless final 
states their action is defined by
\beq
\bbS{i} \, \Rl 
& \equiv & 
- \, \Norm 
\sum_{c\neq i} \sum_{d \neq i,c}
\mc E_{cd}^{(i)} \, \bB^{(icd)}_{cd} 
\, ,
\nnb 
\\ [3pt] \
\bbC{ij} \, \Rl
& \equiv & 
\Norm \,
\frac{P^{\mu\nu}_{ij(r)}}{s_{ij}} \, 
\bB_{\mu\nu}^{(ijr)}
\, ,
\nnb\\ [4pt]
\bbS{i} \, \bbC{ij} \, \Rl
& \equiv & 
2 \, \Norm \, C_{f_j} \, \mc E_{jr}^{(i)} \, 
\bB^{(ijr)}
\, ,
\qquad
r = r_{ij}
\, .
\label{CijR1def}
\eeq
The quantities entering \eq{CijR1def} are defined as follows. We denote by $B$ the 
Born matrix element squared, $B = | {\cal A}_n^{(0)} |^2$, while $B_{\mu \nu}$
is its spin-correlated counterpart, defined by removing the gluon polarisation 
vectors from the matrix element and from its complex conjugate\footnote{If the parent 
parton in the collinear $ij$ splitting is not a gluon, the corresponding kernel is diagonal 
in spin space by helicity conservation, and $B_{\mu\nu}$ reduces to $B$.}. Similarly, $B_{cd}$ 
is the colour-correlated Born, defined in \eq{colcorrBorn}. These three objects are 
evaluated in \eq{CijR1def} with mapped momenta, and are therefore denoted with 
a bar and with a label identifying the specific mapping to be employed. Thus, for 
example, $\bB^{(ilm)} \equiv B(\kkl{ilm})$ is the Born squared matrix element with 
mapped momenta $\kkl{ilm}$. Furthermore, $C_{f_j}$ is the Casimir eigenvalue of 
the colour representation of parton $j$, while the eikonal kernel ${\cal E}_{cd}^{(i)}$ 
and the DGLAP kernels $P^{\mu\nu}_{ij(r)}$ are presented in Eq.~\eqref{eikoker} and
Eq.~\eqref{eq:appsinglecoll} respectively\footnote{We note that, as seen in \appn{app:kernels}, 
these and all other kernels are written in terms of Lorentz invariants and in a manifestly 
flavour-symmetric notation.}. The overall normalisation is given by
\beq
  \Norm \, = \, 8 \pi \as \left( \frac{\mu^2 e^{\gam_E}}{4 \pi} \right)^{\eps} \, .
\label{norm1}
\eeq
Importantly, the improved operators must preserve the correct soft and collinear 
limits of $\Rl$ to ensure the locality of the subtraction procedure: in this case, one 
must verify that
\beq
  \Si \, \bSi \, \Rl \, = \,  \Si \, \Rl \, , \quad \Cj \, \bCj \, \Rl \, = \,  \Cj \, \Rl \, ,
\label{ConsistNLO1}
\eeq
as well as
\beq
  \Si \, \bSi \, \bCj \, \Rl \, = \,  \Si \, \bCj \, \Rl \, , \quad \Cj \, \bSi \, \bCj \, 
  \Rl \, = \,  \Cj \, \bSi \, \Rl \, .
\label{ConsistNLO2}
\eeq
These {\it consistency conditions} are indeed verified by \eq{CijR1def}. 
We also stress that $r=r_{ij}$ is any particle different from $i,j$, 
chosen according to the rule defined in \eq{eq: r_ij} (in this case it means 
that the same $r$ must be chosen for the pair $ij$ and for the pair $ji$). 
In what follows, we will describe the action of the improved operators as realising {\it improved limits}. 
Notice that, at this stage, we have a residual freedom in the definition of 
improved limits of sector functions, subject to the preservation of the 
constraints in \eq{SCsumrules} and in \eq{intij}.
\item
The definition of the improved operators given above contains a subtlety \cite{Bertolotti:2022ohq}, 
which must be analysed with care. The DGLAP kernels $P^{\mu\nu}_{ij(r)}$ reported in 
\appn{app:kernels} are written in terms of the invariants 
\beq
x_i 
\; = \;
\frac{s_{ir}}{s_{ir}+s_{jr}}
\, ,
\qquad
x_j
\; = \;
\frac{s_{jr}}{s_{ir}+s_{jr}}
\, ,
\eeq
as opposed to the energy fractions $e_{i}/(e_{i}+e_{j})$, $e_{j}/(e_{i}+e_{j})$. This 
is a useful choice in view of analytical integration, and a legitimate one since $x_{i}$ 
and $x_{j}$ reduce to $e_i$ and $e_j$ in the collinear limit $\bC{ij}$. This choice, however, 
introduces {\it spurious singularities} in the 
collinear limits $\bC{ir}$ and $\bC{jr}$ in the sectors $\W{ij}$ and $\W{ji}$, so that 
the combination $(1-\bbS{i})(1-\bbC{ij})\Rl\,\W{ij}$ is not integrable in the limits 
$\bC{ir}$ and $\bC{jr}$. This problem can be solved by using our freedom to 
define the action of the improved operators $\bbS{i}$ and $\bbC{ij}$ on sector functions 
$\W{ij}$ ($r=r_{ij}$):
\beq
\label{improlsecf}
\bbS{i} \, \W{ij}
\, \equiv \,
\bS{i} \, \W{ij}
\, = \,
\frac{\frac{1}{w_{ij}}}{\sum\limits_{l \neq i} \frac{1}{w_{il}}} 
\, , 
\qquad
\bbC{ij} \, \W{ij}
\, \equiv \,
\frac{e_jw_{jr}}
{e_i w_{ir} \!+\! e_j w_{jr}} 
\, ,
\qquad
\bbS{i} \, \bbC{ij} \, \W{ij}
\, \equiv \,
1
\, .
\qquad
\eeq
The presence of the angular factors $w_{ir}$ and $w_{jr}$, vanishing in the 
$\bC{ir}$ and $\bC{jr}$ limits respectively, allows to verify the following 
{\it auxiliary consistency conditions}
\beq
\label{auxconsist}
\bC{ir} \, 
\Big\{
\, 1 \, , \; 
\bbS{i} \, , \;
\bbC{ij} \left( 1 - \bbS{i} \right)  
\Big\} \, \Rl \, \W{ij}
\; & \to & \;
\mbox{integrable}
\, ,
\nnb\\
\bC{jr} \, 
\Big\{
\, 1 \, , \;
\bbS{i} \, , \;
\bbC{ij} \, , \;
\bbS{i}\,\bbC{ij}
\Big\} \, \Rl \, \W{ij}
\; & \to & \;
\mbox{integrable}
\, ,
\qquad
\eeq
on top of the standard ones, corresponding to Eqs.~(\ref{ConsistNLO1}) and 
(\ref{ConsistNLO2}), which now need to be written explicitly including the sector 
functions, as
\beq
\bS{i} \, 
\Big\{
\left( 1 - \bbS{i} \right) , \;
\bbC{ij} \left( 1 - \bbS{i} \right)  
\Big\} \, \Rl \, \W{ij}
\; & \to & \;
\mbox{integrable}
\, ,
\nnb\\
\bC{ij} \, 
\Big\{
\left( 1 - \bbC{ij} \right) , \;
\bbS{i} \left( 1 - \bbC{ij} \right) 
\Big\} \, \Rl \, \W{ij} 
\; & \to & \;
\mbox{integrable}
\, .
\eeq
Recall that in \eq{auxconsist} the index $r$ labels the reference vector used to define
the collinear kernel $P^{\mu\nu}_{ij(r)}$: in fact, all collinear projection operators should
properly be labelled with the index $r$, which in general we omit for brevity. Notice also
that our definition of improved limits of sector functions, \eq{improlsecf}, is not symmetric 
under $i \leftrightarrow j$. As a consequence, the two lines of \eq{auxconsist} are not 
identical: in the first line, only the combination $\bbC{ij} ( 1 - \bbS{i} )$ gives an integrable 
result in the $ir$ collinear limit, when acting on $\Rl \, \W{ij}$ (which is sufficient for our 
purposes), while in the second line $\bbC{ij}$ and $\bbS{i} \,  \bbC{ij}$ give separately 
integrable contributions in the $jr$ collinear limit.
\item
With these definitions, our first expressions for the NLO local counterterm is
\beq
\label{eq:RsubW1}
K
\, = \, 
\! \sum_{i, j \neq i} K_{ij}
\, ,
\qquad \quad
K_{ij}
\, = \,
\Big( \bbS{i} + \bbC{ij} - \bbS{i}\,\bbC{ij} \Big) \Rl \, \mc W_{ij}
\, ,
\eeq
so that the subtracted squared matrix element is given by 
\beq
\label{eq:RsubW2}
\Rl_{\,\rm sub}(X) 
\, = \,
\! \sum_{i, j \ne i} \! \Rl^{\,\rm sub}_{\,ij}(X) 
\, ,
\quad
\Rl^{\,\rm sub}_{\,ij}(X) 
\, = \, 
\Rl \, \W{ij}  \, \delta_\npo(X) - K_{ij} \, \delta_n(X)
\, .
\eeq
The counterterm defined in \eq{eq:RsubW1} is sufficient to construct a fully 
functional subtraction algorithm at NLO. There is however some room for optimisation: 
for example, we note that the sector functions $\W{ij}$ are useful to identify the 
improved limits to be defined, and the consistency relations they must satisfy, 
but the stability of numerical integrations will 
improve when sectors involving the same parametrisations are combined. 
To pursue this idea, we introduce {\it symmetrised 
sector functions} defined by 
\beq
\label{eq:NLO_Zij}
\Z{ij} 
& = &
\W{ij} + \W{ji}
\, .
\eeq
The corresponding improved limits read 
\beq
\bbS{i} \Z{ij}
\, = \,
\bbS{i} \W{ij}
\, = \, 
\frac{\frac{1}{w_{ij}}}{\sum\limits_{l \neq i} \frac{1}{w_{il}}} 
\, , 
\qquad
\bbC{ij} \Z{ij}
\, = \, 
1
\, ,
\qquad
\bbS{i} \bbC{ij} \Z{ij}
\, = \, 
\bbS{i} \bbC{ij} \W{ij}
\, = \,
1
\, .
\eeq
This symmetrisation of the sector functions reduces the number of sectors and, 
to some extent, simplifies the scheme in view of an efficient numerical performance. 
In fact, the counterterm $K$, with symmetrised sector functions, can be written as
\beq
\label{eq:RsubZ}
K
\, = \,
\! \sum_{i, j > i} \!\! K_{\{ij\}} 
\, ,
\qquad \quad
K_{\{ij\}}
\, = \,
\left( \bbS{i} + \bbS{j} + \bbHC{ij} \right) \! \Rl \, \Z{ij}
\, ,
\eeq
where we have introduced the hard-collinear improved limit
\beq
\label{eq:RsubZ2}
\bbHC{ij} \, \Rl 
& \equiv &
\bbC{ij} 
\left(1 - \bbS{i} - \bbS{j} \right)
\Rl
\; = \;
\Norm \,
\frac{P_{ij(r)}^{{\rm hc}, \mu\nu}\!\!}{s_{ij}} \,
\bB_{\mu\nu}^{(ijr)} \!
\, ,
\eeq
with the hard-collinear splitting kernel $P_{ij(r)}^{{\rm hc}, \mu\nu}$ defined in 
\appn{app:kernels}. The subtracted squared matrix element is now given by 
\beq
\label{eq:RsubZ3}
\Rl_{\,\rm sub}(X) 
\, = \,
\! \sum_{i,j > i} \!\! \Rl^{\,\rm sub}_{\,\{ij\}}(X) 
\, ,
\qquad
\Rl^{\,\rm sub}_{\,\{ij\}}(X) 
\, = \,
\Rl \, \Z{ij}  \, \delta_\npo(X) - K_{\{ij\}} \, \delta_n(X)
\, .
\eeq
A third expression for the NLO counterterm, important for analytic integration, 
is obtained by summing over all sectors. Using \eq{SCsumrules}, one can then
write
\beq
\label{eq:Rsub no sectors}
\Rl_{\,\rm sub}(X) 
\, = \,
\Rl \; \delta_\npo(X) - K \, \delta_n(X)
\, ,
\qquad \,
K 
\, = \,
\sum_{i} \; \bbS{i} \, \Rl
+
\sum_{i, j > i} \bbHC{ij} \, \Rl
\, .
\qquad
\eeq
This expression for $\Rl_{\,\rm sub}(X)$, though very compact, is not the most 
suited for numerical implementation: the expression in \eq{eq:RsubZ3}, with 
symmetrised sector functions, is to be preferred, since it allows to parallelise 
the contribution of different sectors, and to independently optimise their numerical 
evaluation.

\end{itemize}
As discussed in detail in~\cite{Magnea:2018hab,Magnea:2020trj,Bertolotti:2022ohq},
these definitions enable a straightforward integration of local counterterms, 
and yield an implementation of NLO subtraction that can be extended to 
initial-state radiation as well. We now turn to the case of NNLO massless final 
states.


\subsection{\vspace{-.5mm}Local Analytic Sector Subtraction at NNLO}
\label{FrameNNLO}

\vspace{.5mm}
The NNLO contribution to the differential cross section in \eq{pertexpsig} can be 
written as
\beq
\label{pertO2}
\hspace{-5mm}
\frac{d \sig_\NNLO}{dX} 
= \lim_{d \to 4}  \bigg[
\int d \Phi_n \, VV \, \delta_n(X) 
+
\int d \Phi_\npo \, RV \, \delta_\npo(X) 
+ 
\int d \Phi_\npt \, RR \; \delta_\npt(X) \bigg] \, ,
\eeq
where 
\beq
\hspace{-3mm} 
RR 
\, = \, 
\left| {\cal A}_\npt^{(0)} \right|^2 , 
\quad
RV 
\, = \, 
2 \, {\bf Re} 
\left[ {\cal A}_\npo^{(0) \dag} \, {\cal A}_\npo^{(1)}   \right] ,
\quad
VV 
\, = \, 
\left| {\cal A}_n^{(1)} \right|^2 
+ 
2 \, {\bf Re} 
\left[ {\cal A}_\npo^{(0) \dag} \, {\cal A}_\npo^{(2)} \right] 
.
\label{pertA22}
\eeq  
In this case, the $\overline{\rm MS}$-renormalised double-virtual contribution $VV$ 
displays IR poles up to $\eps^{-4}$, the double-real $RR$ contains up to
four phase-space singularities, and the $\overline{\rm MS}$-renormalised real-virtual 
term $RV$ has poles up to $\eps^{-2}$ and up to 
two phase-space singularities. In order to rewrite \eq{pertO2} as a sum of finite contributions, 
we will define {\it four} local counterterms, which we label 
$K^{\one}$, $K^{\two}$, $K^{\otwo}$ and $K^{\RV}$. The counterterm $K^{\one}$ is 
designed to reproduce all phase-space singularities of $RR$ due to a single particle 
becoming unresolved, while $K^{\two}$ takes care of situations where two particles 
become unresolved at the same rate. The two sets of singularities overlap, and $K^{\otwo}$ is 
responsible for subtracting the double-counted overlap region. Finally, $K^{\RV}$ 
will subtract the phase-space singularities arising from the single-real radiation in $RV$.

In order to integrate these counterterms, we will need to introduce phase-space
parametrisations factorising single and double radiation, in analogy with \eq{phaspafac}.
In this case we will need the factorisations
\beq
d \Phi_{\npt} = \frac{\varsi_\npt}{\varsi_\npo} \, d \Phi_\npo \, d \Phi_{{\rm rad}} 
\, , 
\quad\;
d \Phi_{\npt} = \frac{\varsi_\npt}{\varsi_n} \, d \Phi_n \, d \Phi_{{\rm rad}, 2} 
\, , 
\quad\;
d \Phi_{\npo} = \frac{\varsi_\npo}{\varsi_n} \, d \Phi_n \, d \Phi_{{\rm rad}} 
\, .
\quad
\label{phaspafac2}
\eeq
Once a parametrisation yielding \eq{phaspafac2} is in place, one can define 
{\it integrated counterterms} as
\beq
\label{intcountNNLO}
  I^{\one} \, \equiv \, \int d \Phi_{{\rm rad}} \, K^{\one} \, , 
  && \qquad
  I^{\two} \, \equiv \, \int d \Phi_{{\rm rad}, 2} \, K^{\two} \, ,
  \nnb \\
  I^{\otwo} \, \equiv \, \int d \Phi_{{\rm rad}} \, K^{\otwo} \, , 
  && \qquad
  I^{\RV} \, \equiv \, \int d \Phi_{{\rm rad}} \, K^{\, \RV} \, .
\eeq
We are now ready to write down the master formula for our subtraction at NNLO: 
in the rest of the paper we will precisely define and construct all the necessary 
ingredients, generalising the discussion summarised in \secn{FrameNLO}. We
aim to construct an expression of the form
\beq
\label{eq:sigma NNLO sub}
  \frac{d\sig_\NNLO}{dX} \, = \, 
  \int d \Phi_n \, VV_{\,\rm sub}(X) 
  +
  \int d \Phi_\npo \, RV_{\,\rm sub}(X) 
  + 
  \int d \Phi_\npt \, RR_{\,\rm sub}(X) 
  \, ,
\eeq
where each one of the three contributions is finite in $\eps$ and is free 
of phase-space singularities. 

Using the local counterterms introduced above, and their integrals over the radiative
degrees of freedom, the subtracted matrix elements $VV_{\,\rm sub}$, $RV_{\,\rm sub}$
and $RR_{\,\rm sub}$ are given by
\beq
\label{eq:NNLO_VV_sub}
VV_{\, \rm sub}(X) 
& \equiv &
\left( VV + I^{\two} + I^{\RV} \right) \delta_n (X) 
\, , 
 \\
\label{eq:NNLO_RV_sub}
RV_{\, \rm sub}(X) 
& \equiv &
\left(RV + I^{\one} \right) \delta_\npo(X) 
\, - \, 
\left( K^{\RV} + I^{\otwo} \right) \delta_n(X)
\, , 
\\
\label{eq:NNLO_RR_sub}
RR_{\, \rm sub}(X) 
& \equiv & 
RR \, \delta_\npt (X) 
- 
K^{\one} \, \delta_\npo(X) 
- 
\left( K^{\two} - K^{\otwo} \right) \delta_n (X) 
\, .
\eeq
Once again, Eqs.~\eqref{eq:sigma NNLO sub} and \eqref{eq:NNLO_VV_sub}-\eqref{eq:NNLO_RR_sub} 
provide an identical rewriting of \eq{pertO2}, and their logic is as follows:
\begin{itemize} 
\item in Eq.~\eqref{eq:NNLO_RR_sub}, $RR_{\, \rm sub}(X)$ must be 
integrated in the full phase space $\Phi_\npt$, and it is built out of tree-level 
quantities\footnote{We have implicitly understood the underlying Born reaction to be 
associated with tree-level diagrams; however, in case of loop-induced processes, 
all arguments and techniques presented in this article carry over.}, 
therefore has no explicit IR poles. It has no phase-space singularities either, 
since single-unresolved contributions are subtracted by $K^{\one}$,
double-unresolved contributions are subtracted by $K^{\two}$, and their
double-counted overlap is reinstated by adding back $K^{\otwo}$. 
\item in \eq{eq:NNLO_RV_sub}, $RV$ must be integrated in $\Phi_\npo$, and is 
affected by both explicit IR poles and phase-space singularities. The IR poles
arising from the loop integration in $RV$ are cancelled by the integral 
$I^{\one}$, by virtue of general cancellation theorems; the first parenthesis
is thus finite, but both terms are singular in the phase space of the radiated 
particle. By construction, the phase-space singularities of $I^{\one}$
are cancelled by $I^{\otwo}$, and $K^{\RV}$ is designed to cancel 
the phase-space singularities of $RV$. This however does not guarantee 
that explicit IR poles will cancel in the second parenthesis. Anyway, one can 
fine-tune the definition of $K^{\RV}$, by including explicit IR poles
not associated with the phase-space singularities of $RV$, in order to
make the second parenthesis finite as well. At this point, \eq{eq:NNLO_RV_sub} is 
both finite and integrable. 
\item The complete cancellation of real and virtual 
singularities in \eq{eq:NNLO_RV_sub} and \eq{eq:NNLO_RR_sub} 
guarantees then, 
as a consequence of the KLN theorem, that \eq{eq:NNLO_VV_sub}, to be integrated 
in the Born-level phase space $\Phi_n$, will be free of IR poles.
\end{itemize}
In the next sections we will construct explicit expressions for all counterterms, compute 
their integrals analytically, and finally obtain $RR_{\, \rm sub}(X)$, $RV_{\, \rm sub}(X)$ 
and $VV_{\, \rm sub}(X)$. As was the case at NLO,
this will require identifying the relevant single- and double-unresolved limits,
introducing an appropriate set of NNLO sector functions, and defining flexible
and consistent phase-space mappings. Needless to say, the multiplicity of singular 
configurations and of their overlaps will lead to long and intricate expressions: 
therefore, detailed formulas for NNLO soft and collinear kernels, for the relevant 
mapped limits, and for the required integrals, as well as a number of notational 
shortcuts, will be presented in the Appendices.


\section{The subtracted double-real contribution $RR_{\,\rm sub}$}
\label{RRsub}

In this section we provide a detailed construction of the subtracted squared matrix 
element for double-real radiation, $RR_{\,\rm sub}$. As noted in \eq{eq:NNLO_RR_sub},
this will require the definition of three separate local counterterms. From a combinatorial 
viewpoint, this task represents the most intricate part of the NNLO-subtraction 
programme, due to the large number of overlapping singular limits affecting double-real 
radiation. In analogy to \secn{FrameNLO}, we will proceed as 
follows: first, in \secn{singu2}, we will list and briefly discuss the relevant singular limits, 
which can be single- or double-unresolved; next, in \secn{topo2}, we will introduce a 
set of sector functions, smoothly partitioning the $(\npt)$-particle phase space so as to minimise 
the number of singular configurations to be considered in any given sector. These 
sectors will naturally be grouped into three different {\it topologies}, corresponding to the 
structure of the limits relevant to each sector. Next, in \secn{combising2}, we will identify 
specific combinations of limits that yield integrable contributions in each topology, in 
the spirit of \eq{intij}; we will then construct, in \secn{phaspamap2}, a family of phase-space 
mappings in order to properly factorise the double-radiative phase space in all relevant 
configurations. Finally, in \secn{buildRRsub2}, we will introduce improved limits appropriate 
for each topology, discuss the required consistency conditions, and then use the improved 
limits to compose an expression for the subtracted double-real contribution $RR_{\,\rm sub}$. 
As was the case at NLO for single-real radiation, it is possible to improve upon the 
resulting expression for $RR_{\,\rm sub}$ by introducing symmetrised sector functions 
in order to optimise the subsequent numerical integration. This construction is discussed in 
\secn{RRsubSymSect}. We note that the 
construction presented in this paper differs slightly in some technical choices from the 
one given in Ref.~\cite{Magnea:2018hab}: we will note the differences as we go along.


\subsection{Singular limits for double-real radiation}
\label{singu2}

Double-real radiation matrix elements are characterised by a variety of overlapping 
singular limits. It is important, from the outset, to pick a complete set of limits, in 
order to then study (and subtract) their overlaps, to avoid double counting. Clearly,
single-unresolved soft and collinear limits are relevant also for double radiation,
so our list must include the limits $\bS{i}$ and $\bC{ij}$ introduced in \secn{FrameNLO}.
Next, we need to collect all possible double-unresolved limits. Importantly, when 
two particles become unresolved, one needs to distinguish {\it uniform} limits, 
where the rate at which the two particles become unresolved is the same, and 
{\it strongly-ordered} limits, where one particle becomes unresolved at a higher 
rate with respect to the second one. Obviously, this distinction becomes  
relevant  starting at NNLO. Our set of fundamental uniform limits consists of four 
independent configurations. First, two particles $i$ and $j$ can become soft at the same 
rate, a limit which we denote by $\bS{ij}$; second, a single hard particle can branch into 
three collinear ones, $i$, $j$ and $k$, a limit which we denote by $\bC{ijk}$; third, two 
hard partons can independently branch into two collinear pairs, which we denote by 
$\bC{ijkl}$, with $(i, j)$ and $(k, l)$ labelling the two independent pairs; finally, 
a particle $i$ can become soft while another pair of particles, $j$ and $k$, 
become collinear at the same rate\footnote{In Ref.~\cite{Magnea:2018hab}, 
two strongly-ordered soft-collinear limits were considered, instead of the 
uniform one chosen here.}, which we denote by $\bSC{ijk}$. In these four limits, 
the double-real-radiation squared matrix element factorises, with the relevant kernels 
derived and presented in Ref.~\cite{Catani:1999ss}. Given these uniform limits, 
the strongly-ordered ones can be reached by acting iteratively: for example, the 
strongly-ordered double-soft limit, with particle $i$ becoming soft faster than particle 
$j$, can be reached by computing $\bS{i} \, \bS{ij}$,while the strongly-ordered 
double-collinear limit, with particles $i$ and $j$ becoming collinear faster than 
the third particle $k$, will be given by the combination $\bC{ij} \, \bC{ijk}$. All singular 
configurations can be reached in this way. 

In order to proceed, we need to characterise the limits more precisely, in terms 
of phase-space variables. As was the case at NLO, we choose to define the
limits in terms of Mandelstam invariants, and we pay attention to the fact that
all limits must commute when acting on the double-real radiation squared 
matrix element. Using the variables $e_i$ and $w_{ij}$ given in \eq{sectprelim},
the definitions of the independent limits, both single- and double-unresolved, 
are specified in Table~\ref{tab:limits}. Importantly, our choice of independent
limits is related to our choice of sector functions, which will be tuned so that
only a minimal pre-defined set of the chosen limits will contribute in each sector.
\begin{table}[ht]
\begin{center}
\begin{tabular}{|l|l|}
\hline
$\bS{i}$ & $e_i \to  0$ \quad (soft configuration of parton $i$) \\[2pt]
\hline
$\bC{ij}$ & $w_{ij} \to 0$ \quad (collinear configuration of partons $(i, j)$) \\[2pt]
\hline
$\bS{ij}$ & $e_i, e_j  \to  0$ and $e_i/e_j \to \mbox{constant}$ \\[2pt]
& (uniform double-soft configuration of partons $(i, j)$) \\[2pt]
\hline
$\bC{ijk}$ & $w_{ij}, w_{ik}, w_{jk} \to 0$ and 
$w_{ij}/w_{ik}, w_{ij}/w_{jk}, w_{ik}/w_{jk} \to \mbox{constant}$ \\[2pt]
& (uniform double-collinear configuration of partons $(i, j, k)$) \\[2pt]
\hline
$\bC{ijkl}$ & $w_{ij}, w_{kl} \to 0$ and 
$w_{ij}/w_{kl} \to \mbox{constant}$ \\
& (uniform double-collinear configuration of partons $(i, j)$ and $(k, l)$) \\
\hline
$\bSC{ijk}$ & $e_{i}, w_{jk} \to 0$ and 
$e_{i}/w_{jk} \to \mbox{constant}$ \\
& (uniform soft and collinear configuration for partons $i$ and $(j, k)$) \\
\hline
\end{tabular}
\end{center}
\caption{
Definitions of the single-unresolved singular limits $\bS{i}$, $\bC{ij}$ 
and of our set of basic independent double-unresolved singular limits $\bS{ij}$, 
$\bC{ijk}$, $\bC{ijkl}$, $\bSC{ijk}$.
}
\label{tab:limits}
\end{table}


\subsection{Sector functions and topologies for double-real radiation}
\label{topo2}

We now introduce a smooth unitary partition of the double-real-radiation phase space, 
in the spirit of Ref.~\cite{Frixione:1995ms}. Since at most four particles can be 
involved in singular infrared limits at NNLO, we label the sector functions with
four indices, and denote them by $\W{ijkl}$. We pick the first two indices to
label the single-unresolved configurations assigned to the chosen sector. In 
particular, we will design the sector $(ijkl)$ to contain the limits $\bS{i}$
and $\bC{ij}$ (thus we take $j \neq i$). We then need to distinguish sectors
involving only three distinct particles from sectors involving four distinct particles.
In sectors where only three particles are involved, the double-unresolved limit
$\bC{ijk}$ will be relevant; furthermore, a second particle (besides $i$) may 
become soft, and it can be particle $j$ {\it or} particle $k$. Correspondingly, 
we will have distinct sector functions $\W{ijjk}$ and $\W{ijkj}$, where we 
take the third index to indicate the second particle that can become soft. 
Similarly, if all four indices are distinct, we take $\W{ijkl}$ to select the 
sector where particles $i$ and $k$ can become soft, while the possible 
collinear pairs are $(i,j)$ and $(k,l)$. Notice that in all cases the last three
indices $j$, $k$ and $l$ are distinct from $i$, and $k \neq l$. We will refer 
to the three allowed combinations of sector indices, $(ijjk)$, $(ijkj)$ and 
$(ijkl)$ as {\it topologies}, and we will denote them collectively by $\tau \equiv
abcd \in \{ijjk, ijkj, ijkl\}$.

We now need to introduce a precise definition of NNLO sector functions, which 
will enable us to list all the fundamental limits contributing to each topology.
As was done at NLO (see \eq{sectfunc}), we will define NNLO sector functions
as ratios of the type  
\beq
\W{abcd} 
\; = \;  
\frac{\sigma_{abcd}}{\sigma} 
\, ,
\qquad \quad
\sigma
\; = \;
\!\! \sum_{\substack{a, b \neq a}} 
\sum_{\substack{c \neq a \\ d \neq a, c}}
\sigma_{abcd}
\, ,
\eeq
so that
\beq
\sum_{\substack{a, b \neq a}} 
\sum_{\substack{c \neq a \\ d \neq a, c}} \W{abcd} 
\; = \;
1
\, .
\eeq
Such a partition allows us to rewrite the double-real squared matrix element $RR$ as 
\beq
\label{RRsum}
RR 
& = & \!
\sum_{i, \, j \ne i} \,
\sum_{k \ne i} \,
\sum_{l \ne i, k} \,
RR \; \W{ijkl}
\,\, =  
\sum_{i, \, j \ne i} \,
\sum_{k \ne i, j} \,
\bigg[
RR \; \W{ijjk}
+
RR \; \W{ijkj}
+
\sum_{l \ne i, j, k} \,
RR \; \W{ijkl}
\bigg]
\, .
\qquad
\eeq
Our choice for the functions $\sigma_{abcd}$\footnote{This choice corresponds to setting 
$\alpha = \beta$ in the NNLO sector functions introduced in Ref.~\cite{Magnea:2018hab}.} 
is given by
\beq
\label{eq: sigma_abcd}
\sigma_{abcd} 
\; = \;
\frac{1}{(e_a \, w_{ab})^{\alpha}}
\frac{1}{(e_c + \delta_{bc} \, e_a) \, w_{cd}} 
\, , 
\qquad \quad
\alpha > 1 
\, .
\eeq
Given \eq{eq: sigma_abcd}, we can list which of the fundamental limits discussed 
in \secn{singu2} will affect each topology. One finds that the combination
$RR \; \W{\tau}$ will be singular in the limits listed below.
\begin{alignat}{7}
\label{eq:unbarred_set}
& RR \; \W{ijjk} \, : 
\quad && \bS{i} \, ,
\quad && \bC{ij} \, ,
\quad && \bS{ij} \, ,
\quad && \bC{ijk} \, ,
\quad &&  \bSC{ijk} \, ; 
\nnb
\\
& RR \; \W{ijkj} \, : 
\quad && \bS{i} \, ,
\quad && \bC{ij} \, ,
\quad && \bS{ik} \, ,
\quad && \bC{ijk} \, ,
\quad &&  \bSC{ijk} \, ,
\quad &&  \bSC{kij} \, ;
\\
& RR \; \W{ijkl} \, : 
\quad && \bS{i} \, ,
\quad && \bC{ij} \, ,
\quad && \bS{ik} \, ,
\quad && \bC{ijkl} \, ,
\quad &&  \bSC{ikl} \, ,
\quad &&  \bSC{kij} \, .
\nnb
\end{alignat}
In analogy with the NLO sum-rule requirements in \eq{SCsumrules}, also 
NNLO sector functions which share a given singular configuration must form a unitary 
partition. This is a crucial feature in order to minimise the complexity of the counterterm 
structure in view of analytic integration. The choice of the functions $\sigma_{abcd}$ in 
\eq{eq: sigma_abcd} guarantees that the required partial sums reduce to unity. 
For example, we report the sum rules for the double-unresolved limits in 
Table~\ref{tab:limits}, which read 
\beq
\label{eq:sum_rules_NNLO1}
\bS{ik} \,
\left( \,
\sum_{b \neq i} \sum_{d \neq i,k}
\W{ibkd}
+
\sum_{b \neq k} \sum_{d \neq k,i}
\W{kbid}
\right)
\, = \,
1
\, ,
\\[5pt]
\label{eq:sum_rules_NNLO2}
\bC{ijk} \!\!\!\!\!\!
\sum_{abc \, \in \, \pi(ijk)} \!\!\!\!\!\!
\big(\W{abbc} + \W{abcb} \big)
\, = \,
1
\, ,
\qquad
\bC{ijkl} \!\!\!\!
\sum_{\substack{ab \, \in \, \pi(ij) \\ cd \, \in \, \pi(kl)}} \!\!\!\!
\big(\W{abcd} + \W{cdab} \big)
\, = \,
1
\, ,
\\
\label{eq:sum_rules_NNLO3}
\bSC{ijk} \,
\bigg( \!\!\!\!
\sum_{\substack{d \neq i \\ ab \, \in \, \pi(jk)}} \!\!\!\!\!\!\W{idab}
+ \!\!\!
\sum_{\substack{d \neq i,a \\ ab \, \in \, \pi(jk)}} \!\!\!\!\!\! \W{abid} \,
\bigg)
\, = \,
1
\, ,
\eeq
where by $\pi(ij)$ and $\pi(ijk)$ we denote respectively the sets $\{ij, ji\}$ and 
$\{ijk, ikj, jik, jki, kij, kji\}$.

In order for the double-real contribution to properly combine with the real-virtual
 correction, we require NNLO sector functions to factorise
into NLO-like sector functions under the action of single-unresolved limits. 
As discussed in Ref.~\cite{Magnea:2018hab}, and 
below in \secn{RVsub}, this ensures the local cancellation of integrated phase-space 
singularities with the poles of the real-virtual correction, sector by sector in the 
single-radiative phase space: indeed $RV$ needs to be
partitioned with NLO-like sector functions, since it involves a single-real radiation.
As an example, one may verify that the sector functions for the topology $(ijjk)$ satisfy
\beq
  \bS{i}  \, \W{ijjk} =  \W{jk}  \, \bS{i}  \, \W{ij}^{(\alpha)} , ~~
  \bC{ij}  \, \W{ijjk} =  \W{[ij]k}  \, \bC{ij} \,  \W{ij}^{(\alpha)} , ~~ \,
  \bS{i} \, \bC{ij} \,  \W{ijjk} = \W{jk}  \, \bS{i} \, \bC{ij}  \, \W{ij}^{(\alpha)} ,
  \qquad
\label{propNNLOfact}
\eeq
where $\W{[ij]k}$ is the NLO sector function defined in the $(\npo)$-particle phase 
space including the parent parton $[ij]$ of the collinear pair $(i, j)$, and we introduced
the NLO-like, $\alpha$-dependent sector functions
\beq
\label{NLOtype}
\W{ij}^{(\alpha)}
\, \equiv \,
\frac{\sigma_{ij}^{(\alpha)}}{\sum_{k \neq l} \sigma_{kl}^{(\alpha)}}
\, ,
\qquad \quad
\sigma_{ij}^{(\alpha)}
\, \equiv \,
\frac{1}{(e_i\,w_{ij})^{\alpha}} \, ,
\qquad
\alpha > 1
\, ,
\eeq
so that ordinary NLO sector functions are given by $\W{ij}= \W{ij}^{(1)}$. Similar
relations hold for the other two topologies.


\subsection{Combining singular limits of topologies}
\label{combising2}

As listed in \eq{eq:unbarred_set}, a limited number of products of IR projectors is 
sufficient to collect all singular configurations of the double-real squared matrix 
element in each topology. Since the action of the relevant limits on both $RR$ and 
on the sector functions does not depend on the order they are applied, the following 
combinations are by construction integrable in the whole phase space 
\beq
\label{integrablecomb}
\left( 1 - \bS{i} \right)
\left( 1 - \bC{ij} \right)
\left( 1 - \bS{ij} \right)
\left( 1 - \bC{ijk} \right)
\left( 1 - \bSC{ijk} \right) 
\,
RR \; \W{ijjk}
\; & \to & \;
\mbox{integrable} \, , \qquad
\nnb\\
\left( 1 - \bS{i} \right)
\left( 1 - \bC{ij} \right)
\left( 1 - \bS{ik} \right)
\left( 1 - \bC{ijk} \right)
\left( 1 - \bSC{ijk} \right) 
\left( 1 - \bSC{kij} \right) 
\,
RR \; \W{ijkj}
\; & \to & \;
\mbox{integrable} \, , \qquad
\\
\left( 1 - \bS{i} \right)
\left( 1 - \bC{ij} \right)
\left( 1 - \bS{ik} \right)
\left( 1 - \bC{ijkl} \right)
\left( 1 - \bSC{ikl} \right) 
\left( 1 - \bSC{kij} \right) 
\,
RR \; \W{ijkl}
\; & \to & \;
\mbox{integrable} \, . \qquad
\nnb
\eeq
Note that, in analogy to the definition used for NLO projection operators, if we take 
$\bL{}{}$ to be any one of the singular limits in Table~\ref{tab:limits}, the action 
$\bL{}{} \, RR \, \W{abcd} \equiv (\bL{}{} \, RR) \, (\bL{}{} \,\W{abcd})$  is understood
for all topologies.

Applying directly \eq{integrablecomb} would be quite cumbersome, as the three 
lines generate a total of 160 terms. Fortunately, the resulting combinations of limits 
are not all independent, and several non-trivial relations can be obtained exploiting 
the symmetries of the limits under exchanges of indices, as well as the definitions
of the various limits involved as projection operators on singular terms of $RR$.
Consider for example, in four-particle sector $\W{ijkl}$, the projection $(1 - \bS{ik})\,  
RR \,  \W{ijkl}$. This will contain only terms in $RR$ that are not singular in sector 
$(ijkl)$ when the uniform soft limit is taken for particles $i$ and $k$. As a consequence, 
if further projections involving {\it both} the $i$ and $k$ soft limits are taken, the result 
will be integrable. We conclude, for example, that
\beq
\bSC{ikl} \, \bSC{kij} ( 1 - \bS{ik} ) \, RR \; \W{ijkl}
& \to & 
\mbox{integrable}
\, .
\label{exconstlim}
\eeq
Working in this way, topology by topology, we can write a set of finite relations, which 
help us remove redundant configurations contributing to \eq{integrablecomb}. They read
\beq
\label{eq:footnote relations}
\bC{ij} \, \bSC{ijk} (1-\bS{i})(1-\bS{ij})(1-\bC{ijk}) \, RR \; \W{ijjk}
& \to & 
\mbox{integrable}
\, ,
\nnb\\
\bS{i} \, \bSC{kij} (1-\bS{ik}) (1-\bC{ijk}) \, RR \; \W{ijkj} 
& \to & 
\mbox{integrable}
\, ,
\nnb\\
\bC{ij} \, \bSC{ijk} (1-\bS{i})(1-\bS{ik})(1-\bC{ijk}) \, RR \; \W{ijkj}
& \to & 
\mbox{integrable}
\, ,
\nnb\\
\bC{ij} \, \bS{ik} (1-\bS{i})(1-\bSC{kij})(1-\bC{ijk}) \, RR \; \W{ijkj}
& \to & 
\mbox{integrable}
\, ,
\nnb\\
\bSC{ijk} \, \bSC{kij} ( 1 - \bS{ik} ) \, RR \; \W{ijkj}
& \to & 
\mbox{integrable}
\, ,
\nnb\\
\bS{i} \, \bSC{kij} (1-\bS{ik}) (1-\bC{ijkl}) \, RR \; \W{ijkl} 
& \to & 
\mbox{integrable}
\, ,
\nnb\\
\bS{i}\,\bC{ijkl} (1-\bS{ik}) (1-\bSC{ikl}) \, RR \; \W{ijkl}
& \to & 
\mbox{integrable}
\, ,
\\
\bC{ij} \, \bSC{ikl} (1-\bS{i})(1-\bS{ik})(1-\bC{ijkl}) \, RR \; \W{ijkl}
& \to & 
\mbox{integrable}
\, ,
\nnb\\
\bC{ij} \, \bS{ik} (1-\bS{i})(1-\bSC{kij})(1-\bC{ijkl}) \, RR \; \W{ijkl}
& \to & 
\mbox{integrable}
\, ,
\nnb\\
\bSC{ikl} \, \bSC{kij} ( 1 - \bS{ik} ) \, RR \; \W{ijkl}
& \to & 
\mbox{integrable}
\, ,
\nnb \\
\bC{ijkl} \, \bS{ik} ( 1 - \bSC{ikl} ) \, RR \; \W{ijkl}
& \to & 
\mbox{integrable}
\, ,
\nnb \\
\bC{ijkl} \, \bS{ik} ( 1 - \bSC{kij} ) \, RR \; \W{ijkl}
& \to & 
\mbox{integrable}
\, .
\nnb
\eeq
These finite relations allow us to simplify considerably \eq{integrablecomb}, leading
to the integrable expression
\beq
\label{eq:candidate RRsub}
\RR \, \W{\tau} 
-
\left(
\bL{ij}{\!\one} 
+
\bL{\tau}{\!\two}
-
\bL{\tau}{\!\otwo}
\right)
\RR \, \W{\tau} 
\; & \to & \;
\mbox{integrable}
\, ,
\eeq
which is the NNLO equivalent of \eq{intij} for double-real radiation\footnote{Note 
that there is no ambiguity in the notation: we denote by $(ij)$ the first two indices 
of the sector, which are common to all three topologies.}. In \eq{eq:candidate RRsub}
we distinguished, for each topology $\tau$, the single-unresolved limit $\bL{ij}{\!\one}$, 
the uniform double-unresolved limit $\bL{\tau}{\!\two}$, and the strongly-ordered 
double-unresolved limit $\bL{\tau}{\!\otwo}$. Their explicit expressions for each topology, in terms 
of the projectors discussed in \secn{singu2}, are
\beq
\label{eq:unbarred L}
\bL{ij}{\!\one}
& = &
\bS{i} 
+
\bC{ij} 
\left( 1 - \bS{i} \right)
\, ,
\nnb\\
\bL{ijjk}{\!\two}
& = &
\bS{ij} 
+ 
\bSC{ijk} \left( 1 - \bS{ij} \right)
+ 
\bC{ijk} 
\left( 1 - \bS{ij} \right) 
\left( 1 - \bSC{ijk} \right)
\, ,
\nnb\\
\bL{ijkj}{\!\two}
& = &
\bS{ik} 
+ 
\left( \bSC{ijk} + \bSC{kij} \right)
\left( 1 - \bS{ik} \right)
+
\bC{ijk} 
\left( 1 - \bS{ik} \right)
\left( 1 - \bSC{ijk} - \bSC{kij} \right) 
\, ,
\nnb\\
\bL{ijkl}{\!\two}
& = &
\bS{ik} 
+
\left( \bSC{ikl} + \bSC{kij} \right)
\left( 1 - \bS{ik} \right) 
+
\bC{ijkl} 
\left( 1 + \bS{ik} - \bSC{ikl} - \bSC{kij} \right)
\, ,
\nnb\\
\bL{ijjk}{\!\otwo}
& = &
\bS{i} \,
\Big[
\bS{ij} 
+
\bSC{ijk}
\left( 1 - \bS{ij} \right) 
+
\bC{ijk}
\left( 1 - \bS{ij} \right)
\left(1 - \bSC{ijk} \right)
\Big]
\\ 
&& \hspace{1.5cm}
+ \,
\bC{ij} 
\left( 1 - \bS{i} \right)
\Big[
\bS{ij} 
+
\bC{ijk}
\left( 1 - \bS{ij} \right)
\Big]
\, ,
\nnb\\
\bL{ijkj}{\!\otwo}
& = &
\bS{i} \,
\Big[
\bS{ik} 
+
\bSC{ijk}
\left( 1 - \bS{ik} \right)
+
\bC{ijk}
\left( 1 - \bS{ik} \right)
\left( 1 - \bSC{ijk} \right)
\Big]
\nnb \\ 
&& \hspace{1.5cm}
+ \,
\bC{ij}
\left( 1 - \bS{i} \right)
\Big[
\bSC{kij} 
+
\bC{ijk}
\left( 1 - \bSC{kij} \right)
\Big]
\, ,
\nnb\\
\bL{ijkl}{\!\otwo}
& = &
\bS{i} \,
\Big[
\bS{ik} 
+
\bSC{ikl}
\left( 1 - \bS{ik} \right) 
\Big]
\,+\,
\bC{ij}
\left( 1 - \bS{i} \right)
\Big[
\bSC{kij}
+
\bC{ijkl}
\left( 1 - \bSC{kij} \right)
\Big]
\, .
\nnb
\eeq
The projection operators appearing in \eq{eq:unbarred L} are organised so as to display, 
in order, the soft ($\bf S$), the uniform soft and collinear ($\bf SC$) and the collinear 
($\bf C$) singular contributions. Upon summing over sectors, \eq{eq:candidate RRsub} 
and \eq{eq:unbarred L} build up the equivalent at NNLO of \eq{intij} and \eq{candcount},
for double-real radiation: indeed, applying the limits defined in \eq{eq:unbarred L} on 
$RR$ and on the sector functions gives the starting point to determine the form of the 
counterterms for each sector, since the limits contain all phase-space singularities of 
$RR$ in a given sector, without double counting. In order to promote them to actual 
counterterms, it is now necessary to introduce phase-space mappings, allowing to 
properly factorise the $(\npt)$-body phase space into an $(\npo)$-body phase space 
times a single-radiation phase space for $\bL{}{\!\one}$ and $\bL{}{\!\otwo}$, and into 
an $n$-body phase space times a double-radiation phase space for $\bL{}{\!\two}$, 
as shown in \eq{phaspafac2}. We now turn to the discussion of these mappings.


\subsection{Phase-space mappings for double-real radiation}
\label{phaspamap2}

There is considerable freedom to define phase-space mappings for double-real radiation
(see for example~\cite{DelDuca:2019ctm}). We have chosen to use nested Catani-Seymour 
final-state mappings, which involve a minimal set of the $(\npt)$ momenta, and are built 
in terms of Mandelstam invariants, simplifying both the factorised expression for the 
$(\npt)$-body phase space and the dependence of the counterterms on the integration 
variables of the radiative phase spaces. In this framework, the mappings to factorise 
the $(\npt)$-body phase space into an $(\npo)$-body phase space times a single-radiation 
phase space, necessary for $\bL{}{\!\one}$ and $\bL{}{\!\otwo}$, can be 
constructed with the same procedure followed at NLO, and one is lead to \eq{mapNLO} 
and \eq{momconsCS}, with $i$ running from 1 to $n+2$, and $j$ running from 
1 to $n+1$.

For the construction of an on-shell, momentum conserving $n$-tuple of massless 
momenta in the $(\npt)$-particle phase space, necessary for  $\bL{}{\!\two}$, we 
distinguish the following three possibilities. 
\begin{itemize}
\item
We choose six final-state massless momenta $k_{a}$, $k_{b}$, $k_{c}$, 
$k_{d}$, $k_{e}$, $k_{f}$ (all different) and construct the $n$-tuple 
(without $k_{a}$ and $k_{b}$) 
\beq
\label{sixpart1}
\kkl{acd,bef}
& = & 
\left\{ 
\{ k \}_{\slashed a \slashed b \slashed c \slashed d \slashed e \slashed f}, \, 
\kk{c}{acd,bef},
\kk{d}{acd,bef},
\kk{e}{acd,bef},
\kk{f}{acd,bef}
\right\} \, ,
\eeq
with
\beq
\label{sixpart2}
&&
\kk{c}{acd,bef} 
\, = \, 
k_{a} + k_{c} - \frac{s_{ac}}{s_{[ac]d}} \, k_{d} 
\, , 
\qquad\qquad
\kk{d}{acd,bef} 
\, = \, 
\frac{s_{acd}}{s_{[ac]d}} \, k_{d} 
\, ,
\nnb \\[0.2cm]
&&
\kk{e}{acd,bef} 
\, = \, 
k_{b} + k_{e} - \frac{s_{be}}{s_{[be]f}} \, k_{f} 
\, ,
\qquad\qquad
\kk{f}{acd,bef} 
\, = \,  
\frac{s_{bef}}{s_{[be]f}} \, k_{f}
\, ,
\eeq
while all other momenta are left unchanged 
($\kk{n}{acd,bef} = k_n \, , \, n \neq a,b,c,d,e,f$).
Here and in the following $s_{[ab]c} = s_{ac}+s_{bc}$.
\item
We choose five final-state massless momenta $k_{a}$, $k_{b}$, $k_{c}$, 
$k_{d}$, $k_{e}$ (all different) and construct the $n$-tuple 
(without $k_{a}$ and $k_{b}$) 
\beq
\label{fivepart1}
\kkl{acd,bed}
& = & 
\left\{ 
\{ k \}_{\slashed a \slashed b \slashed c \slashed d \slashed e}, \, 
\kk{c}{acd,bed},
\kk{d}{acd,bed},
\kk{e}{acd,bed}
\right\} \, ,
\eeq
with
\beq
\label{fivepart2}
&&
\kk{c}{acd,bed}
\, = \, 
k_{a} + k_{c} - \frac{s_{ac}}{s_{[ac]d}} \, k_{d} 
\, , 
\qquad\quad
\kk{d}{acd,bed} 
\, = \,
\left(\! 1 + \frac{s_{ac}}{s_{[ac]d}} + \frac{s_{be}}{s_{[be]d}} \right)
k_{d} 
\, ,
\qquad\qquad 
\nnb \\[0.2cm]
&&
\kk{e}{acd,bed} 
\, = \,
k_{b} + k_{e} - \frac{s_{be}}{s_{[be]d}} \, k_{d} 
\, ,
\eeq
while all other momenta are left unchanged 
($\kk{n}{acd,bed} = k_n \, , \, n \neq a,b,c,d,e$).
\item
We choose four final-state massless momenta $k_{a}$, $k_{b}$, $k_{c}$, 
$k_{d}$ (all different) and construct the $n$-tuple (without $k_{a}$ and 
$k_{b}$) 
\beq
\label{fourpart1}
\kkl{acd,bcd}
\; = \;
\kkl{abc,bcd}
\; = \;
\kkl{abcd}
& = & 
\left\{ 
\{ k \}_{\slashed a \slashed b \slashed c \slashed d}, \, 
\kk{c}{abcd},
\kk{d}{abcd}
\right\} \, ,
\eeq
with
\beq
\label{fourpart2}
\kk{c}{abcd} 
\, = \, 
k_{a} + k_{b} + k_{c} - \frac{s_{abc}}{s_{ad} + s_{bd} + s_{cd}} \, k_{d} 
\, , 
\qquad\quad
\kk{d}{abcd} 
\, = \, 
\frac{s_{abcd}}{s_{ad} + s_{bd} + s_{cd}} \, k_{d} 
\, ,
\qquad
\eeq
while all other momenta are left unchanged 
($\kk{n}{abcd} = k_n \, , \, n \neq a,b,c,d$).
\end{itemize}
With these tools, we are now ready to construct improved infrared projectors, with 
a proper factorised structure, and we can use them to define our counterterms.


\subsection{Building $RR_{\,\rm sub}$ with improved singular limits}
\label{buildRRsub2}

To write explicitly the counterterms we introduce {\em improved} versions of the limits 
in Table~\ref{tab:limits} 
\begin{center}
\begin{tabular}{lllllll}
$\bbS{a}$, & $\bbC{ab}$, & $\bbS{ab}$, & $\bbC{abc}$, & $\bbC{abcd}$, & $\bbSC{abc}$ \, .
\end{tabular}
\end{center}
They are to be interpreted as operators which, on top of extracting the corresponding 
singular limit on the objects they act on, convey a specific mapping of momenta, to be 
defined case by case, and may be further refined (for example by tuning their action 
on sector functions) in order to ensure the local cancellation of singularities after the 
implementation of phase-space mappings.

Given the definitions of the improved limits (to be discussed below) we can construct
the expression for $RR_{\,\rm sub}$ in the following way. First, we define the improved 
version of the various ${\bf L}$ operators corresponding to the limits in \eq{eq:unbarred L}, 
for each topology, denoting the improved operators by $\bbL{}$. Next, we define our local 
counterterms, for each topology $\tau= ijjk,ijkj,ijkl$, as
\begin{align}
\label{counttopo} 
K^{\one}_{\tau} 
& = \,
\bbL{ij}^{\one}
\RR \; \W{\tau}
\, ,
&
K^{\two}_{\tau} 
& = \,
\bbL{\tau}^{\two} 
\RR \; \W{\tau}
\, ,
&
K^{\otwo}_{\tau}
& = \,
\bbL{\tau}^{\otwo}
\RR \; \W{\tau}
\, .
\end{align}
The subtracted double-real squared matrix element for topology $\tau$ is 
then given by
\beq
\label{eq:RRsub tau}
\RR^{\,\rm sub}_{\,\tau}(X) 
& = & 
\RR \; \W{\tau}  \, \delta_\npt(X)
- 
K^{\one}_{\tau}  \, \delta_\npo(X)
- 
\Big(
K^{\two}_{\tau} 
-
K^{\otwo}_{\tau}
\Big) 
\delta_n(X)
\, .
\eeq
This allows to build the complete $RR_{\,\rm sub}(X)$ of 
\eq{eq:sigma NNLO sub} by summing the contributions from all sectors according to
\beq
\label{eq:RRsub W}
\RR_{\,\rm sub}(X) 
& = & 
\sum_{i, \, j \ne i}
\sum_{k \ne i, j}
\bigg[
\RR^{\,\rm sub}_{\,ijjk}(X) 
+
\RR^{\,\rm sub}_{\,ijkj}(X) 
+
\sum_{l \neq i,j,k} \!
\RR^{\,\rm sub}_{\,ijkl}(X) 
\bigg]
\, .
\eeq
The structure of \eq{eq:NNLO_RR_sub} is then recovered by using \eq{RRsum}, 
and by defining 
\beq
\label{eq:K1 K2 K12 W}
K^{\one}
& = & 
\sum_{i, \, j \ne i}
\sum_{k \ne i, j}
\bigg[
K^{\one}_{\,ijjk}
+
K^{\one}_{\,ijkj}
+
\!\! \sum_{l \neq i,j,k} \!\!
K^{\one}_{\,ijkl}
\bigg]
\, ,
\nnb\\
K^{\two}
& = & 
\sum_{i, \, j \ne i}
\sum_{k \ne i, j}
\bigg[
K^{\two}_{\,ijjk}
+
K^{\two}_{\,ijkj}
+
\!\! \sum_{l \neq i,j,k} \!\!
K^{\two}_{\,ijkl}
\bigg]
\, ,
\nnb\\
K^{\otwo}
& = & 
\sum_{i, \, j \ne i}
\sum_{k \ne i, j}
\bigg[
K^{\otwo}_{\,ijjk}
+
K^{\otwo}_{\,ijkj}
+
\!\!\! \sum_{l \neq i,j,k} \!\!\!
K^{\otwo}_{\,ijkl}
\bigg]
\, .
\qquad\qquad
\eeq
We emphasise that the definitions of the counterterms
are actually complete only after specifying both the action $\bbL{}{} \, \RR$ of improved 
limits on the double-real matrix element, as well as the action $\bbL{}{} \, \mc{W}_{\tau}$ 
on sector functions. All the improved limits are reported in 
\appn{app:K1,K2,K12}, and are written in terms of the soft and collinear kernels listed 
in \appn{app:kernels}, multiplying appropriate versions of the Born-level probabilities,
expressed in terms of mapped momenta. 

In order to give the reader a feeling for the kind of expressions that emerge from 
this procedure, we reproduce here two representative examples. First, the uniform 
double-unresolved double-soft improved limit $\bbS{ik}$ ($i \neq k$) can be written 
as
\beq
\label{eq:bSik_text}
\bbS{ik} \, \RR
& \equiv &
\frac{\Norm^{\,2}}{2} \!
\sum_{\substack{c \neq i,k \\ d \neq i,k,c}}
\bigg\{ \;
\mc E^{(i)}_{cd} 
\! \sum_{\substack{e \neq i,k,c,d}}
\bigg[
\sum_{\substack{f \neq i,k,c,d,e}} \!
\mc E^{(k)}_{ef} \bB_{cdef}^{(icd,kef)}
+ 
4 \,
\mc E^{(k)}_{ed} \bB_{cded}^{(icd,ked)}
\bigg]
\nnb\\[-3mm]
&&
\hspace{17mm}
+ \,
2 \,
\mc E^{(i)}_{cd} 
\mc E^{(k)}_{cd} 
\bB_{cdcd}^{(icd,kcd)}
+
\mc E^{(ik)}_{cd}
\bB_{cd}^{(ikcd)}
\bigg\}
\, ,
\eeq
where the NLO eikonal kernel $\mc E^{(i)}_{cd}$ and the NNLO eikonal kernel
$\mc E^{(ik)}_{cd}$ are presented in Eqs.~(\ref{eikoker}) and (\ref{eikoker2}), and
we employed six-, five- and four-particle mappings for the 
colour-correlated Born terms, according to the numbers of particles involved.
Note in particular that all eikonal dipoles are mapped differently, which is 
essential for the analytic integration, as discussed in Ref.~\cite{Magnea:2020trj} 
and in \secn{RRint} below.

For the strongly-ordered double-unresolved double-soft improved limit $\bbS{i} \, 
\bbS{ik}$ ($i \neq k$), on the other hand, we write
\beq
\label{eq:bSi_bSik_text}
\bbS{i} \, \bbS{ik} \, \RR 
& \equiv &
\frac{\Norm^{\,2}}{2} \!\!
\sum_{\substack{c \neq i,k \\ d \neq i,k,c}} \!
\Bigg\{ \;
\mc E_{cd}^{(i)} \,
\Bigg[ \;
\sum_{\substack{e \neq i,k,c,d}} \!
\bigg(
\sum_{f \neq i,k,c,d,e} \!\!\!\!\!\!
\bar{\mc E}_{ef}^{(k)(icd)} \,
\bB_{cdef}^{(icd,kef)}
+
2 \,
\bar{\mc E}_{ed}^{(k)(icd)}
\bB_{cded}^{(icd,ked)}
\bigg)
\nnb\\[-2mm]
&&
\hspace{2cm}
+ \,
2 \!\!\!
\sum_{\substack{e \neq i,k,c,d}} \!\!\!
\bar{\mc E}_{ed}^{(k)(idc)}
\bB_{cded}^{(idc,ked)} 
+
2 \;
\bar{\mc E}_{cd}^{(k)(icd)} 
\Big( \bB_{cdcd}^{(icd,kcd)} + C_A\,\bB_{cd}^{(icd,kcd)} \Big)
\Bigg]
\nnb\\
&&
\hspace{2cm}
- \,
2 \, C_A 
\bigg[
\mc E_{kc}^{(i)} \,
\bar{\mc E}_{cd}^{(k)(ick)} \,
\bB_{cd}^{(ick,kcd)} 
+
\mc E_{kd}^{(i)} \,
\bar{\mc E}_{cd}^{(k)(ikd)} \,
\bB_{cd}^{(ikd,kcd)} 
\bigg]
\Bigg\}
\, .
\eeq
As might be expected, the complexity of the kernels has diminished with respect 
to \eq{eq:bSik_text} (indeed the expression solely features NLO eikonal factors), 
but the combinatorics has become more intricate. Notice
that we used mapped momenta also in the eikonal kernels corresponding to the
least-unresolved particle $k$.

It is important to stress that, while there appears to be considerable freedom in 
the choice of the improved limits, there are also stringent constraints that must 
be satisfied. In particular, the improved limits $\bbL{}{} \, \RR$ must carry the same symmetries 
under index exchange as the respective unimproved countertparts, so that the 
improved collections $\bbL{}^{\one}, \bbL{}^{\two}, \bbL{}^{\otwo}$ are still consistent 
with \eq{eq:unbarred L}, whose content is based on the validity of the integrable relations 
listed in \eq{eq:footnote relations}. Within the limitations of this requirement, there 
is still a residual freedom to modify the action of the improved limits acting on 
both $\RR$ and sector functions, with respect to the bare result extracted by their 
unimproved version. However, we must make sure that this procedure does preserve 
the locality of the cancellation of singularities, or, analogously, the finiteness of 
$\RR^{\,\rm sub}_{\,ijjk}$, $\RR^{\,\rm sub}_{\,ijkj}$ and $\RR^{\,\rm sub}_{\,ijkl}$, 
defined in \eq{eq:RRsub tau}. To this end, we checked the consistency of the 
improved limits listed in \appn{app:K1,K2,K12} by analytically verifying that, for 
any topology $\tau$, the corresponding $\RR^{\,\rm sub}_{\,\tau}$ is in fact integrable 
in all singular limits of that tolopogy. Specifically, we verified analytically that 
\beq
\big\{ 
\bS{i}, \; \bC{ij}, \; \bS{ij}, \; \bC{ijk}, \; \bSC{ijk} 
\big\}\,\RR^{\,\rm sub}_{\,ijjk}
& \to & 
\mbox{integrable} 
\, ,
\nnb\\
\big\{
\bS{i}, \; \bC{ij}, \; \bS{ik}, \; \bC{ijk}, \; \bSC{ijk} , \; \bSC{kij} 
\big\}\,\RR^{\,\rm sub}_{\,ijkj}
& \to & 
\mbox{integrable} 
\, ,
\nnb\\
\big\{ 
\bS{i}, \; \bC{ij}, \; \bS{ik}, \; \bC{ijkl}, \; \bSC{ikl} , \; \bSC{kij}
\big\}\,\RR^{\,\rm sub}_{\,ijkl}
& \to & 
\mbox{integrable} 
\, .
\eeq
Furthermore, since the collinear kernels of \appn{app:kernels} display spurious 
collinear singularities involving the reference momentum $k_r$, which are not 
always screened by the sector functions, we verified explicitly that also the 
following relations hold 
\beq
\label{eq:spur_sing}
\big\{ 
\bC{ir}, \; \bC{jr}, \; \bC{ijr} 
\big\}\,\RR^{\,\rm sub}_{\,ijjk}
& \to & 
\mbox{integrable}
\, ,
\nnb\\
\big\{ 
\bC{ir}, \; \bC{kr}, \; \bC{ikr} 
\big\}\,\RR^{\,\rm sub}_{\,ijkj}
& \to & 
\mbox{integrable}
\, ,
\nnb\\
\big\{ 
\bC{ir}, \; \bC{kr}
\big\}\,\RR^{\,\rm sub}_{\,ijkl}
& \to & 
\mbox{integrable} 
\, .
\eeq
Having passed these tests\footnote{
We have also verified that $\RR^{\,\rm sub}_{\,\tau}$ is integrable 
in secondary limits such as $\bS{j}$ for topology $\tau=ijjk$, 
$\bS{k}$ for $\tau=ijkj,ijkl$, $\bC{jk}$ for topologies 
$\tau=ijjk,ijkj$, and $\bC{kl}$ for $\tau=ijkl$.
}, 
the improved limits listed in \appn{app:K1,K2,K12}, when
assembled according to Eqs.~\eqref{counttopo}-\eqref{eq:RRsub W}, provide a fully local
subtraction of phase-space singularities for the double-real-emission contribution to
the cross section, and \eq{eq:RRsub W} is indeed integrable in the $(\npt)$-particle 
phase space. We now go on to illustrate a different construction for $\RR_{\,\rm sub}$ 
based on symmetrised sector functions, similarly to what was done in 
\secn{FrameNLO} at NLO.


\subsection{\vspace{-.5mm}$RR_{\,\rm sub}$ with symmetrised sector functions}
\label{RRsubSymSect}

\vspace{.5mm}
The partition of the $(\npt)$-particle phase space by means of the sector functions 
$\W{abcd}$ that we introduced in \secn{topo2} is not the only possible way forward. 
Analogously to what we did at NLO (see Eqs.~(\ref{eq:RsubZ2})
and (\ref{eq:Rsub no sectors})), this sector structure can be adapted to meet certain 
symmetry conditions that reduce the actual number of sectors: in particular, sectors 
sharing the same double-collinear singularities would naturally be parametrised in the 
same way in a numerical implementation, whence grouping such sectors in a single 
contribution is expected to improve numerical stability. Exploiting the symmetries of 
the improved limit $\bbC{ijk}$, we thus sum up the 6 permutations of $i,j,k$ in sectors 
$\W{ijjk}, \W{ijkj}$ introducing the {\it symmetrised sector functions}
\beq
\Z{ijk} 
& = &
\W{ijjk} + \W{ikkj} + \W{jiik} + \W{jkki} + \W{kiij} + \W{kjji} 
\nnb\\
&+& 
\W{ijkj} + \W{ikjk} + \W{jiki} + \W{jkik} + \W{kiji} + \W{kjij}
\, .
\eeq
Similarly, in the four-particle sectors $\W{ijkl}$, we can exploit the symmetries of the 
improved limit $\bbC{ijkl}$ to sum up the 8 permutations 
$ijkl$, $ijlk$, $jikl$, $jilk$, $klij$, $klji$, $lkij$, $lkji$, and define
\beq
\Z{ijkl}
\; = \;
\W{ijkl} + \W{ijlk} + \W{jikl} + \W{jilk} 
+ 
\W{klij} + \W{klji} + \W{lkij} + \W{lkji}
\, .
\eeq
We also introduce the NLO-type symmetric sector functions 
\beq
\label{NLOtypesym}
\Z{ij}^{(\alpha)}
\; \equiv \;
\W{ij}^{(\alpha)} + \W{ji}^{(\alpha)}
\, ,
\qquad
\Z{ij}
\, \equiv \,
\Z{ij}^{(1)}
\, ,
\eeq
where $\W{ij}^{(\alpha)}$ was defined in \eq{NLOtype}. We will also find it useful
to introduce a notation for the soft limit of the symmetric sector functions
\beq
\label{eq:Zsij}
\Z{{\rm s},ij}^{(\alpha)}
\, \equiv \,
\bS{i} \, \Z{ij}^{(\alpha)}
\, = \,
\bS{i} \, \W{ij}^{(\alpha)} 
\, = \, 
\frac{\frac{1}{w_{ij}^{\alpha}}}
     {\sum\limits_{l \neq i} \frac{1}{w_{il}^{\alpha}}} 
\, ,
\qquad
\Z{{\rm s},ij}
\, \equiv \,
\Z{{\rm s},ij}^{(1)}
\, .
\eeq
The use of $\Z{ijk}$ and $\Z{ijkl}$, upon reducing the number of sectors, simplifies 
the expression of the counterterms. In fact, deriving the action of the generic improved 
limit $\bbL{}{}$ on the new sector functions (which can be directly obtained from the 
$\bbL{}{} \, \W{abcd}$ definitions in \appn{app:K1,K2,K12}), we verify that, thanks 
to their symmetries, any improved limit involving either the operator $\bbC{ijk}$, 
or the operator $\bbC{ijkl}$, when acting on $\Z{ijk}$ and $\Z{ijkl}$ respectively, 
reduces them to unity, according to
\beq
\bbC{ijk} \big( \dots \big) \, RR \, \Z{ijk} \, = \, \bbC{ijk} \big( \dots \big) \, RR
\, ,
\qquad\quad
\bbC{ijkl} \big( \dots \big) \, RR \, \Z{ijkl} \, = \, \bbC{ijkl} \big( \dots \big) \, RR
\, ,
\eeq
where the ellipsis denotes a generic sequence of improved limits. 

In analogy with \eq{counttopo}, we now define our local counterterms with
symmetrised sector functions by
\begin{align}
\label{eq:symmetrised K}
K^{\one}_{\{\sigma\}}
& = \,
\bbL{\{\sigma\}}^{\one} \,
\RR \, \Z{\sigma}
\, ,
&
K^{\two}_{\{\sigma\}}
& = \,
\bbL{\{\sigma\}}^{\two} \,
\RR \, \Z{\sigma}
\, ,
&
K^{\otwo}_{\{\sigma\}}
& = \,
\bbL{\{\sigma\}}^{\otwo} \,
\RR \, \Z{\sigma}
\, ,
\end{align}
where we denote the symmetrised topologies by $\sigma \in \{ijk, ijkl\}$, and
the limits $\bbL{}{}_{\{\sigma\}}$ are symmetrised versions of the limits in
\eq{eq:unbarred L}, to be presented below. The subtracted double-real 
contribution for a given symmetrised sector, in analogy with \eq{eq:RRsub tau},
is then given by
\beq
\RR^{\,\rm sub}_{\{\sigma\}}(X) 
& \equiv &
\RR \; \Z{\sigma} \, \delta_\npt(X) 
- 
K^{\one}_{\{\sigma\}} \, \delta_\npo(X) 
- 
\left( K^{\two}_{\{\sigma\}} - K^{\otwo}_{\{\sigma\}} \right) \delta_n(X) 
\, ,
\eeq
and finally the full expression for $RR_{\,\rm sub}(X)$ of 
\eq{eq:sigma NNLO sub} is obtained by summing the contributions from the 
symmetrised sectors $\Z{ijk}$, $\Z{ijkl}$. It reads 
\beq
\label{eq:RRsub Z}
\RR_{\,\rm sub}(X) 
& = & 
\sum_{i, \, j > i} 
\bigg[
\sum_{k > j} \,
\RR^{\,\rm sub}_{\{ijk\}}(X)
+
\sum_{\substack{k \neq j\\k>i}} 
\sum_{\substack{l \neq i,j\\l>k}}
\RR^{\,\rm sub}_{\{ijkl\}}(X)
\bigg]
\, .
\eeq
This expression can be written in the form of \eq{eq:NNLO_RR_sub} by 
building the complete counterterms $K^{\one}$, $K^{\two}$ and 
$K^{\otwo}$ in terms of symmetrised sector functions, as
\beq
\label{eq:K1 K2 K12 Z}
K^{\one}
& = & 
\sum_{i, \, j > i} 
\bigg[
\sum_{k > j} \,
K^{\one}_{\{ijk\}}
+
\sum_{\substack{k \neq j\\k>i}} 
\sum_{\substack{l \neq i,j\\l>k}}
K^{\one}_{\{ijkl\}}
\bigg]
\, ,
\nnb\\
K^{\two}
& = & 
\sum_{i, \, j > i} 
\bigg[
\sum_{k > j} \,
K^{\two}_{\{ijk\}}
+
\sum_{\substack{k \neq j\\k>i}} 
\sum_{\substack{l \neq i,j\\l>k}}
K^{\two}_{\{ijkl\}}
\bigg]
\, ,
\nnb\\
K^{\otwo}
& = & 
\sum_{i, \, j > i} 
\bigg[
\sum_{k > j} \,
K^{\otwo}_{\{ijk\}}
+
\sum_{\substack{k \neq j\\k>i}} 
\sum_{\substack{l \neq i,j\\l>k}} \!
K^{\otwo}_{\{ijkl\}}
\bigg]
\, .
\qquad\qquad
\eeq
The symmetrised improved limits required to compute the symmetrised counterterms 
defined in \eq{eq:symmetrised K} can be derived from the limits designed for the $\W{abcd}$
sector functions, which were presented in \eq{eq:unbarred L} before improvement. The
symmetrisation must be done carefully, in order not to overcount singular configurations.
We adopt the following procedure. First, we expand all products in \eq{eq:unbarred L}, 
and we express the corresponding \textit{improved} limits as flat sums running over the 
respective sets of relevant singular limits. For example, we write
\beq
\bbL{ab}^{\one}
& = & \!\!
\sum_{\bar{\bf \ell} \, \in \, {\cal L}_{ab}^{\one}} \!\! \bar{\bf \ell}
\, ,
\qquad
\, \mbox{where}
\quad
{\cal L}_{ab}^{\one}
\, = \,
\Big\{ \bbS{a} , \, \bbC{ab}, \, - \, \bbS{a} \, \bbC{ab} \Big\}
\, ,
\nnb\\
\bbL{abbc}^{\two}
& = & \!\!
\sum_{\bar{\bf \ell} \, \in \, {\cal L}_{abbc}^{\two}} \!\! \bar{\bf \ell}
\, ,
\qquad
\mbox{where}
\quad
{\cal L}_{abbc}^{\two}
\, = \,
\Big\{
\bbS{ab} , \,
\bbSC{abc} , \,
- \, \bbSC{abc} \, \bbS{ab} , \,
\bbC{abc} , \,
- \, \bbS{ab} \, \bbC{abc} ,
\nnb\\ [-3mm]
&&
\hspace{5cm}
- \, \bbSC{abc} \, \bbC{abc} , \,
\bbSC{abc} \, \bbS{ab} \, \bbC{abc}
\Big\}
\, ,
\eeq
and similarly for the remaining limits given in \eq{eq:unbarred L}. Next, we introduce
the index sets
\beq
\alpha
& = &
\{ij, ji, ik, ki, jk, kj\}
\, ,
\hspace{27.2mm}
\beta
\, = \,
\{ij, ji, kl, lk\}
\, ,
\nnb\\
\gamma_1
& = &
\{ijjk, ikkj, jkki, jiik, kiij, kjji\}
\, ,
\qquad
\gamma_2
\, = \,
\{ijkj, ikjk, jkik, jiki, kiji, kjij\}
\, ,
\nnb\\
\delta
& = &
\{ijkl, ijlk, jikl, jilk, klij, klji, lkij, lkji\}
\, ,
\eeq
which enumerate the permutations that will need to be summed in order to perform
the required symmetrisations. The limits $\bbL{\{\sigma\}}^{\one}$, $\bbL{\{\sigma\}}^{\two}$
and $\bbL{\{\sigma\}}^{\otwo}$ can now be defined by sums running over unions of 
the sets ${\cal L}$. Specifically, we define
\beq
\label{defsymbarlim}
\bbL{\{ijk\}}^{\one}
& = &
\sum_{\bar{\bf \ell} \, \in \, {\cal L}_{\alpha}^{\one}} \!\! \bar{\bf \ell}
\, ,
\qquad
\mbox{where}
\quad
{\cal L}_{\alpha}^{\one}
\, = \,
\bigcup_{ab \, \in \, \alpha}
{\cal L}_{ab}^{\one}
\, ,
\nnb\\
\bbL{\{ijkl\}}^{\one}
& = &
\sum_{\bar{\bf \ell} \, \in \, {\cal L}_{\beta}^{\one}} \!\! \bar{\bf \ell}
\, ,
\qquad
\mbox{where}
\quad
{\cal L}_{\beta}^{\one}
\, = \,
\bigcup_{ab \, \in \, \beta}
{\cal L}_{ab}^{\one}
\, ,
\nnb\\
\bbL{\{ijk\}}^{\two}
& = &
\sum_{\bar{\bf \ell} \, \in \, {\cal L}_{\gamma}^{\two}} \!\! \bar{\bf \ell}
\, ,
\qquad
\mbox{where}
\quad
{\cal L}_{\gamma}^{\two}
\, = \,
\left[
\bigcup_{abbc \, \in \, \gamma_1} \!\!\!
{\cal L}_{abbc}^{\two}
\right]
\! \cup \!
\left[
\bigcup_{abcb \, \in \, \gamma_2} \!\!\!
{\cal L}_{abcb}^{\two}
\right]
\, ,
\nnb\\
\bbL{\{ijkl\}}^{\two}
& = &
\sum_{\bar{\bf \ell} \, \in \, {\cal L}_{\delta}^{\two}} \!\! \bar{\bf \ell}
\, ,
\qquad
\mbox{where}
\quad
{\cal L}_{\delta}^{\two}
\, = \,
\bigcup_{abcd \, \in \, \delta}
{\cal L}_{abcd}^{\two}
\, .
\eeq
Similarly, the strongly-ordered double-unresolved limits $\bbL{\{\sigma\}}^{\otwo}$ are 
given by analogous sums, where for $\sigma = ijk$ the sum runs over the collection 
${\cal L}_{\gamma}^{\otwo}$, and, for $\sigma = ijkl$, the sum runs over the collection 
${\cal L}_{\delta}^{\otwo}$, defined as in the last two lines of \eq{defsymbarlim}, with 
the replacement $\two \to \otwo$. While assembling the set unions introduced in 
\eq{defsymbarlim}, one must take care to count only once all limits that coincide 
by symmetry: thus, for example, one should use the fact that $\bbC{ij} = \bbC{ji}$,
and $\bbSC{ijk} = \bbSC{ikj}$. To further illustrate the procedure, we note that
the first line of \eq{defsymbarlim} becomes
\beq
\label{expll1ijksym}
\bbL{\{ijk\}}^{\one} & = & \bbS{i} + \bbS{j} + \bbS{k} + \bbC{ij} + \bbC{ik} + \bbC{jk} 
\nnb \\
&& \,
- \, \bbS{i} \, \bbC{ij} - \bbS{j} \, \bbC{ij} - \bbS{i} \, \bbC{ik} - \bbS{k} \, \bbC{ik} 
- \bbS{j} \, \bbC{jk} - \bbS{k} \, \bbC{jk} \nnb \\
& = & 
\bbS{i} + \bbS{j}  + \bbS{k} + \bbHC{ij} + \bbHC{ik} + \bbHC{jk}
\, ,
\eeq
properly including all relevant singular regions without double counting.

The explicit results for the sums in \eq{defsymbarlim} appear rather cumbersome at 
first sight, but in fact they result in relatively compact expressions when the limits 
are evaluated. Indeed, thanks to the symmetry properties of $\Z{ijk} $ and $\Z{ijkl}$, 
it is possible to merge subsets of singular limits which factor identical combinations
of symmetrised sector functions. One finds then that only certain combinations of 
singular limits survive in the result. In detail, all single-unresolved limits can be written
explicitly as sums of single-soft limits $\bbS{a}$ plus hard-collinear combinations 
$\bbHC{ab}$, defined in \eq{eq:RsubZ2}. Furthermore, it is useful to introduce a 
soft-subtracted version of the uniform double-unresolved limit $\bbSC{abc}$, which 
is given by
\beq
\bbSHC{abc} 
& \equiv &
\bbSC{abc}
\left(1 - \bbS{ab}  - \bbS{ac} \right) 
\, .
\eeq
This combination can appear only when attached to either the $\bbS{a}$ or 
$\bbC{abc}$ limits: indeed, in any other case, the operators $\bbSC{abc}$ and 
$\bbS{ab} \, \bbSC{abc}$ do not share the same sector functions in the limit.  
Similarly, considering the double-unresolved improved collinear limit $\bbC{abc}$, 
we can distinguish three useful combinations, defined by
\beq
\label{hc3comb}
\bbHC{abc} 
& \equiv &
\bbC{abc} 
\left(1 - \bbS{ab} - \bbS{bc} - \bbS{ac} \right) 
\, ,
\nnb
\\
\bbHC{abc}^{\;(\bf s)}
& \equiv &
\bbC{abc}
\left(1 - \bbS{ab}  - \bbS{ac} \right) 
\left( 1 -  \bbSC{abc} \right)
\, ,
\nnb
\\
\bbHC{abc}^{\;(\bf c)} 
& \equiv &
\bbC{abc}
\left(1 - \bbS{ab}  -  \bbSC{cab} \right)
\, ,
\eeq
which reflect three different possible strategies for removing soft singularities 
from the collinear kernel. The superscripts $(\bf s)$ and $(\bf c)$ in the second 
and third line of \eq{hc3comb} refer to the fact that the $(\bf s)$ combination can 
appear only in association with a single-soft limit $\bbS{d}$ (with $d \in \{a,b,c\}$),
while the $(\bf c)$ combination can appear only in association with single
hard-collinear limits $\bbHC{de}$, with $de \in \{ab,ac,bc\}$. Finally, for 
the four-particle double-collinear improved limit $\bbC{ijkl}$ we introduce
\beq
\label{hc4comb}
\bbHC{abcd} 
& \equiv &
\bbC{abcd}
\left(1 + \bbS{ac} + \bbS{bc} + \bbS{ad} + \bbS{bd} - \bbSC{acd} - 
\bbSC{bcd} - \bbSC{cab} - \bbSC{dab} \right) 
\, ,
\nnb
\\
\bbHC{abcd}^{\;(\bf c)} 
& \equiv &
\bbC{abcd}
\left(1 - \bbSC{cab}  -  \bbSC{dab} \right)
\, ,
\eeq
where again the notation $(\bf c)$ refers to the fact that the combined limit in the 
second line of \eq{hc4comb} can only appear in association with the hard-collinear
single-unresolved limits $\bbHC{ab}$ and $\bbHC{cd}$.

Using these preliminary definitions, we can write down explicit expressions for the
symmetrised improved limits defined in \eq{defsymbarlim}. They are
\beq
\label{eq:symmetrised L}
\bbL{\{ijk\}}^{\one}
& = &
\bbS{i} + \bbS{j}  + \bbS{k}
+
\bbHC{ij} + \bbHC{jk} + \bbHC{ik}
\, ,
\nnb
\\
\bbL{\{ijkl\}}^{\one}
& = &
\bbS{i} + \bbS{j} + \bbS{k} + \bbS{l}
+
\bbHC{ij} 
+ 
\bbHC{kl}
\, ,
\nnb
\\
\bbL{\{ijk\}}^{\two}
& = &
\bbS{ij} + \bbS{jk} + \bbS{ik}
+
\bbSC{ijk} ( 1 \!-\! \bbS{ij} \!-\! \bbS{ik} )
+
\bbSC{jik} ( 1 \!-\! \bbS{ij} \!-\! \bbS{jk} )
+
\bbSC{kij} ( 1 \!-\! \bbS{ik} \!-\! \bbS{jk} )
\nnb
\\
&&
+ \,
\bbHC{ijk}
-
\bbC{ijk} (\, \bbSHC{ijk} \!+\! \bbSHC{jik} \!+\! \bbSHC{kij} \,)
\, ,
\nnb
\\
\bbL{\{ijkl\}}^{\two}
& = &
\bbS{ik} + \bbS{jk} + \bbS{il} + \bbS{jl}
+
\bbSC{ikl} \left( 1 \!-\! \bbS{ik} \!-\! \bbS{il} \right) 
\!+
\bbSC{jkl} \left( 1 \!-\! \bbS{jk} \!-\! \bbS{jl} \right) 
\nnb\\
&&
+ \,
\bbSC{kij} \left( 1 \!-\! \bbS{ik} \!-\! \bbS{jk} \right) 
\!+
\bbSC{lij} \left( 1 \!-\! \bbS{il} \!-\! \bbS{jl} \right)
+
\bbHC{ijkl}
\, ,
\nnb
\\
\bbL{\{ijk\}}^{\otwo}
& = &
\bbS{i}
\left( \bbS{ij} \!+\! \bbS{ik} \!+\! \bbSHC{ijk} \right)
+
\bbS{j}
\left( \bbS{ij} \!+\! \bbS{jk} \!+\! \bbSHC{jik} \right)
+
\bbS{k}
\left( \bbS{ik} \!+\! \bbS{jk} \!+\! \bbSHC{kij} \right)
\nnb\\[1mm]
&&
+ \,
\left( 
\bbS{i} \!+\! \bbS{j}  \!+\! \bbS{k}
\right)
\bbHC{ijk}^{\;(\bf s)} 
+
\bbHC{ij}
\left( \bbS{ij} \!+\! \bbSC{kij} \!+\! 
\bbHC{ijk}^{\;(\bf c)}
\right) 
\nnb\\[1mm]
&&
+ \,
\bbHC{jk}
\left( \bbS{jk} \!+\! \bbSC{ijk} \!+\! 
\bbHC{ijk}^{\;(\bf c)} 
\right) 
+
\bbHC{ik}
\left( \bbS{ik} \!+\! \bbSC{jik} \!+\! 
\bbHC{ijk}^{\;(\bf c)}
\right) 
\, ;
\nnb
\\
\bbL{\{ijkl\}}^{\otwo}
& = &
\bbS{i}
\left( \bbS{ik} + \bbS{il} \right)
+
\bbS{j}
\left( \bbS{jk} + \bbS{jl} \right)
+
\bbS{k}
\left( \bbS{ik} + \bbS{jk} \right)
+
\bbS{l}
\left( \bbS{il} + \bbS{jl} \right)
\nnb\\[1mm]
&&
+ \,
\bbS{i} \,
\bbSHC{ikl}
+
\bbS{j} \,
\bbSHC{jkl}
+
\bbS{k} \,
\bbSHC{kij}
+
\bbS{l} \,
\bbSHC{lij}
\nnb\\[1mm]
&&
+ \,
\bbHC{ij}
\left( \bbSC{kij} + \bbSC{lij} \right) 
+
\bbHC{kl}
\left( \bbSC{ikl} + \bbSC{jkl} \right) 
+
\left( \bbHC{ij} + \bbHC{kl} \right) \bbHC{ijkl}^{\;(\bf c)}
\, .
\eeq
The actions of all these improved limits on $\RR$ and on the symmetrised sector 
functions $\Z{ijk}$, $\Z{ijkl}$ are reported in \appn{app:K1,K2,K12}.

Comparing the collections of singular projectors relevant to $\W{abcd}$ sector 
functions in \eq{eq:unbarred L} with the ones reported in \eq{eq:symmetrised L} 
for the symmetrised case, it is immediate to notice that the number of different 
non-trivial singular limits contributing to a given sector changes, depending on 
the type of partition we introduce. In particular, this number increases for our 
choice of $\Z{ijk}$ and $\Z{ijkl}$. Despite this, though, the ordered sums in \eq{eq:RRsub Z},
building up the relevant integrable contributions, lead to a significantly more 
compact final expression (in terms of the number of different objects one needs 
to define and evaluate). This is a feature that will translate into a gain in computational 
time and resources in the final numerical implementation.


\section{Integration of the double-real-radiation counterterms}
\label{RRint}

In the previous section we constructed $RR_{\,\rm sub}$ of \eq{eq:NNLO_RR_sub}, a 
combination which is integrable everywhere in the double-radiative phase space, by 
subtracting the local counterterms $K^{\one}$, $K^{\two}$ and $K^{\otwo}$ (given in 
\eq{eq:K1 K2 K12 W}, or equivalently in \eq{eq:K1 K2 K12 Z}) from the double-real 
squared matrix element $RR$. These counterterms must now be added back, after 
integrating out one or two emissions, yielding the integrated counterterms $I^{\one}$, 
$I^{\two}$, $I^{\otwo}$. The integration procedure in the presence of sectors involves 
rather intricate combinatorics, and generates lengthy expressions in the intermediate 
stages. However, all integrals that need to be computed are remarkably simple, and 
in almost all cases have trivial (logarithmic) dependence on the Mandelstam 
invariants~\cite{Magnea:2020trj}.

We will begin, in \secn{phaspapa}, by introducing the relevant phase-space factorisations 
and parameterisations, derived from the nested Catani-Seymour mappings introduced in 
\secn{phaspamap2}. Then, in \secn{InteSec}, we will report the integration of the counterterms 
$K^{\one}$, $K^{\two}$, $K^{\otwo}$, specifying how each singular contribution is treated. 
The resulting expressions can be simplified, by relabelling the momenta and rewriting 
the flavour sums of the original $(n+2)$-body phase space, as explained in \secn{FlaSum}. 
It is then possible to recombine the contributions carrying different mappings, resulting in 
relatively compact collections of integrals for $I^{\one}$, $I^{\two}$, $I^{\otwo}$, presented 
in \secn{AsseCou}. At this stage, the results can be directly employed in the subtraction 
formula, \eq{eq:sigma NNLO sub}.

It is natural to define $I^{\one}$ as the integral of $K^{\one}$ in the single-unresolved 
radiation, and $I^{\two}$ as the integral of $K^{\two}$ in both unresolved 
emissions. For the strongly-ordered counterterm $K^{\otwo}$ both possibilities are 
in principle viable. In our framework, we define $I^{\otwo}$ as the integral
of $K^{\otwo}$ in a single radiation\footnote{We note that in the context of 
the {\tt CoLoRFul} approach to subtraction~\cite{DelDuca:2016csb,DelDuca:2016ily}, 
the strongly-ordered counterterm is integrated directly in both unresolved radiations.}, 
corresponding to the `most unresolved' radiated particle, as explicitly noted 
in \eq{intcountNNLO}. As a consequence, before performing the integrations, 
we rewrite both $K^{\one}$ and $K^{\otwo}$ by summing up the sector functions 
related to the most unresolved radiation (the ones carrying the suffix $\alpha$), 
while keeping the sector functions for the second  (least-unresolved) radiation 
untouched. Note however that these remaining sector functions carry mapped
kinematics. In this way, it will be possible to combine directly the integrated 
counterterms $I^{\one}$ and $I^{\otwo}$ with the real-virtual contribution $RV$,
and with the real-virtual counterterm $K^{\RV}$, in \eq{eq:NNLO_RV_sub}, sector 
by sector in $\Phi_{\npo}$. For the sake of simplicity, in the following all integrations 
are described using the expressions for $K^{\one}$, $K^{\two}$ and $K^{\otwo}$ in 
terms of symmetrised sector functions, as given in \eq{eq:K1 K2 K12 Z}, but the 
resulting expressions for $I^{\one}$, $I^{\two}$ and $I^{\otwo}$ will be given also 
in terms of the $\W{}$ sector functions.


\subsection{Phase-space parametrisations}
\label{phaspapa}

We start by giving precise definitions for the measures of integration in the radiative
phase spaces $d\Phi_{\rm rad}$ and $d\Phi_{\rm rad, 2}$, according to \eq{phaspafac},
but now highlighting the dependence on the chosen mappings (discussed in 
\secn{phaspamap2}), and making specific choices of integration variables.

The single-unresolved counterterm $K^{\one}$ contains just single mappings 
of the type $\kkl{acd}$ ($a,c,d$ all different) and is going to be 
integrated in the corresponding single-radiation phase space. 
On the contrary, $K^{\otwo}$ and $K^{\two}$ are built by means of iterated 
mappings of the type $\kkl{acd,bef}$ ($a,c,d$ all different and $b,e,f$ all different). 
However, while $K^{\otwo}$ needs to be integrated just in the phase space of 
the single radiation corresponding to the first mapping, $K^{\two}$ must 
be integrated in the whole double-radiation phase space.

We start specifying \eq{phaspafac2}, needed for the integration of $K^{\one}$ 
and $K^{\otwo}$. We write
\beq
\label{eq:dPhi rad (acd)}
&&
\int d\Phi_\npt (\{k\})
\; = \;
\frac{\varsi_{n+2}}{\varsi_{n+1}}
\int d\Phi_\npo^{(acd)}
\int d \Phi_{\rm rad}^{(acd)} 
\, ,
\eeq
where we defined
\beq
\label{eq:dPhi rad (acd) 2}
d\Phi_\npo^{(acd)}
\; \equiv \;
d\Phi_\npo(\kkl{acd}) 
\, .
\eeq
The explicit expression for the radiative measure is
\beq
\label{eq:dPhi rad (acd) 3}
\int d \Phi_{\rm rad}^{(acd)} 
\, = \,
N (\eps)  \left(\sk{cd}{acd}\right)^{\!1 - \eps}
\int_0^\pi \!\!\! d\phi \, ( \sin \! \phi )^{- 2 \eps} 
\int_0^1 \!\! dy \, \int_0^1 \!\! dz \,
\Big[ y (1 - y)^2 z (1 - z) \Big]^{- \eps} (1 - y) 
\, ,
\qquad
\eeq
where 
\beq
\quad N(\eps) 
\, \equiv \, 
\frac{(4\pi)^{\eps - 2}}{\sqrt\pi\,\Gam\!\left(\frac12-\eps\right)}  
\, .
\eeq
The invariants composed by the momenta $k_{a}$, $k_{c}$, $k_{d}$ are related 
to the integration variables $y$ and $z$ by
\beq
s_{ac} \, = \, y \, \sk{cd}{acd}
\, , 
\qquad\qquad
s_{ad} \, = \, z (1 - y) \, \sk{cd}{acd}
\, , 
\qquad\qquad
s_{cd} \, = \, (1 - z)(1 - y) \, \sk{cd}{acd}
\, ,
\eeq 
so that $s_{acd} \, = \, s_{ac} + s_{ad} + s_{cd} = \sk{cd}{acd}$.

To parametrise the double-radiative phase space,  needed for $K^{\two}$, we employ 
double mappings of three different types, as discussed in \secn{phaspamap2}. 
We examine them in turn.

The six-particle mapping $\kkl{acd,bef}$ ($a,b,c,d,e,f$ all different) presented in 
Eqs.~(\ref{sixpart1}) and (\ref{sixpart2}) induces the factorisation 
\beq
\int d \Phi_{\npt}(\{k\})
& = &
\frac{\varsi_{n+2}}{\varsi_n} 
\int d\Phi_n^{(acd,bef)}
\int d \Phi_{\rm rad, 2}^{\,(acd,bef)} 
\, ,
\quad\quad
d\Phi_n^{(acd,bef)}
\; \equiv \;
d \Phi_n(\kkl{acd,bef})  
\, ,
\eeq
and the radiative measure of integration is
\beq
\int d \Phi_{\rm rad, 2}^{\,(acd,bef)} 
& = &
N^2 (\eps) \, 
\left(
\sk{cd}{acd,bef} \, \sk{ef}{acd,bef}
\right)^{\!1 - \eps}
\int_0^\pi \!\! d \phi' \, (\sin\phi')^{- 2 \eps}
\int_0^1 \!\! d y'
\int_0^1 \!\! d z'
\int_0^\pi \!\! d \phi \, (\sin\phi)^{- 2 \eps} \!
\nnb\\
&& 
\times\, 
\int_0^1 \!\! d y 
\int_0^1 \!\! d z \,
\Big[ y'(1-y')^2\,z'(1-z')\,y(1-y)^2\,z(1-z) \Big]^{- \eps} 
(1-y')(1-y) 
\, ,
\qquad
\eeq
where the expressions for relevant invariants in terms of the integration variables are
\beq
s_{ac} & = & y' \, \sk{cd}{acd,bef}
\, , 
\qquad
s_{ad} \; = \; z' \, ( 1 - y' ) \, \sk{cd}{acd,bef}
\, , 
\qquad
s_{cd}  \; = \; ( 1 - z' ) ( 1  - y' ) \, \sk{cd}{acd,bef}
\, ,
\nnb \\[2mm]
s_{be} & = & y \, \sk{ef}{acd,bef}
\, , 
\qquad\;
s_{bf} \; = \;  z \, ( 1 - y ) \, \sk{ef}{acd,bef}
\, , 
\qquad\;\;
s_{ef}  \; = \; ( 1  - z ) ( 1 - y ) \, \sk{ef}{acd,bef}
\, ,
\qquad
\eeq
so that $s_{acd} \, = \, s_{ac} + s_{ad} + s_{cd} = \sk{cd}{acd,bef} =  \sk{cd}{acd}$, and $s_{bef} 
\, = \, s_{be} + s_{bf} + s_{ef} = \sk{ef}{acd,bef} =  \sk{ef}{bef}$.

The five-particle mapping $\kkl{acd,bed}$ ($a,b,c,d,e$ all different) presented in 
Eqs.~(\ref{fivepart1}) and (\ref{fivepart2}) induces the factorisation 
\beq
\int d \Phi_{\npt}(\{k\})
& = & 
\frac{\varsi_{n+2}}{\varsi_n} 
\int d\Phi_n^{(acd,bed)} 
\int d \Phi_{\rm rad, 2}^{\,(acd,bed)} 
\, ,
\quad\quad
d\Phi_n^{(acd,bed)}
\; \equiv \;
d \Phi_n(\kkl{acd,bed})  
\, ,
\eeq
and we write
\beq
\int d \Phi_{\rm rad, 2}^{\,(acd,bed)} 
& = & 
N^2 (\eps) \, 
\left(
\sk{cd}{acd,bed} \, \sk{ed}{acd,bed}
\right)^{\!1 - \eps}
\int_0^\pi \!\! d \phi' \, (\sin\phi')^{- 2 \eps} \!
\int_0^1 \!\! d y'
\int_0^1 \!\! d z'
\int_0^\pi \!\! d \phi \, (\sin\phi)^{- 2 \eps} \!
\int_0^1 \!\! d y
\nnb\\
& & 
\quad
\times\, 
\int_0^1 \!\!\! d z \, 
\Big[ y'(1-y')^2\,z'(1-z')\,y(1-y)^3\,z(1-z) \Big]^{- \eps} 
(1-y')(1-y)^2 
\, ,
\eeq
with 
\beq
s_{ac} & = & y' \, ( 1 - y ) \, \sk{cd}{acd,bed}
\,, 
\hspace{30.5mm}
s_{ad} \; = \; z' \, ( 1 - y' ) ( 1 - y ) \, \sk{cd}{acd,bed}
\, , 
\nnb \\[2mm]
s_{be} & = & y \, \sk{ed}{acd,bed}
\, , 
\hspace{42.4mm}
s_{bd} \; = \;  ( 1 - y' ) \, z \, ( 1 - y ) \, \sk{ed}{acd,bed}
\, , 
\nnb \\
\qquad
s_{cd}  & = & ( 1 - y' ) ( 1  - z' ) ( 1 - y ) \, \sk{cd}{acd,bed}
\,,
\hspace{10.5mm}
s_{ed}  \; = \; ( 1 - y' ) ( 1  - z ) ( 1 - y ) \, \sk{ed}{acd,bed} 
\, ,
\qquad
\qquad 
\eeq
so that the five-parton invariant $s_{abcde}=s_{ab}+s_{ac}+s_{ad}+s_{ae}+s_{bc}+
s_{bd}+s_{be}+s_{cd}+s_{ce}+s_{de}$ is equal to $\sk{cde}{acd,bed} = \sk{cd}{acd,bed}+
\sk{ce}{acd,bed}+\sk{de}{acd,bed}$.

Finally, we have the four-particle mapping, $\kkl{acd,bcd} = \kkl{abcd}$, ($a,b,c,d$ all 
different), presented in Eqs.~(\ref{fourpart1}) and (\ref{fourpart2}). This is the most intricate 
mapping, inducing the factorisation 
\beq
\int d \Phi_{\npt}(\{k\})
& = &
\frac{\varsi_{n+2}}{\varsi_n} 
\int d\Phi_n^{(abcd)} 
\int d \Phi_{\rm rad, 2}^{\,(abcd)} 
\, ,
\qquad\qquad
d\Phi_n^{(abcd)}
\; \equiv \;
d \Phi_n(\kkl{abcd})  
\, ,
\eeq
where we write
\beq
\int d \Phi_{\rm rad, 2}^{\,(abcd)} 
& = & 
2^{- 2 \eps} \,
N^2 (\eps) \, 
\left(\sk{cd}{abcd}\right)^{\!2 - 2\eps}
\!\!
\int_0^1 \!\! d w'
\int_0^1 \!\! d y'
\int_0^1 \!\! d z'
\int_0^\pi \!\! d \phi \, (\sin\phi)^{- 2 \eps} \!
\int_0^1 \!\! d y
\int_0^1 \!\! d z
\nnb\\
&& 
\times\,
\Big[ w'(1\!-\!w')\Big]^{-1/2-\eps}  
\Big[ y'(1\!-\!y')^2\,z'(1\!-\!z')\,y^2(1\!-\!y)^2\,z(1\!-\!z) \Big]^{- \eps} 
(1\!-\!y') \, y \, (1\!-\!y) 
\, ,
\nnb
\eeq
with 
\beq
s_{ab} & = & y' \, y \, \sk{cd}{abcd}
\, , 
\qquad\qquad
s_{ac} \; = \; z' ( 1 - y' ) \, y \, \sk{cd}{abcd}
\, , 
\qquad\qquad
s_{bc}  \; = \; ( 1 - y' ) ( 1  - z' ) \, y \, \sk{cd}{abcd}
\, ,
\nnb \\[2mm]
s_{cd}  & = & ( 1 - y' ) ( 1 - y ) ( 1 - z ) \, \sk{cd}{abcd}
\, ,
\nnb \\
s_{ad} 
& = & 
(1-y) 
\left[\, y'(1-z')(1-z) + z'z - 2(1-2w')\sqrt{y'z'(1-z')z(1-z)} \,\right] 
\sk{cd}{abcd}
\, ,
\nnb \\
s_{bd} 
& = & 
(1-y) 
\left[\, y'z'(1-z) + (1-z')z + 2(1-2w')\sqrt{y'z'(1-z')z(1-z)} \,\right] 
\sk{cd}{abcd}
\, ,
\label{invaproco}
\eeq
so that $s_{abcd} \, = \, s_{ab} + s_{ac} + s_{ad} + s_{bc} + s_{bd} + s_{cd} = \sk{cd}{abcd}$.


\subsection{Integration of $K^{\one}$, $K^{\two}$ and $K^{\otwo}$}
\label{InteSec}

We now have all the ingredients to actually perform the required integrations. 
Our task is simplified by the fact that the integrals 
of the azimuthal parts of the collinear kernels (see \eqref{eq:appsinglecoll}) vanish, 
as shown in \appn{app:int Q}. All remaining 
integrals are then computed following the techniques explained in~\cite{Magnea:2020trj}.
We will later recombine the components that were differently 
mapped by relabelling momenta, in order to compose the complete results, which 
will be considerably simpler.

For the single-unresolved counterterm $K^{\one}$ the required integral is 
\beq
\label{intK1}
\int d\Phi_\npt \, K^{\one}
& = &
\int d\Phi_\npt \,
\bigg\{ \;
\sum_{i, j \neq i} \sum_{\substack{k \ne i\\k > j}} \;
\bbS{i} \, RR \, \bZ{jk}
+
\sum_{i, \, j > i} 
\sum_{k \ne i} \,
\sum_{\substack{l \ne i\\l > k}} \;
\bbHC{ij} \, \RR \, \bZ{kl}
\bigg\}
\, .
\qquad 
\eeq
The integrand on the right-hand side has been obtained from $K^{\one}$ of 
\eq{eq:K1 K2 K12 Z} by summing up the NLO sector functions with label 
$\alpha$ of Eqs.~(\ref{eq:bL1 Zijk})-(\ref{eq:bL1 Zijkl}).
As explained in \appn{app:K1,K2,K12}, the mapped sector functions $\bZ{ij}$ are 
understood to carry the same mapping as the matrix elements they multiply. 
Since \eq{intK1} will have to be combined
with the real-virtual contribution $RV$, as part of \eq{eq:NNLO_RV_sub},
we need to express the integral in \eq{intK1} as a sum of terms in which the integration
over the single-particle radiative phase space has been performed, a specific 
parametrisation for the $(\npo)$-particle phase space has been identified, and
the full single-real-radiation squared matrix element $R$ has been factored, and
computed in the chosen variables. The results for the summands of the two terms in
\eq{intK1} take the form
\beq
\int d\Phi_\npt \,
\bbS{i} \, RR \, \bZ{jk}
& = &
- \,
\frac{\varsi_{n+2}}{\varsi_{n+1}}
\sum_{c \neq i}
\sum_{d \neq i,c}
\int \! d\Phi_{\npo}^{(icd)} \, 
J_s^{icd} \,
\bR^{(icd)}_{cd} \,
\bZ{jk}^{(icd)}
\, ,
\\
\label{eq:int HCij}
\int d\Phi_\npt \,
\bbHC{ij} \, RR \, \bZ{kl}
& = &
\frac{\varsi_{n+2}}{\varsi_{n+1}}
\int \! d\Phi_{\npo}^{(ijr)} \,
J_{\rm hc}^{ijr} \,
\bR^{(ijr)} \,
\bZ{kl}^{(ijr)}
\, ,
\qquad
r = r_{ijkl}
\, ,
\eeq
where the measure of integration $d\Phi_{\npo}^{(acd)}$ was defined in \eq{eq:dPhi rad (acd) 2}. 
The integration over the appropriate $d \Phi_{\rm rad}$ has been performed, yielding 
the integrals $J_s^{icd}$ and $J_{\rm hc}^{ijr}$, whose explicit expressions are 
given in \eq{eq:Js^ilm} and in \eq{Jhc^ijr}, respectively. The choice of $r=r_{ijkl} \ne i,j,k,l$, 
according to the rule of \eq{eq: r_ij},  which reflects the choice made for $\bbHC{ij} \, \RR$ in
\eq{eq:bHCij}, causes a dependence of the integrated kernel $J_{\rm hc}^{ijr}$ on the indices 
$k$ and $l$ of the sector function $\bZ{kl}^{(ijr)}$. Notice that the choice $r = r_{ijkl}$ implies 
the need for at least five massless partons in $\Phi_{\npt}$, namely three massless final-state 
partons at Born level. A solution for the case of two massless final-state partons in the Born 
phase space requires minor technical modifications, which have been developed, and will 
be presented elsewhere.

We now turn to the integration of $K^{\two}$, which is the most involved 
part of the calculation. 
In this case, since $I^{\two}$ enters \eq{eq:NNLO_VV_sub}, 
which lives in $\Phi_n$, we start from $K^{\two}$ in \eq{eq:K1 K2 K12 Z}
and perform the complete sum over sector functions, exploiting their sum rules (see for example Eqs.~\eqref{eq:sum_rules_NNLO1}-\eqref{eq:sum_rules_NNLO3}).
This gives 
\beq
\label{intK2}
\int d\Phi_\npt \, K^{\two}
& = &
\int d\Phi_\npt \,
\bigg[ \;
\sum_{i,j > i} \;
\bbS{ij} \, RR
+
\sum_{i, j \neq i}\sum_{\substack{k \ne i\\k > j}} \;
 \bbSHC{ijk} \left(1 - \bbC{ijk} \right) \RR
\nnb\\[-2mm]
&&
\hspace{14mm}
+ \,
\sum_{i, \, j > i} \sum_{k > j} \;
\bbHC{ijk} \RR
+
\sum_{i, \, j > i} \sum_{\substack{k \ne j \\ k > i}}
\sum_{\substack{l \neq j \\ l > k}} \;
\bbHC{ijkl} \, \RR
\bigg]
\, .
\eeq
Each of the four terms in \eq{intK2} must be written as a sum of contributions, where
the double-radiation kernels have been integrated over the parametrised radiative 
phase space, and one is left with a Born-level factor, expressed in terms of 
mapped momenta. To guide the eye of the reader through the following rather 
intricate expressions, we note that, for each one of the limits involved,
the results are of the form
\beq
\label{genintK2}
\int d\Phi_\npt \, \bbL{\, \cdots}^{\two} \, RR 
& = & 
{\rm constant} \, \sum_{\{\mu\}} \, 
\int d \Phi_n^{(\mu)} \,\,
J_{{\rm limit}}^{\mu} \,\, \bB_{{\rm colour}}^{(\mu)} \, , 
\eeq
where the overall constant is related to multiplicities, the sum is over the set $\{\mu\}$ 
of mappings that have been employed, the Born factor may have different colour correlations,
and $J$ will always denote the results of the integration of the kernels appropriate
to the limit being taken\footnote{Note that, since the limit $\bbL{}$ is expressed as a sum
of terms that can be mapped differently, several $J$'s will contribute to each $\bbL{}$.}. 
The relevant $J$'s will be listed in \appn{app:master integrals}. 
Beginning with the integrated double-soft limit in \eq{intK2}, we 
find the explicit expression
\beq
\int d\Phi_\npt \,
\bbS{ij} \, RR 
& = &
\frac{1}{2} \,
\frac{\varsi_{n+2}}{\varsi_{n}} \!
\sum_{\substack{c \neq i,j \\ d \neq i,j,c}} \!
\Bigg\{
\sum_{\substack{e \neq i,j,c,d}}
\bigg[ \;
\sum_{\substack{f \neq i,j,c,d,e}}
\int\!d \Phi_n^{(icd,jef)}  \,
J_{{\rm s} \otimes {\rm s}}^{ijcdef} \,
\bB_{cdef}^{(icd,jef)}
\nnb\\[-3mm]
&&
\hspace{34mm}
+ \,
4 \,
\int\!d \Phi_n^{(icd,jed)}  \,
J_{\rm s \otimes s}^{ijcde} \,
\bB_{cded}^{(icd,jed)}
\bigg]
\nnb\\
&&
\hspace{21mm}
+ \,
\int\!d \Phi_n^{(ijcd)}  \,
\bigg[
2 \,
J_{\rm s \otimes s}^{ijcd} \,
\bB_{cdcd}^{(ijcd)}
+
J_{\rm ss}^{ijcd} \,
\bB_{cd}^{(ijcd)}
\bigg]
\Bigg\}
\, ,
\eeq
where we collected colour correlations involving four, three and two partons, and
each term has been mapped differently, to simplify the corresponding integration.
The integrals relevant for double-soft radiation are presented in \eq{doublesoftint}.
We now turn to the second term in \eq{intK2}, and we find (with $r=r_{ijk}$) 
\beq
&&
\hspace{-3mm} 
\int \! d\Phi_{\!\npt} \,
\bbSHC{ijk} \left(1 - \bbC{ijk} \right) \RR
=
\nnb\\
&&
- \,
\frac{\varsi_{n+2}}{\varsi_{n}} \,
\Bigg\{
\sum_{\substack{c\neq i,j,k,r \\ d\neq i,j,k,r,c}} \!
\int \! d\Phi_n^{(jkr,icd)} \,
J_{\rm s \otimes hc}^{jkricd} 
\bB^{(jkr,icd)}_{cd}
+
2 \!\!
\sum_{\substack{c\neq i,j,k,r}} 
\int \! d\Phi_n^{(jkr,icr)} \,
J_{\rm s \otimes hc}^{jkricr} 
\bB^{(jkr,icr)}_{cr}
\nnb\\
&&
\hspace{12mm}
+ \,
\Bigg[ \;
\sum_{\substack{c\neq i,j,k}} 
\int\!d \Phi_n^{(krj,icj)} 
J_{\rm s \otimes hc}^{krjic}
\bigg(
\rho^{\scriptscriptstyle (C)}_{jk}
\bB^{(krj,icj)}_{[jk]c}
+
\tilde f_{\,jk}^{\,q\bar q} \,
{\cal \bB}^{(krj,icj)}_{[jk]c}
\bigg)
\nnb\\
&&
\hspace{16mm}
+ \,
C_{f_{[jk]}} \,
\rho^{\scriptscriptstyle (C)}_{jk}
\int\!d \Phi_n^{(krj, irj)} 
J_{\rm s \otimes hc}^{krjir} \,
\bB^{(krj,irj)}
+
(j \leftrightarrow k)
\Bigg]
\Bigg\}
\, ,
\eeq
where $[jk]$ represents the parent particle of the pair $(j,k)$, the factors $\rho^{\scriptscriptstyle (C)}_{jk}$, 
involving combinations of Casimir
eigenvalues, are defined in \eq{Cascomb}, the flavour factors such as $\tilde f_{\,jk}^{\,q 
\bar q}$ are presented in \eq{Flavcomb}, and 
${\cal B}_{cd}$
is a colour projection of
the Born contribution involving the symmetric tensor $d_{ABC}$, defined in \eq{deftildeB};
furthermore, 
the phase-space integrals $J_{\rm s \otimes hc}$ are presented in 
\eq{Jsofttimescoll}. 
The remaining contributions to \eq{intK2} are purely 
hard-collinear. 
For the integral of the emission of a cluster of three hard-collinear particles we find
\beq
\int d \Phi_\npt \,
\bbHC{ijk} \RR
& = &
\frac{\varsi_{n+2}}{\varsi_{n}}
\int d \Phi_n^{(ijkr)} \,
J_{\rm hcc}^{ijkr} \,
\bB^{(ijkr)}
\, ,
\qquad
r = r_{ijk}
\, ,
\eeq
while for the emission of two distinct pairs of hard-collinear particles the integral reads
\beq
\int \! d\Phi_{\npt} \,
\bbHC{ijkl} \, \RR
& = &
\frac{\varsi_{n+2}}{\varsi_{n}}
\int \! d\Phi_{n}^{(ijr, klr)} \,
J^{\,ijklr}_{\rm hc \otimes hc} \, \bB^{(ijr, klr)}
\, ,
\qquad
r = r_{ijkl}
\, ,
\eeq
where the integrals $J_{\rm hcc}$ and $J_{\rm hc \otimes hc}$ are reported in 
\eq{Jhcc} and in \eq{Jhchc}, respectively. 

We finally turn to the integration of the strongly-ordered counterterm 
$K^{\otwo}$. 
As announced, we integrate $K^{\otwo}$ only in the phase space of the most unresolved 
radiation, so the integrals involved will be the same that appeared in the case of $K^{\one}$.
Starting from the expression for $K^{\otwo}$ in \eq{eq:K1 K2 K12 Z}, we then sum up the 
NLO sector functions with label $\alpha$ of Eqs.~(\ref{eq:bL12 Zijk})-(\ref{eq:bL12 Zijkl}), 
and we get 
\beq
\label{intK12}
\int d\Phi_\npt \, K^{\otwo}
& = &
\int d\Phi_\npt \,
\Bigg\{ \,
\sum_{i,j \ne i} 
\bbS{i} \,
\bigg[
\sum_{k \ne i,j} \,
\bbS{ij} \, RR \, \bZ{{\rm s},jk}
+
\sum_{\substack{k \ne i\\k > j}} \;
\big( \bbSHC{ijk} + \bbHC{ijk}^{\;(\bf s)} \big) \, \RR
\bigg]
\nnb\\[-2mm]
&&
\hspace{16mm}
+ \,
\sum_{i, \, j > i} \sum_{k \ne i,j} \,
\bbHC{ij} \, 
\bigg[ \;
\bbS{ij} \, \RR \, \bZ{{\rm s},jk}
+
\sum_{l \ne i,k} \, 
\bbSC{kij} \, \RR \, \bZ{{\rm s},kl}
\nnb\\[-2mm]
&&
\hspace{46mm}
+ \,\,
\bbHC{ijk}^{\;(\bf c)} \, \RR
+
\sum_{\substack{l \ne i,j\\l > k}} \,
\bbHC{ijkl}^{\;(\bf c)} \, \RR
\bigg]
\Bigg\}
\, ,
\eeq
where again the mapped sector functions $\bZ{{\rm s},ab}$ carry the same
mapping as the matrix elements they multiply.  No other sector functions appear 
in \eq{intK12}, since the use of symmetrised sector functions has allowed to 
perform the corresponding sector sums, thus replacing sector functions by unity.
Once again, to highlight the general structure of the expressions listed below, we note
that they are all of the form
\beq
\label{genintK12}
\int d\Phi_\npt \, \bbL{\, \cdots}^{\otwo} \, RR 
& = &
{\rm constant} \, \sum_{\{\mu_1,\mu_2\}} \, 
\int d \Phi_\npo^{(\mu_1)} \,\,
J_{{\rm limit}}^{\mu_1} \,\, \bar{\cal K}^{(\mu_1)}_{\mu_2} \,\,  
\bB_{{\rm colour}}^{(\mu_1,\mu_2)} \, . 
\eeq
In this case, the only integrals required for the most unresolved radiation will again be
$J_s^{ilm}$ and $J_{\rm hc}^{ijr}$, given in \eq{eq:Js^ilm} and in \eq{Jhc^ijr} respectively,
and we denoted by $\bar{\cal K}$ a contribution to either a soft or a collinear kernel, associated
with the second radiation, 
which carries mapping $(\mu_1)$, i.e. the first one in the nested 
mapping $(\mu_1,\mu_2)$ of the Born matrix elements.
Proceeding in the order of \eq{intK12}, the integrated
strongly-ordered double-soft limit is given by
\beq
&&
\hspace{-3mm} 
\int \!
d\Phi_{\npt} \,
\bbS{i}\,\bbS{ij} \, \RR \, \bZ{{\rm s},jk}
\; = \;
\\
&&
\Norm \,
\frac{\varsi_{n+2}}{\varsi_{n+1}} \,
\!\!\! \sum_{\substack{c \neq i,j \\ d \neq i,j,c}} \!
\Bigg\{ \;
\int \!
d\Phi_{\npo}^{(icd)} \, 
J_s^{icd} \,
\bigg[ \;
\sum_{\substack{e \neq i,j,c,d}}
\bigg(
\frac{1}{2} \!
\sum_{\substack{f \neq i,j,c,d,e}}
\bar{\mc E}_{ef}^{(j)(icd)} \,
\bB_{cdef}^{(icd,jef)} 
+
\bar{\mc E}_{ed}^{(j)(icd)} \,
\bB_{cded}^{(icd,jed)} 
\bigg)
\nnb\\[-3mm]
&&
\hspace{45mm}
+ \,
\bar{\mc E}_{cd}^{(j)(icd)}
\Big(
\bB_{cdcd}^{(icd,jcd)} 
+
C_A
\bB_{cd}^{(icd,jcd)} 
\Big)
\bigg]
\bZ{{\rm s},\,jk}^{(icd)}
\nnb\\[-1mm]
&&
\hspace{21mm}
+ \,
\int \!
d\Phi_{\npo}^{(idc)} \, 
J_s^{idc} \,
\!\!\! \sum_{\substack{e \neq i,j,c,d}}
\bar{\mc E}_{ed}^{(j)(idc)} \,
\bB_{cded}^{(idc,jed)} \,
\bZ{{\rm s},\,jk}^{(idc)}
\nnb\\[-1mm]
&&
\hspace{21mm}
- \,
C_A
\int \!
d\Phi_{\npo}^{(icj)} \, 
J_s^{icj} \,
\bar{\mc E}_{cd}^{(j)(icj)} \,
\bB_{cd}^{(icj,jcd)} \,
\bZ{{\rm s},\,jk}^{(icj)}
\nnb\\[-1mm]
&&
\hspace{21mm}
- \,
C_A
\int \!
d\Phi_{\npo}^{(ijd)} \, 
J_s^{ijd} \,
\bar{\mc E}_{cd}^{(j)(ijd)} \,
\bB_{cd}^{(ijd,jcd)} \,
\bZ{{\rm s},\,jk}^{(ijd)} \,
\Bigg\}
\, , \nnb
\eeq
and it is entirely expressed in terms of the simple one-loop eikonal kernels given in \eq{eikoker}. 
Next, we need the integral (with $r = r_{ijk}$)
\beq
&&
 \hspace{-3mm} 
\int \!
d\Phi_{\!\npt} \,
\bbS{i} \, \bbSHC{ijk} \, \RR 
\; = \;
\\
&&
- \,
\Norm \,
\frac{\varsi_{n+2}}{\varsi_{n+1}} \,
\! \sum_{\substack{c\neq i,j,k}} 
\Bigg\{ \;
\! \sum_{\substack{d\neq i,j,k,c}} \!
\int \!
d\Phi_\npo^{(icd)} 
J_s^{icd} \,
\frac{\bar P^{(icd){\rm hc},\mu\nu}_{jk}}{\sk{jk}{icd}} \,
\bB^{(icd,jkr)}_{\mu\nu,cd}
\nnb\\
&&
\hspace{27mm}
+ \,
\Bigg[ \,
\int \!
d\Phi_\npo^{(ijc)} 
J_s^{ijc}
\frac{\bar P^{(ijc){\rm hc},\mu\nu\!\!\!\!}_{jk(r)}}{2\sk{jk}{ijc}} \,
\bigg(\!
\rho^{\scriptscriptstyle (C)}_{jk} \,
\bB^{(ijc,krj)}_{\mu\nu,[jk]c}
\!+\!
\tilde f_{\,jk}^{\,q\bar q} \, 
{\cal \bB}^{(ijc,krj)}_{\mu\nu,[jk]c}
\!\bigg)
+
(j \leftrightarrow k)
\Bigg]
\nnb\\
&&
\hspace{27mm}
+ \,
\Bigg[
\int \!
d\Phi_\npo^{(icj)} 
J_s^{icj}
\frac{\bar P^{(icj){\rm hc},\mu\nu\!\!\!\!}_{jk(r)}}{2\sk{jk}{icj}} \,
\bigg(
\rho^{\scriptscriptstyle (C)}_{jk} \,
\bB^{(icj,krj)}_{\mu\nu,[jk]c}
\!+\!
\tilde f_{\,jk}^{\,q\bar q} \,
{\cal \bB}^{(icj,krj)}_{\mu\nu,[jk]c}
\bigg)
+
(j \leftrightarrow k)
\Bigg]
\Bigg\}
\, , \nnb
\eeq
where the hard-collinear kernels are given in \eq{hardcollker}.
We now turn to limits involving triple-collinear configurations. 
First we need
\beq
&&
\hspace{-3mm} 
\int \!
d\Phi_{\!\npt} \,
\bbS{i}\,\bbHC{ijk}^{\;(\bf s)} \, \RR 
\; = \;
\\
&&
\Norm \,
\frac{\varsi_{n+2}}{\varsi_{n+1}} \,
\frac{C_{f_{[jk]}\!}}{2} 
\Bigg\{ \;
\Bigg[
\rho^{\scriptscriptstyle (C)}_{jk} \,
\int \! d\Phi_{\npo}^{(ijr)} \,
J_s^{ijr} \,
\frac{\bar P_{jk(r)}^{(ijr){\rm hc},\mu\nu\!\!\!\!\!}}{\sk{jk}{ijr}} \,
\left(\! \bB^{(ijr,jkr)}_{\mu\nu} \!-\! \bB^{(ijr,krj)}_{\mu\nu} \!\right)\!
+
(j \leftrightarrow k)
\Bigg]
\nnb\\
&&
\hspace{23mm}
+ \,
\Bigg[
\rho^{\scriptscriptstyle (C)}_{jk} \,
\int \! d\Phi_{\npo}^{(irj)} \,
J_s^{irj} \,
\frac{\bar P_{jk(r)}^{(irj){\rm hc},\mu\nu\!\!\!\!\!}}{\sk{jk}{irj}} \,
\left(\! \bB^{(irj,jkr)}_{\mu\nu} \!-\! \bB^{(irj,krj)}_{\mu\nu} \!\right)\!
+
(j \leftrightarrow k)
\Bigg]
\nnb\\
&&
\hspace{23mm}
- \,
\rho^{\scriptscriptstyle (C)}_{[jk]}
\Bigg[
\int \!
d\Phi_\npo^{(ijk)} \, 
J_s^{ijk} \,
\frac{\bar P^{(ijk){\rm hc},\mu\nu\!\!\!}_{jk(r)}}{\sk{jk}{ijk}} \,
\bB^{(ijk,jkr)}_{\mu\nu}
+
(j \leftrightarrow k)
\Bigg]
\Bigg\}
\, ,
\qquad
r = r_{ijk}
\, .
\nnb
\eeq
Next we consider (again with $r = r_{ijk}$)
\beq
\label{HCijSij}
\int \!
d\Phi_{\!\npt} \,
\bbHC{ij} \, \bbS{ij} \, \RR \,
\bZ{{\rm s},\,jk}
& = &
- \,
\Norm \,
\frac{\varsi_{n+2}}{\varsi_{n+1}} \!
\int \! d\Phi_{\npo}^{(ijr)} \,
J_{\rm hc}^{ijr} \,
\!\!\!\!\! \sum_{\substack{c \neq i,j \\ d \neq i,j,c}} \!\!\!\!
\bar{\mc E}^{(j)(ijr)}_{cd} \,
\bB_{cd}^{(ijr,jcd)} \,
\bZ{{\rm s},\,jk}^{(ijr)}
\, ,
\qquad
\eeq
where the choice of $r$ different from $i,j,k$, analogously to the integral 
of $\bbHC{ij}\,\RR$ in \eq{eq:int HCij}, causes a 
dependence of the integrated kernel $J_{\rm hc}^{ijr}$ on the index $k$ 
of the sector function $\bZ{{\rm s},\,jk}^{(ijr)}$. 
Next we have (
$r = r_{ijkl}$, 
$r' = r_{ijk}$
)
\beq
\int \!
d\Phi_{\!\npt} \,
\bbHC{ij} \, \bbSC{kij} \, \RR \,
\bZ{{\rm s},\,kl}
& = &
- \,
\Norm
\frac{\varsi_{n+2}}{\varsi_{n+1}}
\int \! d\Phi_{\npo}^{(ijr)} \,
J_{\rm hc}^{ijr} \,
\Bigg[
\! \sum_{\substack{c \neq i,j,k,r' \\ d\neq i,j,k,r',c}} \!\!\!\!
\bar{\mc E}^{(k)(ijr)}_{cd} 
\bB^{(ijr, kcd)}_{cd} \!
\\
&&
\hspace{5mm}
+ \,
2 \!\!\!
\sum_{c \neq i,j,k,r'} \!\!\!
\bar{\mc E}^{(k)(ijr)}_{cr'} 
\bB^{(ijr, kcr')}_{cr'} 
+ 
2 \!\!
\sum_{\substack{c\neq i,j,k}} \!
\bar{\mc E}^{(k)(ijr)}_{jc} \!
\bB^{(ijr,kcj)}_{jc}
\Bigg]
\,
\bZ{{\rm s},\,kl}^{(ijr)}
\, .
\nnb
\eeq
Finally we need to handle strongly-ordered hard-collinear limits. 
First, with a collinear cluster of three particles we find
($r = r_{ijk}$)
\beq
\int \!
d\Phi_{\!\npt} \,
\bbHC{ij} \, \bbHC{ijk}^{\;(\bf c)} \, \RR
& = &
\Norm \,
\frac{\varsi_{n+2}}{\varsi_{n+1}} \, 
\bigg\{
\int \! d\Phi_{\npo}^{(ijr)} \,
J_{\rm hc}^{ijr} \,
\frac{\bar P_{jk(r)}^{(ijr){\rm hc},\mu\nu\!\!\!\!}}{\sk{jk}{ijr}} \,
\bB^{(ijr,jkr)}_{\mu\nu}
\nnb\\
&&
\hspace{2mm}
- \, 2 \,
C_{f_{[ij]}} \!
\int \! d\Phi_{\npo}^{(ijr)} \,
J_{\rm hc}^{ijr} \,
\bar{\mc E}^{(k)(ijr)}_{jr} \,
\Big(
\bB^{(ijr,krj)} 
-
\bB^{(ijr,kjr)}
\Big)
\bigg\}
\, .
\eeq
Then, with two independent pairs of collinear 
particles, we find
\beq
\int \!
d\Phi_{\!\npt} \,
\bbHC{ij} \, \bbHC{ijkl}^{\;(\bf c)} \, \RR
& = &
\Norm \,
\frac{\varsi_{n+2}}{\varsi_{n+1}}
\int \! d\Phi_{\npo}^{(ijr)} \,
J_{\rm hc}^{ijr} \,
\frac{\bar P_{kl(r)}^{(ijr){\rm hc},\mu\nu\!\!\!\!}}{\sk{kl}{ijr}} \,
\bB^{(ijr,klr)}_{\mu\nu}
\, ,
\qquad
r = r_{ijkl}
\, .
\qquad
\eeq
This concludes the list of all required integrals 
for double-real radiation.


\subsection{Relabelling of momenta and flavour sums}
\label{FlaSum}

Our next step will be to collect the results of the different sectors and combine them
by renaming the mapped momenta in each sector. More precisely, in all $(\npo)$-body 
phase spaces $d\Phi_\npo^{(abc)}$ appearing in the integrals of $K^{\one}$ and 
$K^{\otwo}$, we rename the sets of mapped momenta $\{\bar k^{(abc)}\}_{\npo}$ 
as a unique set of $(n+1)$ momenta $\{k\}_{\npo}$. With this new labelling, the 
indices of the mapped momenta refer directly to the particles of the unique 
$(\npo)$-body phase space, and the reference to the first mapping can be 
simply removed. The relabelling thus leads to
\beq
&&
d\Phi_\npo^{(abc)} \to d\Phi_\npo
\, ,
\qquad
\bZ{\, \cdots}^{(abc)} \to \Z{\, \cdots}
\, ,
\qquad
\bR^{(abc)}_{\, \cdots} \to R_{\, \cdots}
\, ,
\qquad
\bB^{(abc,def)}_{\, \cdots} \to \bB^{(def)}_{\, \cdots}
\, ,
\nnb \\
&&
\hspace{15mm}
\bar s_{ij}^{(abc)} \to s_{ij}
\, ,
\qquad
\bar P_{ij(r)}^{(abc){\rm hc},\mu\nu} \to P_{ij(r)}^{{\rm hc},\mu\nu}
\, ,
\qquad
\bar {\mc E}_{lm}^{(i)(abc)} \to {\mc E}_{lm}^{(i)}
\, .
\qquad
\eeq
Similarly, in the $n$-body phase spaces $d\Phi_n^{(abc,def)}$ appearing in the 
integral of $K^{\two}$, the sets of mapped momenta $\{\bar k^{(abc,def)}\}_{n}$ are 
renamed as a unique set of $n$ momenta $\{k\}_{n}$, which in practice means 
performing the substitutions 
\beq
d\Phi_n^{(abc,def)} \; \to \; d\Phi_n
\, ,
\qquad
&&
\bB^{(abc,def)}_{\, \cdots} \; \to \; B_{\, \cdots}
\qquad
\bar s_{ij}^{(abc,def)} \; \to \; s_{ij}
\, .
\eeq
In particular, in the integral of $\bbSHC{ijk}(1-\bbC{ijk})\RR$, which involves a collinear 
splitting of partons $j$ and $k$, the momenta 
$\kk{k}{jkr,icd}$, $\kk{k}{jkr,icr}$, $\kk{j}{krj,icj}$ and $\kk{j}{krj,ijr}$ 
are all renamed as $k_p$, where $p$ is the parent  particle of $j$ and $k$. 

At this stage, in all integrated counterterms, the only recollection of the particles of 
the original $(\npt)$-body phase space is confined to the flavour factors $f_{i}^{q}$, 
$f_{i}^{\bar q}$, $f_{i}^{g}$. These can be summed up, and, if needed, translated into 
flavour factors for the particles of the $(\npo)$-body and $n$-body phase spaces. 
We now give the rules to perform these sums.

\vspace{3mm}
Let us begin with the simple case in which only one particle is integrated out, which
is what happens for $K^{\one}$ and $K^{\otwo}$. In this case the following rules apply.
\begin{itemize}
\item
When going from an $(\npt)$-body phase space to an $(\npo)$-body phase space by 
discarding particle $i$, which happens when particle $i$ is a soft gluon, the sum over
flavour factors satisfies
\beq
\label{flavsum1soft}
\frac{\varsigma_{n+2}}{\varsigma_{n+1}}
\sum_{i} 
f_{i}^{g}
=
1
\, .
\eeq
For example, if all ($\npt$) particles are gluons, one has $\varsigma_{n+2} = 1/(\npt)!$ 
and $\varsigma_{n+1} = 1/(\npo)!$, and the sum yields the missing factor of $\npt$.
\item
When going from an $(\npt)$-body phase space to an $(\npo)$-body phase space 
by replacing two particles $i,j$ with their parent particle $p$, which happens when $i$ 
and $j$ form a collinear pair, the sum over the flavour factors of particles $i,j$ can 
be written as a sum over flavour factors for particle $p$ according to the rules 
\beq
\label{flavsum1coll}
\frac{\varsigma_{n+2}}{\varsigma_{n+1}}
\sum_{i, j > i}
f_{ij}^{q \bar q}
& = &
N_f \,
\sum_{p} \, f_{p}^{g}
\, ,
\nnb\\[-1mm]
\frac{\varsigma_{n+2}}{\varsigma_{n+1}}
\sum_{i, j > i}
( f_{ij}^{g q} + f_{ij}^{g \bar q} )
& = &
\sum_{p} \, ( f_{p}^{q} + f_{p}^{\bar q} )
\, ,
\nnb\\[-1mm]
\frac{\varsigma_{n+2}}{\varsigma_{n+1}}
\sum_{i, j > i}
f_{ij}^{gg}
& = &
\frac{1}{2} \,
\sum_{p} \, f_{p}^{g}
\, .
\eeq
As an example, consider the production of $n$ gluons and a collinear $q \bar{q}$ pair.
In this case the first line of \eq{flavsum1coll} applies, and one must take into account
the fact that quark flavours must be summed, since the quark pair is integrated out. One
then has $\varsigma_{n+2} = N_f/n!$ and $\varsigma_{n+1} = 1/(\npo)!$, since the new
final state involves $(\npo)$ gluons. For the same reason, the {\it r.h.s.}~yields $N_f (\npo)$.
\end{itemize}
Not surprisingly, the flavour sum rules for the integrated $K^{\two}$ are both more 
varied and more intricate, since one is integrating out two particles, either by removing 
them (when they are soft), or by replacing them with their (grand)parent particles 
when they form collinear sets. We consider the various cases in turn.
\begin{itemize}
\item
When going from an $(\npt)$-body phase space to an $n$-body phase space by
discarding two particles $i,j$, the sum over particles $i,j$ satisfies 
\beq
\frac{\varsigma_{n+2}}{\varsigma_{n}}
\sum_{i} \sum_{j > i} 
f_{ij}^{gg}
=
\frac{1}{2}
\, ,
\qquad\qquad
\frac{\varsigma_{n+2}}{\varsigma_{n}}
\sum_{i} \sum_{j > i} 
f_{ij}^{q \bar q}
=
N_f
\, .
\eeq
As before, the first equality is easily verified when all ($\npt$) particles are gluons, as is the
second one when the final state consists of $n$ gluons and a quark-antiquark pair.
\item
When going from an $(\npt)$-body phase space to an $n$-body phase space by replacing 
two particles $j,k$ with their parent particle $p$, and by discarding particle $i$, the sum 
over particles $i,j,k$ can be written as a sum over $p$ according to the following rules. 
\beq
\frac{\varsigma_{n+2}}{\varsigma_{n}}
\sum_{i, j \neq i}\sum_{\substack{k \ne i\\k > j}}
f_{i}^{g}
f_{jk}^{q \bar q}
& = &
N_f \,
\sum_{p} \, f_{p}^{g}
\, ,
\nnb\\[-1mm]
\frac{\varsigma_{n+2}}{\varsigma_{n}}
\sum_{i, j \neq i}\sum_{\substack{k \ne i\\k > j}}
f_{i}^{g}
( f_{jk}^{g q} + f_{jk}^{g \bar q} )
& = &
\sum_{p} \, ( f_{p}^{q} + f_{p}^{\bar q} )
\, ,
\nnb\\[-1mm]
\frac{\varsigma_{n+2}}{\varsigma_{n}}
\sum_{i, j \neq i}\sum_{\substack{k \ne i\\k > j}}
f_{i}^{g}
f_{jk}^{gg}
& = &
\frac{1}{2} \,
\sum_{p} \, f_{p}^{g}
\, ,
\eeq
where it is important to pay attention to the range of the various sums.
\item
When going from an $(\npt)$-body phase space to an $n$-body phase space by
replacing three particles $i,j,k$ with their grandparent particle $p$, the sum 
over particles $i,j,k$ can be replaced by a sum over $p$ according to the 
following rules. 
\beq
\frac{\varsigma_{n+2}}{\varsigma_{n}}
\sum_{i, \, j > i} \sum_{k > j}
( f_{ijk}^{q \bar q q'} + f_{ijk}^{q \bar q \bar q'} )
& = &
N_f
\sum_{p} \, (f_{p}^{q} \!+\! f_{p}^{\bar q})
\, ,
\nnb\\[-1mm]
\frac{\varsigma_{n+2}}{\varsigma_{n}}
\sum_{i, \, j > i} \sum_{k > j}
( f_{ijk}^{q \bar q q} + f_{ijk}^{q \bar q \bar q} )
& = &
\frac{1}{2}
\sum_{p} \, (f_{p}^{q} \!+\! f_{p}^{\bar q})
\, ,
\nnb\\[-1mm]
\frac{\varsigma_{n+2}}{\varsigma_{n}}
\sum_{i, \, j > i} \sum_{k > j}
f_{ijk}^{q \bar q g}
& = &
N_f
\sum_{p} \, f_{p}^{g}
\, ,
\nnb\\[-1mm]
\frac{\varsigma_{n+2}}{\varsigma_{n}}
\sum_{i, \, j > i} \sum_{k > j}
( f_{ijk}^{gg q} + f_{ijk}^{gg \bar q} )
& = &
\frac{1}{2}
\sum_{p} \, (f_{p}^{q} \!+\! f_{p}^{\bar q})
\, ,
\nnb\\[-1mm]
\frac{\varsigma_{n+2}}{\varsigma_{n}}
\sum_{i, \, j > i} \sum_{k > j}
f_{ijk}^{ggg}
& = &
\frac{1}{6}
\sum_{p} \, f_{p}^{g}
\, ,
\eeq
where one easily recognises in the five lines the five possible partonic channels 
involving the production of a cluster of three collinear particles: in the first line,
the final quark-antiquark pair can have any flavour (including that of the grandparent
(anti)quark, which is the same as that of the final (anti)quark $q'$), while in the
second line all three (anti)quarks have the same flavour.
\item
The most intricate channel for flavour sums arises when going from an $(\npt)$-body 
phase space to an $n$-body phase space by replacing two pairs of particles $i,j$ and 
$k,l$ with their parent particles, $p$ and $t$ respectively. In this case, the sum over 
particles $i,j,k,l$ can be replaced by a sum over $p$ and $t$ according to the following rules. 
\beq
\frac{\varsigma_{n+2}}{\varsigma_{n}}
\sum_{i, \, j > i} \sum_{\substack{k \ne j \\ k > i}}
\sum_{\substack{l \neq j \\ l > k}}
f_{ij}^{q \bar q}
f_{kl}^{q' \bar q'}
& = &
\frac{N_f^{2}}{2} \,
\sum_{p,t\ne p} \, f_{pt}^{gg}
\, ,
\nnb\\[-1mm]
\frac{\varsigma_{n+2}}{\varsigma_{n}}
\sum_{i, \, j > i} \sum_{\substack{k \ne j \\ k > i}}
\sum_{\substack{l \neq j \\ l > k}}
\Big[
f_{ij}^{q \bar q}
( f_{kl}^{g q'} + f_{kl}^{g \bar q'} )
+
( f_{ij}^{g q'} + f_{ij}^{g \bar q'} )
f_{kl}^{q \bar q}
\Big]
& = &
\frac{N_f}{2} \,
\sum_{p,t\ne p} \, 
( f_{pt}^{g q} + f_{pt}^{g \bar q} )
\, ,
\nnb\\[-1mm]
\frac{\varsigma_{n+2}}{\varsigma_{n}}
\sum_{i, \, j > i} \sum_{\substack{k \ne j \\ k > i}}
\sum_{\substack{l \neq j \\ l > k}}
( f_{ij}^{q \bar q}f_{kl}^{gg} + f_{ij}^{gg}f_{kl}^{q \bar q} )
& = &
\frac{N_f}{2} \,
\sum_{p,t\ne p} \, f_{pt}^{gg}
\, ,
\nnb\\[-1mm]
\frac{\varsigma_{n+2}}{\varsigma_{n}}
\sum_{i, \, j > i} \sum_{\substack{k \ne j \\ k > i}}
\sum_{\substack{l \neq j \\ l > k}}
( f_{ij}^{g q} + f_{ij}^{g \bar q} )
( f_{kl}^{g q'} + f_{kl}^{g \bar q'} )
& = &
\frac{1}{2} \,
\sum_{p,t\ne p} \, 
( f_{p}^{q} + f_{p}^{\bar q} )
( f_{t}^{q'} + f_{t}^{\bar q'} )
\, ,
\nnb\\[-1mm]
\frac{\varsigma_{n+2}}{\varsigma_{n}}
\sum_{i, \, j > i} \sum_{\substack{k \ne j \\ k > i}}
\sum_{\substack{l \neq j \\ l > k}}
\Big[
( f_{ij}^{g q} + f_{ij}^{g \bar q} )
f_{kl}^{gg}
+
f_{ij}^{gg}
( f_{kl}^{g q} + f_{kl}^{g \bar q} )
\Big]
& = &
\frac{1}{4} \,
\sum_{p,t\ne p} \, 
( f_{pt}^{g q} + f_{pt}^{g \bar q} )
\, ,
\nnb\\[-1mm]
\frac{\varsigma_{n+2}}{\varsigma_{n}}
\sum_{i, \, j > i} \sum_{\substack{k \ne j \\ k > i}}
\sum_{\substack{l \neq j \\ l > k}}
f_{ij}^{gg}f_{kl}^{gg}
& = &
\frac{1}{8} \,
\sum_{p,t\ne p} \, f_{pt}^{gg}
\, .
\eeq
\end{itemize}
We emphasise that the flavour sum rules listed in this section apply for any final-state
multiplicity and flavour structure. We now have all the tools to assemble the complete
integrated counterterms, which will be naturally organised according to the flavour 
structures of the $(\npo)$-particle and of the $n$-particle phase spaces, as needed.


\subsection{\vspace{-.5mm}Assembling the complete integrated counterterms}
\label{AsseCou}

\vspace{.5mm}
After summing all contributions that were differently mapped, relabelling momenta, 
and making use of the flavour rules listed in \secn{FlaSum}, the resulting integrated 
counterterms do not bear any remaining trace of the original $(\npt)$-body phase space, 
and we can actually get full results for $I^{\one}$, $I^{\two}$, $I^{\otwo}$, as defined 
in \eq{intcountNNLO}.
The simplest case is the integral of the single-unresolved counterterm 
$I^{\one}$, which reads 
\beq
\label{eq:I1 (1)}
I^{\one}
& = & 
\sum_{\substack{i,j \ne i}}
I^{\one}_{ij} \, \W{ij}
\, = \,
\sum_{\substack{i,j > i}}
I^{\one}_{ij} \, \Z{ij}
\, ,
\\
I^{\one}_{ij}
& = &
- \!\!
\sum_{c, d \ne c} \!\!
J_s(s_{cd}) \,
R_{cd}
+
\sum_{k} \,
J_{\rm hc}^{\,k}(s_{kr}) \,
R
\, ,
\qquad
r = r_{ijk}
\, .
\nnb
\eeq
Here $R$ is the full squared matrix element for single-real radiation, defined in \eq{pertA2}, and $R_{cd}$ is its colour-correlated counterpart, 
defined in \eq{colcorrV}. The single-soft integral $J_s$ is given in \eq{eq:Js}, and the 
collinear integral $J_{\rm hc}^{\,k}$ is given in \eq{Jhck}. 
Because of the rule $r=r_{ijk}$, a dependence of $J_{\rm hc}^{\,k}(s_{kr})$ 
on $i$ and $j$ is left, excluding the possibility to sum over sectors in 
the hard-collinear part of $I^{\one}$.

The integral of the double-unresolved counterterm, $I^{\two}$, is more intricate,
and we assemble it according to 
\beq
\label{eq:I2}
I^{\two}
& = & 
I^{\two}_{\sS\sS}
+
I^{\two}_{\sSHC}
+
I^{\two}_{\sHCC}
+
I^{\two}_{\sHCHC}
\, ,
\eeq
distinguishing double-soft, soft-times-hard-collinear and double-hard-collinear 
contributions, the last of which may involve three or four Born-level particles.
For $I^{\two}_{\sS\sS}$ we get contributions containing Born-level 
colour correlations involving four, three and two particles, and we write
\beq
I^{\two}_{\sSS}
& = & 
\frac{1}{4} \!
\sum_{\substack{c,d \neq c}}
\Bigg\{ \;
\sum_{\substack{e \neq c,d}} 
\bigg[
\sum_{f \neq c,d,e} \!
J_{\rm s \otimes s}^{(4)} (s_{cd},s_{ef}) 
B_{cdef}
+
4 \,
J_{\rm s \otimes s}^{(3)} (s_{cd},s_{ed}) 
B_{cded}
\bigg]
\\
&&
\hspace{12mm}
+ \,
2 \,
J_{\rm s \otimes s}^{(2)} (s_{cd}) 
B_{cdcd}
+
2 \, 
\Big[
2 \, 
N_f \, T_R \,
J_{\rm ss}^{(\rm q \bar q)}(s_{cd}) 
-
C_A\,
J_{\rm ss}^{(\rm gg)}(s_{cd}) 
\Big]
B_{cd}
\Bigg\}
\, ,
\nnb
\eeq
where the constituent integrals are given in \eq{varJscrosss}. 
The 
soft-times-hard-collinear
contribution yields
\beq
\label{eq:I2shc}
I^{\two}_{\sSHC}
& = & 
- \,
\sum_{k}
\bigg\{ \;
J_{\rm hc}^{\,k}(s_{kr})
\! \sum_{c, d \ne c} \!
J_{\rm s}(s_{cd})
B_{cd}
+
J_{\rm shc}^{\,k}(s_{kr})
B
+
J_{\rm shc}^{k,\rm A}(s_{kr})
B_{kr}
\\
&&
\hspace{9mm}
+
\sum_{\substack{c\neq k,r}} 
\Big[
J_{\rm shc}^{k,\rm B}(s_{kr},s_{kc})
B_{kc}
+
J_{\rm shc}^{k,\rm B}(s_{kr},s_{cr})
B_{cr}
\Big]
\bigg\}
\, ,
\qquad
r = r_{k}
\, ,
\qquad
\nnb
\eeq
where the rule $r = r_{k}$, as defined in \eq{eq: r_ij}, prevents $r$ from being equal to $k$. 
In \eq{eq:I2shc} we have introduced the following soft-times-hard-collinear 
integrals 
\beq
\label{flavcombshc}
J_{\rm shc}^{\,k}(s)
& = &
( f_{k}^{q} \!+\! f_{k}^{\bar q} ) \,
\Big\{
2 \,
C_F \, 
J_{\rm s \otimes hc}^{gqg}(s)
+
C_A \, 
\Big[ J_{\rm s \otimes hc}^{ggq}(s) - J_{\rm s \otimes hc}^{gqg}(s) \Big]
\Big\}
\nnb\\
&&
\hspace{-4mm}
+ \,
f_{k}^{g} \,
C_A \, 
\Big[ 
2 \,
N_{f} \, 
J_{\rm s \otimes hc}^{gqq}(s)
+
J_{\rm s \otimes hc}^{ggg}(s)
\Big]
\, ,
\nnb\\
J_{\rm shc}^{\,k,\rm A}(s)
& = &
( f_{k}^{q} \!+\! f_{k}^{\bar q} ) \,
\bigg\{
2 \,
J_{\rm s \otimes hc}^{gqg}(s)
+
\frac{C_{A}}{C_{F}}\,
\Big[ J_{\rm s \otimes hc}^{ggq}(s) - J_{\rm s \otimes hc}^{gqg}(s) \Big]
-
2 \,
J_{\rm s \otimes hc}^{\, 4 \rm(2g)}\!(s,s)
\bigg\}
\nnb\\
&&
\hspace{-4mm}
+ \,
f_{k}^{g} \,
\Big\{ 
2 \, N_{f} \,
\Big[
J_{\rm s \otimes hc}^{gqq}(s)
-
J_{\rm s \otimes hc}^{\, 4 \rm(1g)}\!(s,s)
\Big]
+
J_{\rm s \otimes hc}^{ggg}(s)
-
J_{\rm s \otimes hc}^{\, 4 \rm(3g)}\!(s,s)
\Big\}
\, ,
\nnb\\
J_{\rm shc}^{\,k,\rm B}(s,s')
& = &
( f_{k}^{q} \!+\! f_{k}^{\bar q} ) \,
\Big[
2 \,
J_{\rm s \otimes hc}^{\, 3 \rm(2g)}\!(s,s')
-
2 \,
J_{\rm s \otimes hc}^{\, 4 \rm(2g)}\!(s,s')
\Big]
\nnb\\
&&
\hspace{-4mm}
+ \,
f_{k}^{g} \,
\Big\{
2 \,
N_{f} \, 
\Big[
J_{\rm s \otimes hc}^{\, 3 \rm(1g)}\!(s,s')
-
J_{\rm s \otimes hc}^{\, 4 \rm(1g)}\!(s,s')
\Big]
+
J_{\rm s \otimes hc}^{\, 3 \rm(3g)}\!(s,s')
-
J_{\rm s \otimes hc}^{\, 4 \rm(3g)}\!(s,s')
\Big\}
\, ,
\eeq
whose constituent integrals can be found in \eq{constintshc}. 
Next, we turn to the double-hard-collinear
integral involving three Born-level particles, which reads
\beq
I^{\two}_{ \sHCC}
& = &
\sum_{k} \, 
\bigg\{ \;
( f_{k}^{q} \!+\! f_{k}^{\bar q} ) \,
\bigg[
N_f \, 
J_{\rm hcc}^{\zg} (s_{kr})
+ 
\frac{1}{2} \, 
J_{\rm hcc}^{(\rm 0g, id)} (s_{kr})
+
\frac{1}{2} \, 
J_{\rm hcc}^{\tg} (s_{kr})
\bigg]
\nnb\\
&&
\hspace{6mm}
+ \, 
f_{k}^{g} \,
\bigg[
N_f \, 
J_{\rm hcc}^{\og} (s_{kr})
+
\frac{1}{6} \, 
J_{\rm hcc}^{(\rm 3g)} (s_{kr})
\bigg]
\bigg\}
\, B
\, ,
\qquad
r = r_{k}
\, ,
\qquad
\nnb
\eeq
where 
the relevant constituent integrals are given in \eq{constinthcc}. 
Finally, we come to the double-hard-collinear integral involving 
four Born-level particles, which reads 
\beq
I^{\two}_{\sHCHC}
& = &
\frac{1}{2} 
\sum_{j,l\ne j} 
\bigg\{ \;
( f_{j}^{q} + f_{j}^{\bar q} )
( f_{l}^{q'} + f_{l}^{\bar q'} )
J_{\rm hc \otimes hc}^{\rm qgqg} \big(s_{jr}s_{lr}\big)
\\
&&
\hspace{10mm} 
+ \,
( f_{jl}^{g q} \!+\! f_{jl}^{g \bar q} )
\bigg[
N_f \,
J_{\rm hc \otimes hc}^{\rm qqqg} \big(s_{jr}s_{lr}\big)
+
\frac{1}{2} \,
J_{\rm hc \otimes hc}^{\rm qggg} \big(s_{jr}s_{lr}\big)
\bigg]
\nnb\\
&&
\hspace{10mm}
+ \,
f_{jl}^{gg}
\bigg[
N_f^{2} 
J_{\rm hc \otimes hc}^{\rm qqqq}\big(s_{jr}s_{lr}\big)
+
N_f 
J_{\rm hc \otimes hc}^{\rm qqgg}\big(s_{jr}s_{lr}\big)
+
\frac{1}{4} 
J_{\rm hc \otimes hc}^{\rm gggg}\big(s_{jr}s_{lr}\big)
\bigg]
\bigg\}
B
\nnb
\, ,
\qquad
r = r_{jl}
\, ,
\eeq
where the constituent integrals are given in \eq{constinthchc}.
Similarly to $I^{\one}$, for the integral of the strongly-ordered 
counterterm, $I^{\otwo}$, we provide expressions with both unsymmetrised 
and symmetrised sector functions, so as to make it straightforward to prove that 
$I^{\otwo}$ compensates sector by sector the phase-space singularities of 
$I^{\one}$. 
Beginning with the expression involving
the original sector functions $\W{ij}$, we write
\beq
\label{eq:I12 Wij}
I^{\otwo}
& = & 
\sum_{\substack{i,j \ne i}}
I^{\otwo}_{ij}
\, ,
\qquad\qquad
I^{\otwo}_{ij}
\; = \;
I^{\otwo}_{{\rm S},ij} \, \W{{\rm s},ij}
+ 
I^{\otwo}_{{\rm C},ij} \,
-
I^{\otwo}_{{\rm SC},ij}
\, ,
\eeq
where 
the soft limit of sector functions $\W{{\rm s},ij}$ is given in \eq{eq:Wsij}.
The soft integral $I^{\otwo}_{{\rm S},ij}$ can again be organised 
in terms of quadruple, triple and simple Born-level colour correlations, 
which in this case will be multiplied times eikonal kernels for the second radiation, 
and NLO-type soft and hard-collinear integrals. 
We find ($r=r_{ik}$, $r'=r_{ij}$, $r''=r_{ijk}$)
\beq
\label{eq:I12_components_S}
I^{\otwo}_{{\rm S},ij}
& = &
\Norm \!
\sum_{\substack{c\neq i \\ d\neq i,c}} \!
\mc E^{(i)}_{cd}
\bigg\{ \;
\frac{1}{2}
\sum_{\substack{e\neq i \\ f\neq i,e}} \!\!
J_s (s_{ef}) \,
\bB^{(icd)}_{cdef}
+
\sum_{\substack{e\neq i,d}} \!\!
J_s (s_{de}) \,
\Big( \bB^{(icd)}_{cded} - \bB^{(idc)}_{cded} \Big)
\\
&&
\hspace{17mm}
- \,
C_A \,
\Big[
J_s(s_{ic}) + J_s(s_{id}) - J_s(s_{cd})
\Big]
\bB^{(icd)}_{cd}
-
J_{\rm hc}^{\,i}(s_{ir'}) 
\bB_{cd}^{(icd)}
\bigg\} 
\nnb\\
&&
\hspace{-3mm}
- \,
\Norm \,
\sum_{k \neq i} \,
J_{\rm hc}^{\,k}(s_{kr''}) 
\bigg[
\sum_{\substack{c \neq i,k,r \\ d \neq i,c,k,r}} \!\!\!
\mc E^{(i)}_{cd} \,
\bB^{(icd)}_{cd}
+
2 \!\!
\sum_{\substack{c \neq i,k,r}} \!\!
\mc E^{(i)}_{cr} \,
\bB^{(icr)}_{cr}
+
2 \!
\sum_{\substack{c\neq i,k}} \!
\mc E^{(i)}_{kc} \,
\bB^{(ick)}_{kc}
\bigg]
\, ,
\qquad
\nnb
\eeq
where the component integrals are given in \eq{eq:Js} and in \eq{Jhck}. 
We notice that the expression contains 
three
different reference particles 
$r$, $r'$ and $r''$, all
built according to the rule in \eq{eq: r_ij}. 
In particular 
$r'=r_{ij}$ and $r''=r_{ijk}$ introduce
a dependence in $I^{\otwo}_{{\rm S},ij}$ on 
the particle $j$ of the soft sector function $\W{{\rm s},ij}$.
The collinear integral
$I^{\otwo}_{ {\rm C},{ij} }$ is expressed in terms of spin-correlated Born-level squared matrix 
elements, which in this case are multiplied times LO collinear kernels for the least-unresolved
collinear splitting, and times suitable combinations of the same constituent integrals as in 
\eq{eq:I12_components_S}. We find (with $r=r_{ij}$, $r'=r_{ijk}$)
\beq
\label{eq:I12_components_C}
I^{\otwo}_{{\rm C},ij}
& = & 
- \,
\Norm \,
\frac{P^{\mu\nu}_{ij(r)}}{s_{ij}}
\bigg\{ \,
\sum_{\substack{c\neq i,j}}
\sum_{\substack{d\neq i,j,c}} \!
J_s(s_{cd}) \,
\bB^{(ijr)}_{\mu\nu,cd}
+
C_{f_{[ij]}} \,
\rho^{\scriptscriptstyle (C)}_{[ij]} \,
J_s(s_{ij}) \,
\bB^{(ijr)}_{\mu\nu}
\\
&&
\hspace{18mm}
+ \,
\bigg[ \;
\sum_{\substack{c\neq i,j}}
J_s(s_{ic}) \,
\Big(
\rho^{\scriptscriptstyle (C)}_{ij} \,
\bB^{(jri)}_{\mu\nu,[ij]c}
+
\tilde f_{\,ij}^{\,q\bar q} \,
{\cal \bB}^{(jri)}_{\mu\nu,[ij]c}
\Big)
\nnb\\
&&
\hspace{22mm}
+ \,
C_{f_{[ij]}} \,
\rho^{\scriptscriptstyle (C)}_{ij} \,
J_s(s_{ir}) \,
\Big(
\bB^{(jri)}_{\mu\nu}
-
\bB^{(ijr)}_{\mu\nu}
\Big)
+
(i \leftrightarrow j)
\bigg]
\bigg\}
\, \W{{\rm c},ij(r)}
\nnb\\
&&
+ \,
\Norm \,
\frac{P_{ij(r)\!}^{\mu\nu\!\!}}{s_{ij}} \,
\Big[
J_{\rm hc}^{\,i}(s_{ir})
+
J_{\rm hc}^{\,j}(s_{jr})
\Big]
\bB^{(ijr)}_{\mu\nu}
\, \W{{\rm c},ij(r)}
+
\Norm \!
\sum_{k \neq i,j} \!
\frac{P_{ij(r')\!}^{\mu\nu\!\!}}{s_{ij}} \,
J_{\rm hc}^{\,k}(s_{kr'}) \,
\bB^{(ijr')}_{\mu\nu}
\, \W{{\rm c},ij(r')}
\, ,
\nnb
\eeq
where the collinear limit of sector functions $\W{{\rm c},ij}$ is given in \eq{eq:Wcij}, and 
again two reference particles have to be introduced. Finally, the soft-collinear integral has 
a similar structure and reads (with $r=r_{ij}$, $r'=r_{ijk}$)
\beq
\label{eq:I12_components_SC}
I^{\otwo}_{{\rm SC},ij}
& = & 
- \,
2 \,
\Norm \,
\mc E^{(i)}_{jr}
\bigg\{ \,
C_{f_{j}} \!
\sum_{\substack{c\neq i,j}}
\sum_{\substack{d\neq i,j,c}} \!
J_s(s_{cd}) \,
\bB^{(ijr)}_{cd}
-
C_{f_{j}}  
C_A \,
J_s(s_{ij}) \,
\bB^{(ijr)}
\\
&&
\hspace{17mm}
+ \,
C_A \,
\bigg[ \;
\sum_{\substack{c\neq i,j}}
J_s(s_{ic}) \,
\bB^{(jri)}_{[ij]c}
+
C_{f_{j}} \,
J_s(s_{ir}) 
\Big(
\bB^{(jri)} \!
-
\bB^{(ijr)}
\Big)
\bigg]
\nnb\\
&&
\hspace{17mm}
+ \,
(2 C_{\!f_{j}} \!-\! C_{\!A})
\bigg[
\sum_{\substack{c\neq i,j}} \!
J_s(s_{jc}) \,
\bB^{(irj)}_{[ij]c}
+
C_{f_{j}} \,
J_s(s_{jr}) 
\Big(
\bB^{(irj)} \!
-
\bB^{(ijr)}
\Big)
\bigg]
\bigg\}
\nnb\\
&&
\hspace{-3mm}
+ \,
2 \,
\Norm 
C_{f_{j}} 
\mc E^{(i)}_{jr}
\Big[
J_{\rm hc}^{\,i}(s_{ir})\,
\bB^{(ijr)}
+
J_{\rm hc}^{\,j}(s_{jr}) \,
\bB^{(irj)}
\Big]
+
2 \,
\Norm 
C_{f_{j}} 
\mc E^{(i)}_{jr'} \!\!
\sum_{k \neq i,j} \!\!
J_{\rm hc}^{\,k}(s_{kr'}) \,
\bB^{(ijr')} \!
\, .
\qquad
\nnb
\eeq
As already noted, a more compact expression can be obtained using symmetrised 
sector functions. We can write
\beq
\label{eq:I12}
I^{\otwo}
& = & 
\sum_{\substack{i,j > i}}
I^{\otwo}_{\{ij\}}
\, ,
\qquad\qquad
I^{\otwo}_{\{ij\}}
\; = \;
I^{\otwo}_{{\rm S},ij} \,
\Z{{\rm s},ij}
+ 
I^{\otwo}_{{\rm S},ji} \,
\Z{{\rm s},ji}
+ 
I^{\otwo}_{ {\rm HC},{ij} }
\, ,
\eeq
where the soft contributions are given by \eq{eq:I12_components_S} and the 
hard-collinear contribution $I^{\otwo}_{{\rm HC},{ij} }$ reads 
($r=r_{ij}$, $r'=r_{ijk}$) 
\beq
I^{\otwo}_{ {\rm HC},{ij} }
& = & 
I^{\otwo}_{ {\rm C},{ij} }
+
I^{\otwo}_{ {\rm C},{ji} }
-
I^{\otwo}_{ {\rm SC},{ij} }
-
I^{\otwo}_{ {\rm SC},{ji} }
\\
& = &
- \,
\Norm \,
\frac{P^{{\rm hc},\mu\nu}_{ij(r)}}{s_{ij}}
\bigg\{ \,
\sum_{\substack{c\neq i,j}}
\sum_{\substack{d\neq i,j,c}} \!
J_s(s_{cd}) \,
\bB^{(ijr)}_{\mu\nu,cd}
+
C_{f_{[ij]}} \,
\rho^{\scriptscriptstyle (C)}_{[ij]} \,
J_s(s_{ij}) \,
\bB^{(ijr)}_{\mu\nu}
\nnb\\
&&
\hspace{20mm}
+ \,
\bigg[ \;
\sum_{\substack{c\neq i,j}}
J_s(s_{ic}) \,
\Big(
\rho^{\scriptscriptstyle (C)}_{ij} \,
\bB^{(jri)}_{\mu\nu,[ij]c}
+
\tilde f_{\,ij}^{\,q\bar q} \,
{\cal \bB}^{(jri)}_{\mu\nu,[ij]c}
\Big)
\nnb\\
&&
\hspace{24mm}
+ \,
C_{f_{[ij]}} \,
\rho^{\scriptscriptstyle (C)}_{ij} \,
J_s(s_{ir}) \,
\Big(
\bB^{(jri)}_{\mu\nu}
-
\bB^{(ijr)}_{\mu\nu}
\Big)
+
(i \leftrightarrow j)
\bigg]
\bigg\}
\nnb\\
&&
\hspace{-3mm}
+ \,
\Norm \,
\frac{P_{ij(r)}^{{\rm hc},\mu\nu\!\!}}{s_{ij}} \,
\Big[
J_{\rm hc}^{\,i}(s_{ir}) \,
+
J_{\rm hc}^{\,j}(s_{jr}) \,
\Big]
\bB^{(ijr)}_{\mu\nu}
+
\Norm \!
\sum_{k \neq i,j} \!
\frac{P_{ij(r')}^{{\rm hc},\mu\nu\!\!}}{s_{ij}} \,
J_{\rm hc}^{\,k}(s_{kr'}) \,
\bB^{(ijr')}_{\mu\nu}
\nnb\\
&&
\hspace{-3mm}
- \,
2 \,
\Norm 
\bigg[
C_{f_{i}} 
\mc E^{(j)}_{ir} \,
J_{\rm hc}^{\,i}(s_{ir}) \,
\Big(
\bB^{(jri)}
-
\bB^{(jir)}
\Big)
+
C_{f_{j}} 
\mc E^{(i)}_{jr} \,
J_{\rm hc}^{\,j}(s_{jr}) \,
\Big(
\bB^{(irj)}
-
\bB^{(ijr)}
\Big)
\bigg]
\, . \;\;\;\;
\nnb
\eeq
This concludes the list of integrated counterterms
for double-real radiation. We now turn to the treatment of real-virtual contributions.


\section{The subtracted real-virtual contribution $RV_{\,\rm sub}$}
\label{RVsub}

Let us take stock of what we have achieved so far. After subtracting the appropriate 
combination of the local counterterms $K^{\one}$, $K^{\two}$ and $K^{\otwo}$ from 
the double-real squared matrix element $\RR$, and after adding back the corresponding
integrated counterterms, $I^{\one}$, $I^{\two}$ and $I^{\otwo}$, we can write a partially
subtracted expression for the differential distribution in \eq{pertO2}. It reads
\beq
\label{eq:sigma NNLO (1)}
\frac{d\sig_\NNLO}{dX} 
& = &
\int d \Phi_n \, \Big[ VV + I^{\two} \Big] \, \delta_n(X)
\\
&+ &
\int d \Phi_\npo \, 
\Big[
\left(RV + I^{\one} \right) \delta_\npo(X) - I^{\otwo} \,\delta_n(X)
\Big]\nnb
\\
& + &
\int d \Phi_\npt \, RR_{\,\rm sub}(X) 
\, .
\nnb
\eeq
Notice that no approximations have been made in reaching \eq{eq:sigma NNLO (1)}, 
since all local terms that were subtracted from \eq{pertO2} were added back exactly
in integrated form. 
At this stage, $RR_{\,\rm sub}$, given in \eq{eq:RRsub W} 
or in \eq{eq:RRsub Z}, is free of phase-space singularities in 
$\Phi_\npt$, and (evidently) does not contain explicit poles in $\eps$. 
Therefore it can be directly integrated in four dimensions, as desired. 
We now focus on the second line of \eq{eq:sigma NNLO (1)}. 
While the introduction of the integrated counterterm $I^{\one}$ exactly 
cancels the $\eps$ poles of $RV$ (in the same way as, at next-to-leading order, $I$ 
cancels the poles of $V$), new poles in $\eps$ are introduced through $I^{\otwo}$; 
on top of this, the combination in square brackets is still affected by 
phase-space singularities in $\Phi_\npo$. To be more precise, 
the second line of \eq{eq:sigma NNLO (1)} verifies now two crucial properties 
that follow from general cancellation theorems and from the definitions given so far. 
Specifically
\beq
\label{eq:RVproperties1and2}
(1)
& \quad & 
\big(RV + I^{\one} \big)\,\delta_\npo(X)
\; \to \;
\mbox{finite}
\, ,
\nnb\\
(2)
& \quad & 
I^{\one} \, \delta_\npo(X) - I^{\otwo} \, \delta_n(X)
\; \to \;
\mbox{integrable}
\, ,
\eeq
where the integrated couterterms are defined in \eq{eq:I1 (1)} and \eq{eq:I12 Wij}. 
The first property is expected from the KLN theorem: indeed, $I^{\one}$ is the
integral over the most unresolved radiation of $RR$, and its IR poles must 
compensate the virtual poles arising when one of the two unresolved particles
becomes virtual, while the other one is unaffected. These are precisely the 
poles of $RV$. To check this, which provides a test of the results obtained so far, 
it is  sufficient to perform the $\eps$ expansion of $I^{\one}$, as given in 
\eq{eq:I1 (1)}, writing
\beq
\label{eq:I1 (2)}
I^{\one}
& = &
I^{\one}_{\rm poles}
+
I^{\one}_{\rm fin}
+
\mc O(\eps)
\, .
\eeq
Performing the sum over sectors in $I^{\one}_{\rm poles}$, we get
\beq
\label{eq:I1 poles}
I^{\one}_{\rm poles}
& = &
\frac{\alpha_s}{2\pi} \,
\bigg[
\frac{1}{\eps^2} \,
\Sigma_{\!_C} \, 
R
+
\frac{1}{\eps} \, 
\bigg(
\Sigma_{\gamma} \, R
+
\sum_{c,d \neq c} \!
L_{cd} \, 
R_{cd}
\bigg)
\bigg]
\; = \;
- \, RV_{\rm poles}
\, .
\eeq
Keeping the complete dependence on sector functions in $I^{\one}_{\rm fin}$, 
we have
\beq
\label{eq:I1 fin}
I^{\one}_{\rm fin}
& = &
\sum_{i,j\neq i}
I^{\one}_{{\rm fin},ij} \, \W{ij}
\, = \,
\sum_{i,j > i}
I^{\one}_{{\rm fin},ij} \, \Z{ij}
\, ,
\\
I^{\one}_{{\rm fin},ij}
& = &
\frac{\alpha_s}{2\pi} \,
\bigg[ 
\bigg(
\Sigma_{\phi}
-
\sum_{k}
\gamma^{\rm hc}_{k} \, 
L_{kr}
\bigg) 
R
+
\! \sum_{c,d \neq c} \!
L_{cd}  
\bigg( 2 - \frac{1}{2} \, L_{cd} \bigg) 
R_{cd}
\bigg]
\, ,
\qquad
r = r_{ijk}
\, .
\nnb
\eeq
In Eqs.~(\ref{eq:I1 poles})-(\ref{eq:I1 fin}), $L_{ab} = \log(s_{ab}/\mu^2)$, and the 
numerical coefficients are given in Eqs.~(\ref{Cascomb})-(\ref{hccoeff}).
One easily verifies that $I^{\one}_{\rm poles}$
matches the explicit poles of the real-virtual matrix element 
$RV_{\rm poles}$, which have the well-known universal NLO structure (see 
for example \cite{Giele:1993dj,Catani:1998bh}), upon replacing the $n$-point 
amplitude with the $(\npo)$-point amplitude.

In order to prove the second property in \eq{eq:RVproperties1and2}, we start 
from the decompositions of Eqs.~(\ref{eq:I1 (1)})-(\ref{eq:I12 Wij}) in terms of the 
sector fuctions $\W{ij}$ and write 
\beq
I^{\!\one} \delta_{\npo\!}(X) - I^{\!\otwo} \delta_{n}(X)
\, = \,
\!\!\! \sum_{\substack{i,j \ne i}} \!
\Big\{
I^{\one}_{ij} \W{ij} \, 
\delta_{\npo\!}(X)
- 
\Big[
I^{\!\otwo}_{{\rm S},ij} \, \W{{\rm s},ij}
+
I^{\!\otwo}_{{\rm C},ij} 
-
I^{\!\otwo}_{{\rm SC},ij}
\Big]
\delta_n(X)
\Big\}
,
\qquad
\eeq
where the NLO sector functions $\W{ij}$ and $\W{{\rm s},ij}$ are defined in 
\eq{sectfunc} and \eq{eq:Wsij} respectively.
The second property in \eq{eq:RVproperties1and2} is thus satisfied at the 
level of single sectors $\W{ij}$ owing to the relations 
\beq
\label{I1-I12-integrable}
\bS{i}\,
\Big[
I^{\one}_{ij} \, \W{ij}
-
I^{\otwo}_{{\rm S},ij} \, \W{{\rm s},ij}
\Big]
& \to &
\mbox{integrable}
\, ,
\qquad \, \hspace{10mm}
\bS{i}\,
\Big[
I^{\otwo}_{{\rm C},ij}
-
I^{\otwo}_{{\rm SC},ij}
\Big]
\; \to \;
\mbox{integrable}
\, , \qquad
\nnb \\
\bC{ij}\,
\Big[
I^{\one}_{ij} \, \W{ij}
-
I^{\otwo}_{{\rm C},ij}
\Big]
& \to &
\mbox{integrable}
\, ,
\qquad
\bC{ij}\,
\Big[
I^{\otwo}_{{\rm S},ij} \, \W{{\rm s},ij}
-
I^{\otwo}_{{\rm SC},ij}
\Big]
\; \to \;
\mbox{integrable}
\, . \qquad
\eeq
For concreteness, consider the first relation. Under soft limit, the 
$(\npo)$-particle matrix element in $I^{\one}_{ij}$ returns a sum of products 
of eikonal factors and Born-level, colour-correlated matrix elements, and its 
sector function $\W{ij}$ becomes equal to $\W{{\rm s},ij}$. 
At the same time, when the operator $\bS{i}$ acts on 
$I^{\otwo}_{{\rm S},ij}$, it effectively removes the phase-space 
mappings, so that \eq{eq:I12_components_S} tends to the $\bS{i}$ limit of 
the square parenthesis in \eq{eq:I1 (1)}, up to the overall sign.
Similar steps show the validity of the other relations in 
\eq{I1-I12-integrable}.

At this point, on the one hand we have shown that the combination $\big(RV + 
I^{\one}\big)\,\delta_\npo(X)$ is free of explicit poles, but it still contains phase-space 
singularities. On the other hand, we have proven that $I^{\one} \, \delta_\npo(X) 
- I^{\otwo} \, \delta_n(X)$ is integrable in $\Phi_\npo$, but may still contain 
poles in $\eps$. In order to build a fully subtracted real-virtual matrix element 
$RV_{\,\rm sub}$, free of poles in $\eps$ and integrable in the whole $(\npo)$-body 
phase space, we need to define, in each sector $ij$, a real-virtual counterterm 
$K^{\RV}_{ij}$ satisfying the two further properties 
\beq
\label{eq: prop KRV}
(3)
& \quad & 
K^{\RV}_{ij} + I^{\otwo}_{ij}
\; \to \;
\mbox{finite}
\, ,
\nnb\\
(4)
& \quad & 
RV \, \W{ij} \, \delta_\npo(X) - K^{\RV}_{ij} \, \delta_n(X)
\; \to \;
\mbox{integrable}
\, .
\eeq
With a real-virtual counterterm satisfying the two properties in \eq{eq: prop KRV}, the 
subtracted real-virtual contribution to the cross section, defined in \eq{eq:NNLO_RV_sub}, 
is manifestly finite and integrable in $\Phi_\npo$. To
construct $RV_{\,\rm sub}$ explicitly, we rewrite it here as a sum over sectors:
\beq
\label{eq:RVsub W}
RV_{\,\rm sub}(X) 
& = & 
\sum_{\substack{i,j \ne i}}
\bigg[
\Big( RV + I^{\one}_{ij} \Big) \W{ij} \,
\delta_\npo(X)
-
\Big( K^{\RV}_{ij} + I^{\otwo}_{ij} \Big) \,
\delta_n(X)
\bigg]
\, .
\eeq
Thanks to the presence of sector functions, the second condition of \eq{eq: prop KRV}
actually simplifies to
\beq
RV \, \W{ij} \, \delta_\npo(X) - K^{\RV}_{ij} \, \delta_n(X)
\; \to \;
\mbox{integrable in the limits $\bS{i}$, $\bC{ij}$}
\, .
\eeq
In order to find a suitable definition for $K^{\RV}_{ij}$, satisfying the required properties, 
we start by introducing soft and collinear improved limits, $\bbS{i}$ and $\bbC{ij}$, for the
real-virtual squared matrix element. On the one hand, these limits must reproduce the 
singular behaviour of $RV$, so that
\beq
\bS{i}\,
\Big[
\left( 1 - \bbS{i} \right)
RV \, \W{ij}
\Big]
& \to &
\mbox{integrable}
\, ,
\quad \hspace{2mm}
\bS{i}\,
\Big[
\bbC{ij} \left( 1 - \bbS{i} \right)
RV \, \W{ij}
\Big]
\; \to \;
\mbox{integrable}
\, , \qquad
\nnb \\
\bC{ij}\,
\Big[
\left( 1 - \bbC{ij} \right)
RV \, \W{ij}
\Big]
& \to &
\mbox{integrable}
\, ,
\quad
\bC{ij}\,
\Big[
\bbS{i} \left( 1 - \bbC{ij} \right)
RV \, \W{ij}
\Big]
\; \to \;
\mbox{integrable}
\, . \qquad
\eeq
On the other hand, the improved limits must feature appropriate mappings, such
that they fulfil momentum conservation and on-shell conditions for the Born-level 
particles, and, at the same time, they simplify as much as possible the 
analytic integration over the radiation phase space. Following the discussion 
presented at NLO, and the choices made in Ref.~\cite{Magnea:2020trj}, 
we introduce
\beq
\label{eq:barred limits RV}
\bbS{i} \, RV \, \W{ij}
& \equiv &
- \, 
\Norm
\sum_{\substack{c \neq i \\ d \neq i,c}}
\bigg[
\mc E_{cd}^{(i)} \, 
\bV_{cd}^{(icd)}
- 
\frac{\as}{2\pi} 
\bigg(
\tilde{\mc E}_{cd}^{(i)} \, 
+
\mc E_{cd}^{(i)} \,
\frac{\beta_0}{2\eps}
\bigg)
\bB_{cd}^{(icd)}
+
\as \!\!\!
\sum_{\substack{e \neq i,c,d}} \!\!
\tilde{\mc E}_{cde}^{(i)} \, 
\bB_{cde}^{(icd)}
\;
\bigg]
\,
\W{{\rm s},ij}
\, ,
\nnb \\
\bbC{ij} \, RV \, \W{ij}
& \equiv &
\frac{\Norm}{s_{ij}} \,
\bigg[
P_{ij(r)}^{\mu\nu}\,\bV_{\mu\nu}^{(ijr)}
+
\frac{\as}{2\pi} \, 
\bigg(
\tilde P_{ij(r)}^{\mu\nu} \!
-
P_{ij(r)}^{\mu\nu}
\frac{\beta_0}{2\eps}
\bigg) 
\bB_{\mu\nu}^{(ijr)}
\bigg]
\W{{\rm c},ij}
\, ,
\nnb\\
\bbS{i} \, \bbC{ij} \, RV \, \W{ij}
& \equiv &
2 \,
\Norm \,
C_{f_j} \,
\bigg[
\mc E^{(i)}_{jr}
\bV^{(ijr)}
-
\frac{\as}{2\pi}
\bigg(
\tilde{\mc E}^{(i)}_{jr} \, 
+
\mc E^{(i)}_{jr} \,
\frac{\beta_0}{2\eps}
\bigg)
\bB^{(ijr)}
\bigg]
\, ,
\qquad\qquad
r = r_{ij}
\, .
\eeq
The kernels $\mc E^{(i)}_{cd}$ and $P_{ij(r)}^{\mu\nu}$ are the eikonal and 
collinear kernels from tree-level factorisation, introduced already at NLO, 
and given in \eq{eikoker} and in \eq{eq:appsinglecoll}, respectively. 
In addition, $\tilde{\mc E}^{(i)}_{cd}$, $\tilde{\mc E}_{cde}^{(i)}$ 
and $\tilde P_{ij(r)}^{\mu\nu}$ are the genuine real-virtual soft and 
collinear kernels \cite{Bern:1999ry,Catani:2000pi}, presented here 
in \eq{tildeik} and in \eq{tildAP}, respectively.

Since the combination $ ( 1 - \bS{i}) ( 1 - \bC{ij} ) \, RV \, \W{ij} $ is integrable 
everywhere in $\Phi_\npo$, one would expect to define the counterterm $K^{\RV}_{ij}$ 
simply as an NLO-like  combination of improved limits, namely 
\beq
\label{KRVexpected}
K^{\RV}_{ij , \, {\rm  expected}}
\, = \, 
\Big[ \bbS{i} + \bbC{ij} \left( 1 -  \bbS{i} \right) \Big] 
\, 
RV \, 
\W{ij} 
\, .
\eeq
Although such a choice preserves the minimal structure of the real-virtual counterterm, 
and automatically fulfils the condition (4) of \eq{eq: prop KRV}, explicit computations show 
that it spoils the condition (3) of \eq{eq: prop KRV}. In principle, it would have been natural 
to expect that the poles of \eq{KRVexpected} would cancel those of $I^{\otwo}_{ij}$. 
Indeed, the poles of \eq{KRVexpected} are designed to match the poles of $RV$ that 
are accompanied by phase-space singularities. At the same time, $I^{\otwo}_{ij}$ is the 
result of integrating the strongly-ordered counterterm over the phase space of the most 
unresolved radiation: thus, it collects precisely terms that have phase-space singularities 
in the remaining radiation, as well as poles that should match their virtual counterpart, 
given by $RV$. On the other hand, there are subtleties that prevent the poles of  $I^{\otwo}_{ij}$ 
from matching exactly those of $K^{\RV}_{ij,{\rm  expected}}$. The first subtlety stems from 
the specific phase-space mappings one has to adopt in order to define the improved limits 
in \eq{eq:barred limits RV}. Since such contributions are affected by both double poles in 
$\eps$ and by phase-space singularities, they feature single poles in $\eps$ with coefficients 
depending on kinematic invariants. This generates a mismatch: in fact, we notice that in 
Eqs.~(\ref{eq:I12_components_S})-(\ref{eq:I12_components_SC}) the residues of the poles 
in $I^{\otwo}_{ij}$ that depend on kinematics are proportional to logarithms of Lorentz invariants 
constructed with {\it unmapped} momenta, \emph{i.e.}~with $(\npo)$-body kinematics. In 
contrast, the residues of the poles in the real-virtual improved limits of \eq{eq:barred limits RV} 
can also depend on logarithms of {\it mapped} invariants, obtained via momentum mappings 
from the $(n+1)$- to the $n$-particle phase space. This is the case, for instance, for the 
virtual component of the soft limit: the pole content of $\bV_{cd}^{(icd)}$ includes terms 
of the type $\log \big(\bar{s}_{ef}^{(icd)}/\mu^2 \big)$, which cannot appear in $I^{\otwo}_{ij}$. 
More involved mismatches occur in the collinear sector, where the kinematics of the poles 
of $I^{\otwo}_{ij}$ fails to match that of $K^{\RV}_{ij,{\rm  expected}}$ out of the collinear 
region, irrespectively of mappings.

The fact that all discrepancies in the single pole in $\eps$ disappear in the singular regions 
of phase space, as they must, gives us the possibility to refine the definition of $K^{\RV}_{ij,{\rm  
expected}}$, by adding back precisely the mismatched terms, thus obtaining the desired cancellation 
of the $I^{\otwo}_{ij}$ poles, without introducing new phase-space singularities. Schematically, 
we define
\beq
\label{eq:KRVij}
K^{\RV}_{ij} 
\, \equiv \, 
K^{\RV}_{ij , \, {\rm  expected}}
+ 
\Delta_{ij} 
\, = \, 
\Big[ \,
\bbS{i} + \bbC{ij} \left( 1 - \bbS{i} \right)
\Big]
RV \, \W{ij}
+
\Delta_{ij}
\,.
\eeq
The extra term $\Delta_{ij}$ appearing in \eq{eq:KRVij} is required not to spoil condition 
(4) of \eq{eq: prop KRV}, and therefore cannot have any phase-space singularity in the 
limits $\bS{i}$ and $\bC{ij}$. Thus we impose that
\beq
\label{eq:Delta_finite_limits}
\bS{i}\, \Delta_{ij}
\; \to \;
\mbox{integrable}
\, ,
\qquad
\bC{ij}\, \Delta_{ij}
\; \to \;
\mbox{integrable}
\, .
\eeq
At the same time, $\Delta_{ij}$ has the crucial role of matching the explicit 
$\eps$ poles of $I^{\otwo}_{ij}$, implying the finiteness of the 
combination $K^{\RV}_{ij} + I^{\otwo}_{ij}$, in agreement with condition (3) 
of \eq{eq: prop KRV}. 
In practice, we introduce for $\Delta_{ij}$ a decomposition into soft, 
collinear and soft-collinear components, along the lines discussed for 
$I^{\otwo}_{ij}$ in Eq.\eqref{eq:I12 Wij}, and 
we write
\beq
\label{eq:structDeltaij}
\Delta_{ij}
& \equiv &
\Delta_{{\rm S},i} \, \W{{\rm s},ij}
+
\Delta_{{\rm C},ij}
-
\Delta_{{\rm SC},ij}
\, .
\eeq
Using this decomposition, 
the properties \eq{eq:Delta_finite_limits} can be better detailed, and read 
\beq
\label{eq:Delta_finite_limits 2}
\bS{i} \, 
\Delta_{{\rm S},i} \, \W{{\rm s},ij}
\; \to \;
\mbox{integrable}
\, ,
&\qquad& \hspace{8.6mm}
\bS{i} \big( \Delta_{{\rm C},ij} - \Delta_{{\rm SC},ij} \big)
\; \to \;
\mbox{integrable}
\, ,
\nnb\\
\bC{ij} \, \Delta_{{\rm C},ij} 
\; \to \;
\mbox{integrable}
\, ,
&\qquad&
\bC{ij} \big( \Delta_{{\rm S},i} \, \W{{\rm s},ij} - \Delta_{{\rm SC},ij} \big)
\; \to \;
\mbox{integrable}
\, .
\eeq
Furthermore, 
we can enforce the desired cancellation between $K^{\RV}_{ij}$ 
and $I^{\otwo}_{ij}$ for each component. 
Specifically, we require that
\beq
\label{eq:Delta_poles}
\Big[ 
\bbS{i} \, RV \, \W{ij} 
+
\Big( \Delta_{{\rm S},i} + I^{\otwo}_{{\rm S},ij} \Big)
\, \W{{\rm s},ij} \Big]_{\rm poles} & = & 0 \, ,
\nnb \\
\Big[ 
\bbC{ij} \, RV \, \W{ij} 
+
\Big( \Delta_{{\rm C},ij} + I^{\otwo}_{{\rm C},ij} \Big)
\Big]_{\rm poles} & = & 0 \, ,
\nnb \\
\Big[ 
\bbS{i} \, \bbC{ij} \, RV \, \W{ij} 
+
\Big( \Delta_{{\rm SC},ij} + I^{\otwo}_{{\rm SC},ij} \Big)
\Big]_{\rm poles}
& = &
0
\, .
\eeq
Since the pole parts of both $I^{\otwo}_{ij}$ and $K^{\RV}_{ij , \, {\rm  expected}}$ are 
explicitly known, the necessary compensating terms are easily determined. 
An expression for the three components of $\Delta_{ij}$ can be constructed 
in a minimal way by considering all and only the single poles of $I^{\otwo}_{ij}$ with 
mismatching kinematics. Since they consist in differences of logarithms, or differences
of Born matrix elements (which vanish in the soft or collinear limit), we decided to 
promote the differences of logarithms to ratios of scales, raised to a power vanishing 
with $\eps$. This non-minimal structure simplifies subsequent integrations, and it 
only affects finite parts, without introducing new phase-space singularities.
Beginning with the soft term $\Delta_{{\rm S},i}$, we define 
\beq
\label{eq:DeltaSi}
\Delta_{{\rm S},i}
& = &
- \,
\frac{\alpha_s}{2 \pi} \,
\Norm \!\!
\sum_{\substack{c\neq i \\ d\neq i,c}} \!
\mc E^{(i)}_{cd} \,
\Bigg\{ \;\;
\frac{1}{2\eps^{2}} \!\!\!
\sum_{\substack{e\neq i,c \\ f\neq i,c,e}} \!
\bigg[ \bigg(\frac{s_{ef}}{\sk{ef}{icd}}\bigg)^{\!\!-\eps} \!\! -1 \bigg] \,
\bB^{(icd)}_{efcd}
+
\frac{1}{\eps^{2}} \!
\sum_{\substack{e\neq i,d}} 
\bigg[ \bigg(\frac{s_{ed}}{\sk{ed}{icd}}\bigg)^{\!\!-\eps} \!\! -1 \bigg] \,
\bB^{(icd)}_{edcd}
\nnb\\[-2mm]
&&
\hspace{26mm}
+ \,
\bigg[
\bigg( 
\frac{1}{\eps^2} 
+ 
\frac{2}{\eps} 
\bigg)
2 \, C_{\!f_{c}} 
+
\frac{\gamma^{\rm hc}_{c}}{\eps}
\bigg]
\Big( \bB^{(icd)}_{cd} - \bB^{(idc)}_{cd} \Big)
\Bigg\}
\nnb\\
&&
- \,
\frac{\as}{2\pi}\,
\Norm \!\!
\sum_{\substack{k\neq i \\ c\neq i,k,r}} \!
\mc E^{(i)}_{cr} \,
\frac{\gamma^{\rm hc}_{k}}{\eps}
\Big(
\bB^{(irc)}_{cr}
-
\bB^{(icr)}_{cr}
\Big)
\, ,
\qquad\qquad\qquad\qquad\qquad
r = r_{ik}
\, .
\eeq
Thanks to the fact that in the soft limit the mapped momenta coincide 
with the unmapped ones, the first \eq{eq:Delta_finite_limits 2} is 
fulfilled in a straightforward way. 
The first relation in \eq{eq:Delta_poles} is less evident, but can be 
proven by simply performing the $\eps$ expansion of 
$\bbS{i} \, RV$, $\Delta_{{\rm S},i}$ and 
$I^{\otwo}_{{\rm S},ij}$.
For the collinear component, we define 
($r = r_{ij}$, $r' = r_{ijk}$)
\beq
\label{eq:DeltaCij}
\Delta_{{\rm C},ij}
& = &
\frac{\as}{2\pi}\,
\Norm \,
\frac{P_{ij(r)}^{\mu\nu}}{s_{ij}} \,
\frac{1}{\eps^2} 
\sum_{\substack{c\neq i,j}}
\Bigg\{ \,
\sum_{\substack{d\neq i,j,c}} \!
\bigg[ \bigg(\frac{s_{cd}}{\sk{cd}{ijr}}\!\bigg)^{\!\!-\eps} - 1 \bigg] \,
\bB_{\mu\nu,cd}^{(ijr)}
+
2 \,
\bigg[
1 -
\bigg(\!\frac{\sk{jc}{ijr}}{s_{[ij]r}}\!\bigg)^{\!\!-\eps} \, 
\bigg]
\bB_{\mu\nu,[ij]c}^{(ijr)}
\nnb\\
&&
\hspace{33mm} 
+ \,
\bigg\{\;
\rho^{\scriptscriptstyle (C)}_{ij} \,
\bigg[ 
\bigg(\!\frac{\sk{ic}{jri}}{\sk{ir}{jri}}\!\bigg)^{\!\!-\eps} 
-
\bigg(\!\frac{s_{ir}\sk{ic}{jri}}{\sk{ir}{jri}s_{ic}}\!\bigg)^{\!\!-\eps} \,
\bigg]
\bB_{\mu\nu,[ij]c}^{(jri)} \nnb \\
&&
\hspace{37mm} 
+ \,
\tilde f_{\,ij}^{\,q\bar q} \,
\bigg[ 
\bigg(\!\frac{\sk{ic}{jri}}{\mu^2}\!\bigg)^{\!\!-\eps} 
-
\bigg(\!\frac{\sk{ic}{jri}}{s_{ic}}\!\bigg)^{\!\!-\eps} \,
\bigg]
{\cal \bB}_{\mu\nu,[ij]c}^{(jri)}
+ 
(i \leftrightarrow j)
\bigg\} \!
\Bigg\}
\,
\W{{\rm c},ij(r)}
\nnb\\
& + &
\frac{\as}{2\pi} \,
\Norm \!
\sum_{\substack{k \neq i,j}}
\bigg(
\frac{\gamma^{\rm hc}_{k}}{\eps}
+
\phi^{\rm hc}_{k}
\bigg)
\Bigg[
\frac{P_{ij(r)}^{\mu \nu}}{s_{ij}} \,
\bB^{(ijr)}_{\mu \nu}
\,
\W{{\rm c},ij(r)}
-
\frac{P_{ij(r')}^{\mu \nu}}{s_{ij}} \,
\bB^{(ijr')}_{\mu \nu}
\,
\W{{\rm c},ij(r')}
\Bigg]
\, ,
\eeq
where $\rho^{\scriptscriptstyle (C)}_{ij}$, $\tilde f_{\,ij}^{\,q\bar q}$, 
$\gamma^{\rm hc}_{k}$, $\phi^{\rm hc}_{k}$ and ${\cal \bB}$ are defined in 
\appn{app:notations}, and $\W{{\rm c},ij(r)}$ is given in \eq{eq:Wcij}.
The third \eq{eq:Delta_finite_limits 2} can be verified by considering that in 
the collinear limit $\bC{ij}$ we have
\beq
\kk{j}{ijr},
\kk{i}{jri}
\xrightarrow{\bC{ij}}
k_{[ij]}
,
\qquad
\kk{r}{ijr},
\kk{r}{jri}
\xrightarrow{\bC{ij}}
k_{r}
,
\qquad
\kk{c}{ijr},
\kk{c}{jri}
\xrightarrow{\bC{ij}}
k_{c}
.
\eeq
Again the second \eq{eq:Delta_poles} can be proven upon expansion in $\eps$. 
Finally for the soft-collinear component we introduce (with $r = r_{ij}$, $r' = r_{ijk}$)
\beq
\label{eq:DeltaSCij}
\Delta_{{\rm SC},ij}
& = &
\frac{\as}{2\pi} \,
2 \,
\Norm \,
C_{f_j} \,
\mc E^{(i)}_{jr} \,
\frac{1}{\eps^2} 
\sum_{\substack{c\neq i,j}}
\Bigg\{ \hspace{-1.7pt}
\sum_{\substack{d\neq i,j,c}} \!
\bigg[ \bigg( \frac{s_{cd}}{\sk{cd}{ijr}}\!\bigg)^{\!\!-\eps} - 1 \bigg] \,
\bB_{cd}^{(ijr)}
+
 2 \,
\bigg[ 
 \bigg( \! \frac{s_{jr}}{s_{[ij]r}} \! \bigg)^{\!\!-\eps}
\!-\! 
\bigg( \! \frac{\sk{jc}{ijr}}{s_{[ij]r}} \! \bigg)^{\!\!-\eps} \,
\bigg]
\bB_{[ij]c}^{(ijr)}
\nnb \\
&&
\hspace{5mm}
+ \,
\frac{C_{\!A}}{C_{\!f_{j}}} 
\bigg[ 
\bigg( \frac{\sk{ic}{jri}}{\sk{ir}{jri}}\!\bigg)^{\!\!\!-\eps} \!\!\!
-
\!\bigg( \frac{s_{ir}\sk{ic}{jri}}{\sk{ir}{jri}s_{ic}}\!\bigg)^{\!\!\!-\eps}
\bigg]
\bB_{[ij]c}^{(jri)}
+ \,
\frac{2 \, C_{\!f_{j}} \!\!-\! C_{\!A}}{C_{f_{j}}} 
\bigg[ 
\bigg( \frac{\sk{jc}{irj}}{\sk{jr}{irj}}\!\bigg)^{\!\!\!-\eps} \!\!\!
-
\!\bigg( \frac{s_{jr}\sk{jc}{irj}}{\sk{jr}{irj}s_{jc}}\!\bigg)^{\!\!\!-\eps} 
\bigg]
\bB_{[ij]c}^{(irj)}
\Bigg\}
\nnb\\
&&
\hspace{-3mm}
+ \,
\frac{\as}{2\pi} \,
2 \,
\Norm \,
C_{f_j} 
\bigg[
J_{\rm hc}^{\,j}(s_{jr}) \,
\mc E^{(i)}_{jr}
\Big(
\bB^{(ijr)} \!
-
\bB^{(irj)} 
\Big)
\!+
\!\!\!
\sum_{\substack{k\neq i,j}} \!
\bigg(
\frac{\gamma^{\rm hc}_{k}}{\eps}
+
\phi^{\rm hc}_{k}
\bigg) \!
\bigg(
\mc E^{(i)}_{jr} 
\bB^{(ijr)}
-
\mc E^{(i)}_{jr'} 
\bB^{(ijr')} 
\bigg) \!
\bigg]
 .
\nnb\\
\eeq
With the latter definition we are able to prove the second and the fourth  
\eq{eq:Delta_finite_limits 2}, by exploiting the colour algebra of the 
colour-connected matrix elements, and the cancellation of the $\eps$ poles 
in the third \eq{eq:Delta_poles}.
The explicit expression of the components of $\Delta_{ij}$ in \eq{eq:structDeltaij}
completes the list of definitions required to implement the subtracted real-virtual 
squared  matrix element $RV_{\,\rm sub}$. 
Given its finiteness in $d=4$, we can now rewrite \eq{eq:RVsub W} as
\beq
\label{eq:RVsubW_fin}
RV_{\,\rm sub}(X) 
& = & 
\sum_{\substack{i,j \ne i}}
\bigg[
\Big( RV_{\rm fin} + I^{\one}_{{\rm fin},ij} \Big) \W{ij} \,
\delta_\npo(X)
-
\Big( K^{\RV}_{{\rm fin},ij} + I^{\otwo}_{{\rm fin},ij} \Big) \,
\delta_n(X)
\bigg]
\, ,
\eeq
where the subscript emphasises that, at this stage, all the explicit poles 
have already been cancelled. 
The finite component $I^{\one}_{{\rm fin},ij}$ is given in \eq{eq:I1 fin}, 
while $I^{\otwo}_{{\rm fin},ij}$ can easily be derived 
from Eqs.~\eqref{eq:I12_components_S}-\eqref{eq:I12_components_SC}.
Finally, we obtain the finite contribution $K^{\RV}_{{\rm fin},ij}$ by computing the
expansion in powers of $\eps$ of the sum of Eqs.~\eqref{eq:barred limits RV} and 
\eqref{eq:DeltaSi}-\eqref{eq:DeltaSCij}. 
We refrain from giving here the explicit expression for the quantities in 
\eq{eq:RVsubW_fin}, as we will derive a more compact result for $RV_{\,\rm sub}(X)$, 
in terms of symmetrised sector functions in the next section.


\subsection{\vspace{-.5mm}$RV_{\,\rm sub}$ with symmetrised sector functions}

\vspace{.5mm}
In the previous section we presented the construction of the subtracted real-virtual 
matrix element. We started by introducing the general properties of $RV_{\,\rm sub}$,
and we discussed the main steps necessary to provide an explicit form for all the 
terms that contribute to its definition, according to \eq{eq:RVsub W}. We then
proved that $RV_{\,\rm sub}$ is free of both explicit poles and phase-space 
singularities in each $\W{ij}$ sector separately. 
As was mentioned in \secn{FrameNLO} and in \secn{RRsubSymSect}, however, 
one can improve the numerical performance of the scheme by appropriately 
symmetrising the sector functions.
In this section we present explicit expressions for $RV_{\,\rm sub}$ in terms of
symmetrised sector functions.

In analogy to the procedure applied at NLO in \eq{eq:RsubZ}, and later generalised 
to $RR_{\,\rm sub}$ in \secn{RRsubSymSect}, 
we rewrite the real-virtual counterterm $K^{\RV}$ in terms of the symmetrised  
sector counterterms $K^{\RV}_{\{ij\}}$, defined as 
\beq
\label{eq:KRV}
K^{\RV}_{\{ij\}} 
\; = \;
K^{\RV}_{ij} 
+
K^{\RV}_{ji} 
\, ,
\qquad\qquad
K^{\RV}
\; = \;
\sum_{\substack{i,j \ne i}}
K^{\RV}_{ij}
\; = \;
\sum_{\substack{i,j > i}}
K^{\RV}_{\{ij\}}
\, .
\eeq
Starting from  Eq.~\eqref{eq:RVsubW_fin}, it is then straightforward to obtain 
\beq
\label{eq:RVsub}
RV_{\,\rm sub}(X) 
& = & 
\sum_{\substack{i,j > i}}
\bigg\{
\Big[
RV_{\rm fin}  + I^{\one}_{{\rm fin},ij}
\Big]
\Z{ij}
\,
\delta_\npo(X)
+
\Big[
K^{\RV}_{{\rm fin},\{ij\}} + I^{\otwo}_{{\rm fin},\{ij\}}
\Big] \,
\delta_n(X)
\bigg\}
\, ,
\eeq
with $I^{\one}_{{\rm fin},ij}$ given in \eq{eq:I1 fin}.
To present explicitly the other finite terms featuring in 
Eq.~\eqref{eq:RVsub}, we organise them in terms of soft and hard-collinear 
components, writing
\beq
K^{\RV}_{{\rm fin},\{ij\}} + I^{\otwo}_{{\rm fin},\{ij\}}
& = &
\IK^{\otwoRV}_{{\rm S},ij} \,
\Z{{\rm s},ij}
+
\IK^{\otwoRV}_{{\rm S},ji}
\Z{{\rm s},ji}
+
\IK^{\otwoRV}_{{\rm HC},{ij}}
\, ,
\eeq
where the soft limit of the symmetrised sector functions, $\Z{{\rm s},ij}$, 
is defined in \eq{eq:Zsij}. 
The finite soft counterterm 
$\IK^{\otwoRV}_{{\rm S},ij}$
is obtained 
by combining \eq{eq:I12_components_S} with Eqs.~(\ref{eq:barred limits RV}) 
and (\ref{eq:DeltaSi}), dropping the explicit poles. 
The result is extremely compact, and, except for the process-dependent finite 
part of the single-virtual squared matrix element, it displays only simple 
logarithmic dependence on the kinematics. 
We find 
($r = r_{ik}$, $r' = r_{ij}$, $r''=r_{ijk}$)
\beq
\IK^{\otwoRV}_{{\rm S},ij}
& = &
4 \, \as^{2} \!
\sum_{\substack{c \neq i \\ d\neq i,c}} \!
\mc E^{(i)}_{cd}
\Bigg\{
\sum_{\substack{e \neq i \\ f\neq i,e}} \!\!
\bigg( L_{ef} - \frac14 L_{ef}^{2} \bigg) \,
\bB^{(icd)}_{cdef}
+
2 \!
\sum_{\substack{e \neq i,d}} \!\!
\bigg( L_{ed} - \frac14 L_{ed}^{2} \bigg)
\Big( \bB^{(icd)}_{cded} - \bB^{(idc)}_{cded} \Big)
\nnb \\[-2mm]
&&
\hspace{21mm}
+ 
\sum_{\substack{e\neq i,d}} 
\ln^{2} \frac{\sk{de}{icd}}{s_{de}} \,
\bB^{(icd)}_{cded}
-
\frac12
\ln^{2} \frac{\sk{cd}{icd}}{s_{cd}} \,
\bB^{(icd)}_{cdcd}
-
2 \pi \!\!\!
\sum_{\substack{e \neq i,c,d}} \!\!
\ln\frac{s_{id}s_{ie}}{\mu^2\,s_{de}} \,
\bB_{cde}^{(icd)}
\nnb\\
&&
\hspace{21mm}
+ \,
\bigg[
\bigg( 6 - \frac{7}{2}\,\zeta_{2} \bigg)
\big(
\Sigma_{\!_C} \!+\! 2 C_{f_{d}} \!-\! 2C_{f_{c}}
\big)
+
\sum_{k} \,
\phi^{\rm hc}_{k}
-
\sum_{k \ne i} \,
\gamma^{\rm hc}_{k} \, L_{kr''}
-
\gamma^{\rm hc}_{i} \, L_{ir'}
\nnb\\[-2mm]
&&
\hspace{26mm}
+ \,
C_A
\bigg(
6 - \zeta_{2}
-
\ln\frac{s_{ic}}{s_{cd}}\ln\frac{s_{id}}{s_{cd}}
- 
2 \,
\ln \frac{s_{ic}s_{id}}{\mu^{2}s_{cd}}
\bigg)
\bigg] \, 
\bB_{cd}^{(icd)}
\Bigg\} 
\nnb \\
&&
\hspace{-4mm}
+ \,
4 \, \as^{2} \,
\sum_{k \neq i} 
\big( \phi^{\rm hc}_{k} \!- \gamma^{\rm hc}_{k} L_{kr''} \big) 
\bigg[ \;
\! \sum_{\substack{c\neq i,k}} \!
\mc E^{(i)}_{kc} 
\Big( \bB^{(ick)}_{kc} \!- \bB^{(ikc)}_{kc} \Big)
+
\!\! \sum_{\substack{c\neq i,k,r}} \!\!
\mc E^{(i)}_{cr} 
\Big( \bB^{(icr)}_{cr} \!- \bB^{(irc)}_{cr} \Big)
\bigg]
\nnb\\[-1mm]
&&
\hspace{-4mm}
+ \, 
8\pi \, \as \!
\sum_{\substack{c \neq i\\ d \neq i,c}}
\mc E_{cd}^{(i)} \, 
\bar{V}_{{\rm fin},cd}^{(icd)}
\, ,
\eeq
where $\bar{V}_{{\rm fin},cd}^{(icd)}$ is the finite part of the 
colour-correlated, single-virtual squared matrix element, expressed 
in the mapped kinematics.
We notice that, as happened for $I^{\otwo}_{{\rm S},ij}$, the presence of the 
reference particle $r'=r_{ij}$ introduces a dependence on 
the particle $j$ of the soft sector function $\Z{{\rm s},ij}$ which 
multiplies $\IK^{\otwoRV}_{{\rm S},ij}$. 

To conclude this section, we also report the finite hard-collinear 
counterterm 
$\IK^{\otwoRV}_{{\rm HC},ij}$, 
which is the result 
of summing Eqs.~(\ref{eq:I12_components_C}) and (\ref{eq:I12_components_SC}) 
with Eqs.~(\ref{eq:barred limits RV}), (\ref{eq:DeltaCij}), and (\ref{eq:DeltaSCij}). 
We find (with $r=r_{ij}$, $r'=r_{ijk}$) 
\beq
\label{eq:IK(12-RV)(HC)}
\IK^{\otwoRV}_{{\rm HC},ij}
& = &
4 \, \as^{2} \,
\frac{P_{ij(r)}^{{\rm hc},\mu\nu}}{s_{ij}}
\Bigg\{ \;
\! \sum_{\substack{c\neq i,j}}
\Bigg[
\ln^{2}\frac{\sk{jc}{ijr}}{s_{[ij]r}} \,
\bB_{\mu\nu,[ij]c}^{(ijr)}
-  
\frac12 \!
\sum_{d\neq i,j,c} \!\!
\big( 4 L_{cd} \!-\! L_{cd}^{2} \big) \,
\bB^{(ijr)}_{\mu\nu,cd}
\Bigg]
\nnb\\
&&
\hspace{2cm}
- \,
\!\! \sum_{\substack{c\neq i,j,r}}
\Bigg[
\ln^{2} \frac{\sk{cr}{ijr}}{s_{cr}} \,
\bB_{\mu\nu,cr}^{(ijr)}
+
\frac{\rho^{\scriptscriptstyle (C)}_{ij}}{2} \,
{\mc L}_{ijcr} \,
\bB_{\mu\nu,[ij]c}^{(jri)}
+
\frac{\rho^{\scriptscriptstyle (C)}_{ji}}{2} \,
{\mc L}_{jicr} \,
\bB_{\mu\nu,[ij]c}^{(irj)}
\Bigg]
\nnb\\
&&
\hspace{2cm}
- \,
\frac{1}{2} 
\sum_{\substack{c\neq i,j}}
\tilde f_{\,ij}^{\,q\bar q} \,
\Big(
\tilde{\mc L}_{ijcr} \,
{\cal \bB}_{\mu\nu,[ij]c}^{(jri)}
-
\tilde{\mc L}_{jicr} \,
{\cal \bB}_{\mu\nu,[ij]c}^{(irj)}
\Big)
\nnb\\
&&
\hspace{17mm}
- \,
\bigg[
\bigg( 6 - \frac{7}{2}\,\zeta_{2} \bigg)
\Big( \Sigma_{\!_C} \!-\! C_{f_{[ij]}} \rho^{\scriptscriptstyle (C)}_{[ij]} \Big)
+
C_{f_{[ij]}} \frac{\rho^{\scriptscriptstyle (C)}_{[ij]}}{2} \,
\big( 4 L_{ij} \!-\! L_{ij}^{2} \big)
\nnb\\
&&
\hspace{22mm}
- \,
C_{f_{[ij]}} \frac{\rho^{\scriptscriptstyle (C)}_{ij}}{2} \,
\big( 4 L_{ir} \!-\! L_{ir}^{2} \big)
-
C_{f_{[ij]}} \frac{\rho^{\scriptscriptstyle (C)}_{ji}}{2} \,
\big( 4 L_{jr} \!-\! L_{jr}^{2} \big)
+
\Sigma_{\phi}^{\rm hc}
\bigg]
\bB^{(ijr)}_{\mu\nu}
\Bigg\}
\nnb\\
& - & 
4 \, \as^{2} \,
\bigg[ 
2\,
C_{f_j} \,
\mc E^{(i)}_{jr} \,
C_{f_{[ij]}}
\ln^{2}\!\frac{s_{jr}}{s_{[ij]r}} \, 
\bB^{(ijr)}
+
(i \!\leftrightarrow\! j)
\bigg]
\nnb\\
&+&
4 \, \as^{2} \,
\Bigg[
\frac{P_{ij(r)}^{{\rm hc},\mu \nu}}{s_{ij}}
\Big( \gamma^{\rm hc}_{i} L_{ir} + \gamma^{\rm hc}_{j} L_{jr} \Big) \,
\bB^{(ijr)}_{\mu \nu}
+
\sum_{\substack{k \neq i,j}}
\frac{P_{ij(r')}^{{\rm hc},\mu \nu}}{s_{ij}} \,
\gamma^{\rm hc}_{k} \, L_{kr'} \,
\bB^{(ijr')}_{\mu \nu}
\Bigg]
\nnb\\
& - &
4 \, \as^{2} \,\,
\frac{\tilde P_{{\rm fin},ij(r)}^{{\rm hc},\mu\nu}}{s_{ij}} \,\,
\bB^{(ijr)}_{\mu \nu}
-
8 \pi \, \as \,
\frac{P_{ij(r)}^{{\rm hc},\mu\nu}}{s_{ij}} \,
\bar{V}_{{\rm fin},\mu \nu}^{(ijr)}
\, ,
\eeq
where we introduced the shorthand notation
\beq
{\mc L}_{ijcr}
\, = \,
2
\ln \frac{s_{ic}}{s_{ir}}
\left[
2
-
L_{ic}
+
\ln \frac{\sk{ic}{jri}}{\sk{ir}{jri}} \,
\right]
\, ,
\qquad
\tilde{\mc L}_{ijcr}
\, = \,
2
L_{ic}
\left[
2
-
L_{ic}
+
\ln \frac{\sk{ic}{jri}}{\mu^2} \,
\right]
\, .
\eeq
Notice that also in \eq{eq:IK(12-RV)(HC)} the kinematic dependence is expressed
only in terms of simple logarithms. Our next step is now to integrate the real-virtual 
counterterm, and add back the result to complete \eq{eq:NNLO_VV_sub}.


\section{\vspace{-.5mm}Integration of the real-virtual counterterm}
\label{RVint}

\vspace{.5mm}
In Eqs.~\eqref{eq:KRVij}, \eqref{eq:KRV} we have defined the counterterm $K^{\RV}$, 
that enabled us to build the subtracted real-virtual squared matrix element $RV_{\,\rm sub}$, 
integrable in the whole $(\npo)$-body phase space, and free of poles in $\eps$. The 
$K^{\RV}$ counterterm needs to be integrated in $d=4-2\eps$ dimensions in the 
radiation phase space, and then the result must be added back, according to the 
subtraction structure given in Eqs.~(\ref{eq:sigma NNLO sub})-(\ref{eq:NNLO_RV_sub}).
In order to compute the integrated counterterm, $I^{\RV}$, as defined in \eq{intcountNNLO}, 
we proceed by summing over all sectors $\W{ij}$, so that sector functions drop out of the calculation, 
owing to the sum rules they satisfy (like for example those in \eqref{SCsumrules}). 
We then perform the integration over the radiative 
phase space, with the measure $d \Phi_{\rm rad}^{(acd)}$, naturally induced by the mapping
$(acd)$, according to
\beq
&&
\int d\Phi_\npo (\{k\})
\; = \;
\frac{\varsi_{n+1}}{\varsi_{n}}
\int d\Phi_n^{(acd)}
\int d \Phi_{\rm rad}^{(acd)} 
\, ,
\qquad\qquad
d\Phi_n^{(acd)}
\; \equiv \;
d\Phi_n(\kkl{acd}) 
\, ,
\eeq
where $d \Phi_{\rm rad}^{(acd)}$ is defined in \eq{eq:dPhi rad (acd) 3}.
The integration of $K^{\RV}$ is carried out following the methods described in
Ref.~\cite{Magnea:2020trj}, and using the fact that the spin-correlated contributions
proportional to the kernels $Q_{ij(r)}^{\mu\nu}$ and $\tilde Q_{ij(r)}^{\mu\nu}$ vanish 
upon integration, as discussed in \appn{app:int Q}. The formal expression for the 
integrated version of $K^{\RV}$ can be written as
\beq
\label{eq:integKRV}
\int d\Phi_\npo \, K^{\RV}
& = &
\int d\Phi_\npo \,
\bigg[
\sum_{i} \;
\Big( \bbS{i} \, RV + \Delta_{{\rm S},{i}} \Big)
+
\sum_{i, \, j > i} 
\Big( \bbHC{ij} \, RV + \Delta_{{\rm HC},ij} \Big)
\bigg]
\, ,
\eeq
where the integrands are defined in Eqs.~\eqref{eq:barred limits RV} and 
\eqref{eq:DeltaSi}-\eqref{eq:DeltaSCij} and we use the shorthand notations
(see \eq{eq:RsubZ2})
\beq
\bbHC{ij} \, RV \equiv \bbC{ij} ( 1 - \bbS{i} - \bbS{j} ) \, RV \, , \qquad \quad
\Delta_{{\rm HC},ij} \equiv \Delta_{{\rm C},ij} + \Delta_{{\rm C},ji} - 
\Delta_{{\rm SC},ij} - \Delta_{{\rm SC},ji} \, .
\eeq
Before integrating, we can further simplify the expressions for $ \Delta_{{\rm S},i}$ 
and $\Delta_{{\rm C},ij}$, given in \eqref{eq:DeltaSi}-\eqref{eq:DeltaCij}. In fact, since 
$\sk{ef}{icd} = s_{ef}$ for  $e,f \neq i,c,d$, and $\sk{cd}{ijr} = s_{cd}$ for  $c,d \neq i,j,r$, 
one finds that
\beq
\label{somerels}
&&
\frac{1}{2} \!\!
\sum_{\substack{e\neq i,c \\ f\neq i,c,e}} \!\!
\Bigg[ \bigg(\frac{s_{ef}}{\sk{ef}{icd}}\!\bigg)^{\!\!-\eps} \!\! -1 \Bigg] \,
\bB^{(icd)}_{efcd}
+
\sum_{\substack{e\neq i,d}} \!
\Bigg[ \bigg( \frac{s_{ed}}{\sk{ed}{icd}} \bigg)^{\!-\eps} - 1 \Bigg] \,
\bB^{(icd)}_{edcd} \, = 
\nnb \\
&&
\hspace{3cm} 
\, = \, 
2 \!\!
\sum_{\substack{e\neq i,c,d}} \!
\Bigg[ \bigg(\frac{s_{ed}}{\sk{ed}{icd}}\!\bigg)^{\!\!-\eps} - 1 \Bigg]
\bB^{(icd)}_{edcd}
+
\Bigg[ \bigg(\frac{s_{cd}}{\sk{cd}{icd}}\!\bigg)^{\!\!-\eps} - 1 \Bigg]
\bB^{(icd)}_{cdcd}
\, ,
\eeq
as well as
\beq
\sum_{\substack{c\neq i,j \\ d\neq i,j,c}} \!
\Bigg[ \bigg(\frac{s_{cd}}{\sk{cd}{ijr}}\!\bigg)^{\!\!-\eps} - 1 \Bigg] \,
\bB_{\mu\nu,cd}^{(ijr)}
\, = \,
2 \!\!
\sum_{\substack{c\neq i,j,r}} \!
\Bigg[ \bigg(\frac{s_{cr}}{\sk{cr}{ijr}}\!\bigg)^{\!\!-\eps} - 1 \Bigg] \,
\bB_{\mu\nu,cr}^{(ijr)}
\, .
\eeq
After integration, the soft contributions to \eq{eq:integKRV} read
\beq
\label{softintRVcount}
\int d\Phi_\npo \,
\bbS{i} \, RV
& = &
- \,
\frac{\varsi_{n+1}}{\varsi_{n}} \,
\sum_{\substack{c \neq i \\ d \neq i,c}} \,
\int \! d \Phi_n^{(icd)} 
\bigg[ \;
J_{\rm s}^{icd} \,
\bar V_{cd}^{(icd)}
\, - \,
\frac{\as}{2\pi} \,
\bigg(
\tilde J_{\rm s}^{icd}
+
J_{\rm s}^{icd} \,
\frac{\beta_0}{2\eps}
\bigg)
\bar B_{cd}^{(icd)} \qquad
\\[-3mm]
&&
\hspace{33mm}
+ \,\,
\as \!\!
\sum_{\substack{e \neq i,c,d}} \!\!\!
\tilde J_{\rm s}^{\,icde} \, 
\bar B_{cde}^{(icd)} \,
\bigg]
\, , 
\nnb
\eeq
while 
($r = r_{ik}$)
\beq
\int \! d\Phi_\npo \,
\Delta_{{\rm S},i}
& = &
- \,
\frac{\alpha_s}{2 \pi} \,
\frac{\varsi_{n+1}}{\varsi_{n}} \!\!
\sum_{\substack{c \neq i \\ d \neq i,c}} \!
\Bigg\{ \;
\int \! d \Phi_n^{(icd)} \,
\bigg[
\sum_{e \neq i,c,d} \!\!
J_{\!_{\Delta \rm s}}^{\, icd(e)} \,
\bB^{(icd)}_{edcd}
+
J_{\!_{\Delta \rm s}}^{\, icd} \,
\bB^{(icd)}_{cdcd}
\bigg]
\\[-3mm]
&&
\hspace{21mm}
+ \,
\bigg[
2 \, C_{f_{c}}
\bigg( 
\frac{1}{\eps^2} 
\!+\! 
\frac{2}{\eps} 
\bigg)
\!+\!
\frac{\gamma^{\rm hc}_{c}}{\eps}
\bigg]
\bigg[
\int \! d \Phi_n^{(icd)} \!
J_{\rm s}^{\, icd} 
\bB^{(icd)}_{cd}
\!-\!
\int \! d \Phi_n^{(idc)} \!
J_{\rm s}^{\, idc} 
\bB^{(idc)}_{cd}
\bigg]
\!\Bigg\}
\nnb\\
&&
- \,
\frac{\as}{2\pi}\,
\frac{\varsi_{n+1}}{\varsi_{n}} \!\!\!
\sum_{\substack{k\neq i \\ c\neq i,k,r}} \!\!\!
\frac{\gamma^{\rm hc}_{k}}{\eps} \, 
\bigg[
\int \! d \Phi_n^{(irc)} 
J_{\rm s}^{irc} \,
\bB^{(irc)}_{cr}
-
\int \! d \Phi_n^{(icr)} 
J_{\rm s}^{icr} \,
\bB^{(icr)}_{cr}
\bigg]
\, .
\nnb
\eeq
Explicit expressions for the constituent integrals $\tilde J_s^{\, icd}$, $\tilde 
J_{\rm s}^{\,icde}$, $J_{\!_{\Delta \rm s}}^{\, icd(e)}$ and $J_{\!_{\Delta \rm s}}^{\, 
icd}$ are given in \eq{softRVintker}, while the NLO integral $J_{\rm s}^{\, icd}$ is 
given in \eq{eq:Js^ilm}. We notice that the soft integrated real-virtual counterterm in 
\eq{softintRVcount} receives contributions from the triple-colour-correlated 
squared matrix element $\bar{B}_{cde}$. However, the pole content of such term 
vanishes upon performing the appropriate colour sums (see Ref.~\cite{Magnea:2020trj} 
for further details). This cancellation represents a strong test for the method: it is 
protected by the fact that no singular contributions proportional to colour tripoles 
can arise from double-virtual nor from double-real corrections. On the other hand, 
integrating the tripole contribution to the soft real-virtual kernel requires the non-trivial 
procedure described in Ref.~\cite{Magnea:2020trj}, which is necessary in order to 
verify the pole cancellation, and to compute the finite remainder. To complete the 
discussion we also report the integrated hard-collinear component,
\beq
\int \! d\Phi_\npo \,
\bbHC{ij} \, RV
& = &
\frac{\varsi_{n+1}}{\varsi_{n}} \!
\int \! d\Phi_n^{(ijr)} \, 
\bigg[  
J_{\rm hc}^{\, ijr} \,
\bar{V}^{(ijr)}
+ 
\frac{\as}{2\pi}
\bigg(
\tilde J_{\rm hc}^{\, ijr} \!
- 
J_{\rm hc}^{\, ijr}
\frac{\beta_0}{2\eps}
\bigg)
\bB^{(ijr)}
\bigg]
\, ,
\qquad
r = r_{ij}
\, ,
\qquad
\eeq
while the compensating hard-collinear term integrates to ($r = r_{ij}$, $r' = r_{ijk}$) 
\beq
\label{eq:integDeltaHCij}
\int d\Phi_\npo \,
\Delta_{{\rm HC},ij}
& = &
\frac{\as}{2\pi}
\frac{\varsi_{n+1}}{\varsi_{n}} \,
\Bigg\{ \;
\int d \Phi_n^{(ijr)} \,
\bigg[
\sum_{\substack{c\neq i,j,r}}
J_{\!_{\Delta {\rm hc}}}^{\, ijr} \,
\bB_{cr}^{(ijr)}
+
\sum_{\substack{c\neq i,j}}
J_{\!_{\Delta {\rm hc}}}^{\,ijrc} \,
\bB_{[ij]c}^{(ijr)}
\bigg]
\\
&&
\hspace{14mm}
+ \,
\! \sum_{\substack{c\neq i,j,r}}
\bigg[
\int d\Phi_n^{(jri)} \,
J_{\!_{\Delta \rm hc}}^{\, jri,c} \,
\bB_{[ij]c}^{(jri)}
+
\int d\Phi_n^{(irj)} \,
J_{\!_{\Delta \rm hc}}^{\, irj,c} \,
\bB_{[ij]c}^{(irj)}
\bigg]
\nnb \\
&&
\hspace{14mm}
+ \,
\! \sum_{k \ne i,j} \!
\bigg(
\frac{\gamma^{\rm hc}_{k}}{\eps}
+
\phi^{\rm hc}_{k}
\bigg)
\bigg[
\int d\Phi_n^{(ijr)} 
J_{\rm hc}^{\, ijr} \,
\bB^{(ijr)}
-
\int  d\Phi_n^{(ijr')} 
J_{\rm hc}^{\, ijr'} \,
\bB^{(ijr')}
\bigg]
\nnb\\
&&
\hspace{14mm}
+ \,
\tilde f_{ij}^{q\bar q} \,
\sum_{\substack{c\neq i,j}} 
\bigg[
\int d\Phi_n^{(jri)} \,
\tilde J_{\!_{\Delta \rm hc}}^{\, jri,c} \,
{\cal \bB}_{[ij]c}^{(jri)}
-
\int d\Phi_n^{(irj)} \,
\tilde J_{\!_{\Delta \rm hc}}^{\, irj,c} \,
{\cal \bB}_{[ij]c}^{(irj)}
\bigg]
\nnb
\Bigg\}
\, .
\eeq
Explicit expressions for the hard-collinear constituent integrals 
$\tilde J_{\rm hc}^{\, ijr}$, $J_{\!_{\Delta {\rm hc}}}^{\, ijr}$, 
$J_{\!_{\Delta {\rm hc}}}^{\, ijrc}$, $J_{\!_{\Delta \rm hc}}^{\, jri,c}$, and 
$\tilde J_{\!_{\Delta \rm hc}}^{\, jri,c}$  
are given in \eq{Jhc^ijr tilde}, while the NLO hard-collinear integral 
$J_{\rm hc}^{\, ijr}$ is given in \eq{Jhc^ijr}.

Having computed all relevant integrals, we now recombine them, 
following a procedure analogous to the one described at the end of \secn{InteSec}.
We rename the sets of mapped momenta $\{\bar k^{(abc)}\}_{n}$ to the same set 
of Born-level momenta $\{k\}_{n}$ by means of the replacements 
\beq
d\Phi_n^{(abc)} \, \to \, d\Phi_n
\, ,
\qquad
\bB^{(abc)}_{\cdots} \, \to \, B_{\cdots}
\, ,
\qquad
{\cal \bB}^{(abc)}_{\cdots} \, \to \, {\cal \bB}_{\cdots}
\, ,
\qquad
\bar s_{lm}^{(abc)} \, \to \, s_{lm}
\, ,
\eeq
where the ellipsis in the Born-level matrix element stands for a generic colour 
correlation. 
In particular, in the integral of $\Delta_{{\rm HC},ij}$, all momenta 
$\kk{j}{ijr}$, $\kk{i}{jri}$, $\kk{j}{irj}$, and $\kk{j}{ijr'}$ are renamed as $k_p$, 
where $p$ is the label of the parent particle splitting into $i$ and $j$.
As a consequence of this renaming, the integrals involving 
${\cal \bB}_{[ij]c}$ 
can be recombined, and do not contribute to the integrated counterterm. 
Indeed
\beq
&&
\int d\Phi_n^{(jri)} \,
\tilde J_{\!_{\Delta \rm hc}}^{\, jri,c} \,
{\cal \bB}_{[ij]c}^{(jri)}
-
\int d\Phi_n^{(irj)} \,
\tilde J_{\!_{\Delta \rm hc}}^{\, irj,c} \,
{\cal \bB}_{[ij]c}^{(irj)}
\, = \,
\\
&&
\hspace{1.5cm}
= \,
\int d \Phi_n^{(jri)} \,
\tilde J_{\!_{\Delta \rm hc}}^{\, c} \! \Big( \sk{ir}{jri}, \sk{ic}{jri} \Big) \,
{\cal \bB}_{[ij]c}^{(jri)}
-
\int d\Phi_n^{(irj)} \,
\tilde J_{\!_{\Delta \rm hc}}^{\, c} \! \Big( \sk{jr}{irj}, \sk{jc}{irj} \Big) \,
{\cal \bB}_{[ij]c}^{(irj)}
\nnb\\
&&
\hspace{1.5cm}
\to \;
\int d \Phi_n \,
\tilde J_{\!_{\Delta \rm hc}}^{\, c} \! \Big( s_{pr}, s_{pc} \Big) \,
{\cal B}_{pc}
-
\int d \Phi_n \,
\tilde J_{\!_{\Delta \rm hc}}^{\, c} \! \Big( s_{pr}, s_{pc} \Big) \,
{\cal B}_{pc}
\; = \;
0
\, . 
\nnb
\eeq
The dependence on the $(\npo)$-body phase-space particles is 
now limited to the flavour factors $f_{i}^{q}$, $f_{i}^{\bar q}$ and 
$f_{i}^{g}$, which can be 
translated into flavour factors for the $n$-body-phase-space particles, as 
was done in \secn{FlaSum} for the double-real contribution. 
In particular, when going from an $(\npo)$-body phase space to an $n$-body 
phase space the relations in \eq{flavsum1soft} and \eq{flavsum1coll} apply, 
with the formal replacement $n\to n-1$.
After performing the flavour sums, no dependence on the original $(\npo)$-body phase 
space remains. Simplifying the colour correlations where possible, we finally get 
\beq
\label{eq:IRV}
I^{\RV}
& = & 
-
\sum_{\substack{c,d \neq c}} 
\bigg[
J_{\rm s}(s_{cd}) \, 
V_{cd}
+
J_{\sRV}(s_{cd}) \, 
B_{cd}
+
J_{\sRV}^{^{(2)}}(s_{cd}) \, 
B_{cdcd}
+
\! \sum_{\substack{e \neq c,d}}
J_{\sRV}^{cde} \,
B_{cde} \, 
\bigg]
\nnb\\
&&
+ \,
\sum_{j}
\Bigg\{ \,
J_{\rm hc}^{\,j}(s_{jr}) \, V
+
J_{\hcRV}^{\,j}(s_{jr}) \, B
+
J_{\hcRV}^{\,j,\rm A}(s_{jr}) \, B_{jr}
+
\! \sum_{c\neq j,r} \!
\bigg[
J_{\hcRV}^{\,j,\rm B}(s_{jc}) \, B_{jc}
+
J_{\hcRV}^{\,j,\rm C}(s_{jr}) \, B_{cr}
\bigg]
\nnb\\
&&
\hspace{8mm}
+ \,
\frac{\as}{2\pi} \,
\sum_{k \ne j} \,
\bigg(
\frac{\gamma^{\rm hc}_{k}}{\eps}
+
\phi^{\rm hc}_{k}
\bigg)
\Big[
J_{\rm hc}^{\,j}(s_{jr}) 
-
J_{\rm hc}^{\,j}(s_{jr'}) 
\Big]
\Bigg\}
\, ,
\eeq
where we introduced the following combinations of constituent integrals:
\beq
J_{\sRV}(s)
& = &
- \,
\frac{\as}{2\pi} \,
\bigg[
C_{A} \,
\tilde J_{\rm s}(s)
+
\frac{\beta_0}{2\eps} \,
J_{\rm s}(s)
+
2 \,
C_{f_{d}} \,
J_{\!_{\Delta \rm s}}^{(3)}(s)
\bigg]
\, ,
\eeq
\beq
J_{\sRV}^{^{(2)}}(s)
& = &
\frac{\as}{2\pi} \,
\Big[
J_{_{\Delta \rm s}}^{(2)}(s)
-
J_{_{\Delta \rm s}}^{(3)}(s)
\Big]
\, ,
\eeq
\beq
\label{tripoleint}
J_{\sRV}^{cde}
& = &
- \,
\frac{\as^{2}}{2\pi} \,
\bigg[ \;
\frac{1}{2} \,
\ln \frac{s_{ce}}{s_{de}} \,
\ln^2 \frac{s_{cd}}{\mu^2}
\, + \, 
\frac{1}{6} \ln^3 \frac{s_{ce}}{s_{de}}
\, + \, 
\Li_3 \bigg(\!\!-\frac{s_{ce}}{s_{de}} \bigg)
+
\mc O(\eps)
\bigg]
\, ,
\eeq
\beq
J_{\hcRV}^{\, j}(s)
& = &
\frac{\as}{2\pi} \,
\Bigg\{ \;
\big( f_{j}^{q} \!+\! f_{j}^{\bar q} \big) \,
\bigg[
\tilde J_{\rm hc}^{\og}(s)
-
\frac{\beta_0}{2\eps} \,
J_{\rm hc}^{\og}(s)
-
C_F \,
J_{\!_{\Delta \rm hc,A}}^{\og}(s)
-
C_F \,
J_{\!_{\Delta \rm hc,A}}^{\rm qg}(s)
-
C_F \,
J_{\!_{\Delta \rm hc,A}}^{\rm gq}(s)
\bigg]
\nnb\\
&&
\hspace{8mm}
+ \,
f_{j}^{g} \, 
\bigg[ \,
\frac12 \, 
\bigg(
\tilde J_{\rm hc}^{\tg}(s)
-
\frac{\beta_0}{2\eps} \,
J_{\rm hc}^{\tg}(s)
-
C_A \,
J_{\!_{\Delta \rm hc,A}}^{\tg}(s)
-
2 \, C_A \,
J_{\!_{\Delta \rm hc,A}}^{\rm gg}(s)
\bigg)
\nnb\\
&&
\hspace{18mm}
+ \,
N_{f} \, 
\bigg(
\tilde J_{\rm hc}^{\zg}(s)
-
\frac{\beta_0}{2\eps} \,
J_{\rm hc}^{\zg}(s)
-
C_A \,
J_{\!_{\Delta \rm hc,A}}^{\zg}(s)
-
2 \, C_A \,
J_{\!_{\Delta \rm hc,A}}^{\rm qq}(s)
\bigg)
\bigg]
\Bigg\}
\, ,
\eeq
\beq
J_{\hcRV}^{\,j,\rm A}(s)
& = &
\frac{\as}{2\pi} \,
\Bigg\{ \;
\big( f_{j}^{q} \!+\! f_{j}^{\bar q} \big) \,
\Big(
J_{\!_{\Delta \rm hc,B}}^{\og}(s)
-
J_{\!_{\Delta \rm hc,A}}^{\rm qg}(s)
-
J_{\!_{\Delta \rm hc,A}}^{\rm gq}(s)
\Big)
\nnb \\
&&
\hspace{8mm}
+ \,
f_{j}^{g} \, 
\bigg[ \,
\frac12 \, 
\Big(
J_{\!_{\Delta \rm hc,B}}^{\tg}(s)
-
2 \,
J_{\!_{\Delta \rm hc,A}}^{\rm gg}(s)
\Big)
+
N_{f} \, 
\Big(
J_{\!_{\Delta \rm hc,B}}^{\zg}(s)
-
2 \,
J_{\!_{\Delta \rm hc,A}}^{\rm qq}(s)
\Big)
\bigg]
\Bigg\}
\, ,
\eeq
\beq
J_{\hcRV}^{\,j,\rm B}(s)
& = &
\frac{\as}{2\pi} \,
\Bigg\{ \,
\big( f_{j}^{q} + f_{j}^{\bar q} \big) \,
\Big(
J_{\!_{\Delta \rm hc,B}}^{\og}(s)
+
J_{\!_{\Delta \rm hc,B}}^{\rm qg}(s)
+
J_{\!_{\Delta \rm hc,B}}^{\rm gq}(s)
\Big)
\nnb\\
&&
\hspace{8mm}
+ \,
f_{j}^{g} \, 
\bigg[ \,
\frac12 \, 
\Big(
J_{\!_{\Delta \rm hc,B}}^{\tg}(s)
+
2 \,
J_{\!_{\Delta \rm hc,B}}^{\rm gg}(s)
\Big)
+
N_{f} \,
\Big(
J_{\!_{\Delta \rm hc,B}}^{\zg}(s)
+
2 \,
J_{\!_{\Delta \rm hc,B}}^{\rm qq}(s)
\Big)
\bigg]
\Bigg\}
\, ,
\eeq
\beq
J_{\hcRV}^{\,j,\rm C}(s)
& = &
\frac{\as}{2\pi} \,
\Bigg\{ \,
\big( f_{j}^{q} + f_{j}^{\bar q} \big) \,
J_{\!_{\Delta \rm hc}}^{\og}(s)
+
f_{j}^{g} \, 
\bigg[ \,
\frac12 \, J_{\!_{\Delta \rm hc}}^{\tg}(s)
+
N_{f} \, J_{\!_{\Delta \rm hc}}^{\zg}(s)
\bigg]
\Bigg\}
\, .
\eeq
All new constituent integrals appearing in the above results are listed in 
\appn{app:master integrals}: the soft integrals are presented in \eq{somesoftintrv}, 
the hard-collinear integrals in \eq{tildehcint}, and the integrals arising from the 
compensating $\Delta_{ij}$ terms in Eqs.~(\ref{somedeltaint})-(\ref{yetsomeotherdeltaint}).
We note once again that all integrals involved are single-scale, and thus involve only
simple logarithms. Interestingly, the only exception is \eq{tripoleint}, a uniform-weight-three
function featuring three scales and a single trilogarithm: this integral 
arises as 
a finite remainder of the non-trivial integration of the tripole term.

The integrated counterterm $I^{\RV}$ given in \eq{eq:IRV}, which features Born-level 
kinematics, contains explicit poles in $\eps$, that must be combined with those of the 
integrated counterterm $I^{\two}$, and must, together, cancel the singularities of the 
double-virtual squared matrix element. In the next section we turn to the proof of this 
statement, which provides a highly non-trivial test of all our calculations, and completes
the subtraction programme for generic massless final states.


\section{The subtracted double-virtual contribution $VV_{\,\rm sub}$}
\label{VVsub}

Finally, we turn our attention to the first line in \eq{eq:NNLO_VV_sub}, which we 
rewrite here as
\beq
\label{eq:VVsub}
VV_{\,\rm sub}(X)
& = &
\Big[ VV + I^{\two} + I^{\RV} \Big] \, \delta_n(X)
\, .
\eeq
It is our task to show that the equation above is free of $\eps$ poles. To verify this, 
we first explicitly derive the $\eps$ poles of $VV$, and then we provide the complete 
$\eps$ expansion of $I^{\two} + I^{\RV}$, including ${\cal O} (\eps^0)$ terms, 
obtained by combining \eq{eq:I2} and \eq{eq:IRV}.


\subsection{The pole part of the double-virtual matrix element $VV$}
\label{dovipo}

All infrared poles of gauge-theory scattering amplitudes can be expressed in a factorised 
form through the formula~\cite{Catani:1998bh,Sterman:2002qn,Gardi:2009qi,Becher:2009qa,
Feige:2014wja}
\beq
  {\cal A} \left( \frac{k_i}{\mu}, \alpha_s (\mu), \eps \right) \, = \, {\bf Z} 
  \left( \frac{k_i}{\mu}, \alpha_s (\mu), \eps \right) {\cal H}
  \left( \frac{k_i}{\mu}, \alpha_s (\mu), \eps \right) \, ,
\label{factoramp}
\eeq
where ${\cal H}$ is finite as $\eps \to 0$, and ${\bf Z}$ is a colour operator with a
universal form, to be discussed below. The infrared operator ${\bf Z}$ obeys a (matrix) 
renormalisation-group equation, which can be 
solved in exponential form, with a trivial initial condition, in terms of an 
anomalous-dimension matrix ${\bf \Gamma}$. One may write
\beq
  {\bf Z} \left( \frac{k_i}{\mu}, \alpha_s (\mu), \eps \right) \, = \,
  {\cal P} \exp \left[ \int_0^{\mu} \frac{d \lambda}{\lambda} \,\,
  {\bf \Gamma} \left( \frac{k_i}{\lambda}, \alpha_s (\lambda), \eps \right) \right] \, ,
\label{defGamma}
\eeq
where the integral converges at $\lambda = 0$ in dimensional regularisation thanks 
to the behaviour of the $\beta$ function in $d = 4 - 2 \eps$, for $\eps < 0$ ($d >4$). 
Indeed, in dimensional  regularisation one has
\beq
  \mu \frac{d \alpha_s}{d \mu}
  \, \equiv \,
  \beta \left( \eps, \alpha_s \right)
  \, = \,
  - \, 2 \eps \, \alpha_s
  \, - \,
  \frac{\alpha_s^2}{2 \pi} \, \beta_0 \, + \, {\cal O} \left( \alpha_s^3 \right) \, ,
\label{beta4}
\eeq
whose solution implies~\cite{Magnea:1990zb} that the $d$-dimensional running coupling
$\alpha_s (\mu, \eps)$ vanishes at $\mu = 0$ for $\eps < 0$, so that the 
corresponding initial condition is ${\bf Z} (\mu = 0) ={\bf 1}$, leading 
to \eq{defGamma}. For the purposes of NNLO subtraction (and thus at
two loops for virtual amplitudes), ${\bf \Gamma}$ is given by the dipole formula 
\cite{Gardi:2009qi,Becher:2009qa}
\beq
  {\bf \Gamma} \left( \frac{p_i}{\lambda}, \alpha_s (\lambda), \eps \right) \, = \,
  \frac{1}{2} \, \widehat{\gamma}_K \left( \alpha_s (\lambda, \eps) \right)
  \sum_{i,j>i} \ln \left( \frac{s_{ij} \, e^{{\rm i} \pi \sigma_{ij}}}{\lambda^2} 
  \right) {\bf T}_i  \cdot {\bf T}_j \, + \, \sum_{i} \gamma_i \left( 
  \alpha_s (\lambda, \eps) \right) \, .
\label{dipole}
\eeq
In \eq{dipole}, the phases $\sigma_{ij}$ are given by $\sigma_{ij} = +1$ if partons
$i$ and $j$ are either both in the initial state or both in the final state, while $\sigma_{ij} 
= 0$ otherwise.
For our present final-state application, we can thus henceforth replace all phase
factors using $e^{{\rm i} \pi \sigma_{ij}} = -1$, with the understanding that 
the logarithm is taken above the cut.

The anomalous dimensions appearing in \eq{dipole} are the cusp anomalous 
dimension $\widehat{\gamma}_K \left( \alpha_s \right)$ and the collinear anomalous 
dimensions $\gamma_i \left( \alpha_s \right)$. More precisely, in the derivation of 
\eq{dipole} it has been assumed that the (light-like) cusp anomalous dimension 
$\gamma_K^{(r)} (\alpha_s)$, in colour representation $r$, obeys `Casimir scaling', 
\emph{i.e.}~it can be written as
\beq
  \gamma_K^{(r)} (\alpha_s) \, = \, C_r \, \widehat{\gamma}_K (\alpha_s) \, ,
\label{Casscal}
\eeq
where $C_r$ is the quadratic Casimir eigenvalue for colour representation $r$, while
$ \widehat{\gamma}_K (\alpha_s)$ is a universal (representation-independent) function.
This assumption is known to fail at four loops \cite{Grozin:2017css,Moch:2018wjh}.
The collinear anomalous dimensions $\gamma_i (\alpha_s)$ are related to the
anomalous dimensions of quark and gluon fields, and can be derived from essentially 
colour-singlet calculations such as those of form factors. 

One important consequence of the dipole formula is that the scale integration
in \eq{defGamma} can be performed without affecting the colour structure (which
is scale-independent): one may therefore omit the path-ordering in \eq{defGamma},
simplifying considerably the necessary calculations. Expanding the various ingredients 
perturbatively as
\beq
  \widehat{\gamma}_K (\alpha_s) \, = \, \sum_{n = 1}^\infty
  \widehat{\gamma}_K^{(n)} \left( \frac{\alpha_s}{2\pi} \right)^n \, , \quad
  \gamma_i (\alpha_s) \, = \, \sum_{n = 1}^\infty
  \gamma_i^{(n)} \left( \frac{\alpha_s}{2\pi} \right)^n \, , \quad
  {\bf \Gamma} (\alpha_s) \, = \, \sum_{n = 1}^\infty
  {\bf \Gamma}^{(n)} \left( \frac{\alpha_s}{2\pi} \right)^n \, ,
\label{expGam}
\eeq
one gets at NLO
\beq
{\bf \Gamma}^{(1)}
  & = & \frac{1}{4} \, \widehat{\gamma}_K^{(1)}
  \sum_{i,j \neq i} \ln \left( \frac{- s_{ij} + {\rm i} \eta }{\mu^2} \right)
  {\bf T}_i \cdot {\bf T}_j \, + \, \sum_{i} \gamma_i^{(1)} -
  \frac{1}{4} \, \widehat{\gamma}_K^{(1)} \ln \left(\frac{\mu^2}{\lambda^2}\right)
  \sum_{i} C_{f_i} \, ,
\label{Gamma1}
\eeq
and consequently
\beq
{\bf Z}^{(1)} \left( \frac{p_i}{\mu}, \eps \right)
\, = \,
- \frac{1}{\eps^2}
\frac{\widehat{\gamma}_K^{(1)}}{8} \Sigma_{\!_C} 
- \frac{1}{\eps}
\left(
\frac{\widehat{\gamma}_K^{(1)}}{8}
\sum_{i,j\neq i} L_{ij}\, {\bf T}_i \cdot {\bf T}_j
+ \frac{1}{2} \Sigma_{\gamma}
\right) 
+ \, {\rm i} \pi \, \frac{\gamma_K^{(1)}}{8 \eps} \, \Sigma_{\!_C} \, ,
\label{NLOres_reim}
\eeq
where $L_{ij} = \ln (s_{ij}/\mu^2)$. \eq{NLOres_reim} is in agreement with~\cite{Catani:1998bh,
Becher:2009qa}, with the one-loop anomalous-dimension coefficients given by
\beq
\hspace{-5mm}
\widehat{\gamma}_K^{(1)}
\, = \,
4
\, ,
\quad 
\gamma^{(1)}_i
\, \equiv \,
\gamma_i
\, = \,
\frac{3}{2} \, C_F \, (f_{i}^{q} \!+\! f_{i}^{\bar q})
+
\frac{1}{2} \, \beta_{0} \, f_{i}^{g}
\, ,
\quad
\Sigma_{\!_C}
\, = \,
\sum_i C_{f_i}
\, ,
\quad
\Sigma_{\gamma}
\, = \,
\sum_i \gamma_i
\, ,
\label{oloandin}
\eeq
where we noted that in the text we have sometimes used the notation $\gamma_i$ for the 
one-loop coefficient denoted here by $\gamma_i^{(1)}$. ~Expanding the anomalous
dimensions to two loops and performing the relevant integrals, the NNLO result for
the ${\bf Z}$ factor is
\beq
{\bf Z}^{(2)}
& = &
\frac{1}{\eps^4} \,
\frac{\Big(\widehat{\gamma}_K^{(1)}\Big)^2}{128} 
\Sigma_{\!_C}^2
\nnb\\
& + &
\frac{1}{\eps^3} \, \,
\frac{\widehat{\gamma}_K^{(1)}}{64}
\Sigma_{\!_C}
\Bigg[
3 \beta_0 \, + \, 4 \Sigma_{\gamma}  \, + \, 
\widehat{\gamma}_K^{(1)}
\sum_{i,j \neq i} \ln \bigg( \frac{- s_{ij} + {\rm i} \eta}{\mu^2} \bigg) \, 
{\bf T}_i \cdot {\bf T}_j 
\Bigg]
\nnb\\
& + &
\frac{1}{\eps^2} \,
\frac18 \, 
\Bigg[
\frac{\beta_0 \, \widehat{\gamma}_K^{(1)}}{4} 
\sum_{i,j \neq i} \ln \bigg( \frac{- s_{ij} + {\rm i} \eta}{\mu^2} \bigg) \,  
{\bf T}_i \cdot {\bf T}_j \, + \, 
\beta_0 \Sigma_{\gamma} \, - \, 
\frac{\widehat{\gamma}_K^{(2)}}{4} \, \Sigma_{\!_C}
\nnb\\
&&
\hspace{1cm}
+ \,\, 
\Sigma_{\gamma}^2  \, + \,
\frac{\widehat{\gamma}_K^{(1)}}{2} \, \Sigma_{\gamma} 
\! \sum_{i,j \neq i} \ln \bigg( \frac{- s_{ij} + {\rm i} \eta}{\mu^2} \bigg) \,
{\bf T}_i \cdot {\bf T}_j
\nnb\\
&&
\hspace{1cm}
+ \, \frac{\Big(\widehat\gamma_K^{(1)}\Big)^2}{16}
\sum_{\substack{i,j \neq i\\k,l \neq k}} 
\ln \bigg( \frac{- s_{ij} + {\rm i} \eta}{\mu^2} \bigg) 
\ln \bigg( \frac{- s_{kl} + {\rm i} \eta}{\mu^2} \bigg) \,
{\bf T}_i \cdot {\bf T}_j \, {\bf T}_k \cdot {\bf T}_l
\Bigg]
\nnb\\
& - &
\frac{1}{\eps} \,
\frac 14 \,
\Bigg[
\frac{\widehat{\gamma}_K^{(2)}}{4} 
\sum_{i,j \neq i} \ln \bigg( \frac{- s_{ij} + {\rm i} \eta}{\mu^2} \bigg) \, 
{\bf T}_i \cdot {\bf T}_j \, + \,
\Sigma_{\gamma}^{(2)}
\Bigg]
\, ,
\label{Zexp}
\eeq
which agrees with~\cite{Becher:2009qa}, with the anomalous dimension coefficients 
given in \eq{twoloandim}, 
and where we defined $\Sigma_{\gamma}^{(2)} = \sum_i \gamma_i^{(2)}$. 
Having deduced the $\bf{Z}$ elements up to the needed order, we can now 
interfere the double-virtual amplitude with the Born, and extract the poles. 
The perturbative expansion of \eqref{factoramp} yields
\beq
  {\cal A}^{(0)} & = & {\cal H}^{(0)} \, , \nonumber \\
  {\cal A}^{(1)} & = & \frac{\alpha_s}{2\pi} \, \Big[ {\cal H}^{(1)} + {\bf Z}^{(1)} 
  {\cal H}^{(0)} \Big] \, \equiv \, \frac{\alpha_s}{2\pi} \, A^{(1)} \, , \nonumber \\
  {\cal A}^{(2)} & = & \bigg( \frac{\alpha_s}{2\pi} \bigg)^2 \Big[ {\cal H}^{(2)} + 
  {\bf Z}^{(1)} {\cal H}^{(1)} + {\bf Z}^{(2)} {\cal H}^{(0)} \Big] \, \equiv \, \bigg(
  \frac{\alpha_s}{2\pi} \bigg)^2 A^{(2)} \, ,
\label{ampcoeff}
\eeq
implying
\beq
\label{squaredA}  
  \left| {\cal A} \right|^2  & = & \left| {\cal H}^{(0)} \right|^2 + \frac{\alpha_s}{2\pi} \, 
  2 \, {\bf Re} \left[ \left( {\cal H}^{(0)} \right)^\dagger {\cal H}^{(1)} +
  \left( {\cal H}^{(0)} \right)^\dagger {\bf Z}^{(1)} {\cal H}^{(0)} \right]  \\
  && + \, \bigg( \frac{\alpha_s}{2\pi} \bigg)^2
  \left[ 2 \, {\bf Re} \left( \left( {\cal H}^{(0)} \right)^\dagger {\cal H}^{(2)} +
  \left( {\cal H}^{(0)} \right)^\dagger {\bf Z}^{(1)} {\cal H}^{(1)} + 
  \left( {\cal H}^{(0)} \right)^\dagger {\bf Z}^{(2)} {\cal H}^{(0)} \right) \right. \nonumber \\
  && \left. \hspace{18mm} + \, \left| {\cal H}^{(1)} \right|^2 + 
  \left( {\cal H}^{(0)} \right)^\dagger \left( {\bf Z}^{(1)} \right)^\dagger
  {\bf Z}^{(1)} {\cal H}^{(0)} + 2 \, {\bf Re} \left( \left( {\cal H}^{(1)} \right)^\dagger 
  {\bf Z}^{(1)} {\cal H}^{(0)} \right) \right] + {\cal O} (\alpha_s^3) \, . \nonumber 
\eeq
We are interested in the divergent contributions to \eq{squaredA} at ${\cal  O} (\alpha_s^2)$: 
we extract them in turn. First, the direct contribution of the two-loop ${\bf Z}$ matrix is given by
\beq
\label{hz2h}
2 \, {\bf Re} \bigg(
\left( {\cal H}^{(0)} \right)^\dagger {\bf Z}^{(2)} {\cal H}^{(0)}
\bigg)
& = &
{{\cal H}^{(0)}}^\dagger \big({\bf Z}^{(2)}+{{\bf Z}^{(2)}}^\dag\big) {\cal H}^{(0)}
\nnb\\
& = &
\frac{1}{\eps^4} \,
\frac1{4}
\Sigma_{\!_C}^2 \, B
+
\frac{1}{\eps^3} \,
\frac12
\Sigma_{\!_C}
\Bigg[ 
\Big( \frac{3}{4} \, \beta_0 \, + \, \Sigma_{\gamma} \! \Big) \, B
+ \, \sum_{i, j\neq i} L_{ij} \, B_{ij}
\Bigg]
\nonumber \\
& + &
\frac{1}{\eps^2} \,
\frac14\,
\Bigg[
\Big( \beta_0 \Sigma_{\gamma} \, - \, 
\frac{\widehat{\gamma}_K^{(2)}}{4} \Sigma_{\!_C}
+  \Sigma_{\gamma}^2 \Big) \, B
\, + \,\left(\beta_0 + 2 \, \Sigma_{\gamma}\right)
\sum_{i,j \neq i} L_{ij} \, B_{ij}
\nonumber \\
& &
\hspace{1cm}
+ \, \frac{1}{2} \,
\sum_{\substack{i,j\neq i\\k,l\neq k}} 
\Big( L_{ij} \, L_{kl} -\pi^2 \Big)
\, B_{ijkl}
\Bigg]
\nonumber \\
& - &
\frac{1}{\eps} \,
\frac18\,
\left[4\,\Sigma_{\gamma}^{(2)} \, B
\, + \,\widehat{\gamma}_K^{(2)}
\sum_{i,j\neq i} L_{ij} \, B_{ij} \right] \, ,
\eeq
where again $L_{ij} = \ln (s_{ij}/\mu^2)$, and the colour-correlated Born amplitudes $B_{ij}$
and $B_{ijkl}$ are defined in \eq{colcorrBorn}. The square of the one-loop ${\bf Z}$ matrix
contributes
\beq
\label{hz1z1h}
{{\cal H}^{(0)}}^\dag {{\bf Z}^{(1)}}^\dag {\bf Z}^{(1)} {\cal H}^{(0)}
& = &
\frac1{\eps^4}\,
\frac1{4}\,
\Sigma_{\!_C}^2 B
+
\frac1{\eps^3}\,
\frac12\,
\Sigma_{\!_C}
\Bigg[
\Sigma_{\gamma} \, B
+
\sum_{i,j \neq i} L_{ij} \, B_{ij}
\Bigg]
\nnb\\
&+&
\frac1{\eps^2}\,
\frac14\,
\Bigg[
\Sigma_{\gamma}^2 \, B
+
2 \, \Sigma_{\gamma}
\sum_{i,j \neq i} L_{ij} \, B_{ij}
+
\frac1{2}
\sum_{\substack{i,j \neq i\\k,l\neq k}}
\Big( L_{ij} \, L_{kl} + \pi^2 \Big) \, B_{ijkl}
\Bigg] \, .
\eeq
Note that in \eq{hz2h} and in \eq{hz1z1h}, for simplicity, we already substituted $\widehat
\gamma_K^{(1)}=4$. Finally, terms involving the product of the one-loop hard part and the 
one-loop ${\bf Z}$ matrix give
\beq
\label{hz1h1}
2 \, {\bf Re}
\left(
{{\cal H}^{(0)}}^\dag {\bf Z}^{(1)} {\cal H}^{(1)}
+
{{\cal H}^{(1)}}^\dag {\bf Z}^{(1)} {\cal H}^{(0)}
\right)
& = &
{{\cal H}^{(0)}}^\dag \Big( {\bf Z}^{(1)} + {{\bf Z}^{(1)}}^\dag \Big) {\cal H}^{(1)} 
\nnb \\ && 
+ \, 
{{\cal H}^{(1)}}^\dag \Big( {\bf Z}^{(1)} + {{\bf Z}^{(1)}}^\dag \Big) {\cal H}^{(0)}
\, .
\eeq
In order to make use in practice of \eq{hz1h1}, it is useful to rewrite $\mc H^{(1)}$ in 
terms of the full virtual amplitude $A^{(1)}$, using
\beq
{\cal H}^{(1)} & = & A^{(1)} - {\bf Z}^{(1)} {\cal H}^{(0)} \, .
\eeq
\eq{hz1h1} then becomes
\beq
\label{hz1h11}
2 \, {\bf Re}
\left(
{{\cal H}^{(0)}}^\dag {\bf Z}^{(1)} {\cal H}^{(1)}
+
{{\cal H}^{(1)}}^\dag {\bf Z}^{(1)} {\cal H}^{(0)}
\right)
& = &
{{\cal H}^{(0)}}^\dag \Big( {\bf Z}^{(1)} + {{\bf Z}^{(1)}}^\dag \Big) A^{(1)}
+
{A^{(1)}}^\dag \Big( {\bf Z}^{(1)} + {{\bf Z}^{(1)}}^\dag \Big) {\cal H}^{(0)} \nnb \\
&& - \,  
{{\cal H}^{(0)}}^\dag \Big( {{\bf Z}^{(1)}}^2 + 2\,{{\bf Z}^{(1)}}^\dag{{\bf Z}^{(1)}}
+ {{{\bf Z}^{(1)}}^\dag}^2 \Big) {\cal H}^{(0)}
\, .
\eeq
The term on the second line of \eq{hz1h11} is easily computed using \eq{NLOres_reim} 
and yields
\beq
- \,
{{\cal H}^{(0)}}^\dag \Big( {{\bf Z}^{(1)}}^2 + 2\,{{\bf Z}^{(1)}}^\dag{{\bf Z}^{(1)}}
+ {{{\bf Z}^{(1)}}^\dag}^2 \Big) {\cal H}^{(0)}
& = &
- \, 
\frac1{\eps^4}\,
\Sigma_{\!_C}^2 B
-
\frac1{\eps^3}
2 \, \Sigma_{\!_C}
\Bigg[
\Sigma_{\gamma} \, B
+
\sum_{i,j \neq i} L_{ij} \, B_{ij}
\Bigg]
\nnb \\
&& \hspace{-3cm}
- \, \frac1{\eps^2}
\Bigg[
\Sigma_{\gamma}^2 B
+
2 \, \Sigma_{\gamma}
\sum_{i,j \neq i} L_{ij} \, B_{ij}
+
\frac12
\sum_{\substack{i,j \neq i\\k,l\neq k}}
L_{ij} \, L_{kl} \, B_{ijkl}
\Bigg]
\, .
\eeq
The first two terms on the {\it r.h.s.}~of \eq{hz1h11} can be expressed in terms of the
one-loop virtual correction to the cross section. One finds
\beq
\label{h0z1a1}
&&\hspace{-15mm}
\frac{\alpha_s}{2 \pi} \left[ 
{{\cal H}^{(0)}}^\dag \Big( {\bf Z}^{(1)} + {{\bf Z}^{(1)}}^\dag \Big) A^{(1)}
+
{A^{(1)}}^\dag \Big( {\bf Z}^{(1)} + {{\bf Z}^{(1)}}^\dag \Big) {\cal H}^{(0)} \right]
\nnb \\
&=&
{{\cal H}^{(0)}}^\dag
\Bigg[
- \frac{1}{\eps^2}\frac{\widehat{\gamma}_K^{(1)}}{4} \Sigma_{\!_C} 
- \frac{1}{\eps} \left( \frac{\widehat{\gamma}_K^{(1)}}{4}
\sum_{i, j\neq i} L_{ij} \, 
{\bf T}_i \cdot {\bf T}_j + \Sigma_{\gamma}\right)
\Bigg]
{{\cal A}^{(1)}} \, + \, {\rm h.\,c.}
\nnb\\
&=&
- \frac{1}{\eps^2}\,
\Sigma_{\!_C} \, V
- \frac1\eps \Sigma_{\gamma} \, V
- \frac{1}{\eps}\,
\sum_{i,j \neq i} L_{ij} \, V_{ij}
\, ,
\eeq
where the colour-correlated virtual correction $V_{ij}$ is defined in \eq{colcorrV}.
Combining \eq{hz2h} with \eq{hz1z1h} and \eq{h0z1a1}, we get a complete and explicit
expression for the pole part of the double-virtual contribution to the cross section,
\beq
\label{VVpolesfin}
VV \Big|_{\rm poles}
& = &
\bigg( \frac{\alpha_s}{2\pi} \bigg)^2
\Bigg\{
- \, \frac{1}{\eps^4} \,
\frac1{2}
\, \Sigma_{\!_C}^2 \, B
+ \,
\frac{1}{\eps^3} \,
\Sigma_{\!_C} \bigg[ 
\bigg( \frac{3}{8} \, \beta_0 \, - \, \Sigma_{\gamma} \! \bigg) \, B
- \, \sum_{i, j\neq i} L_{ij} \, B_{ij} \bigg]
\nonumber \\
&&
\hspace{1.5cm}
+ \,\, 
\frac{1}{\eps^2} \,
\frac14\,
\bigg[
\bigg( \beta_0 \Sigma_{\gamma} \, - \, 
\frac{\widehat{\gamma}_K^{(2)}}{4} \Sigma_{\!_C}
-\,2 \, \Sigma_{\gamma}^2 \bigg) \, B
\nonumber \\
& &
\qquad\qquad\qquad\qquad
\, + \,\bigg(\beta_0 - \,4 \, \Sigma_{\gamma}\bigg)
\sum_{i,j \neq i} L_{ij} \, B_{ij}
- \,
\sum_{\substack{i,j\neq i\\k,l\neq k}} 
L_{ij} \, L_{kl} \, B_{ijkl} \bigg]
\nonumber \\
&&
\hspace{1.5cm}
- \,\,
\frac{1}{\eps} \,
\frac18\,
\bigg[4\,\Sigma_{\gamma}^{(2)} \, B
\, + \,\widehat{\gamma}_K^{(2)}
\sum_{i,j\neq i} L_{ij} \, B_{ij} \bigg]
\Bigg\}
\nnb\\
&& - \, \, 
\frac{\alpha_s}{2\pi} \,\,
\bigg[
\frac{1}{\eps^2}\,
\Sigma_{\!_C} \, V
+\frac1\eps \Sigma_{\gamma} \, V
+ \frac{1}{\eps}\,
\sum_{i,j \neq i} L_{ij} \, V_{ij}
\bigg]
\, .
\eeq
\eq{VVpolesfin} can now be combined with the integrals of the double-radiative
and the real-virtual counterterms to form the subtracted double-virtual contribution
to the cross section, $VV_{\,\rm sub}$, given in \eq{eq:VVsub}.


\subsection{\vspace{-.5mm}Integrated counterterms for double-virtual poles}
\label{VVcount}

\vspace{.5mm}
The expressions for the relevant integrated counterterms, $I^{\two}$ and $I^{\RV}$, 
were given in \eq{eq:I2} and in \eq{eq:IRV}, respectively. We only need to expand
these expressions in powers of $\eps$, including terms up to ${\cal O} (\eps^0)$.
We define
\beq
I^{\two} + I^{\RV}
\, \equiv \,
I^{\twoRV}_{\rm poles}
+
I^{\twoRV}_{\rm fin} + {\cal O} (\eps)
\, .
\eeq
As expected, the pole part $I^{\twoRV}_{\rm poles}$ exactly cancels \eq{VVpolesfin}:
\beq
I^{\twoRV}_{\rm poles}
& = & - \,
VV \Big|_{\rm poles}
\, .
\eeq
We note in particular that \textit{it is not necessary} to compute NLO virtual corrections up
to ${\cal O} (\eps^2)$, since the last term in \eq{VVpolesfin}, containing virtual corrections
multiplied times explicit poles up to $\eps^{-2}$, is exactly reproduced by $I^{\twoRV}_{\rm 
poles}$, so that ${\cal O} (\eps)$ contributions to NLO corrections never appear in our
subtraction formula\footnote{This understands the technical capability by a two-loop 
provider to turn off the ${\cal O}(\eps)$ NLO virtual contribution in the computation of 
$VV$. Were this is not the case, the evaluation of $I^{\two}$ as well would have to be 
performed with such a contribution turned on.}. This was anticipated in Ref.~\cite{Weinzierl:2011uz} 
and emerges clearly in our approach thanks to the factorisation properties of the one-loop 
amplitude, and the minimal scheme we adopt for the factorisation of virtual corrections.
The finite part of the integrated counterterms can be written as ($r = r_{j}$, $r' = r_{jl}$)
\beq
\label{itworv}
I^{\twoRV}_{\rm fin}
& = &
\bigg( \frac{\alpha_s}{2\pi} \bigg)^{\!\!2}
\Bigg\{ \;
\bigg[ \;
I^{(0)}
+
\sum_{j} 
I^{(1)}_{j} \, 
L_{jr}
+
\sum_{j} 
I^{(2)}_{j} \, 
L_{jr}^{2}
+
\frac{1}{2} \,
\!\! \sum_{j, l\ne j} \!
\gamma^{\rm hc}_{j} \, \gamma^{\rm hc}_{l} \,
L_{jr'} L_{lr'}
\bigg]
\,
B
\\
&&
\hspace{14mm}
+ \, 
\sum_{j}
\Big[ 
I^{(0)}_{jr}
+
I^{(1)}_{jr} \, L_{jr}
\Big]
\,
B_{jr}
-
2 \,
\big( 1 - \zeta_2 \big) \,
\!\! \sum_{j,c\neq j,r} \!
\gamma^{\rm hc}_{j} \,
\big( 2 - L_{cr} \big)
\,
B_{cr}
\nnb\\
&&
\hspace{14mm}
+ \!
\sum_{c,d \neq c} \! L_{cd}
\bigg[ 
I^{(0)}_{cd}
+
I^{(1)}_{cd} \, L_{cd}
+
\frac{\beta_{0}}{12} \, L_{cd}^{2}
-
\frac{1}{2} 
\big( 4 - L_{cd} \big)
\sum_{j} \,
\gamma^{\rm hc}_{j} \,
L_{jr}
\bigg]
\,
B_{cd}
\nnb\\
&&
\hspace{14mm}
+ \!
\sum_{c,d \neq c} 
\bigg[ 
- 2 + \zeta_2 + 2\,\zeta_3 - \frac{5}{4}\,\zeta_4
+
2 \big(1-\zeta_3\big) \, L_{cd}
\bigg] 
\,
B_{cdcd}
\nnb\\
&&
\hspace{14mm}
+ \,
\big(1-\zeta_2\big)
\! \sum_{\substack{c,d\neq c\\e\neq d}} \!
L_{cd} \, L_{ed} \, 
B_{cded}
+
\! \sum_{\substack{c,d\neq c\\e,f\neq e}} \!
L_{cd} \, L_{ef}
\bigg[
1
-
\frac{1}{2} \, L_{cd}
\bigg(\! 1 \!-\! \frac{1}{8} L_{ef} \!\bigg) 
\bigg]
\,
B_{cdef} 
\nnb\\
&&
\hspace{14mm}
+ \,
\pi
\!\sum_{\substack{c,d\neq c\\e\neq c,d}} \!
\bigg[ 
\ln\!\frac{s_{ce}}{s_{de}} \,
L_{cd}^2
+
\frac{1}{3}\ln^3\!\frac{s_{ce}}{s_{de}}
+
2 \, \Li_3\!\left(\!-\frac{s_{ce}}{s_{de}}\!\right)
\bigg]
B_{cde}
\Bigg\}
\nnb\\
&&
+ \,\,
\frac{\alpha_s}{2\pi} \,
\bigg[
\bigg(
\Sigma_{\phi}
-
\sum_{j}
\gamma^{\rm hc}_{j} \, 
L_{jr}
\bigg)
\,
V^{\rm fin}
+
\! \sum_{c,d \neq c} \!
L_{cd} \, 
\bigg(\! 2 \!-\! \frac{1}{2}L_{cd} \!\bigg) 
\,
V_{cd}^{\rm fin}
\bigg]
\, ,
\nnb
\eeq
where $V^{\rm fin}$ and $V_{cd}^{\rm fin}$ are the ${\cal O} (\eps^0)$ terms in
the virtual and colour-correlated virtual contributions, which are obtained from the
full virtual contributions $V$ and $V_{cd}$ by subtracting the IR poles given explicitly
by \eq{NLOres_reim}. We emphasise that the kinematic dependence of \eq{itworv}
is only through simple logarithms of kinematic invariants, with the single exception 
of the trilogarithm multiplying the tripole Born-level colour correlation $B_{cde}$ on
the one-but-last line of \eq{NLOres_reim}. All the integral coefficients appearing in
\eq{NLOres_reim} are pure numbers, and they are given by
\beq
\label{intcoeff_fin}
I^{(0)}
& = &
N_q^{2} C_{\!_F}^{2} \,
\bigg[
\frac{101}{8} \!-\! \frac{141}{8} \zeta_{2} \!+\! \frac{245}{16} \zeta_{4}  
\bigg]
\, + \, 
N_g 
N_q C_{\!_F} \, 
\bigg[
C_{\!\!_A} 
\bigg(
\frac{13}{3} \!-\! \frac{125}{6} \zeta_{2} \!+\! \frac{245}{8} \zeta_{4} \!
\bigg)
+
\beta_{0} 
\bigg(
\frac{77}{12} \!-\! \frac{53}{12} \zeta_{2} \!
\bigg)
\bigg]
\nnb \\
&+&
N_g^{2} \, 
\bigg[
C_{\!\!_A}^{2} 
\bigg(
\frac{20}{9} - \frac{13}{3} \zeta_{2} + \frac{245}{16} \zeta_{4} 
\bigg)
+
\beta_{0}^{2} 
\bigg(
\frac{73}{72} - \frac{1}{8} \zeta_{2}
\bigg)
+
C_{\!\!_A} \beta_{0} 
\bigg(
- \frac{1}{9} - \frac{11}{3} \zeta_{2}
\bigg)
\bigg]
\nnb\\
&+&
N_q C_{\!_F} \, 
\bigg[
C_{\!_F} 
\bigg(
\frac{53}{32} \!-\! \frac{57}{8}\zeta_{2} 
\!+\! 
\frac{1}{2}\zeta_{3} \!+\! \frac{21}{4}\zeta_{4} \!
\bigg)
+
C_{\!\!_A} 
\bigg(
\frac{677}{432} \!+\! \frac{5}{3}\zeta_{2} 
\!-\! \frac{25}{2}\zeta_{3} \!+\! \frac{47}{8}\zeta_{4} \!
\bigg)
\nnb\\
&&
\hspace{11mm}
+ \,
\beta_{0}
\bigg(
\frac{5669}{864} \!-\! \frac{85}{24}\zeta_{2} \!-\! \frac{11}{12}\zeta_{3} \!
\bigg)
\bigg]
\nnb\\
&+&
N_g \, 
\bigg[
C_{\!_F} C_{\!\!_A}
\bigg(
- \frac{737}{48} + 11\zeta_{3}
\bigg)
+
C_{\!_F} \beta_{0}
\bigg(
\frac{67}{16} - 3\zeta_{3}
\bigg)
+
\beta_{0}^{2} 
\bigg(
\frac{73}{72} - \frac{3}{8}\zeta_{2}
\bigg)
\nnb\\
&&
\hspace{7mm}
+ \,
C_{\!\!_A}^{2} 
\bigg(
- \frac{4289}{216} + \frac{15}{2}\zeta_{2} - 14\zeta_{3} + \frac{89}{8}\zeta_{4} 
\bigg)
+
C_{\!\!_A} \beta_{0} 
\bigg(
\frac{647}{54} - \frac{53}{8}\zeta_{2} - \frac{11}{12}\zeta_{3}
\bigg)
\bigg]
\, ,
\nnb
\eeq
\beq
I^{(1)}_{j}
& = &
(f_j^q+f_j^{\bar q}) \, C_{\!_F} \,
\bigg[
N_q C_{\!_F}
\bigg(
\frac{5}{2} - \frac{7}{4}\zeta_{2}
\bigg)
+
N_g 
C_{\!\!_A} 
\bigg(
\frac{1}{3} - \frac{7}{4}\zeta_{2}
\bigg)
+
\frac{2}{3} 
N_g 
\beta_{0}
\nnb\\
&&
\hspace{18mm}
+ \,
C_{\!_F} 
\bigg(
- \frac{3}{8} - 4\zeta_{2} + 2\zeta_{3}
\bigg)
+
C_{\!\!_A} 
\bigg(
\frac{25}{12} - 3\zeta_{2} + 3\zeta_{3}
\bigg)
+
\beta_{0} 
\bigg(
\frac{1}{24} + \zeta_{2}
\bigg)
\bigg]
\nnb\\
&+&
f_j^g \,
\bigg[
N_q C_{\!_F} 
C_{\!\!_A}
\big(
10 \!-\! 7\zeta_{2}
\big)
-
N_q C_{\!_F} 
\beta_{0}
\bigg(
\frac{5}{2} \!-\! \frac{7}{4}\zeta_{2} \!
\bigg)
+
N_g 
C_{\!\!_A}^{2} 
\bigg(
\frac{4}{3} \!-\! 7\zeta_{2} \!
\bigg)
+
N_g 
C_{\!\!_A} 
\beta_{0}
\bigg(
\frac{7}{3} \!+\! \frac{7}{4}\zeta_{2} \!
\bigg)
\nnb\\
&&
\hspace{9mm}
- \,
\frac{2}{3} 
(N_g+1) 
\beta_{0}^{2} 
+
\frac{11}{4} 
C_{\!_F} C_{\!\!_A}
- 
\frac{3}{4}
C_{\!_F} \beta_{0}
+
C_{\!\!_A}^{2} 
\bigg(
\frac{28}{3} \!-\! \frac{23}{2}\zeta_{2} \!+\! 5\zeta_{3} \!
\bigg)
-
C_{\!\!_A} \beta_{0}
\bigg(
\frac{2}{3} \!-\! \frac{5}{2}\zeta_{2} \!
\bigg)
\bigg]
\, ,
\nnb\\
I^{(2)}_{j}
& = &
\frac{1}{8}\big( 15\,C_{A} - 7\,\beta_{0} \big)
C_{f_{j}} 
-
\frac{1}{4}\big( 5\,C_{A} - 2\,\beta_{0} \big)
\gamma_{j}
+ \frac{1}{8} \big(
16\,\zeta_2\, - 15 \big) C_{f_{j}}^{2}
\, ,
\nnb\\
I^{(0)}_{jr}
& = &
\big( - 1 + 3\zeta_2 - 2\zeta_3\big) C_{A} 
-
\frac{1}{2}\big( 13 + 10\zeta_2 + 2\zeta_3 \big) C_{f_{j}} 
+ 
\big( 5 + 2\zeta_3 \big) \gamma_{j}
\, ,
\nnb\\
I^{(1)}_{jr}
& = &
\big( 1 - \zeta_2 \big) C_{A} 
+ 
\frac{1}{2}\big( 4 + 7\zeta_2 \big) C_{f_{j}} 
- 
\big( 2 + \zeta_2 \big) \gamma_{j}
\, ,
\nnb\\
I^{(0)}_{cd}
& = &
\bigg( \frac{20}{9} - 2\zeta_2 - \frac{7}{2}\zeta_3 \bigg) C_{A}
+
\frac{31}{9}\,\beta_{0}
+
2\,\Sigma_{\phi}
+
8 \, \big( 1 - \zeta_2 \big) \, C_{f_{d}}
\, ,
\nnb\\
I^{(1)}_{cd}
& = &
-
\bigg( \frac{1}{3} - \frac{1}{2}\zeta_2\bigg) C_{A}
- 
\frac{11}{12}\,\beta_{0}
- 
\frac{1}{2}\,\Sigma_{\phi}
\, .
\eeq
We stress that, as expected, the pole part $I^{\twoRV}_{\rm poles}$ does not depend 
on the reference momenta $r, \, r'$; conversely, the dependence on $r, \, r'$ arising in 
the finite part $I^{\twoRV}_{\rm fin}$ is necessary to cancel the explicit one in the 
counterterms $K^{\two}$ and $K^{\RV}$.


\section{\vspace{-.5mm}An explicit example of cancellation}
\label{Explicit}

\begin{figure}[h]
\hspace{5mm}
\includegraphics[width=0.3\linewidth]{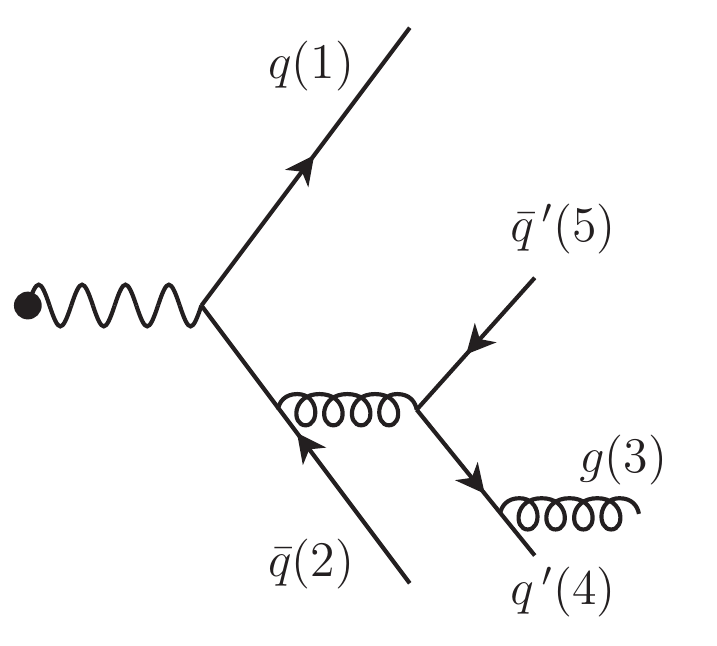}
\quad
\includegraphics[width=0.3\linewidth]{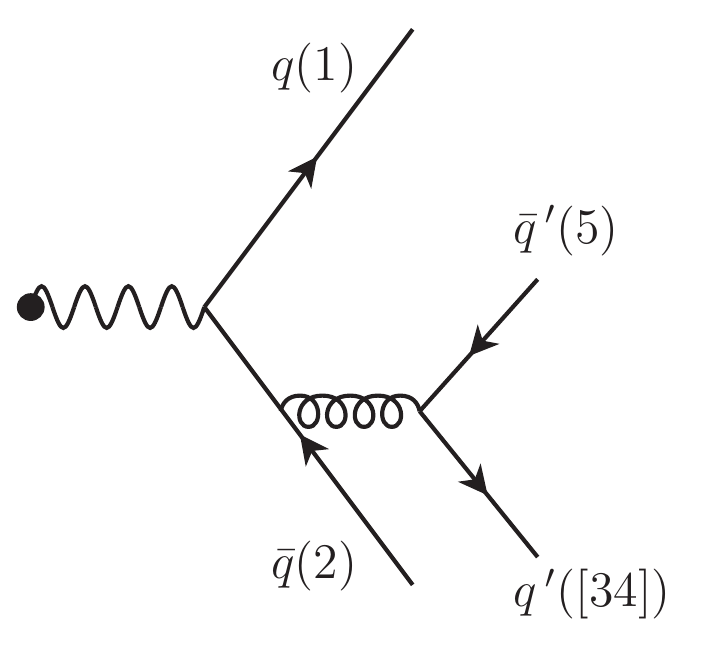}
\quad
\includegraphics[width=0.3\linewidth]{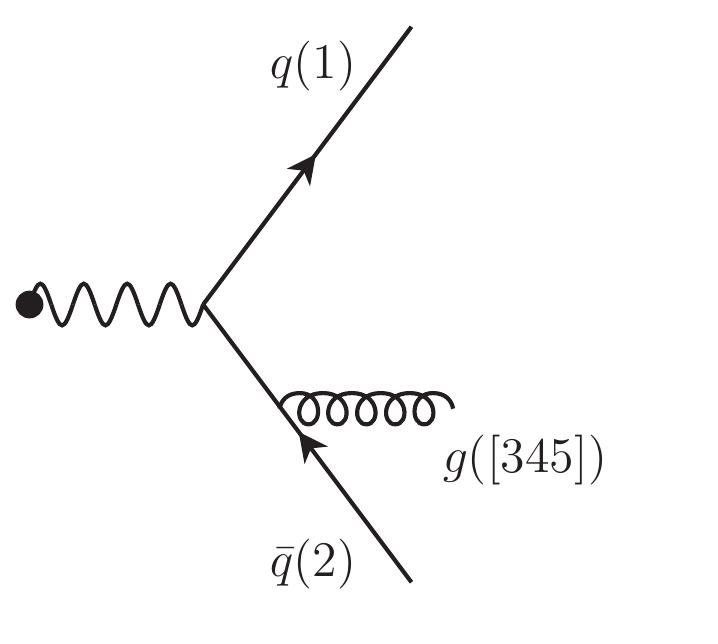}
\caption{
Sample Feynman diagrams for three-jet production in leptonic collisions: NNLO 
double real emission (left), NLO single real emission (middle), Born process (right).}
\label{fig:graphs}
\end{figure}

In this section we work out in detail the cancellation of IR singularities for the process 
$e^+ e^- \to q(1) \bar q(2) g([345])$ at NNLO, focusing on the double-real-emission 
channel where an extra quark-antiquark pair is emitted, $e^+ e^- \to q(1) \bar q(2) g(3) 
\q'(4) \bar q'(5)$ (with $q$ and $q'$ being different quark flavours), and picking the sector 
$\W{4353}$ as a test case. The sector function selects $\bbC{43}$ as single-unresolved 
limit, hence the only possible underlying single-real-emission channel is $e^+ e^- \to 
q(1) \bar q(2)$ $q'([34]) \bar q'(5)$. Moreover, the only double-unresolved configurations 
allowed are $\bbS{45}$ and $\bbC{435}$. In Fig.~\ref{fig:graphs} we show a sample 
NNLO double-real-emission Feynman diagram contributing to this sector (left), 
together with its underlying NLO single-real-emission diagram (middle), as well as 
the LO Born diagram (right). 

The expression for the relevant kernels, according to Eqs.~\eqref{eq:bSik} and 
\eqref{eq:bCijCijk} is rather simple. Since partons 4 and 5 are quarks, the eikonal 
kernels $\mc E^{(4)}_{ab}$ and $\mc E^{(5)}_{ab}$ vanish; moreover, since the 
parent parton [34] of the 34 pair is a quark, there is no azimuthal dependence in 
the collinear splitting kernels, hence $Q^{\mu\nu}_{43(r)} = 0$, and $P^{\mu\nu}_{43(r)} 
\bR_{\mu\nu}^{(43r)} = P_{43(r)} \bR^{(43r)}$. The required limits are then given 
by (see Appendix~\ref{app:kernels})
\beq
\label{eq:kernel_explicit_ex}
\bbC{43} \, RR 
& \equiv &
\frac{\Norm}{s_{43}} \, P_{43(r)} \, 
\bR^{(43r)}
\, ,
\hspace{30mm}
\bbC{435} \, \RR 
\, \equiv \,
\frac{\Norm^{\,2}}{s_{435}^2} \,
P_{435(r)}^{\mu\nu} \,\bB_{\mu\nu}^{(435r)} 
\, ,
\nonumber  \\
\bbS{45} \, \RR
& \equiv &
\frac{\Norm^{\,2}}{2} \!
\sum_{\substack{c \neq 4,5 \\ d \neq 4,5,c}}
\mc E^{(45)}_{cd}
\bB_{cd}^{(45cd)}
\, ,
\hspace{14.6mm}
\bbS{45} \, \bbC{435} \, \RR 
\, \equiv \,
- \,
\Norm^{\,2} \,
C_{f_3} \,
\mc E_{3r}^{(45)}
\bB^{(453r)} 
\, ,
\nonumber  \\
\bbC{43} \, \bbC{435} \, \RR 
& \equiv &
\Norm^{\,2} \, 
\frac{P_{43(r)}}{s_{43}} \,
\frac{\bar P_{[34]5(r)}^{(43r)\mu\nu}}{\sk{[34]5}{43r}} \,
\bB_{\mu\nu}^{(43r,[34]5r)}
\, .
\eeq
In the case we are considering, the reference index $r$ (used for collinear limits) could 
be either $r=1$ or $r=2$. The subtracted double-real contribution in this sector is thus
\beq
RR^{\, \rm sub}_{4353}
& = &
RR \, \W{4353} -
K^{\one}_{4353} -
\Big( K^{\two}_{4353} - K^{\otwo}_{4353} \Big)
\, ,
\eeq
where the NNLO counterterms read
\beq
K^{\one}_{4353}
& \equiv &
\bbC{43} \, RR \, \W{4353}
\nnb\\
& = &
\Norm \, \frac{P_{43(r)}}{s_{43}} \, 
\bR^{(43r)} \,
\bW{5[34]}^{\, (43r)} \,
\W{{\rm c},43(r)}^{\, (\alpha)}
\, ,
\nnb \\ [5pt]
K^{\two}_{4353}
& \equiv &
\Big[
\bbS{45} +
\bbC{435} \, \big(1-\bbS{45}\big)
\Big]
\, RR \, \W{4353}
\nnb \\
& = &
\frac{\Norm^{\,2}}{2} \!
\sum_{\substack{c \neq 4,5 \\ d \neq 4,5,c}}
\mc E^{(45)}_{cd} \,
\bB_{cd}^{(45cd)} \, 
\big( \bbS{45} \, \W{4353} \big)
+
\frac{\Norm^{\,2}}{s_{435}^2} \,
P_{435(r)}^{\mu\nu} \,\bB_{\mu\nu}^{(435r)} 
\big(\bbC{435}\,\W{4353}\big)
\nnb \\
&&
+ \,
\Norm^{\,2} \,
C_{f_3} \,
\mc E_{3r}^{(45)} \,
\bB^{(453r)} \,
\big( \bbS{45} \, \bbC{435} \, \W{4353} \big)
\, ,
\nnb \\ [5pt]
K^{\otwo}_{4353}
& \equiv &
\bbC{43} \, \bbC{435}
\, RR \, \W{4353}
\nnb \\
& = &
\Norm^{\,2} \,
\frac{P_{43(r)}}{s_{43}} \,
\frac{\bar P_{[34]5(r)}^{(43r) \mu \nu}}{\sk{[34]5}{43r}} \,
\bB_{\mu \nu}^{(43r,[34]5r)}
\, \bW{{\rm c},5[34](r)}^{\, (43r)} \,
\W{{\rm c},43(r)}^{\, (\alpha)}
\, .
\eeq
Let us begin by showing the cancellation of the singular behaviour in the $\bC{43}$ 
limit. We note that particle momenta in this limit obey $\bC{43} \{ k \} = \bC{43}
\{\bar k\}^{(43r)} = \{k\}_{\slashed4 \slashed3, [34]}$, which implies that the limit
taken on sector functions gives $\bC{43} \, \bW{5[34]}^{(43r)} = \W{5[34]}$, and 
$\bC{43} \, \W{{\rm c},43(r)}^{(\alpha)} = \bC{43} \, \W{43}^{(\alpha)}$. This, 
together with the relation $\bC{43} \W{4353} = \bC{43} \W{5[34]} \W{43}^{(\alpha)}$, 
and with the known $\bC{43}$ limit of $RR$, is sufficient to show that
\beq
\bC{43} \,
\Big[1-\bbC{43}\Big] \, RR \, \W{4353}
& \to & {\rm integrable}
\, .
\eeq
Next, we need to show that
\beq
\label{eq:explicit_consistency_1}
\bC{43} \,
\Big[K^{\two}_{4353}-K^{\otwo}_{4353}\Big]
& \to & {\rm integrable}
\, .
\eeq
To this end, let us note that the double soft kernel factorised in $\bbS{45}\, RR$ is 
not singular in the $\bC{43}$ limit, since the denominator of $\mc E^{(45)}$ (see 
\eq{eq:kernel_explicit_ex}) features the sum $s_{[34]5} \equiv s_{34} + s_{35}$. 
The same is true for the $\bbC{435} \, \bbS{45}\, RR$ limit, which is constructed 
with the same double-soft kernel. 
%
%
%
As a consequence, checking the requirement in \eq{eq:explicit_consistency_1} 
reduces to verifying that
\beq
\label{C43integrable}
\bC{43} \, \bbC{435} \,
\Big[1-\bbC{43}\Big] \, RR \, \W{4353}
& \to & {\rm integrable}
\, .
\eeq
As far as sector functions are concerned, it is straightforward to check that
$\bC{43} \, \bbC{435} \, \W{4353} = 
\bC{43} \, \bbC{43} \, \bbC{435} \, \W{4353} = 
\bC{43} \,\bW{{\rm c},5[34](r)}^{(43r)} \, \W{{\rm c},43(r)}^{(\alpha)} = 
\bC{[34]5} \, \W{5[34]} \, \bC{43} \, \W{43}^{(\alpha)}$.
The mapped kinematics is such that the identity $\{\bar k\}^{(435r)} = 
\{\bar k\}^{(43r,[34]5r)}$ holds also far from the collinear limit. Finally, 
the double-collinear kernel can be written as
\beq
\frac{P^{\mu\nu}_{435(r)}}{s_{345}^2}
& = &
\frac1{s_{345}^2}
\bigg[
-
P_{453(r)}^{\og} \, g^{\mu\nu}
+
\sum_{a=3,4,5}Q_{453(r)}^{\og,a} \, d_a^{\, \mu\nu} \,
\bigg]
\, .
\eeq
It is easy to show that in the collinear limit $\bC{43}$ the non-abelian contribution to this kernel is non-singular\footnote{
To this end we use the following equivalence relations in the $\bC{43}$ limit 
($\kt^{\mu}_{5} = - \kt^{\mu}_{3} - \kt^{\mu}_{4}$): 
\[
\bC{43} \,
\frac{s_{35}}{s_{45}}
=
\bC{43} \,
\frac{s_{3r}}{s_{4r}}
=
\bC{43} \,
\frac{z_{3}}{z_{4}}
\, ,
\qquad
\bC{43} \,
\kt^{\mu}_{3}
=
- \,
\bC{43} \,
\frac{z_{3}}{z_{34}} \, \kt^{\mu}_{5}
\, ,
\qquad
\bC{43} \,
\kt^{\mu}_{4}
=
- \,
\bC{43} \,
\frac{z_{4}}{z_{34}} \, \kt^{\mu}_{5}
\, .
\]
}, 
while the abelian part, owing to the relations $x_3 = 
s_{3r}/s_{[34]r}$ and $x_4 = 1 - x_3$, becomes 
\beq
\bC{43} \, 
\frac{P^{\mu\nu}_{435(r)}}{s_{345}^2}
& = &
\bC{43} \,
\frac{C_F \, T_R}{s_{[34]5}^2} \, 
\frac{2 x_4+(1-\eps)x_3^2}{x_3} \,
\frac{s_{[34]5}}{s_{43}} \,
\bigg[
\!\!
-g^{\mu\nu}
+
4 \, z_5(1-z_5)\frac{\kt_5^\mu \kt_5^\nu}{\kt_5^2}
\bigg]
\nnb\\
& = & 
\bC{43}\,
\frac{P_{43(r)}}{s_{43}}
\,
\frac{P_{[34]5(r)}^{\mu\nu}}{s_{[34]5}}
\, = \,
\bC{43} \, 
\frac{P_{43(r)}}{s_{43}}
\frac{\bar P_{[34]5(r)}^{(43r)\mu\nu}}{\sk{[34]5}{43r}}
\, ,
\eeq 
which shows that singular terms in the $\bC{43}$ limit cancel in \eq{C43integrable}.

Moving on to the double-soft $\bS{45}$ limit, we note that its action on kinematics 
is such that $\bS{45} \, \{k\} = \bS{45} \, \{\bar k\}^{(45cd)} = \{k\}_{\slashed4\slashed5}$. 
This consideration, together with the relation $\bS{45} \, \W{4353} = \bS{45} \, \bbS{45} \, 
\W{4353}$, and the known double-soft limit of the double-real matrix element, immediately 
implies that
\beq
\bS{45} \,
\Big[1-\bbS{45}\Big] \, RR \, \W{4353}
& \to & {\rm integrable}
\, .
\eeq
On the other hand, the single-unresolved kernel $K^{\one}_{4353}$ is non-singular 
in the $\bS{45}$ double-soft limit, as it does not feature any $1/s_{45}$ enhancement, 
so $\bS{45} \, K^{\one}_{4353} = 0$. The same holds for the strongly-ordered collinear 
kernel $ \bbC{43} \, \bbC{435} \, RR$, whence $\bS{45} \, K^{\otwo}_{4353} = 0$. 
We thus are left with the requirement
\beq
\label{eq:explicit_consistency_2}
\bS{45} \,
\bbC{435} \, \big(1-\bbS{45}\big) RR \, \W{4353}
& \to & {\rm integrable}
\, .
\eeq
As far as sector functions are concerned, it is straightforward to verify that $\bS{45} \, 
\bbC{435} \, \W{4353} = \bS{45} \, \bbS{45} \, \bbC{435} \,\W{4353}$, using \eq{eq:bCijkW} 
and \eq{eq:bSikCijkW}. As to matrix elements, the relevant kernel is
\beq
\bbC{435} \, \big(1-\bbS{45}\big) RR
& = &
\frac{\Norm^{\,2}}{s_{435}^2} \,
P_{435(r)}^{\mu\nu} \,\bB_{\mu\nu}^{(435r)} 
+
\Norm^{\,2} \,
C_{f_3} 
\mc E_{3r}^{(45)}
\bB^{(453r)}
\, .
\eeq
Here, using the kinematic condition $\bS{45} \, \{\bar k\}^{(435r)} = \bS{45} \, 
\{\bar k\}^{(453r)}$, one can show that the second term on the right-hand side 
precisely removes from the first one all double-soft enhancements proportional 
to $1/s_{45}$, as was the case for unimproved limits. Thus we verify that
\beq
\bS{45} \, \bbC{435} \, \big(1-\bbS{45}\big) RR \to {\rm integrable} .
\eeq
Next, we consider the behaviour of the counterterm $K^{\two}_{4353}$ under the 
double-collinear limit $\bC{435}$. First, we notice that $\bC{435} \, \{ k\} = \bC{435} \, 
\{\bar k\}^{(435r)} = \bC{435} \, \{\bar k\}^{(453r)} = \{k\}_{\slashed 3 \slashed 4 
\slashed 5[345]}$, and that, using \eq{eq: sigma_abcd} and \eq{eq: hatsigma}, one 
has $\bC{435} \, \bbC{435} \, \W{4353} = \bC{435} \, \W{4353}$. These relations, 
together with the known $\bC{435}$ limit of $RR$, are sufficient to show that 
$\bC{435} \, \bbC{435} \, RR \, \W{4353} = \bC{435} \, RR \, \W{4353}$. 
As for the remaining part of $K^{\two}_{4353}$, we have
\beq
\bC{435} \, \bbS{45}\,\W{4353}
& = &
\bC{435} \, \bbS{45}\,\bbC{435}\,\W{4353}
\, = \,
\frac{\sigma_{4353}}{\sigma_{4353}+\sigma_{5343}+\sigma_{4553}+\sigma_{5443}}
\, ,
\eeq
as can be deduced from the definitions in \eq{eq: hatsigma} and \eq{eq:bSikCijkW}. 
The action of the double-soft kernel on the matrix element, on the other hand, gives
\beq
\label{C435matr}
\bC{435} \,
\sum_{\substack{c \neq 4,5 \\ d \neq 4,5,c}}
\mc E^{(45)}_{cd}
\bB_{cd}^{(45cd)}
\, = \,
2 \, \mc E^{(45)}_{3r} \!
\sum_{\substack{d \neq 4,5,3}}
\bC{435} \,
\bB_{3d}^{(453d)}
\, = \,
2 \, \mc E^{(45)}_{3r} \!
\sum_{\substack{d \neq 4,5,3}}
\bC{435} \,
\bB_{3d}^{(453r)}
\, ,
\eeq
where, in the last step, we used $\bC{435} \, \{\bar k\}^{(453d)} = \bC{435} \, 
\{\bar k\}^{(453r)}$. By performing the sum over colours, \eq{C435matr} becomes
\beq
2 \, \mc E^{(45)}_{3r} \!
\sum_{\substack{d \neq 4,5,3}}
\bC{435} \,
\bB_{3d}^{(453r)}
& = &
-2 \, C_{f_3} \, \mc E^{(45)}_{3r}
\bC{435} \,
\bB^{(453r)}
\, ,
\eeq
which matches (with opposite sign) the kernel in $\bbC{435} \, \bbS{45} \, RR$, 
finally showing that
\beq
\bC{435} \,
\bbS{45} \,
\Big[
1 - \bbC{435}
\Big]
\, RR \, \W{4353}
& \to &
{\rm integrable}
\, .
\eeq
In order to complete the proof of the cancellation of singular contributions in the 
$\bC{435}$ limit, it is finally necessary to show that
\beq
\bC{435}
\Big[K^{\one}_{4353} - K^{\otwo}_{4353}\Big]
& \to & {\rm integrable}
\, .
\eeq
The sector functions appearing in $K^{\one}_{4353}$ and $K^{\otwo}_{4353}$ approach 
the same value under the double-collinear limit, since $\bC{435} \, \bW{{\rm c},5[34](r)} = 
\bC{435} \, \bW{5[34]}$. As for the kernels, one just needs to check that
\beq
\bC{435} \,
\bR^{(43r)}\,
& = &
\bC{435} \,
\frac{\bar P_{[34]5(r)}^{(43r)\mu\nu}}{\sk{[34]5}{43r}} \,
\bB_{\mu\nu}^{(43r,[34]5r)}
\, ,
\eeq
which is indeed the case, since the $\bC{435}$ double-collinear limit acts on the mapped 
kinematics as a single-collinear limit between parton 5 and the parent parton $[34]$.

After proving the local cancellation of all phase-space singularities in sector $\W{4353}$, 
we conclude this section by showing that the functional form chosen for the sector 
functions is also capable of eliminating \emph{spurious} singularities, arising from 
collinear kernels, as detailed in \eq{eq:spur_sing}. 
To this end, we consider the $\bC{3r}$ 
limit (in fact, the $\bC{4r}$ limit does not generate any spurious singularities), we note 
that the double-real $RR \, \W{4353}$, the double-soft $\bbS{45} \, RR \, \W{4353}$, 
and the double-soft-collinear $\bbS{45} \, \bbC{435} \,RR \, \W{4353}$ contributions 
are non-singular, and therefore we will not considered any further. As for kernels 
involving the single-collinear contribution $\bbC{43} \, RR$, they are always 
accompanied by the $\W{{\rm c},43(r)}$ collinear sector functions. Making use 
of the definition in \eq{eq:Wcij}, one verifies that $\bC{3r} \, \W{{\rm c},43(r)} \, 
\propto \, w_{3r}^\alpha$, which (over)compensates the $1/s_{3r}$ spurious 
divergence stemming from the $1/z_3$ factor in $P_{43(r)}$. An analogous 
compensating mechanism is at work for $\bbC{43} \, \bbC{435} \, RR \, \W{4353}$. 
We see that introducing an $r$-dependence in the definition of sector functions enables 
us to achieve a non-singular behaviour in all relevant limits. The double-collinear 
kernel in $\bbC{435} \, RR$ similarly features enhanced $1/s_{3r}$ terms stemming 
from $1/z_3$ denominators: in this case, the numerator of the $\bbC{435} \, \W{4353}$ 
sector function is finite in the $\bC{3r}$ limit, while its denominator contains terms 
like $\hat \sigma_{3545}$ that diverge when $w_{3r}^\alpha$ vanishes, eventually 
achieving $\bC{3r} \, \bbC{435} \, \W{4353} = 0$. We notice once again that such a 
dampening is crucially induced by the choice of including angles involving the 
recoiler $r$ in the definition of collinear sector functions.\\
\indent Finally, we comment on the $\bC{45r}$ spurious singularity. 
We introduce a common 
scaling parameter $\lambda$ for the invariants vanishing in this limit, $s_{45}$, $s_{4r}$
and $s_{5r}$. The $K^{\one}_{4353}$ counterterm is non-singular in this limit, since 
it does not feature any isolated $1/s_{4r}$ denominator, so we focus on $K^{\two}_{4353}$ 
and $K^{\otwo}_{4353}$. The kernel $\bbS{45} \, RR$ diverges as $\lambda^{-2}$ in 
the limit, due to denominators of type $1/s^2_{45}$, or  $1/s_{45} s_{[45]r}$, however 
the corresponding sector function $\bbS{45} \, \W{4353}$ scales as $\lambda^2$, 
thus compensating the singularity. Analogously, the counterterm $\bbC{43} \, \bbC{435}
\, RR \, \W{4353}$ is non-singular in the $\bC{45r}$ limit. As for the remaining counterterms, 
we have
\beq
\bC{45r} \, 
\bbC{435} \, RR
& = &
\bC{45r} \, 
\Norm^{\,2} \,
2 \, C_A T_R \,
\bigg[
\frac{1}{s_{45}^2}
\left(
\frac{s_{43}s_{5r}-s_{4r}s_{53}}{s_{[45]3} s_{[45]r}}
\right)^2
- 
\frac{s_{3r}}{s_{45} s_{[45]3} s_{[45]r}}
\bigg]
\, \sim \, 
\lambda^{-2}
\nnb\\
& = &
\bC{45r} \, 
\bbS{45} \, \bbC{435} \, RR
\, ,
\nnb \\
\bC{45r} \, 
\bbC{435} \, \W{4353}
& = &
\bC{45r} \,
\frac{\hat \sigma_{4353}}{\hat \sigma_{4553}+\hat \sigma_{5443}}
\, \sim \,
\lambda^{\alpha}
\nnb\\
& = &
\bC{45r} \, 
\bbS{45} \, \bbC{435} \, \W{4353}
\, ,
\eeq
where the dependence on the parameter $\alpha$ emerges from the definition
of the sector function, see \eq{eq: sigma_abcd}. In this case, both $\bbC{435} \, 
RR \, \W{4353}$ and $\bbS{45} \, \bbC{435} \, RR \, \W{4353}$ display at most 
an integrable $\lambda^{\alpha -2}$ singularity, which is ultimately due to the 
chosen sector.
However, even in a sector in which the two contributions are separately 
non-integrable (for instance in $\W{4553}$ or $\W{5443}$), the fact that both 
kernels and sector functions tend to become identical prevents a singular 
scaling of the double-hard collinear contribution 
$\bbC{435} (1 - \bbS{45} ) RR \, \W{4353}$. 
This completes our analysis of this example, showing that the subtracted 
double-real emission contribution under consideration is completely free of 
phase-space singularities.


\section{\vspace{-.5mm}Status and perspective}
\label{Persp}

\vspace{.5mm}
We have presented a complete analytic solution to the NNLO subtraction problem for 
general massless coloured final states, within the framework of Local Analytic Sector 
Subtraction, which can be implemented in conjunction with any numerical code providing 
the appropriate one- and two-loop matrix elements, and an efficient phase-space 
integrator.

The main ingredients for our construction are the following. Beginning with the double-radiative
contribution, we introduce a smooth partition of the radiative phase space into {\it sectors}, 
each containing a minimal number of soft and collinear singularities, following the basic 
logic of Ref.~\cite{Frixione:1995ms}. Next, we list 
all {\it uniform} soft and collinear limits, with up to two particles becoming unresolved, that 
contribute to each sector. Denoting these limits by ${\bf \ell}_{{\rm sect}, \, i}$, we then follow 
the strategy of Ref.~\cite{Frixione:2004is}, and construct combinations of the form $\prod_i 
(1 \, - \, {\bf \ell}_{{\rm sect}, \, i})$, which are guaranteed to be integrable in the relevant 
phase spaces, and for which double-counted nested limits have been properly subtracted.
Crucially, in all cases we define {\it commuting} limits, which significantly simplify subsequent 
manipulations. Exploiting the soft- and collinear-factorisation properties of matrix elements, all 
relevant limits can be expressed as products of known splitting kernels times lower-multiplicity 
matrix elements. In order to properly exploit this factorised structure for matrix elements, 
we then introduce a flexible set of phase-space {\it mappings}, which lead to the complete
factorisation of the phase-space integration, separating the on-shell Born-level configuration
from radiative factors. 
Using these mappings, we construct {\it improved limits} 
$\bar{\bf \ell}_{{\rm sect}, \, i}$, which 
are highly optimised, with different choices of mappings
for different limits, different sectors and different terms in each sector; furthermore, the 
action of the improved
limits on sector functions is tuned, when needed. Importantly, these 
optimisations must pass stringent {\it consistency conditions}, ensuring that nested 
improved limits with different mappings remain aligned with the underlying physical 
soft and collinear limits. An analogous procedure is followed for the real-virtual contribution,
where the radiative phase-space structure is much simpler, but splitting kernels (and thus
improved limits) are more intricate.

This lengthy optimisation work pays off when the local counterterms, thus obtained, are 
integrated over the radiative degrees of freedom. All counterterms can be analytically 
integrated, and all singular contributions to the integrated counterterms are given by
single-scale integrals, with trivial logarithmic dependence on Born-level kinematic 
invariants. When integrated counterterms are properly combined with the singular
part of the double-virtual contribution to the cross section, all poles are analytically 
shown to cancel. All finite contributions can also be obtained analytically, and they are 
of similar simplicity, with just a single contribution (proportional to a colour tripole) displaying 
a weight-three polylogarithm depending on two physical scales. In a sense, the existing
tension between the remarkable simplicity of double-virtual singularities and the increasing
intricacy of real-virtual and double-real radiative contributions is resolved by a judicious
choice of sectors and mappings. Indeed, the simplicity of the integrals associated with 
both double-real and real-virtual counterterms kindles hopes that a generalisation to
N$^3$LO subtraction with the same degree of generality might be achievable. On the other 
hand, our approach is undeniably costly from the combinatorial viewpoint, and requires
a fast-growing number of consistency checks, which would be challenging to tackle at 
higher orders.

The future developments of our work are clearly outlined. First of all, the formalism must
be numerically implemented and tested for efficiency. This work is under way, and was
completed at NLO in Ref.~\cite{Bertolotti:2022ohq}. In that paper, the NLO formalism 
was also extended to initial-state coloured particles, without significantly raising the 
technical difficulties. Obviously, the inclusion of non-trivial initial states is a high-priority
goal at NNLO as well, in view of LHC applications. Also in this case, this generalisation
is not expected to involve new major technical obstacles: as observed at NLO, new 
classes of mappings are needed, and collinear factorisation must be consistently 
implemented, but all of these developments are expected to be comparatively straightforward. 
Importantly, new phase-space integrals are expected to be of the same level of complexity 
as those presented here, so that a completely analytic result is expected to be within 
reach. Work is in progress also on this front. In the longer run, an important further 
ingredient to achieve complete generality for NNLO subtraction is the inclusion of 
massive particles in the final state. This task is going to be simplified by the fact that 
the number and type of singular limits associated with massive coloured particles are 
limited, since collinear limits for real radiation are non-singular in this case. Since our 
approach is combinatorially intensive, this is expected to be a significant advantage. 
On the other hand, massive particles will require adjustments in phase-space mappings, 
and will likely involve new classes of integrals, with a more intricate scale dependence. 
We are, nonetheless, confident that a complete analytic expression can be derived also 
in that case.

Finally, we believe that, notwithstanding the simplicity of our analytic results, there is
further room for optimisation, which would be very important in view of a future generalisation
of our approach to N$^3$LO. We note for example that the minimal interference between 
soft and collinear singularities which is suggested by factorisation at amplitude level still 
emerges in our formalism as an output rather than being introduced from the outset. 
We hope that a more detailed understanding of the factorisation structure for real radiation,
in particular for strongly-ordered limits, along the lines of Refs.~\cite{Magnea:2018ebr,
Magnea:2022twu} will provide further insights in this direction. Simplifications in the 
structure of nested infrared limits would likely improve significantly the combinatorial
challenges of our approach, and open the way to higher orders.

In summary, we believe that our results bring the goal of establishing a completely 
general, local, analytic and efficient NNLO-subtraction formalism one step closer.


\vspace{1cm}


\section*{Acknowledgments}

We thank Ezio Maina for collaboration in the early stages of this project, and Francesco Tramontano for useful discussions. 
We thank Zolt\'an Tr\'ocs\'anyi, \'Ad\'am Kardos and Sven-Olaf Moch for useful discussions on technical aspects of the method.
This research was partially supported by the Deutsche Forschungsgemeinschaft 
(DFG, German Research Foundation) under grant 396021762 - TRR 257, and
by the Italian Ministry of University and Research (MIUR), grant PRIN 20172LNEEZ. 
The work of PT has received support from Compagnia di San Paolo, grant 
n.~TORP\_S1921\_EX-POST\_21\_01.
\\
\\


\appendix


\section{\vspace{-.5mm}General notation}
\label{app:notations}

\vspace{.5mm}
We denote by $s$ the squared centre-of-mass energy and by
$\q^\mu = (\sqrt s,\vec 0\,)$ the centre-of-mass four-momentum. 
Given two final-state momenta $k_i^\mu$ and $k_j^\mu$, we define
\beq
s_{\q i} & = & 2 \, \q \cdot k_i 
\, ,
\qquad 
s_{ij} \; = \; 2 \, k_i \cdot k_j 
\, , 
\qquad
L_{ij} \; = \; \ln\frac{s_{ij}}{\mu^2}
\, ,
\nnb\\[3pt] 
e_i & = & \frac{s_{\q i}}{s} 
\, , 
\qquad\qquad
w_{ij} \; = \; \frac{s \, s_{ij}}{s_{\q i} \, s_{\q j}} 
\, .
\eeq
In addition, given four final-state momenta $k_a^\mu$, $k_b^\mu$, 
$k_c^\mu$  and $k_d^\mu$, we define
\beq
&&
s_{abc} \; = \; s_{ab} + s_{bc} + s_{ac}
\, ,
\qquad
s_{[ab]c} \; = \; s_{ac} + s_{bc}
\, ,
\qquad
k_{[ab]}^{\mu} \; = \; k_{a}^{\mu} + k_{b}^{\mu}
\, ,
\nnb\\
&&
s_{abcd} \; = \; s_{ab} + s_{ac} + s_{ad} + s_{bc} + s_{bd} + s_{cd}
\, ,
\qquad
s_{[abc]d} \; = \; s_{ad} + s_{bd} + s_{cd}
\, .
\eeq
For the sake of compactness, we define the following flavour structures: 
\[
f_{i}^{q} 
=
\left\{
\begin{array}{ll}
1 & \mbox{ if $i$ is a quark} \\
0 & \mbox{ if $i$ is not a quark}
\end{array}
\right.
\quad
f_{i}^{\bar q} 
=
\left\{
\begin{array}{ll}
1 & \mbox{ if $i$ is an antiquark} \\
0 & \mbox{ if $i$ is not an antiquark}
\end{array}
\right.
\quad
f_{i}^{g} 
=
\left\{
\begin{array}{ll}
1 & \mbox{ if $i$ is a gluon} \\
0 & \mbox{ if $i$ is not a gluon}
\end{array}
\right.
\]
\beq
\label{Flavcomb}
f_{ij}^{q \bar q} 
= 
f_{i}^{q}f_{j}^{\bar q} + f_{i}^{\bar q}f_{j}^{q}
\, ,
\qquad
f_{ij}^{gg} 
= 
f_{i}^{g}f_{j}^{g}
\, ,
\qquad
f_{ijk}^{ggg} 
= 
f_{i}^{g}f_{j}^{g}f_{k}^{g}
\, ,
\qquad
\tilde f_{\,ij}^{\,q\bar q}
=
f_{i}^{q}f_{j}^{\bar q} - f_{i}^{\bar q}f_{j}^{q}
\, ,
\eeq
which are special cases of the general rule 
\beq
f_{i_{1}\dots i_{n}}^{{f_{1}\dots f_{n}}} = 
\sum_{\substack{g_{1},\dots, g_{n} = \\ P(f_1,\dots,f_n)}} \!\!\!
f_{i_{1}}^{g_{1}} \cdots f_{i_{n}}^{g_{n}}
\, ,
\qquad\qquad
\tilde f_{i_{1}\dots i_{n}}^{{f_{1}\dots f_{n}}} = 
\sum_{\substack{g_{1},\dots, g_{n} = \\ P(f_1,\dots,f_n)}} \!\!\!
{\rm sign}(P)\,
f_{i_{1}}^{g_{1}} \cdots f_{i_{n}}^{g_{n}}
\, ,
\eeq
where $P(f_1,\dots,f_n)$ is a generic permutation of indices $f_1,\dots,f_n$. \\
We introduce a compact notation for Born-level colour correlations:
\beq
\label{colcorrBorn}
B_{cd} 
\,\equiv\,
{\cal A}_n^{(0)\dagger} \,
{\bf T}_c\cdot{\bf T}_d \,
{\cal A}_n^{(0)}
\, ,
\qquad
B_{cdef} 
\, \equiv \,
{\cal A}_n^{(0)\dagger} \,
\{ {\bf T}_c\cdot{\bf T}_d , {\bf T}_e\cdot {\bf T}_f \} \,
{\cal A}_n^{(0)}
\, ,
\eeq
\beq
\label{deftildeB}
{\cal B}_{cd}
\, \equiv \,
f_{c}^{g}\,
{\cal A}_n^{(0) \dagger} \,
{\cal T}_c \cdot {\bf T}_d \,
{\cal A}_n^{(0)}
\, ,
\qquad\qquad
({\cal T}_A)_{BC}
\; = \;
d_{ABC}
\, .
\eeq
Analogously, the colour-correlated real and virtual matrix elements are defined as 
\beq
\label{colcorrV}
V_{cd} \, &\equiv \, 2 \, {\bf Re} \left[ {\cal A}_n^{(1)\dagger} \, {\bf T}_c\cdot{\bf T}_d \,
  {\cal A}_n^{(0)} \right] \, ,\qquad
R_{cd} \, \equiv \,  {\cal A}_{\npo}^{(0)\dagger} \,\, {\bf T}_c\cdot{\bf T}_d \,\,
{\cal A}_{\npo}^{(0)} \, ,
\eeq
which are of relative order $\alpha_s$ with respect to the corresponding Born-level terms. 
\\
We define the following combinations of Casimir operators,
\beq
\label{Cascomb}
\rho^{\scriptscriptstyle (C)}_{ab}
\; = \;
\frac{ C_{f_{[ab]}} + C_{f_a} - C_{f_b} }{ C_{f_{[ab]}} }
\, ,
\qquad\qquad
\rho^{\scriptscriptstyle (C)}_{[ab]}
\; = \;
\frac{ C_{f_{[ab]}} - C_{f_a} - C_{f_b} }{ C_{f_{[ab]}} }
\, ,
\qquad
\Sigma_{\!_C}
\; = \;
\sum_{a} C_{f_a}
\, ,
\eeq
and 
\beq
\gamma_a
\; = \; 
\frac{3}{2} \, C_F \, (f_{a}^{q} \!+\! f_{a}^{\bar q})
+
\frac{1}{2} \, \beta_{0} \, f_{a}^{g}
\, ,
\qquad
\Sigma_{\gamma}
\; = \;
\sum_{a} \gamma_{a}
\, ,
\qquad
\gamma^{\rm hc}_{a}
\; = \;
\gamma_{a} - 2 C_{f_{a}}
\, ,
\eeq
\beq
\phi_a
\; = \;
\frac{13}{3} \, C_F \, (f_{a}^{q} \!+\! f_{a}^{\bar q})
+ 
\frac{4}{3} \, \beta_{0} \, f_{a}^{g}
+
\bigg( \frac{2}{3} \!-\! \frac{7}{2}\zeta_2 \!\bigg)
C_{f_{a}}
\, ,
\qquad
\Sigma_{\phi}
\; = \;
\sum_{a} \phi_{a}
\, ,
\eeq
\beq
\label{hccoeff}
\qquad
\phi^{\rm hc}_{a}
\; = \;
\frac{13}{3} \, C_F \, (f_{a}^{q} \!+\! f_{a}^{\bar q})
+
\frac{4}{3} \, \beta_{0} \, f_{a}^{g}
- 
\frac{16}{3} \, C_{f_{a}}
\, ,
\qquad
\Sigma^{\rm hc}_{\phi}
\; = \;
\sum_{a} \phi^{\rm hc}_{a}
\, ,
\eeq
where the sums run over all final-state QCD partons and
\beq
\beta_{0} = \frac{11 C_A - 4 \,T_R\, N_f}{3}
\, .
\eeq
The two-loop anomalous dimensions are given by
\beq
\widehat{\gamma}_{\!_K}^{(2)}
& = & 4 \, \bigg \{\left( \frac{67}{18} - \zeta_2 \right) C_A \, - \, 
\frac{10}{9} T_R \, N_f \bigg\} \, = \, 
\bigg( \frac{8}{3} - 4\zeta_2 \bigg) C_A + \frac{10}{3} \, \beta_{0} \, , \nnb \\
\gamma_i^{(2)}
& = &
(f_{i}^{q} \!+\! f_{i}^{\bar q}) \,
C_F \, 
\bigg[
3 \left( \frac{1}{8} - \zeta_{2} + 2 \zeta_{3} \right) C_F 
+
\left( \frac{41}{36} - \frac{13}{2} \zeta_{3} \right) C_A
+
\left( \frac{65}{72} + \frac{3}{4} \zeta_{2} \right) \beta_{0}
\bigg] \nnb \\
&&
\hspace{-4mm}
+ \,
f_{i}^{g}
\bigg\{
C_A \, 
\bigg[
-
\frac{11}{4} \, C_F 
+
\left( - \frac{1}{9} - \frac{1}{2} \zeta_{3} \right) C_A
\bigg]
+
\beta_{0} \, 
\bigg[
\frac{3}{4} \, C_F 
+
\left( \frac{16}{9} - \frac{1}{4} \zeta_{2} \right) C_A
\bigg]
\bigg\} \, . 
\label{twoloandim}
\eeq 
As for the labelling of particles we introduce the notation 
\beq
\label{eq: r_ij}
r_{i_{1} \dots i_{n}} = R_{n}(i_{1},\dots,i_{n}) \ne i_{1},\dots,i_{n}
\, ,
\eeq
to indicate a generic particle label different from $i_{1},\dots,i_{n}$, defined following a 
specific rule $R_{n}$. Such a rule is arbitrary to some extent, and could for instance assign 
$r_{i_{1} \dots i_{n}}$ as the smallest label different from all $i_{1},\dots,i_{n}$, or the largest, 
and so on. A crucial feature, however, is that $R_{n}$ must be symmetric under permutations 
of all indices $i_{1},\dots,i_{n}$, and must be the same for all $r_{i_{1} \dots i_{n}}$ with the 
same $n$. As a consequence, the notation $r_{i_{1} \dots i_{n}}$ always refers to the 
rule $R_{n}(i_{1},\dots,i_{n})$, which is a symmetric function of its indices $i_{1},\dots,i_{n}$, 
and just depends on $n$.


\section{Infrared kernels}
\label{app:kernels}


\subsection{Soft kernels at tree level}
\label{app:softkernels}

We introduce the kernels associated with the real emission of one or two soft partons,
as given in Ref.~\cite{Magnea:2020trj}, relevant for both NLO (with the emission of just 
one parton) and NNLO corrections (with the emission of either one or two partons). We 
express all kernels in terms of Lorentz-invariant quantities, and using the flavour structures
defined in \appn{app:notations}. The resulting expressions are
\beq
\mc I^{(i)}_{cd}
\; = \;
f_{i}^{g} \,
\frac{s_{cd}}{s_{ic}\,s_{id}}
\, ,
\qquad\qquad
\mc I^{(ij)}_{cd}
\; = \;
f_{ij}^{q \bar q} \, 2 \, T_R \;  \mc I^{(q \bar q)(ij)}_{cd}
- 
f_{ij}^{gg} \, 2 \,C_A \;  \mc I^{(gg)(ij)}_{cd}
\, ,
\eeq
where
\beq
\mc I_{cd}^{(q\bar q)(ij)}
& = &
\frac{s_{ic}s_{jd}+s_{id}s_{jc}-s_{ij}s_{cd}}{s_{ij}^2\,s_{[ij]c}\,s_{[ij]d}}
\, ,
\label{eq:soft_double_kernel}
\\[3mm]
\mc I_{cd}^{(gg)(ij)}
& = &
\frac{(1-\eps)(s_{ic}s_{jd}+s_{id}s_{jc})-2s_{ij}s_{cd}}
     {s_{ij}^2\,s_{[ij]c}\,s_{[ij]d}}
+
s_{cd} \,\,
\frac{s_{ic}s_{jd}+s_{id}s_{jc}-s_{ij}s_{cd}}{s_{ij}s_{ic}s_{jd}s_{id}s_{jc}} \,
\bigg[
1
-
\frac12 \,
\frac{s_{ic}s_{jd}+s_{id}s_{jc}}{s_{[ij]c}\,s_{[ij]d}}
\bigg]\,.
\nnb
\eeq
We also define the combinations of eikonal kernels
\beq
\label{eikoker}
\mc E^{(i)}_{cd}
& \equiv &
\mc I^{(i)}_{cd}
\; = \;
f_{i}^{g} \, 
\frac{s_{cd}}{s_{ic}\,s_{id}}
\, ,
\nnb\\
\mc E^{(ij)}_{cd}
& \equiv &
\mc I_{cd}^{(ij)} 
-
\frac12\,\mc I_{cc}^{(ij)} 
-
\frac12\,\mc I_{dd}^{(ij)} 
\; = \;
f_{ij}^{q \bar q} \, 2 \, T_R \;  \mc E^{(q \bar q)(ij)}_{cd}
- 
f_{ij}^{gg} \, 2 \,C_A \;  \mc E^{(gg)(ij)}_{cd}
\, ,
\eeq
with 
\beq
\label{eikoker2}
\mc E_{cd}^{(q \bar q)(ij)} 
& = &
\frac{1}{s_{ij}^{2}}
\left[
\frac{s_{ic}s_{jd}+s_{id}s_{jc}}{s_{[ij]c}s_{[ij]d}}
-
\frac{s_{ic}s_{jc}}{s_{[ij]c}^{2}}
-
\frac{s_{id}s_{jd}}{s_{[ij]d}^{2}}
\right]
-
\frac{s_{cd}}{s_{ij}s_{[ij]c}s_{[ij]d}}
\, ,
\nnb\\
\mc E_{cd}^{(gg)(ij)} 
& = &
\frac{1-\eps}{s_{ij}^{2}}
\left[
\frac{s_{ic}s_{jd}+s_{id}s_{jc}}{s_{[ij]c}s_{[ij]d}}
-
\frac{s_{ic}s_{jc}}{s_{[ij]c}^{2}}
-
\frac{s_{id}s_{jd}}{s_{[ij]d}^{2}}
\right]
-
2 \,
\frac{s_{cd}}{s_{ij}s_{[ij]c}s_{[ij]d}}
\nnb\\
&&
+ \,
s_{cd} \,
\frac{s_{ic}s_{jd}+s_{id}s_{jc}-s_{ij}s_{cd}}{s_{ij}s_{ic}s_{jd}s_{id}s_{jc}}
\bigg[
1
-
\frac12 \,
\frac{s_{ic}s_{jd}+s_{id}s_{jc}}{s_{[ij]c}s_{[ij]d}}
\bigg]
\, .
\eeq


\subsection{Soft kernels at one loop}
\label{app:softkernels2}

We introduce kernels associated to the emission of a
single-soft gluon at one-loop level, relevant for the soft part
of the real-virtual counterterm at NNLO,
\beq
\label{tildeik}
\tilde{\mc E}_{cd}^{(i)}
& \equiv &
f_{i}^{g} \,
C_A \,
\frac{\Gamma^3(1+\eps)\Gamma^4(1-\eps)}{\eps^2 \,
\Gamma(1+2\eps)\Gamma^2(1-2\eps)} \, 
\frac{s_{cd}}{s_{ic}s_{id}}
\left(
\frac{e^{\euler}\mu^2 s_{cd}}{s_{ic}s_{id}}
\right)^{\!\eps}
\nnb\\
& = &
C_A \,
{\mc E}_{cd}^{(i)} \,
\bigg[ 
\frac{1}{\eps^{2}} 
-
\frac{1}{\eps} 
\ln\frac{s_{ic}s_{id}}{\mu^2\,s_{cd}}
-
\frac{5}{2}\,\zeta_{2}
+
\frac{1}{2} 
\ln^{2}\frac{s_{ic}s_{id}}{\mu^2\,s_{cd}}
+
\mc O(\eps)
\bigg]
\, ,
\nnb\\
\tilde{\mc E}_{cde}^{(i)}
& \equiv &
f_{i}^{g} \,
\frac{\Gamma(1+\eps)\Gamma^2(1-\eps)}{\eps\,\Gamma(1-2\eps)} \,
\frac{s_{cd}}{s_{ic}s_{id}} \,
\bigg(
\frac{e^{\euler}\mu^2\,s_{de}}{s_{id}s_{ie}}
\bigg)^{\eps}
\nnb\\
& = &
{\mc E}_{cd}^{(i)} \,
\bigg[ 
\frac{1}{\eps} 
-
\ln\frac{s_{id}s_{ie}}{\mu^2\,s_{de}}
+
\mc O(\eps)
\bigg]
\, ,
\eeq
where $\eps$ is the dimensional regulator ($d=4-2\eps$).


\subsection{Collinear and hard-collinear kernels at tree level}\label{app:collkernels}

In order to define the kernel associated to the tree-level emission
of two collinear final-state particles $i$ and $j$
(labelled \emph{single-collinear}), we choose a reference momentum $k_{r}$, 
with $r\neq i,j$, and introduce the following kinematic structures:
\beq
x_i 
\; = \;
\frac{s_{ir}}{s_{[ij]r}}
\, ,
\qquad
x_j
\; = \;
\frac{s_{jr}}{s_{[ij]r}}
\, ,
\qquad
\kt_{i}
\; = \;
x_i\,k_j - x_j \,k_i - (1\!-\!2x_j) \frac{s_{ij}}{s_{[ij]r}} \, k_{r}
\, .
\eeq
Then, the collinear (Altarelli-Parisi) kernels $P_{ij(r)}^{\mu\nu}$ are defined as
\beq\label{eq:appsinglecoll}
&&
P_{ij(r)}^{\mu\nu}
\; = \; 
- \,
P_{ij(r)} \,
g^{\mu\nu}
+
Q_{ij(r)}^{\mu\nu}
\, ,
\qquad
Q_{ij(r)}^{\mu\nu}
\; = \;
Q_{ij(r)} \,
d_{i}^{\,\mu\nu}
\, ,
\eeq
where the azimuthal tensor reads
\beq
d_i^{\mu\nu} 
& = & 
- g^{\mu\nu} + (d-2)\,\frac{\kt_i^\mu\kt_i^\nu}{\kt_i^2}
\,,
\eeq
and
\beq
P_{ij(r)} 
& = &
P_{ij(r)}^{\zg} \,f_{ij}^{q \bar q}
+ 
P_{ij(r)}^{\og}\,f_{i}^{g}(f_{j}^{q} \!+\! f_{j}^{\bar q})
+  
P_{ji(r)}^{\og}\,(f_{i}^{q} \!+\! f_{i}^{\bar q})f_{j}^{g}
+
P_{ij(r)}^{\tg}\,f_{ij}^{gg} 
\, ,
\\
Q_{ij(r)} 
& = &
T_R \, \frac{2 x_i x_j}{1 - \eps} \,f_{ij}^{q \bar q}
-
2 \, C_A \,  x_i x_j \, f_{ij}^{gg}
\, ,\nnb\\
P_{ij(r)}^{\zg}
&=&
T_R \bigg( 1 - \frac{2 x_i x_j}{1 - \eps} \bigg) 
\, ,
\quad
P_{ij(r)}^{\og}
=
C_F \bigg[ 2\,\frac{x_j}{x_i} + (1-\eps)x_i \bigg]
\, ,
\quad
P_{ij(r)}^{\tg}
=
2 \, C_A \bigg( \!\frac{x_i}{x_j} + \frac{x_j}{x_i} + x_i x_j \!\bigg) 
\, .
\nnb
\eeq
The hard-collinear kernels $P_{ij(r)}^{{\rm hc}, \mu\nu}$ are defined as 
\beq
\label{hardcollker}
P_{ij(r)}^{{\rm hc}, \mu\nu}
\; \equiv \; 
P_{ij(r)}^{\mu\nu}
+
s_{ij}
\Big[
2 \, C_{f_{j}} \, \mc E^{(i)}_{jr}
+
2 \, C_{f_{i}} \, \mc E^{(j)}_{ir}
\Big]
g^{\mu\nu}
\; \equiv \;
- \,
P_{ij(r)}^{{\rm hc}} \,
g^{\mu\nu}
+
Q_{ij(r)}^{\mu\nu}
\, ,
\eeq
where
\beq
P_{ij(r)}^{{\rm hc}} 
& = &
P_{ij(r)}^{{\rm hc},\zg} \,f_{ij}^{q \bar q}
+ 
P_{ij(r)}^{{\rm hc},\og}\,f_{i}^{g}(f_{j}^{q} \!+\! f_{j}^{\bar q})
+  
P_{ji(r)}^{{\rm hc},\og}\,(f_{i}^{q} \!+\! f_{i}^{\bar q})f_{j}^{g}
+
P_{ij(r)}^{{\rm hc},\tg}\,f_{ij}^{gg} \, , \\
P_{ij(r)}^{{\rm hc},\zg}
&=&
P_{ij(r)}^{\zg}
=
T_R \bigg( 1 - \frac{2 x_i x_j}{1 - \eps} \bigg) 
\, ,
\quad
P_{ij(r)}^{{\rm hc},\og}
=
C_F (1-\eps)x_i
\, ,
\quad
P_{ij(r)}^{{\rm hc},\tg}
=
2 \, C_A \, x_i x_j
\, .\nnb
\eeq
The kernel associated to the emission of three collinear final-state partons $i$, $j$ 
and $k$ (labelled \emph{double-collinear}) relies on the choice of a reference momentum
$k_{r}$, with $r\neq i,j,k$, and on the following kinematic structures,
\beq
z_a
& = &
\frac{s_{ar}}{s_{[ijk]r}}
\, ,
\qquad
\qquad
z_{ab}
\, = \,
z_a + z_b
\, ,
\qquad
\qquad
a, b
\, = \,
i, j, k
\,\\
\kt_{a}^\mu 
&=& 
k_a^\mu - z_a (k_i^\mu + k_j^\mu + k_k^\mu) 
- 
( s_{[ijk]a} - 2\,z_a s_{ijk} ) \frac{k_r^\mu}{s_{[ijk]r}}
\, ,
\quad\qquad
a,b,c = i,j,k
\, ,
\nnb\\
\tk_a^2 
&=&
z_a( z_a s_{ijk} - s_{[ijk]a} )
\, = \,
z_a( s_{bc} - z_{bc} s_{ijk}) 
\, .\nnb
\eeq
The double-collinear kernels $P_{ijk(r)}^{\mu\nu}$ are defined as 
\beq\label{eq:appdoublecoll}
P_{ijk(r)}^{\mu\nu}
& \equiv &
- 
P_{ijk(r)}\,g^{\mu\nu}
+
Q_{ijk(r)}^{\mu\nu}
\, ,
\qquad
Q_{ijk(r)}^{\mu\nu}
\; = \;
\sum_{a=i,j,k}
Q_{ijk(r)}^{a} \,
d_a^{\mu\nu} 
\, .
\eeq
The $P_{ijk(r)}$ kernels, organised by flavour structures, are given by 
\beq
\label{eq:Pijk-Qijk-1}
P_{ijk(r)} 
& = & 
  P_{ijk(r)}^{\zg}\,f_{ij}^{q \bar q}\,(f_{k}^{q'} \!+\! f_{k}^{\bar q'})
+ P_{jki(r)}^{\zg}\,f_{jk}^{q \bar q}\,(f_{i}^{q'} \!+\! f_{i}^{\bar q'})
+ P_{kij(r)}^{\zg}\,f_{ik}^{q \bar q}\,(f_{j}^{q'} \!+\! f_{j}^{\bar q'})
\nnb\\
& + &
P_{ijk(r)}^{(\rm{0g},\rm id)}
(f_{i}^{q}f_{j}^{q}f_{k}^{\bar q} \!+\! f_{i}^{\bar q}f_{j}^{\bar q}f_{k}^{q})
+
P_{jki(r)}^{(\rm{0g},\rm id)}
(f_{j}^{q}f_{k}^{q}f_{i}^{\bar q} \!+\! f_{j}^{\bar q}f_{k}^{\bar q}f_{i}^{q})
+
P_{kij(r)}^{(\rm{0g},\rm id)}
(f_{i}^{q}f_{k}^{q}f_{j}^{\bar q} \!+\! f_{i}^{\bar q}f_{k}^{\bar q}f_{j}^{q})
\nnb\\
& + &
P_{ijk(r)}^{\og}\,f_{ij}^{q \bar q}\,f_{k}^{g}
+
P_{jki(r)}^{\og}\,f_{jk}^{q \bar q}\,f_{i}^{g}
+
P_{kij(r)}^{\og}\,f_{ik}^{q \bar q}\,f_{j}^{g}
\nnb\\
& + &
P_{ijk(r)}^{\tg}\,f_{ij}^{gg}\,(f_{k}^{q} \!+\! f_{k}^{\bar q})
+
P_{jki(r)}^{\tg}\,f_{jk}^{gg}\,(f_{i}^{q} \!+\! f_{i}^{\bar q})
+
P_{kij(r)}^{\tg}\,f_{ik}^{gg}\,(f_{j}^{q} \!+\! f_{j}^{\bar q})
\nnb\\
& + &
P_{ijk(r)}^{({\rm 3g})}\,f_{ijk}^{ggg} 
\, ,
\eeq
where $q'$ is a quark of flavour equal to or different from that of $q$; 
similarly, the azimuthal tensor kernel can be written as
\beq
\label{eq:Pijk-Qijk-2}
Q_{ijk(r)}^{a}
& = &
Q_{ijk(r)}^{\og,a}\,f_{ij}^{q \bar q}\,f_{k}^{g}
+ 
Q_{jki(r)}^{\og,a}\,f_{jk}^{q \bar q}\,f_{i}^{g}
+ 
Q_{kij(r)}^{\og,a}\,f_{ik}^{q \bar q}\,f_{j}^{g}
+ 
Q_{ijk(r)}^{({\rm{3g}}), a}\,f_{ijk}^{ggg} 
\, .
\eeq
The expressions for $P_{ijk(r)}^{\zg}$, $P_{ijk(r)}^{(\rm{0g},\rm id)}$, 
$P_{ijk(r)}^{\og}$, $P_{ijk(r)}^{\tg}$, and $P_{ijk(r)}^{(\rm 3g)}$ read:
\beq
\label{Pijk0g}
P_{ijk(r)}^{\zg}
& = &
C_F T_R \,
\Bigg\{
\!\!
- 
\frac{s_{ijk}^2}{2s_{ij}^2}\,
\bigg(
\frac{s_{jk}}{s_{ijk}} - \frac{s_{ik}}{s_{ijk}} + \frac{z_i\!-\!z_j}{z_{ij}}
\bigg)^2
\!\!
+
\frac{s_{ijk}}{s_{ij}}\,
\bigg[
2 \, \frac{z_k \!-\! z_iz_j}{z_{ij}}
+
(1-\eps) z_{ij}
\bigg]
\!
-
\!\frac12
+
\eps
\Bigg\}
\, ,
\qquad\quad
\eeq
\beq
\label{Pijk0gid}
P_{ijk(r)}^{(\rm{0g},\rm id)}
& = &
C_F(2C_F\!-\!C_A)\,
\Bigg\{ \!
- 
\frac{s_{ijk}^2\, z_k}{2s_{jk}s_{ik}}  \,
\bigg[
\frac{1+z_k^2}{z_{jk}z_{ik}}
-
\eps\,
\bigg(
\frac{z_{ik}}{z_{jk}}
+
\frac{z_{jk}}{z_{ik}}
+
1
+
\eps
\bigg)
\bigg]
+ \,
(1-\eps)\bigg[ \frac{s_{ij}}{s_{jk}} + \frac{s_{ij}}{s_{ik}} - \eps \bigg]
\nnb \\
&&
\hspace{26mm}
+ \,
\frac{s_{ijk}}{2s_{jk}}\,
\bigg[
\frac{1 + z_k^2 - \eps z_{jk}^2}{z_{ik}}
-
2(1-\eps)\frac{z_j}{z_{jk}}
-
\eps(1+z_k)
-
\eps^2\,z_{jk}
\bigg]
\nnb \\
&&
\hspace{26mm}
+ \,
\frac{s_{ijk}}{2s_{ik}}\,
\bigg[
\frac{1 + z_k^2 - \eps z_{ik}^2}{z_{jk}}
-
2(1-\eps)\frac{z_i}{z_{ik}}
-
\eps(1+z_k)
-
\eps^2\,z_{ik}
\bigg]
\Bigg\}
\, ,
\eeq
%
%
\beq
\label{Pijk1g}
P_{ijk(r)}^{\og}
& = &
C_F T_R \,
\bigg[
\frac{2 s_{ijk} s_{ij}}{s_{ik}s_{jk}}
+
(1-\eps)\left( \frac{s_{ik}}{s_{jk}} + \frac{s_{jk}}{s_{ik}} + 2 \right)
-
2
\bigg]
\nnb \\
&&
\hspace{-4.2mm}
+ \,\,
C_A T_R \,\,
\bigg[
- 
\frac{s_{ijk}^2}{2s_{ij}^2}\,
\bigg(
\frac{s_{jk}}{s_{ijk}} - \frac{s_{ik}}{s_{ijk}} + \frac{z_i-z_j}{z_{ij}}
\bigg)^2
-
\frac{s_{ijk}^2}{s_{ik}s_{jk}}
+
\frac{s_{ijk}^2}{2s_{ij}}\,\frac{1 - 2 z_k}{z_k z_{ij}}
\bigg( \frac{z_i}{s_{ik}} + \frac{z_j}{s_{jk}} \bigg)
\nnb \\
&&
\hspace{12.5mm}
+ \,
\frac{s_{ijk}}{2 z_k z_{ij}} 
\bigg( \frac{z_{ik}}{s_{ik}} + \frac{z_{jk}}{s_{jk}} \bigg)
+
\frac{s_{ijk}}{s_{ij}}\,\frac{1 - z_k + 2 z_k}{z_kz_{ij}}
-
\frac12
+
\eps
\bigg]
\nnb \\
&&
\hspace{-4.2mm}
- \,
\sum_{a=i,j,k}
Q_{ijk(r)}^{a}
\, ,
\eeq
\beq
\label{Pijk2g}
P_{ijk(r)}^{\tg}
& = &
C_F^2\,
\Bigg\{
\frac{s_{ijk}^2\,z_k}{2s_{ik}s_{jk}}\,
\bigg[
\frac{1+z_k^2-\eps z_{ij}^2}{z_iz_j}
+
\eps(1-\eps)
\bigg]
-
(1-\eps)^2 \, \frac{s_{jk}}{s_{ik}}
+
\eps(1-\eps)
\nnb \\
&&
\hspace{10mm}
+ \,
\frac{s_{ijk}}{s_{ik}}\,
\bigg[
\frac{z_k z_{jk} + z_{ik}^3 - \eps z_{ik} z_{ij}^2}{z_iz_j}
+
\eps\,z_{ik}
+
\eps^2\,(1+z_k)
\bigg]
\Bigg\}
\nnb \\
&& \hspace{-3.5mm}
+ \,\,
C_F C_A\,\,
\Bigg\{
(1-\eps)\frac{s_{ijk}^2}{4s_{ij}^2}
\bigg(
\frac{s_{jk}}{s_{ijk}} - \frac{s_{ik}}{s_{ijk}} + \frac{z_i-z_j}{z_{ij}}
\bigg)^2
- 
\frac{s_{ijk}^2\,z_k}{4s_{ik}s_{jk}}\,
\bigg[
\frac{z_{ij}^2(1-\eps) + 2z_k}{z_iz_j}
+
\eps(1-\eps)
\bigg]
\nnb \\
&&
\hspace{15mm}
+ \,
\frac{s_{ijk}^2}{2s_{ij}s_{ik}}\,
\bigg[
\frac{z_{ij}^2(1-\eps) + 2z_k}{z_j}
+
\frac{z_j^2(1-\eps) + 2z_{ik}}{z_{ij}}
\bigg]
+
\frac14(1-\eps)(1-2\eps)
\nnb \\
&&
\hspace{15mm}
+ \,
\frac{s_{ijk}}{2s_{ik}}\,
\bigg[
(1-\eps)
\frac{z_{ik}^3+z_k^2-z_j}{z_j z_{ij}}
-
2\eps \,
\frac{z_{ik}(z_j-z_k)}{z_j z_{ij}}
\nnb\\
&&
\hspace{28mm}
- \,
\frac{z_k z_{jk}+z_{ik}^3}{z_iz_j}
+ 
\eps\,z_{ik}
\frac{z_{ij}^2}{z_iz_j}
-
\eps(1+z_k)
-
\eps^2 z_{ik}
\bigg]
\nnb \\
&&
\hspace{15mm}
+ \,
\frac{s_{ijk}}{2s_{ij}}\,
\bigg[
(1-\eps)
\frac{z_i(2z_{jk}+z_i^2)-z_j(6z_{ik}+z_j^2)}{z_j z_{ij}}
+
2\eps \,
\frac{z_k(z_i-2z_j)-z_j}{z_j z_{ij}}
\bigg]
\Bigg\}
\nnb \\
&& \hspace{-3.5mm}
+ \,\,
( i \leftrightarrow j)
\, ,
\eeq
\beq
\label{Pijk3g}
P_{ijk(r)}^{(\rm 3g)}
& = &
C_A^2 \,
\Bigg\{
(1-\eps)\frac{s_{ijk}^2}{4s_{ij}^2}
\bigg(
\frac{s_{jk}}{s_{ijk}} - \frac{s_{ik}}{s_{ijk}} + \frac{z_i-z_j}{z_{ij}}
\bigg)^2
+
\frac34(1-\eps)
\nnb \\
&&
\hspace{8mm}
+ \,
\frac{s_{ijk}^2}{2s_{ij}s_{ik}}
\bigg[
\frac{2 z_i z_j z_{ik}(1\!-\!2z_k)}{z_k z_{ij}}
+
\frac{1\!+\!2z_i\!+\!2z_i^2}{z_{ik} z_{ij}}
+ 
\frac{1\!-\!2z_iz_{jk}}{z_j z_k}
+
2 z_j z_k
+
z_i(1\!+\!2z_i)
-
4
\bigg]
\nnb\\
&&
\hspace{8mm}
+ \, 
\frac{s_{ijk}}{s_{ij}}\,
\bigg[
4\,\frac{z_iz_j-1}{z_{ij}}
+
\frac{z_iz_j-2}{z_k}
+
\frac{(1-z_kz_{ij})^2}{z_iz_k z_{jk}}
+
\frac52\,z_k
+
\frac32
\bigg]
\Bigg\}
\nnb\\
&&
+ \, \,
\mbox{( 5 permutations )}
\, .
\eeq
\\
The azimuthal kernels $Q_{ijk(r)}^{\og,a}$ and 
$Q_{ijk(r)}^{({\rm{3g}}), a}$ are defined according to the following expressions:
\beq
Q_{ijk(r)}^{\og,i} \!
& = &
T_R
\frac{\tk_i^2}{1\!-\!\eps}
\frac{s_{ijk}}{s_{ik}s_{jk}}
\bigg\{
C_A
\bigg[
1
\!-\!
\frac{2z_j}{z_k}
\frac{s_{ij} \!+\! 2s_{jk}}{s_{ij}^2}
s_{ik}
\!+\!
\frac{z_is_{jk} \!+\! z_js_{ik}}{z_{ij}s_{ij}}
\!+\!
\bigg(\! \frac{z_iz_j}{z_k z_{ij}} \!-\! \frac{1 \!-\! \eps}{2} \!\bigg)
\frac{s_{ik}\!-\!s_{jk}}{s_{ij}}
\bigg] \!
- \!
2 C_F 
\!\bigg\}
,
\nnb\\
Q_{ijk(r)}^{\og,j} \!
& = &
T_R
\frac{\tk_j^2}{1\!-\!\eps}
\frac{s_{ijk}}{s_{ik}s_{jk}}
\bigg\{
C_A
\bigg[
1
\!-\!
\frac{2z_i}{z_k}
\frac{s_{ij} \!+\! 2s_{ik}}{s_{ij}^2}
s_{jk}
\!+\!
\frac{z_is_{jk} \!+\! z_js_{ik}}{z_{ij}s_{ij}}
\!+\!
\bigg(\! \frac{z_iz_j}{z_k z_{ij}} \!-\! \frac{1 \!-\! \eps}{2} \!\bigg)
\frac{s_{jk}\!-\!s_{ik}}{s_{ij}}
\bigg] \!
- \!
2 C_F 
\!\bigg\}
,
\nnb\\
Q_{ijk(r)}^{\og,k} \!
& = &
T_R
\frac{\tk_k^2}{1\!-\!\eps}
\frac{s_{ijk}}{s_{ik}s_{jk}}
\bigg\{
C_A
\bigg[
\frac{z_iz_j}{z_k z_{ij}}
\frac{4 s_{ik}s_{jk} \!+\! s_{ij}s_{[ij]k}}{s_{ij}^{2}}
\!+\!
\frac{z_i\!-\!z_j}{2z_{ij}}\frac{s_{ik}\!-\!s_{jk}}{s_{ij}}
\!-\!
\eps \,
\frac{s_{ijk}\!+\!s_{ij}}{2s_{ij}}
\bigg]
+ 
2 C_F \, \eps
\bigg\}
,
\nnb
\eeq
\beq
\sum_{a=i,j,k} \!\!\!
Q_{ijk(r)}^{({\rm{3g}}), a}
d_a^{\mu\nu} 
& = &
C_A^2 \,
\frac{s_{ijk}}{s_{ij}}
\Bigg\{ \;
\bigg[
\frac{2z_j}{z_k}
\frac{1}{s_{ij}}
+
\bigg(\! \frac{z_jz_{ik}}{z_k z_{ij}} - \frac32 \!\bigg)
\frac{1}{s_{ik}}
\bigg]
\,
\tk_i^2\,d_i^{\mu\nu}
\\
&&
\hspace{12.5mm}
+ \,
\bigg[
\frac{2z_i}{z_k}
\frac{1}{s_{ij}}
-
\bigg(\! 
\frac{z_jz_{ik}}{z_k z_{ij}} - \frac32 - \frac{z_i}{z_k} + \frac{z_i}{z_{ij}} 
\!\bigg)
\frac{1}{s_{ik}}
\bigg]
\,
\tk_j^2 \,d_j^{\mu\nu}
\nnb \\
&&
\hspace{12.5mm}
- \, 
\bigg[
\frac{2z_iz_j}{z_{ij}z_k}
\frac{1}{s_{ij}}
\!+\!
\bigg(\! 
\frac{z_jz_{ik}}{z_k z_{ij}} \!-\! \frac32 \!-\! \frac{z_i}{z_j} \!+\! \frac{z_i}{z_{ik}} 
\!\bigg)
\frac{1}{s_{ik}}
\bigg]
\,
\tk_k^2 \,d_k^{\mu\nu}
\Bigg\} \,
+ \,
\mbox{(5 permutations)}
\, .
\nnb
\eeq
\\
The hard-double-collinear kernels $P_{ijk(r)}^{{\rm hc}, \mu\nu}$ are defined as 
\beq
P_{ijk(r)}^{{\rm hc}, \mu\nu}
\; \equiv \; 
- \,
P_{ijk(r)}^{{\rm hc}} \,
g^{\mu\nu}
+
Q_{ijk(r)}^{\mu\nu}
\, ,
\eeq
where $Q_{ijk(r)}^{\mu\nu}$ is given in \eq{eq:appdoublecoll} and
\beq
P_{ijk(r)}^{{\rm hc}} 
& \equiv &
P_{ijk(r)}
-
s_{ijk}^{2}
\Big[
C_{f_k} 
\Big( 4\,C_{f_k}\mc E_{kr}^{(i)} \mc E_{kr}^{(j)} - \mc E_{kr}^{(ij)} \Big)
+
(i \leftrightarrow k)
+
(j \leftrightarrow k)
\Big]
\, .
\eeq


\subsection{\vspace{-.5mm}Collinear and hard-collinear kernels at one loop}
\label{app:collkernels2}

\vspace{.5mm}
The collinear contribution to the real-virtual counterterm at NNLO
depends on the one-loop, single-collinear kernel which reads 
($r\ne i,j$):
\beq
\label{tildAP}
\tilde P^{\mu\nu}_{ij(r)} 
& \equiv &
\frac{\Gamma^2(1\!+\!\eps)\Gamma^3(1\!-\!\eps)}
     {\Gamma(1\!+\!2\eps)\Gamma^2(1\!-\!2\eps)}
\!\left(\! \frac{e^{\euler\!}\mu^2}{s_{ij}} \!\right)^{\!\!\eps\!}
\bigg\{
\frac{C_{\!f_{[ij]}\!}}{\eps^2}
\Big[
\rho^{\scriptscriptstyle (C)}_{[ij]}
+
\rho^{\scriptscriptstyle (C)}_{ij} \!
F(x_i)
+
\rho^{\scriptscriptstyle (C)}_{ji} \!
F(x_j)
\Big]
P^{\mu\nu}_{ij(r)}
+
\hat P^{\mu\nu}_{ij(r)} 
\bigg\}
,
\qquad\quad
\eeq
where the function $F(x)$ is defined by  
\beq
F(x) 
& \equiv &
1 - {}_2F_1 \left(1,-\eps;1-\eps;\frac{x-1}{x} \right)
\, = \,
\eps\,\ln x
+
\sum_{n=2}^{+\infty} \, \eps^n \, \Li_n\left(\frac{x-1}{x} \right)
\, ,
\eeq
and $\hat P^{\mu\nu}_{ij(r)}$ reads 
\beq
\hat P^{\mu\nu}_{ij(r)}
& = &
\bigg[
-
g_{\mu\nu}
+
4 x_i x_j \,
\frac{\tk_i^\mu\tk_i^\nu}{\tk_i^2}
\bigg]
\,
\frac{T_R}{1\!-\!2\eps} \,
\bigg[
\frac1\eps \big( \beta_0 - 3\,C_F \big)
+
C_A - 2C_F
+
\frac{C_A+4\,T_R\,N_f}{3(3-2\eps)}
\bigg]
f_{ij}^{q \bar q}
\nnb\\
&&
- \,
g_{\mu\nu} \,
C_F \,
\frac{C_A\!-\!C_F}{1-2\eps}
\Big[
\big(1-\eps x_i \big)
f_{i}^{g}(f_{j}^{q} \!+\! f_{j}^{\bar q})
+
\big(1-\eps x_j \big)
(f_{i}^{q} \!+\! f_{i}^{\bar q})f_{j}^{g}
\Big]
\nnb\\
&&
+ \,
\frac{\tk_i^\mu\tk_i^\nu}{\tk_i^2} \,
4\,C_A\,
\frac{2\,T_R\,N_f-C_A(1-\eps)}{(1-2\eps)(2\!-\!2\eps)(3\!-\!2\eps)}\,
\big(1-2\eps x_i x_j \big)
f_{ij}^{gg} 
\, .
\eeq
The expansion of $\tilde P^{\mu\nu}_{ij(r)}$ in the dimensional regulator $\eps$ gives
\beq
\tilde P^{\mu\nu}_{ij(r)} 
& = &
P^{\mu\nu}_{ij(r)} \;
C_{\!f_{[ij]}}
\bigg\{ \;
\rho^{\scriptscriptstyle (C)}_{[ij]}
\bigg[
\frac{1}{\eps^2}
-
\frac{1}{\eps} \, \ln\frac{s_{ij}}{\mu^2}
-
\frac{1}{2} 
\bigg(
7\,\zeta_{2}
-
\ln^{2}\frac{s_{ij}}{\mu^2}
\bigg)
\bigg]
\nnb\\
&&
\hspace{18mm}
+ \,
\bigg[ \frac{1}{\eps} - \ln\frac{s_{ij}}{\mu^2} \bigg]
\left(
\rho^{\scriptscriptstyle (C)}_{ij}
\ln x_{i}
+
\rho^{\scriptscriptstyle (C)}_{ji}
\ln x_{j}
\right)
+
\rho^{\scriptscriptstyle (C)}_{ij}
\Li_2\!\left(\frac{-x_{j}}{x_{i}} \right)
+
\rho^{\scriptscriptstyle (C)}_{ji} 
\Li_2\!\left(\frac{-x_{i}}{x_{j}} \right)
\bigg\}
\nnb\\
& + &
\bigg[
-
g_{\mu\nu}
+
4 x_i x_j \,
\frac{\tk_i^\mu\tk_i^\nu}{\tk_i^2}
\bigg]
\;
f_{ij}^{q \bar q} \;
T_R \,
\bigg[
\left( \frac{1}{\eps} - \ln\frac{s_{ij}}{\mu^2} \right) \!
\big( \beta_0 - 3\,C_F \big)
+
\frac{7}{3} C_A 
+
\frac{5}{3} \beta_{0}
- 
8C_F
\bigg]
\nnb\\
& - &
g_{\mu\nu} \;
(f_{ij}^{gq} \!+\! f_{ij}^{g \bar q}) \;
C_F \,
(C_A\!-\!C_F)
+
\frac{\tk_i^\mu\tk_i^\nu}{\tk_i^2} \;
f_{ij}^{gg} \;
C_A \, (3C_A-\beta_{0})
+
\mc O(\eps)
\, .
\eeq
The one-loop collinear kernel $\hat P_{ij(r)}^{\mu\nu}$ can be rewritten according to
the same structure as in \eq{eq:appsinglecoll},
\beq
&&
\hat P_{ij(r)}^{\mu\nu}
\; = \; 
- \,
\hat P_{ij(r)} \,
g^{\mu\nu}
+
\hat Q_{ij(r)}^{\mu\nu}
\, ,
\qquad
\hat Q_{ij(r)}^{\mu\nu}
\; = \;
\hat Q_{ij(r)} \,
d_{i}^{\,\mu\nu}
\, ,
\eeq
where we have introduced
\beq
\hat P_{ij(r)}
& = &
\frac{T_R}{1\!-\!2\eps} \,
\bigg[
1
-
\frac{2 x_i x_j}{1\!-\!\eps} \,
\bigg]
\bigg[
\frac1\eps \big( \beta_0 - 3\,C_F \big)
+
C_A - 2C_F
+
\frac{C_A+4\,T_R\,N_f}{3(3-2\eps)}
\bigg]
f_{ij}^{q \bar q}
\nnb\\
&&
+ \,
C_F \,
\frac{C_A\!-\!C_F}{1-2\eps}
\Big[
\big(1-\eps x_i \big)
f_{i}^{g}(f_{j}^{q} \!+\! f_{j}^{\bar q})
+
\big(1-\eps x_j \big)
(f_{i}^{q} \!+\! f_{i}^{\bar q})f_{j}^{g}
\Big]
\nnb\\
&&
+ \,
4\,C_A\,
\frac{C_A(1-\eps)-2\,T_R\,N_f}{(1-2\eps)(2\!-\!2\eps)^{2}(3\!-\!2\eps)}\,
\big(1-2\eps x_i x_j \big)
f_{ij}^{gg} 
\, ,
\nnb\\
\hat Q_{ij(r)}
& = &
2 x_i x_j \,
\frac{T_R}{(1\!-\!2\eps)(1\!-\!\eps)} \,
\bigg[
\frac1\eps \big( \beta_0 - 3\,C_F \big)
+
C_A - 2C_F
+
\frac{C_A+4\,T_R\,N_f}{3(3-2\eps)}
\bigg]
f_{ij}^{q \bar q}
\nnb\\
&&
+ \,
4\,C_A\,
\frac{2\,T_R\,N_f-C_A(1-\eps)}{(1-2\eps)(2\!-\!2\eps)^{2}(3\!-\!2\eps)}\,
\big(1-2\eps x_i x_j \big)
f_{ij}^{gg} 
\,.
\eeq
Analogously, the $\eps$ expansion $\tilde P^{\mu\nu}_{ij(r)}$ can be recast in the same form, as
\beq
&&
\tilde P_{ij(r)}^{\mu\nu}
\; = \; 
- \,
\tilde P_{ij(r)} \,
g^{\mu\nu}
+
\tilde Q_{ij(r)}^{\mu\nu}
\, ,
\qquad
\tilde Q_{ij(r)}^{\mu\nu}
\; = \;
\tilde Q_{ij(r)} \,
d_{i}^{\,\mu\nu}
\, ,
\eeq
where $\tilde P_{ij(r)}$ and $\tilde Q_{ij(r)}$ are given by 
($\mc F = P,Q$) 
\beq
\tilde{\mc F}_{ij(r)} 
& = &
\frac{\Gamma^2(1\!+\!\eps)\Gamma^3(1\!-\!\eps)}
     {\Gamma(1\!+\!2\eps)\Gamma^2(1\!-\!2\eps)}
\!\left(\! \frac{e^{\euler\!}\mu^2}{s_{ij}} \!\right)^{\!\!\eps\!}
\bigg\{
\frac{C_{\!f_{[ij]}\!}}{\eps^2}
\Big[
\rho^{\scriptscriptstyle (C)}_{[ij]}
+
\rho^{\scriptscriptstyle (C)}_{ij} \!
F(x_i)
+
\rho^{\scriptscriptstyle (C)}_{ji} \!
F(x_j)
\Big]
\mc F_{ij(r)}
+
\hat{\mc F}_{ij(r)} 
\bigg\}
\, . \qquad
\eeq
The hard-collinear real-virtual kernel, expanded in the regulator $\eps$, reads
\beq
\tilde P_{ij(r)}^{{\rm hc},\mu\nu}
& \equiv &
\tilde P_{ij(r)}^{\mu\nu}
-
s_{ij}
\Big[
2\,C_{f_{j}}\,
\tilde{\mc E}_{jr}^{(i)}
+
2\,C_{f_{i}}\,
\tilde{\mc E}_{ir}^{(j)}
\Big]
g^{\mu\nu}
\\
& = &
\tilde P_{{\rm fin},ij(r)}^{{\rm hc},\mu\nu}
+
C_{\!f_{[ij]}}
\bigg[
\rho^{\scriptscriptstyle (C)}_{[ij]}
\bigg(
\frac{1}{\eps^2}
-
\frac{1}{\eps} \, \ln\frac{s_{ij}}{\mu^2}
\bigg)
+
\frac{1}{\eps}
\Big(
\rho^{\scriptscriptstyle (C)}_{ij}
\ln x_{i}
+
\rho^{\scriptscriptstyle (C)}_{ji} 
\ln x_{j}
\Big)
\bigg]
\,
P^{{\rm hc},\mu\nu}_{ij(r)}
\nnb\\
&&
\hspace{-3mm}
- \,
\frac{4}{\eps} \, 
\bigg[
f_{i}^{g} 
C_{\!f_{j}}^{2} \, 
\frac{x_{j}}{x_{i}}
\ln x_{j}
+
f_{j}^{g} 
C_{\!f_{i}}^{2} \, 
\frac{x_{i}}{x_{j}}
\ln x_{i}
\bigg] \,
g^{\mu\nu}
-
\frac{T_R}{\eps}
\big( \beta_0 \!-\! 3C_F \big) \,
f_{ij}^{q \bar q} 
\bigg[
g_{\mu\nu}
-
4 x_i x_j 
\frac{\tk_i^\mu\tk_i^\nu}{\tk_i^2}
\bigg]
+
\mc O(\eps)
\, ,
\nnb
\eeq
where 
\beq
\tilde P_{{\rm fin},ij(r)}^{{\rm hc},\mu\nu}
& = &
P^{{\rm hc},\mu\nu}_{ij(r)} \;
C_{\!f_{[ij]}}
\bigg\{ \;
\rho^{\scriptscriptstyle (C)}_{[ij]}
\bigg[
\bigg(
\frac{1}{2} 
\ln^{2}\frac{s_{ij}}{\mu^2}
-
\frac{7}{2} \, \zeta_{2}
\bigg)
\bigg]
\nnb\\
&&
\hspace{19mm}
+ \,
\rho^{\scriptscriptstyle (C)}_{ij}
\bigg[
\Li_2\!\left(\frac{-x_{j}}{x_{i}} \right)
- 
\ln\frac{s_{ij}}{\mu^2}
\ln x_{i}
\bigg]
+
\rho^{\scriptscriptstyle (C)}_{ji} 
\bigg[
\Li_2\!\left(\frac{-x_{i}}{x_{j}} \right)
- 
\ln\frac{s_{ij}}{\mu^2}
\ln x_{j}
\bigg]
\bigg\}
\nnb\\
&-&
g^{\mu\nu} \,
2 \,
f_{i}^{g} \,
C_{f_{j}} \, 
\frac{x_{j}}{x_{i}} \,
\bigg\{ \;
C_A
\bigg[
\ln^{2}\!x_{j}
+
2\,\Li_2(x_{i})
\bigg]
+
2C_{f_{j}}
\bigg[
\Li_2\!\left(\frac{-x_{i}}{x_{j}} \right)
- 
\ln\frac{s_{ij}}{\mu^2}
\ln x_{j}
\bigg]
\bigg\}
\nnb\\
&-&
g^{\mu\nu} \,
2 \,
f_{j}^{g} \,
C_{f_{i}} \, 
\frac{x_{i}}{x_{j}} \,
\bigg\{ \;
C_A
\bigg[
\ln^{2}\!x_{i}
+
2\,\Li_2(x_{j})
\bigg]
+
2C_{f_{i}}
\bigg[
\Li_2\!\left(\frac{-x_{j}}{x_{i}} \right)
- 
\ln\frac{s_{ij}}{\mu^2}
\ln x_{i}
\bigg]
\bigg\}
\nnb\\
& - &
\bigg[
g_{\mu\nu}
-
4 x_i x_j \,
\frac{\tk_i^\mu\tk_i^\nu}{\tk_i^2}
\bigg]
\;
f_{ij}^{q \bar q} \;
T_R \,
\bigg[
\ln\frac{s_{ij}}{\mu^2} 
\big( 3\,C_F - \beta_0 \big)
+
\frac{7}{3} C_A 
+
\frac{5}{3} \beta_{0}
- 
8C_F
\bigg]
\nnb\\
&-&
g_{\mu\nu} \;
(f_{ij}^{gq} \!+\! f_{ij}^{g \bar q}) \;
C_F \,
(C_A\!-\!C_F)
+
\frac{\tk_i^\mu\tk_i^\nu}{\tk_i^2} \;
f_{ij}^{gg} \;
C_A \, (3C_A-\beta_{0})
\, ,
\eeq
and 
\beq
\Li_2\!\left(\!-\frac{x_{i}}{x_{j}} \!\right)
& = &
- \,
\Li_2(x_{i})
-
\frac12 \,
\ln^{2} x_{j}
=
\Li_2(x_{j})
+
\ln x_{i}
\ln x_{j}
-
\frac12 \,
\ln^{2} x_{j}
-
\zeta_{2}
\, ,
\nnb\\
\Li_2\!\left(\!-\frac{x_{j}}{x_{i}} \!\right)
& = &
- \,
\Li_2(x_{j})
-
\frac12 \,
\ln^{2} x_{i}
=
\Li_2(x_{i})
+
\ln x_{i}
\ln x_{j}
-
\frac12 \,
\ln^{2} x_{i}
-
\zeta_{2}
\, .
\eeq
Equivalently we can write $\tilde P_{ij(r)}^{{\rm hc},\mu\nu}$ in the form 
\beq
&&
\tilde P_{ij(r)}^{{\rm hc},\mu\nu}
\; = \; 
- \,
\tilde P_{ij(r)}^{{\rm hc}} \,
g^{\mu\nu}
+
\tilde Q_{ij(r)}^{\mu\nu}
\, ,
\eeq
with 
\beq
\tilde P_{ij(r)}^{{\rm hc}}
& \equiv &
\tilde P_{ij(r)}
+
s_{ij}
\Big[
2\,C_{f_{j}}\,
\tilde{\mc E}_{jr}^{(i)}
+
2\,C_{f_{i}}\,
\tilde{\mc E}_{ir}^{(j)}
\Big]
\\
& = &
\frac{\Gamma^2(1\!+\!\eps)\Gamma^3(1\!-\!\eps)}
     {\Gamma(1\!+\!2\eps)\Gamma^2(1\!-\!2\eps)}
\!\left(\! \frac{e^{\euler\!}\mu^2}{s_{ij}} \!\right)^{\!\!\eps\!}
\bigg\{
\frac{C_{\!f_{[ij]}\!}}{\eps^2}
\Big[
\rho^{\scriptscriptstyle (C)}_{[ij]}
+
\rho^{\scriptscriptstyle (C)}_{ij} \!
F(x_i)
+
\rho^{\scriptscriptstyle (C)}_{ji} \!
F(x_j)
\Big]
P_{ij(r)}
+
\hat P_{ij(r)} 
\nnb\\
&&
\hspace{43mm}
+ \,
2\,C_A 
\frac{\Gamma(1\!+\!\eps)\Gamma(1\!-\!\eps)}{\eps^2} 
\bigg[
f_{i}^{g} 
C_{f_{j}}\!\!
\left(
\frac{x_{j}}{x_{i}}
\right)^{\!\!1+\eps}
+
f_{j}^{g} 
C_{f_{i}}\!\!
\left(
\frac{x_{i}}{x_{j}}
\right)^{\!\!1+\eps}
\bigg]
\bigg\}
\, .
\nnb
\eeq


\section{Improved limits}
\label{app:K1,K2,K12}

In this Appendix we provide three Sections collecting the building blocks for the construction 
of our local counterterms, namely we explicitly define the action of 
\begin{itemize}
\item{improved limits on the double-real matrix element $\RR$ (Section \ref{app:LRR});}
\item{improved limits on sector functions $\W{ijjk}$, $\W{ijkj}$, $\W{ijkl}$ (Section \ref{app:LW});}
\item{improved limits on symmetrised sector functions $\Z{ijk}$, $\Z{ijkl}$ (Section \ref{app:LZ}).}
\end{itemize}
The content of each section is organised according to the nature of the singular limits involved, 
which can be single-unresolved, uniform double-unresolved, and strongly-ordered double-unresolved.
The action of improved limits $\bbL{}$ on matrix elements times sector functions is specified by 
$\bbL{}{} \, \RR \, \W{abcd} \equiv \left(\bbL{}{} \, \RR \right) \left(\bbL{}{} \, \W{abcd} \right)$, and 
similarly for $\Z{}{}$ functions. When acting on sector functions, single-unresolved and 
strongly-ordered improved limits imply the latter to be evaluated with mapped kinematics. 
Mapped sector functions are indicated generically as $\bW{}{}$ or $\bZ{}{}$ with no mapping 
labels in Sections \ref{app:LW}, \ref{app:LZ}, understanding that the actual mapping to be 
used must be adapted to the one of the matrix elements the sector function is associated to. 
To be more precise, for each term of an improved limit, the mapping of 
$\bW{}{}$ or $\bZ{}{}$ is always the same as the first mapping of matrix elements in that 
term.

To give an explicit example, let us apply this rule to the $\bbS{i} \, \bbS{ik} \, \RR \, \W{ijkl}$ 
contribution to $K^{\otwo}_{ijkl}$ counterterm. Starting with the definitions
\beq
&&
\hspace{-3mm} 
\bbS{i} \, \bbS{ik} \, \RR 
\; \equiv \;
\nnb\\
&&
\frac{\Norm^{\,2}}{2} \!\!
\sum_{\substack{c \neq i,k \\ d \neq i,k,c}} \!
\bigg\{ \;
\mc E_{cd}^{(i)} \,
\bigg[ \;
\sum_{\substack{e \neq i,k,c,d}} \!
\bigg(
\sum_{f \neq i,k,c,d,e} \!\!\!\!\!\!
\bar{\mc E}_{ef}^{(k)(icd)} \,
\bB_{cdef}^{(icd,kef)}
+
2 \,
\bar{\mc E}_{ed}^{(k)(icd)}
\bB_{cded}^{(icd,ked)}
\bigg)
\nnb\\[-3mm]
&&
\hspace{24mm}
+ \,
2 \!\!\!
\sum_{\substack{e \neq i,k,c,d}} \!\!\!
\bar{\mc E}_{ed}^{(k)(idc)}
\bB_{cded}^{(idc,ked)} 
+
2 \;
\bar{\mc E}_{cd}^{(k)(icd)} 
\Big( \bB_{cdcd}^{(icd,kcd)} + C_A\,\bB_{cd}^{(icd,kcd)} \Big)
\bigg]
\nnb\\
&&
\hspace{15mm}
- \,
2 \, C_A 
\Big[
\mc E_{kc}^{(i)} \,
\bar{\mc E}_{cd}^{(k)(ick)} \,
\bB_{cd}^{(ick,kcd)} 
+
\mc E_{kd}^{(i)} \,
\bar{\mc E}_{cd}^{(k)(ikd)} \,
\bB_{cd}^{(ikd,kcd)} 
\Big]
\bigg\}
\, ,
\eeq
and 
\beq
\bbS{i} \, \bbS{ik} \, \W{ijkl}
& \equiv &
\bW{\!{\rm s},\,kl} \,
\W{\!{\rm s},\,ij}^{(\alpha)}
\, ,
\eeq
according to the procedure detailed above, the explicit expression for $\bbS{i} \, \bbS{ik} \, 
\RR \, \W{ijkl}$ results in
\beq
&&
\hspace{-3mm} 
\bbS{i} \, \bbS{ik} \, \RR \, \W{ijkl}
\; \equiv \;
\nnb\\
&&
\frac{\Norm^{\,2}}{2} \!\!
\sum_{\substack{c \neq i,k \\ d \neq i,k,c}} \!
\bigg\{ \;
\mc E_{cd}^{(i)} \,
\bigg[ \;
\sum_{\substack{e \neq i,k,c,d}} \!
\bigg(
\sum_{f \neq i,k,c,d,e} \!\!\!\!\!\!
\bar{\mc E}_{ef}^{(k)(icd)} \,
\bB_{cdef}^{(icd,kef)}
+
2 \,
\bar{\mc E}_{ed}^{(k)(icd)}
\bB_{cded}^{(icd,ked)}
\bigg)
\bW{\!{\rm s},\,kl}^{(icd)}
\nnb\\[-3mm]
&&
\hspace{24mm}
+ \,
2 \!\!\!
\sum_{\substack{e \neq i,k,c,d}} \!\!\!
\bar{\mc E}_{ed}^{(k)(idc)}
\bB_{cded}^{(idc,ked)} \,
\bW{\!{\rm s},\,kl}^{(idc)}
+
2 \;
\bar{\mc E}_{cd}^{(k)(icd)} 
\Big( \bB_{cdcd}^{(icd,kcd)} + C_A\,\bB_{cd}^{(icd,kcd)} \Big)
\bW{\!{\rm s},\,kl}^{(icd)}
\bigg]
\nnb\\
&&
\hspace{15mm}
- \,
2 \, C_A 
\Big[
\mc E_{kc}^{(i)} \,
\bar{\mc E}_{cd}^{(k)(ick)} \,
\bB_{cd}^{(ick,kcd)} \,
\bW{\!{\rm s},\,kl}^{(ick)}
+
\mc E_{kd}^{(i)} \,
\bar{\mc E}_{cd}^{(k)(ikd)} \,
\bB_{cd}^{(ikd,kcd)} \,
\bW{\!{\rm s},\,kl}^{(ikd)}
\Big]
\bigg\}
\W{\!{\rm s},\,ij}^{(\alpha)}
\, ,
\eeq 
where it is evident that each $\bW{ab}$ contribution is mapped according to the first 
mapping of the Born matrix element it accompanies.

Finally, we introduce a shorthand notation to simplify the treatment in Section 
\ref{app:LW}: we define single-unresolved improved limits on NLO sector functions as
\beq
\label{eq:Wsij}
\W{\!{\rm s},\,ij}^{(\alpha)}
& \equiv &
\bbS{i} \, \W{ij}^{(\alpha)}
\, \equiv \,
\frac{\frac{1}{w_{ij}^{\alpha}}}
     {\sum\limits_{l \neq i} \frac{1}{w_{il}^{\alpha}}} 
\, , 
\qquad
\W{\!{\rm s},\,ij}
\, \equiv \,
\W{\!{\rm s},\,ij}^{(1)}
\, ,
\\
\label{eq:Wcij}
\W{\!{\rm c},\,ij(r)}^{(\alpha)}
& \equiv &
\bbC{ij} \, \W{ij}^{(\alpha)}
\, \equiv \,
\frac{e_j^{\alpha}w_{jr}^{\alpha}}
{e_i^{\alpha}w_{ir}^{\alpha} + e_j^{\alpha}w_{jr}^{\alpha}} 
\, ,
\qquad
\W{\!{\rm c},\,ij(r)}
\, \equiv \,
\W{\!{\rm c},\,ij(r)}^{(1)}
\, ,
\eeq
depending on a reference particle $r \ne i,j$, whose choice will be 
specified case by case;
as for NNLO sector functions, we introduce 
\beq
\label{eq: hatsigma}
\hat\sigma_{abcd (r)} 
& = &
\frac{1}{(e_a\,w_{ab}\,w_{ar})^{\alpha}}
\frac{1}{(e_c\,w_{cr} + \delta_{bc} \, e_a\,w_{ar}) \, w_{cd}} 
\, , 
\eeq
and
\beq
\hat\sigma_{\{ijk\}(r)}
& =& 
\hat\sigma_{ijjk(r)} \!+\! \hat\sigma_{ikjk(r)}
\!+\! 
\hat\sigma_{jiik(r)} \!+\! \hat\sigma_{jkik(r)}
\!+\! 
\hat\sigma_{ijkj(r)} \!+\! \hat\sigma_{ikkj(r)}
\nnb
\\
& + &
\hat\sigma_{kiij(r)} \!+\! \hat\sigma_{kjij(r)}
\!+\! 
\hat\sigma_{jiki(r)} \!+\! \hat\sigma_{jkki(r)}
\!+\! 
\hat\sigma_{kiji(r)} \!+\! \hat\sigma_{kjji(r)}
\, .
\qquad\qquad
\eeq


\subsection{Improved limits of $\RR$}
\label{app:LRR}


\subsubsection*{Single-unresolved improved limits}

For the single-unresolved improved limits we have ($j \ne i$) 
\beq
\label{eq:bSi}
\bbS{i} \, \RR 
& \equiv &
- \,
\Norm
\sum_{\substack{c \neq i \\ d \neq i,c}}
\mc E_{cd}^{(i)} \, \bR^{(icd)}_{cd} 
\, ,
\\
\label{eq:bCij}
\bbC{ij} \, RR 
& \equiv &
\Norm \frac{P_{ij(r)}^{\mu\nu}}{s_{ij}} \, 
\bR_{\mu\nu}^{(ijr)}
\, ,
\\
\label{eq:bSiCij}
\bbS{i} \, \bbC{ij} \, \RR 
& \equiv &
\bbS{i} \, \bbC{ji} \, \RR 
\; \equiv \;
\Norm \, 2 \, C_{f_j} \, \mc E^{(i)}_{jr} \, 
\bR^{(ijr)}
\, ;
\eeq
\beq
\label{eq:bHCij}
\bbHC{ij} \, \RR
& \equiv &
\bbC{ij}\left( 1 - \bbS{i} - \bbS{j} \right) \RR
\, = \,
\Norm \,
\frac{P_{ij(r)}^{{\rm hc},\mu\nu}}{s_{ij}} \, 
\bR_{\mu\nu}^{(ijr)}
\, .
\eeq
In these equations $r$ must be chosen according  to 
the rule of \eq{eq: r_ij} as $r = r_{ijkl} \ne i,j,k,l$, where $i,j,k,l$ are 
the indices appearing in the NNLO sector functions multiplying the improved 
limits $\bbC{ij}$, $\bbS{i}\,\bbC{ij}$, 
$\bbHC{ij}$. 
This means that in the topologies $\W{ijjk}$, $\W{ijkj}$ the index 
$r = r_{ijk}$ is different from the three indices of the sector, while 
for the topology $\W{ijkl}$ ($i,j,k,l$ all different) the index 
$r = r_{ijkl}$ is different from the four indices of the sector.
We stress that, having defined $r = r_{ijkl}$, one needs at least five
massless partons in $\Phi_{\npt}$, namely three massless final-state partons
at Born level. We work under this assumption throughout the paper.


\subsubsection*{Uniform double-unresolved improved limits}

The double-soft improved limit is given by ($k \ne i$) 
\beq
\label{eq:bSik}
\bbS{ik} \, \RR
& \equiv &
\frac{\Norm^{\,2}}{2} \!
\sum_{\substack{c \neq i,k \\ d \neq i,k,c}}
\bigg\{ \;
\mc E^{(i)}_{cd} 
\! \sum_{\substack{e \neq i,k,c,d}}
\bigg[
\sum_{\substack{f \neq i,k,c,d,e}} \!
\mc E^{(k)}_{ef} \bB_{cdef}^{(icd,kef)}
+ 
4 \,
\mc E^{(k)}_{ed} \bB_{cded}^{(icd,ked)}
\bigg]
\nnb\\[-3mm]
&&
\hspace{17mm}
+ \,
2 \,
\mc E^{(i)}_{cd} 
\mc E^{(k)}_{cd} 
\bB_{cdcd}^{(icd,kcd)}
+
\mc E^{(ik)}_{cd}
\bB_{cd}^{(ikcd)}
\bigg\}
\, .
\eeq
The soft-collinear improved limit $\bbSC{ikl}$ 
and its double-soft version $\bbS{ik}\,\bbSC{ikl}$
read ($k \ne i$, $l \ne i,k$, and $r = r_{ikl} \ne i,k,l$ 
defined with the rule of \eq{eq: r_ij}) 
\beq
\label{eq:bSCikl}
\bbSC{ikl} \, \RR
& \equiv & 
- \,
\Norm^{\,2} \,
\frac{P_{kl(r)}^{\mu\nu}}{s_{kl}} \,
\bigg\{
\sum_{\substack{c\neq i,k,l,r}}
\bigg[
\sum_{\substack{d\neq i,k,l,r,c}} \!\!\!
\mc E^{(i)}_{cd} \, 
\bB^{(klr,icd)}_{\mu\nu,cd}
+
2 \,
\mc E^{(i)}_{cr} \, 
\bB^{(klr,icr)}_{\mu\nu,cr}
\bigg]
\nnb\\
&&
\hspace{20mm}
+ \!\!
\sum_{\substack{c\neq i,k,l}}
\left[
\mc E^{(i)}_{kc}
\left(
\rho^{\scriptscriptstyle (C)}_{kl}
\bB^{(lrk,ick)}_{\mu\nu,[kl]c}
+
{\cal \bB}^{(lrk,ick)}_{\mu\nu,[kl]c}
\tilde f_{\,kl}^{\,q\bar q}
\right)
+
(k \leftrightarrow l)
\right]
\bigg\}
\, ,
\qquad
\eeq
\vspace{-5mm}
\beq
\label{eq:bSikSCikl}
\bbS{ik} \,\bbSC{ikl} \, \RR 
& \equiv &
\bbS{ki} \,\bbSC{ikl} \, \RR 
\; \equiv \;
\bbS{ik} \,\bbSC{ilk} \, \RR 
\nnb
\\
& \equiv &
- \, 
2 \,
\Norm^{\,2} \,
\mc E^{(k)}_{lr} \,
\bigg\{ \;\;
C_{f_{l}} \!\!
\sum_{\substack{c\neq i,k,l,r}}
\bigg[
\sum_{d\neq i,k,l,r,c} \!\!
\mc E^{(i)}_{cd} \, 
\bB^{(klr,icd)}_{cd}
+
2 \, 
\mc E^{(i)}_{cr} \, 
\bB^{(klr,icr)}_{cr}
\bigg]
\nnb \\
&&
\hspace{19mm}
+ 
\sum_{\substack{c\neq i,k,l}}
\Big[
C_{A}\,
\mc E^{(i)}_{kc} \,
\bB^{(lrk,ick)}_{[kl]c}
+ 
(2C_{f_l}\!-\!C_{A}) \,
\mc E^{(i)}_{lc} \,
\bB^{(krl,icl)}_{[kl]c}
\Big]
\bigg\}
\, .
\qquad
\eeq
The improved limits $\bbSC{ijk}$, $\bbSC{kij}$, $\bbS{ij}\,\bbSC{ijk}$, 
$\bbS{ik}\,\bbSC{ijk}$, $\bbS{ik}\,\bbSC{kij}$ can be obtained 
from these limits with a renaming of indices.
For the uniform double-unresolved limits involving $\bbC{ijk}$, 
we have ($j \ne i$, $k \ne i,j$ and $r = r_{ijk} \ne i,j,k$)
\beq
\label{eq:bCijk}
\bbC{ijk} \, \RR 
& \equiv & 
\frac{\Norm^{\,2}}{s_{ijk}^2} \,
P_{ijk(r)}^{\mu\nu} \,\bB_{\mu\nu}^{(ijkr)} 
\, ;
\eeq
\beq
\label{eq:bSijCijk}
\bbS{ij} \, \bbC{ijk} \, \RR 
& \equiv & 
\bbS{ij} \, \bbC{ikj} \, \RR 
\; \equiv \;
\bbS{ij} \, \bbC{kij} \, \RR 
\; \equiv \;
\Norm^{\,2} \,
C_{f_k} 
\bigg[
4 \, C_{f_k} \,
\mc E_{kr}^{(i)} \, \mc E_{kr}^{(j)}
- \mc E_{kr}^{(ij)}
\bigg] 
\bB^{(ijkr)} 
\, ,
\qquad
\eeq
\vspace{-5mm}
\beq
\bbHC{ijk} \, \RR
& \equiv &
\bbC{ijk}
\left( 1 - \bbS{ij} - \bbS{ik} - \bbS{jk} \right) \RR
\, = \,
\frac{\Norm^{\,2}}{s_{ijk}^2} \,
P_{ijk(r)}^{{\rm hc},\mu\nu} \,\bB_{\mu\nu}^{(ijkr)}
\, ;
\eeq
\vspace{-5mm}
\beq
\label{eq:bCijkSCijk}
\bbC{ijk}\,\bbSC{ijk} \, \RR 
& \equiv &
\bbC{jki}\,\bbSC{ijk} \, \RR 
\nnb\\
& \equiv &
\Norm^{\,2} \,
C_{f_{[jk]}} \,
\frac{P_{jk(r)}^{\mu\nu}}{s_{jk}}
\bigg[
\rho^{\scriptscriptstyle (C)}_{jk}
\mc E^{(i)}_{jr}
\bB^{(krj,irj)}_{\mu\nu}
+
\rho^{\scriptscriptstyle (C)}_{kj}
\mc E^{(i)}_{kr}
\bB^{(jrk, irk)}_{\mu\nu}
\bigg]
\, ,
\eeq
\vspace{-5mm}
\beq
\label{eq:bSijCijkSCijk}
\bbS{ij} \, \bbC{ijk} \, \bbSC{ijk} \, \RR 
& \equiv & 
\bbS{ij} \, \bbC{ikj} \, \bbSC{ikj} \, \RR 
=
\bbS{ji} \, \bbC{jki} \, \bbSC{ijk} \, \RR
\nnb\\
& \equiv &
2 \, 
\Norm^{\,2} \,
C_{f_{k}} \,
\mc E^{(j)}_{kr} \,
\Big[
C_{A}\,
\mc E^{(i)}_{jr} \,
\bB^{(krj,irj)}
+ 
(2C_{f_k}\!-\!C_{A}) \,
\mc E^{(i)}_{kr} \,
\bB^{(jrk,irk)}
\Big]
\, ,
\qquad
\eeq
\beq
\bbC{ijk} \, \bbSHC{ijk} \, \RR
& \equiv &
\bbC{ijk} \, \bbSC{ijk}
\left( 1 - \bbS{ij} - \bbS{ik} \right) \RR
\nnb\\
& \equiv &
\Norm^{\,2} \,
C_{f_{[jk]}} \,
\frac{P_{jk(r)}^{{\rm hc},\mu\nu}}{s_{jk}}
\bigg[
\rho^{\scriptscriptstyle (C)}_{jk}
\mc E^{(i)}_{jr}
\bB^{(krj,irj)}_{\mu\nu}
+
\rho^{\scriptscriptstyle (C)}_{kj}
\mc E^{(i)}_{kr}
\bB^{(jrk,irk)}_{\mu\nu}
\bigg]
\, ,
\eeq
\beq
\bbSHC{ijk} \left( 1 \!-\! \bbC{ijk} \right) \, \RR
& \equiv & 
\bbSC{ijk}
\left( 1 - \bbC{ijk} \right)
\left( 1 - \bbS{ij} - \bbS{ik} \right) \RR
\\
& \equiv &
- \,
\Norm^{\,2} \,
\frac{P_{jk(r)}^{{\rm hc},\mu\nu}}{s_{jk}}
\bigg\{
\sum_{\substack{c\neq i,j,k,r}}
\bigg[
\sum_{\substack{d\neq i,j,k,r,c}} \!\!\!
\mc E^{(i)}_{cd} \, 
\bB^{(jkr,icd)}_{\mu\nu,cd}
+
2 \,
\mc E^{(i)}_{cr} \, 
\bB^{(jkr,icr)}_{\mu\nu,cr}
\bigg]
\nnb\\
&&
\hspace{20mm}
+ \!
\sum_{\substack{c\neq i,j,k}} 
\bigg[
\mc E^{(i)}_{jc}
\Big(
\rho^{\scriptscriptstyle (C)}_{jk}
\bB^{(krj,icj)}_{\mu\nu,[jk]c}
+
{\cal \bB}^{(krj,icj)}_{\mu\nu,[jk]c}
\tilde f_{\,jk}^{\,q\bar q}
\Big)
\!+\!
(j \leftrightarrow k)
\bigg]
\nnb\\
&&
\hspace{20mm}
+ \,
C_{f_{[jk]}}
\bigg[
\rho^{\scriptscriptstyle (C)}_{jk}
\mc E^{(i)}_{jr} \,
\bB^{(krj,irj)}_{\mu\nu}
+
\rho^{\scriptscriptstyle (C)}_{kj}
\mc E^{(i)}_{kr} \,
\bB^{(jrk,irk)}_{\mu\nu}
\bigg]
\bigg\}
\, .
\nnb
\eeq
Finally, the limits involving $\bbC{ijkl}$ are given by 
($j \ne i$, $k \ne i,j$, $l \ne i,j,k$ and $r = r_{ijkl} \ne i,j,k,l$) 
\beq
\label{eq:bCijkl}
\bbC{ijkl}\, \RR
& \equiv &
\Norm^{\,2}\, 
\frac{P_{ij(r)}^{\mu \nu}}{s_{ij}} \,
\frac{P_{kl(r)}^{\rho \sigma}}{s_{kl}} \,
\bB^{(ijr, klr)}_{\mu \nu \rho \sigma} 
\, ,
\\
\label{eq:bSikCijkl}
\hspace{-30mm}
\bbS{ik} \, \bbC{ijkl} \, \RR 
& \equiv &
\bbS{ik} \, \bbC{jikl} \, \RR 
\; \equiv \;
\bbS{ik} \, \bbC{ijlk} \, \RR 
\; \equiv \;
\bbS{ik} \, \bbC{jilk} \, \RR 
\nnb\\
& \equiv &
4 \,
\Norm^{\,2} \, 
C_{f_j} \,
C_{f_l} \,
\mc E^{(i)}_{jr} \, 
\mc E^{(k)}_{lr} \, 
\bB^{(ijr,klr)} 
\, ;
\\
\label{eq:bSCiklCijkl}
\bbSC{ikl}\,\bbC{ijkl} \, \RR 
& \equiv &
\bbSC{ikl}\,\bbC{klij} \, \RR 
\; \equiv \;
\bbSC{ikl}\,\bbC{jikl} \, \RR 
\; \equiv \;
\bbSC{ikl}\,\bbC{klji} \, \RR 
\nnb\\
& \equiv &
2 \,
\Norm^{\,2} \, 
C_{f_j} \,
\mc E^{(i)}_{jr} \,
\frac{P_{kl(r)}^{\mu\nu}}{s_{kl}}
\bB_{\mu\nu}^{(ijr,klr)} 
\, ;
\eeq
\vspace{-5mm}
\beq
\bbHC{ijkl} \, \RR 
& \equiv &
\bbC{ijkl}
\left(
1
+
\bbS{ik}
+
\bbS{jk}
+
\bbS{il}
+
\bbS{jl}
-
\bbSC{ikl}
-
\bbSC{jkl}
-
\bbSC{kij}
-
\bbSC{lij}
\right)
\RR
\nnb
\\
& = &
\Norm^{\,2}\, 
\frac{P_{ij(r)}^{{\rm hc},\mu \nu}}{s_{ij}} \,
\frac{P_{kl(r)}^{{\rm hc},\rho \sigma}}{s_{kl}} \,
\bB^{(ijr, klr)}_{\mu \nu \rho \sigma}
\, .
\qquad\qquad
\eeq


\subsubsection*{Strongly-ordered double-unresolved improved limits}

The improved limit $\bbS{i}\,\bbS{ik}$ is given by ($k \ne i$) 
\beq
\bbS{i} \, \bbS{ik} \, \RR 
& \equiv &
\frac{\Norm^{\,2}}{2} \!\!\!
\sum_{\substack{c \neq i,k \\ d \neq i,k,c}} \!\!
\bigg\{ \;
\mc E_{cd}^{(i)} \,
\bigg[ \;
\sum_{\substack{e \neq i,k,c,d}} \!
\bigg(
\sum_{f \neq i,k,c,d,e} \!\!\!\!\!\!
\bar{\mc E}_{ef}^{(k)(icd)} \,
\bB_{cdef}^{(icd,kef)}
+
2 \,
\bar{\mc E}_{ed}^{(k)(icd)}
\bB_{cded}^{(icd,ked)}
\bigg)
\nnb\\[-3mm]
&&
\hspace{23mm}
+ \,
2 \!\!\!
\sum_{\substack{e \neq i,k,c,d}} \!\!\!\!
\bar{\mc E}_{ed}^{(k)(idc)}
\bB_{cded}^{(idc,ked)} \!
+
2 \,
\bar{\mc E}_{cd}^{(k)(icd)} \!
\Big( \bB_{cdcd}^{(icd,kcd)} + C_{\!A}\,\bB_{cd}^{(icd,kcd)} \Big)
\bigg]
\quad
\nnb\\
&&
\hspace{15mm}
- \,
2 \, C_A 
\Big[
\mc E_{kc}^{(i)} \,
\bar{\mc E}_{cd}^{(k)(ick)} \,
\bB_{cd}^{(ick,kcd)} 
+
\mc E_{kd}^{(i)} \,
\bar{\mc E}_{cd}^{(k)(ikd)} \,
\bB_{cd}^{(ikd,kcd)} 
\Big]
\bigg\}
\, .
\eeq
For $\bbS{i}\,\bbSC{ikl}$ and $\bbS{i}\,\bbS{ik}\,\bbSC{ikl}$ we have 
($k \ne i$, $l \ne i,k$, and $r = r_{ikl} \ne i,k,l$) 
\beq
\label{eq:bSiSCikl}
\bbS{i}\,\bbSC{ikl} \, \RR 
& \equiv &
- \,
\Norm^{\,2} \!\!\!\!
\sum_{\substack{c\neq i,k,l}} \!
\Bigg\{ \;
\sum_{\substack{d\neq i,k,l,c}} \!\!\!
\mc E^{(i)}_{cd} \, 
\frac{\bar P_{kl(r)}^{(icd)\mu\nu}}{\sk{kl}{icd}} \,
\bB^{(icd,klr)}_{\mu\nu,cd}
\\
&&
\hspace{16mm}
+ \,
\Bigg[
\mc E^{(i)}_{kc} 
\frac{\bar P_{kl(r)}^{(ikc)\mu\nu}}{2\,\sk{kl}{ikc}} \!
\Big(
\rho^{\scriptscriptstyle (C)}_{kl}
\bB^{(ikc,lrk)}_{\mu\nu,[kl]c}
\!+\!
{\cal \bB}^{(ikc,lrk)}_{\mu\nu,[kl]c}
\tilde f_{\,kl}^{\,q\bar q}
\Big)
+
(k \leftrightarrow l)
\!\Bigg]
\nnb\\
&&
\hspace{16mm}
+ \,
\Bigg[
\mc E^{(i)}_{kc} 
\frac{\bar P_{kl(r)}^{(ick)\mu\nu}}{2\,\sk{kl}{ick}} \!
\Big(
\rho^{\scriptscriptstyle (C)}_{kl}
\bB^{(ick,lrk)}_{\mu\nu,[kl]c}
\!+\!
{\cal \bB}^{(ick,lrk)}_{\mu\nu,[kl]c}
\tilde f_{\,kl}^{\,q\bar q}
\Big)
+
(k \leftrightarrow l)
\!\Bigg]
\Bigg\}
\, ,
\nnb
\eeq
\vspace{-2mm}
\beq
\label{eq:bSiSikSCikl}
\bbS{i} \, \bbS{ik} \, \bbSC{ikl} \, \RR 
& \equiv &
\bbS{i} \, \bbS{ik} \, \bbSC{ilk} \, \RR 
\nnb\\
& \equiv &
- \,
\Norm^{\,2} \!
\sum_{\substack{c\neq i,k,l}} \!
\bigg[ \;
2 \,
C_{f_l} \!\!\!
\sum_{\substack{d\neq i,k,l,c}} \!\!\!
\mc E^{(i)}_{cd} \, 
\mc{\bar E}^{(k)(icd)}_{lr}
{\bar B}^{(icd,klr)}_{cd}
\\
&&
\hspace{17mm}
+ \, 
C_A \,
\mc E^{(i)}_{kc} 
\Big(
\bar{\mc E}^{(k)(ikc)}_{lr} 
{\bar B}^{(ikc,lrk)}_{lc}
+
\bar{\mc E}^{(k)(ick)}_{lr} 
{\bar B}^{(ick,lrk)}_{lc}
\Big)
\nnb\\
&&
\hspace{17mm}
+ \, 
(2C_{f_l} \!-\! C_{A} ) \,
\mc E^{(i)}_{lc} 
\Big(
\bar{\mc E}^{(k)(ilc)}_{lr}
{\bar B}^{(ilc,krl)}_{lc}
+
\bar{\mc E}^{(k)(icl)}_{lr}
{\bar B}^{(icl,krl)}_{lc}
\Big)
\bigg]
\, .
\nnb
\eeq
Combining the previous definitions we have 
($j \ne i$, $k \ne i, j$, and $r = r_{ijk} \ne i,j,k$) 
\beq
\bbS{i}\,\bbSHC{ijk} \, \RR 
& \equiv &
\bbS{i}\,\bbSC{ijk} 
\left( 1 - \bbS{ij} - \bbS{ik} \right)  \RR
\\
& \equiv &
- \,
\Norm^{\,2} \!\!\!\!
\sum_{\substack{c\neq i,j,k}} \!
\Bigg\{ \;
\sum_{\substack{d\neq i,j,k,c}} \!\!\!
\mc E^{(i)}_{cd} \, 
\frac{\bar P_{jk(r)}^{(icd){\rm hc},\mu\nu\!\!}}{\sk{jk}{icd}} \,
\bB^{(icd,jkr)}_{\mu\nu,cd}
\nnb\\
&&
\hspace{16mm}
+ \,
\Bigg[
\mc E^{(i)}_{jc} \,
\frac{\bar P_{jk(r)}^{(ijc){\rm hc},\mu\nu\!\!}}{2\,\sk{jk}{ijc}} 
\Big(
\rho^{\scriptscriptstyle (C)}_{jk}
\bB^{(ijc,krj)}_{\mu\nu,[jk]c}
\!+\!
{\cal \bB}^{(ijc,krj)}_{\mu\nu,[jk]c}
\tilde f_{\,jk}^{\,q\bar q}
\Big)
+
(j \leftrightarrow k)
\!\Bigg]
\nnb\\
&&
\hspace{16mm}
+ \,
\Bigg[
\mc E^{(i)}_{jc} \,
\frac{\bar P_{jk(r)}^{(icj){\rm hc},\mu\nu\!\!}}{2\,\sk{jk}{icj}} 
\Big(
\rho^{\scriptscriptstyle (C)}_{jk}
\bB^{(icj,krj)}_{\mu\nu,[jk]c}
\!+\!
{\cal \bB}^{(icj,krj)}_{\mu\nu,[jk]c}
\tilde f_{\,jk}^{\,q\bar q}
\Big)
+
(j \leftrightarrow k)
\!\Bigg]
\Bigg\}
\, .
\nnb
\eeq
For the strongly-ordered double-unresolved limits involving 
$\bbS{i}\,\bbC{ijk}$, we have ($j \ne i$, $k \ne i,j$, 
$r = r_{ijk} \ne i,j,k$) 
\beq
&&
\hspace{-3mm} 
\bbS{i}\,\bbC{ijk} (1\!-\!\bbSC{ijk}) \RR 
\; \equiv \;
\nnb\\
&&
\Norm^{\,2}
\frac{C_{\!f_{[jk]}}}{2}
\Bigg\{ \;
\rho^{\scriptscriptstyle (C)}_{jk} \,
\mc E^{(i)}_{jr} 
\Bigg[\!
\frac{\bar P_{jk(r)}^{(ijr)\mu\nu}}{\sk{jk}{ijr}}
\!\left(\! \bB^{(ijr,jkr)}_{\mu\nu} \!-\! \bB^{(ijr,krj)}_{\mu\nu} \!\right)
+
\frac{\bar P_{jk(r)}^{(irj)\mu\nu}}{\sk{jk}{irj}}
\!\left(\! \bB^{(irj,jkr)}_{\mu\nu} \!-\! \bB^{(irj,krj)}_{\mu\nu} \!\right)
\!\Bigg]
\nnb\\
&&
\hspace{15mm}
+ \,
\rho^{\scriptscriptstyle (C)}_{kj} 
\mc E^{(i)}_{kr} 
\Bigg[\!
\frac{\bar P_{jk(r)}^{(ikr)\mu\nu}}{\sk{jk}{ikr}}
\!\left(\! \bB^{(ikr,jkr)}_{\mu\nu} \!\!-\! \bB^{(ikr,jrk)}_{\mu\nu} \!\right)
+
\frac{\bar P_{jk(r)}^{(irk)\mu\nu}}{\sk{jk}{irk}}
\!\left(\! \bB^{(irk,jkr)}_{\mu\nu} \!\!-\! \bB^{(irk,jrk)}_{\mu\nu} \!\right)
\!\Bigg]
\nnb\\
&&
\hspace{15mm}
- \,
\rho^{\scriptscriptstyle (C)}_{[jk]} \,
\mc E^{(i)}_{jk} 
\Bigg[
\frac{\bar P_{jk(r)}^{(ijk)\mu\nu}}{\sk{jk}{ijk}}
\bB^{(ijk,jkr)}_{\mu\nu} 
+
\frac{\bar P_{jk(r)}^{(ikj)\mu\nu}}{\sk{jk}{ikj}}
\bB^{(ikj,jkr)}_{\mu\nu}
\Bigg]
\Bigg\}
\, ,
\qquad
\eeq
\beq
&&
\hspace{-3mm} 
\bbS{i} \, \bbS{ij} \, \bbC{ijk} (1\!-\!\bbSC{ijk}) \RR 
\; \equiv \;
\bbS{i} \, \bbS{ij} \, \bbC{ikj} (1\!-\!\bbSC{ikj}) \RR 
\nnb\\
&&
\equiv
\Norm^{\,2} \, 
C_{f_k}
\bigg\{ \;
C_A \,
\mc E_{jr}^{(i)} 
\Big[ 
\bar{\mc E}_{kr}^{(j)(ijr)}
\!\left( \bB^{(ijr,jkr)} \!-\! \bB^{(ijr,krj)} \right)
+
\bar{\mc E}_{kr}^{(j)(irj)}
\!\left( \bB^{(irj,jkr)} \!-\! \bB^{(irj,krj)} \right)
\Big]
\nnb\\
&&
\hspace{16mm}
+ \,
(2C_{\!f_k} \!\!-\! C_{\!A} ) \,
\mc E_{kr}^{(i)} \,
\Big[
\bar{\mc E}_{kr}^{(j)(ikr)}
\!\left( \bB^{(ikr,jkr)} \!-\! \bB^{(ikr,jrk)} \right)\!
+
\bar{\mc E}_{kr}^{(j)(irk)}
\!\left( \bB^{(irk,jkr)} \!-\! \bB^{(irk,jrk)} \right)\!
\Big]
\nnb\\
&&
\hspace{16mm}
+ \,
C_A \,
\mc E_{jk}^{(i)} 
\Big[ 
\bar{\mc E}_{kr}^{(j)(ijk)}
\bar B^{(ijk,jkr)}
+
\bar{\mc E}_{kr}^{(j)(ikj)}
\bar B^{(ikj,jkr)}
\Big]
\bigg\}
\, ,
\eeq
\beq
&&
\hspace{-3mm} 
\bbS{i} \, \bbHC{ijk}^{\;(\bf s)} \, \RR 
\; \equiv \;
\bbS{i} \,
\bbC{ijk}
\left( 1 - \bbS{ij} - \bbS{ik} \right) \!
\left( 1 - \bbSC{ijk} \right)
\RR 
\\
&&
=
\Norm^{\,2}
\frac{C_{\!f_{[jk]}}}{2}
\Bigg\{ \;
\rho^{\scriptscriptstyle (C)}_{jk} \,
\mc E^{(i)}_{jr} 
\Bigg[\!
\frac{\bar P_{jk(r)}^{(ijr){\rm hc},\mu\nu\!\!\!\!}}{\sk{jk}{ijr}}
\left(\! \bB^{(ijr,jkr)}_{\mu\nu} \!-\! \bB^{(ijr,krj)}_{\mu\nu} \!\right)\!
+
\frac{\bar P_{jk(r)}^{(irj){\rm hc},\mu\nu\!\!\!\!}}{\sk{jk}{irj}}
\left(\! \bB^{(irj,jkr)}_{\mu\nu} \!-\! \bB^{(irj,krj)}_{\mu\nu} \!\right)\!
\!\Bigg]
\nnb\\
&&
\hspace{19mm}
+ \,
\rho^{\scriptscriptstyle (C)}_{kj} 
\mc E^{(i)}_{kr} 
\Bigg[\!
\frac{\bar P_{jk(r)}^{(ikr){\rm hc},\mu\nu\!\!\!\!}}{\sk{jk}{ikr}}
\left(\! \bB^{(ikr,jkr)}_{\mu\nu} \!\!-\! \bB^{(ikr,jrk)}_{\mu\nu} \!\right)\!
+
\frac{\bar P_{jk(r)}^{(irk){\rm hc},\mu\nu\!\!\!\!}}{\sk{jk}{irk}}
\left(\! \bB^{(irk,jkr)}_{\mu\nu} \!\!-\! \bB^{(irk,jrk)}_{\mu\nu} \!\right)\!
\!\Bigg]
\nnb\\
&&
\hspace{19mm}
- \,
\rho^{\scriptscriptstyle (C)}_{[jk]} \,
\mc E^{(i)}_{jk} 
\Bigg[
\frac{\bar P_{jk(r)}^{(ijk){\rm hc},\mu\nu}}{\sk{jk}{ijk}}
\bB^{(ijk,jkr)}_{\mu\nu}
+
\frac{\bar P_{jk(r)}^{(ikj){\rm hc},\mu\nu}}{\sk{jk}{ikj}}
\bB^{(ikj,jkr)}_{\mu\nu}
\Bigg]
\Bigg\}
\, .
\nnb
\eeq
For $\bbC{ij}\,\bbSC{kij}$ and $\bbS{i}\,\bbC{ij}\,\bbSC{kij}$ we have 
($j \ne i$, $k \ne i,j$, 
$r = r_{ijkl} \ne i,j,k,l$, 
$r' = r_{ijk} \ne i,j,k$
in sector $\W{ijkl}$
) 
\beq
\label{eq:bCijSCkij}
\bbC{ij} \, \bbSC{kij} \, \RR 
& \equiv & 
- \,
\Norm^{\,2} \,
\frac{P_{ij(r)}^{\mu\nu}}{s_{ij}} 
\bigg\{ \;
\sum_{c \neq i,j,k,r'} \!
\bigg[
\sum_{d\neq i,j,k,r',c} \!\!\!\!
\bar{\mc E}^{(k)(ijr)}_{cd} 
\bB^{(ijr, kcd)}_{\mu\nu,cd}
+
2 \,
\bar{\mc E}^{(k)(ijr)}_{cr'} 
\bB^{(ijr, kcr')}_{\mu\nu,cr'}
\bigg]
\nnb\\
&&
\hspace{19mm}
+ \,
2 \!\!\!
\sum_{\substack{c\neq i,j,k}} \!
\bar{\mc E}^{(k)(ijr)}_{jc} 
\bB^{(ijr,kcj)}_{\mu\nu,jc}
\bigg\}
\, ,
\\
\label{eq:bSiCijSCkij}
\bbS{i} \, \bbC{ij} \, \bbSC{kij} \, \RR 
& \equiv & 
\bbS{i} \, \bbC{ji} \, \bbSC{kji} \, \RR 
\nnb\\
& \equiv & 
- \,
2 \,
\Norm^{\,2} \,
C_{f_j}\, 
\mc E^{(i)}_{jr} 
\bigg\{ 
\sum_{c \neq i,j,k,r'} \!
\bigg[
\sum_{d\neq i,j,k,r',c} \!\!\!\!\!\!
\bar{\mc E}^{(k)(ijr)}_{cd} \, \bB^{(ijr, kcd)}_{cd}
+
2 \,
\bar{\mc E}^{(k)(ijr)}_{cr'} \, \bB^{(ijr, kcr')}_{cr'}
\bigg]
\nnb\\
&&
\hspace{24mm}
+ \,
2 \!\!
\sum_{\substack{c\neq i,j,k}} \!\!
\bar{\mc E}^{(k)(ijr)}_{jc} \, \bB^{(ijr, kcj)}_{jc}
\bigg]
\, ,
\\
\bbHC{ij}\,\bbSC{kij} \, \RR 
& \equiv &
\bbC{ij} \left( 1 - \bbS{i} - \bbS{j} \right) \bbSC{kij} \, \RR 
\nnb\\
& = & 
- \,
\Norm^{\,2} \,
\frac{P_{ij(r)}^{{\rm hc},\mu\nu\!\!}}{s_{ij}} 
\bigg\{ \;
\sum_{c \neq i,j,k,r'} \!
\bigg[
\sum_{d\neq i,j,k,r',c} \!\!\!\!
\bar{\mc E}^{(k)(ijr)}_{cd} 
\bB^{(ijr, kcd)}_{\mu\nu,cd}
+
2 \,
\bar{\mc E}^{(k)(ijr)}_{cr'} 
\bB^{(ijr, kcr')}_{\mu\nu,cr'}
\bigg]
\nnb\\
&&
\hspace{16mm}
+ \,
2 \!\!\!
\sum_{\substack{c\neq i,j,k}} \!
\bar{\mc E}^{(k)(ijr)}_{jc} \!
\bB^{(ijr,kcj)}_{\mu\nu,jc}
\bigg\}
\, .
\eeq
The improved limits $\bbC{ij}\,\bbS{ij}\,\RR$,  
$\bbS{i}\,\bbC{ij}\,\bbS{ij}\,\RR$ and their combination 
$\bbHC{ij} \, \bbS{ij} \, \RR$ appear in the sector topology 
$\W{ijjk}$ only, and are given by ($j \ne i$ and $r = r_{ijk} \ne i,j,k$) 
\beq
\!\!
\bbC{ij} \, \bbS{ij} \RR 
& \equiv &
- \,
\Norm^{2} \!\!\!\!
\sum_{\substack{c \neq i,j \\ d \neq i,j,c}} \!\!\!
\Bigg\{ \!
\frac{P_{ij(r)\!}}{s_{ij}} \,
\bar{\mc E}^{(j)(ijr)}_{cd} \!
+
\frac{Q_{ij(r)\!}^{\mu\nu}}{s_{ij}}
\Bigg[\!
\frac{\kk{c,\mu}{ijr}}{\sk{jc}{ijr}}
-
\frac{\kk{d,\mu}{ijr}}{\sk{jd}{ijr}}
\!\Bigg]
\Bigg[\!
\frac{\kk{c,\nu}{ijr}}{\sk{jc}{ijr}}
-
\frac{\kk{d,\nu}{ijr}}{\sk{jd}{ijr}}
\!\Bigg]
\Bigg\}
\bB_{cd}^{(ijr,jcd)} 
\! ,
\qquad\quad
\eeq
\beq
\bbS{i} \, \bbC{ij} \, \bbS{ij} \, \RR 
& \equiv &
\bbS{i} \, \bbC{ji} \, \bbS{ji} \, \RR 
\; \equiv \;
- \, 2 \, \Norm^{\,2}  \, 
 C_{f_j} \, \mc E_{jr}^{(i)} \!
\sum_{\substack{c \neq i,j \\ d \neq i,j,c}} \!  \bar{\mc E}_{cd}^{(j)(ijr)} \,
\bB_{cd}^{(ijr,jcd)} 
\, ,
\eeq
\beq
\bbHC{ij} \, \bbS{ij} \, \RR 
& \equiv &
\bbC{ij} \! \left( 1 - \bbS{i} - \bbS{j} \right) \bbS{ij} \RR 
\\
& = &
- \,
\Norm^2 \!\!
\sum_{\substack{c \neq i,j \\ d \neq i,j,c}} \!
\Bigg[
\frac{P_{ij(r)}^{\rm hc}}{s_{ij}} \,
\bar{\mc E}^{(j)(ijr)}_{cd}
+
\frac{Q_{ij(r)}^{\mu\nu}}{s_{ij}}
\Bigg(
\frac{\kk{c,\mu}{ijr}}{\sk{jc}{ijr}}
-
\frac{\kk{d,\mu}{ijr}}{\sk{jd}{ijr}}
\Bigg)
\Bigg(
\frac{\kk{c,\nu}{ijr}}{\sk{jc}{ijr}}
-
\frac{\kk{d,\nu}{ijr}}{\sk{jd}{ijr}}
\Bigg)
\Bigg] \,
\bB_{cd}^{(ijr,jcd)} 
\! .
\nnb
\eeq
For the strongly-ordered double-unresolved limits involving 
$\bbC{ij}\,\bbC{ijk}$, we have ($j \ne i$, $k \ne i,j$, 
$r = r_{ijk} \ne i,j,k$) 
\beq
\label{eq:bCijCijk}
\bbC{ij} \, \bbC{ijk} \, \RR 
& \equiv & 
\Norm^{\,2} 
\Bigg\{ \;
\frac{P_{ij(r)}}{s_{ij}}
\frac{\bar P_{jk(r)}^{(ijr)\mu\nu}}{\sk{jk}{ijr}}
\bB^{(ijr,jkr)}_{\mu\nu}
+
2 \, C_A \, 
\bar{\mc E}^{(k)(ijr)}_{jr}
\frac{Q_{ij(r)}^{\mu\nu}}{s_{ij}} \, 
\bB^{(ijr,jkr)}_{\mu\nu}
\\
&& 
\hspace{6mm}
- \,
2 \, 
C_{f_k}\,
\bar{\mc E}^{(j)(ijr)}_{kr}
\frac{Q_{ij(r)}^{\mu\nu}}{s_{ij}} \, 
\frac{\tilde{\bar k}^{(ijr)}_{\mu}\tilde{\bar k}^{(ijr)}_{\nu}}
     {\big(\tilde{\bar k}^{(ijr)}\big)^2}
\bB^{(ijr,jkr)}
\Bigg\}
\, ,
\nnb\\
\label{eq:bSiCijCijk}
\bbS{i} \, \bbC{ij} \, \bbC{ijk} \, \RR 
& \equiv & 
\bbS{i} \, \bbC{ji} \, \bbC{jik} \, \RR 
\; \equiv \;
2 \, \Norm^{\,2} \, 
C_{f_j} \, 
\mc E_{jr}^{(i)} \,
\frac{\bar P_{jk(r)}^{(ijr)\mu\nu}}{\sk{jk}{ijr}} \,
\bB^{(ijr,jkr)}_{\mu\nu}
\, ,
\eeq
\beq
\label{eq:bCijSijCijk}
\bbC{ij} \, \bbS{ij} \, \bbC{ijk} \RR 
& \equiv &
2 \, 
\Norm^{\,2} \,
C_{f_k} \, 
\bar{\mc E}^{(j)(ijr)}_{kr}
\Bigg\{
\frac{P_{ij(r)}}{s_{ij}}
-
\frac{Q_{ij(r)}^{\mu\nu}}{s_{ij}} \, 
\frac{\tilde{\bar k}^{(ijr)}_{\mu}\tilde{\bar k}^{(ijr)}_{\nu}}
     {\big(\tilde{\bar k}^{(ijr)}\big)^2}
\Bigg\} 
\bB^{(ijr,jkr)} 
\, ,
\qquad\quad
\\
\label{eq:bSiCijSijCijk}
\bbS{i} \, \bbC{ij} \, \bbS{ij} \, \bbC{ijk}\, \RR 
& \equiv & 
\bbS{i} \, \bbC{ji} \, \bbS{ji} \, \bbC{jik}\, \RR 
\; \equiv \;
4 \, \Norm^{\,2} \,
C_{f_j} \, C_{f_k} \, 
\mc E_{jr}^{(i)}\,
\bar{\mc E}_{kr}^{(j)(ijr)}\,
\bB^{(ijr,jkr)}
\, ,
\eeq
\vspace{-5mm}
\beq
\label{eq:bCijCijkSCkij}
\bbC{ij} \, \bbC{ijk} \, \bbSC{kij} \, \RR
& \equiv &
2 \,
\Norm^{\,2} \, 
C_{f_{[ij]}} \,
\bar{\mc E}^{(k)(ijr)}_{jr} \,
\frac{P_{ij(r)}^{\mu\nu}}{s_{ij}} \,
\bB^{(ijr,krj)}_{\mu\nu} 
\, ,
\\
\label{eq:bSiCijCijkSCkij}
\bbS{i}\, \bbC{ij}\, \bbC{ijk}\, \bbSC{kij}\, \RR
& \equiv & 
\bbS{i}\, \bbC{ji}\, \bbC{jik}\, \bbSC{kji}\, \RR
\; \equiv \;
4 \, 
\Norm^{\,2} \, 
C_{f_j}^{2} \, 
\mc E^{(i)}_{jr}  \,
\bar{\mc E}^{(k)(ijr)}_{jr} \, 
\bB^{(ijr,krj)}
\, ,
\eeq
\beq
\bbHC{ij} \, \bbHC{ijk}^{\;(\bf c)} \, \RR \, \Z{ijk}
& \equiv &
\bbC{ij} \! \left( 1 - \bbS{i} - \bbS{j} \right) 
\bbC{ijk} 
\left( 1 - \bbS{ij} - \bbSC{kij} \right) 
\RR 
\\
& = &
\Norm^{\,2} \,
\frac{P_{ij(r)}^{\rm hc}}{s_{ij}} \,
\frac{\bar P_{jk(r)}^{(ijr){\rm hc},\mu\nu\!\!\!\!}}{\sk{jk}{ijr}} \,
\bB^{(ijr,jkr)}_{\mu\nu}
\nnb\\
&&
- \,
2 \,
\Norm^{\,2} \, 
C_{f_{[ij]}} \,
\bar{\mc E}^{(k)(ijr)}_{jr} \,
\frac{P_{ij(r)}^{{\rm hc},\mu\nu}}{s_{ij}} \,
\Big(
\bB^{(ijr,krj)}_{\mu\nu} 
-
\bB^{(ijr,kjr)}_{\mu\nu} 
\Big)
\, .
\nnb
\eeq
Finally the limits involving $\bbC{ij}\,\bbC{ijkl}$ are given by 
($j \ne i$, $k \ne i,j$, $l \ne i,j,k$ and 
$r = r_{ijkl} \ne i,j,k,l$) 
\beq
\label{eq:bCijCijkl}
\bbC{ij}\, \bbC{ijkl}\, \RR 
& \equiv &
\Norm^{\,2} \, 
\frac{P_{ij(r)}^{\mu \nu}}{s_{ij}}  \; 
\frac{\bar P_{kl(r)}^{(ijr)\rho \sigma}}{\sk{kl}{ijr}} \; 
\bB^{(ijr, klr)}_{\mu \nu \rho \sigma}
\, ,
\\
\label{eq:bSiCijCijkl}
\bbS{i}\, \bbC{ij}\, \bbC{ijkl}\, \RR 
& \equiv & 
\bbS{i}\, \bbC{ji}\, \bbC{jikl}\, \RR 
\; \equiv \;
2\,
\Norm^{\,2} \,
C_{f_j} \,
\mc E^{(i)}_{jr} \, 
\frac{\bar P_{kl(r)}^{(ijr)\rho \sigma}}{\sk{kl}{ijr}} 
\bB^{(ijr, klr)}_{\rho \sigma} 
\, ,
\\
\label{eq:bCijSCkijCijkl}
\bbC{ij}\, \bbSC{kij}\, \bbC{ijkl}\, \RR
& \equiv &
\bbC{ij}\, \bbSC{kij}\, \bbC{ijlk}\, \RR
\; \equiv \;
2\,
\Norm^{\,2} \, 
C_{f_l}\,
\frac{P_{ij(r)}^{\mu \nu}}{s_{ij}} \, 
\bar{\mc E}^{(k) (ijr)}_{lr} \,
\bB^{(ijr, klr)}_{\mu \nu} 
\, ,
\\
\label{eq:bSiCijSCkijCijkl}
\bbS{i}\, \bbC{ij}\, \bbSC{kij}\, \bbC{ijkl}\, \RR 
& \equiv &
\bbS{i}\, \bbC{ji}\, \bbSC{kji}\, \bbC{jikl}\, \RR 
\, \equiv \,
\bbS{i}\, \bbC{ij}\, \bbSC{kij}\, \bbC{ijlk}\, \RR 
\, \equiv \,
\bbS{i}\, \bbC{ji}\, \bbSC{kji}\, \bbC{jilk}\, \RR 
\nnb\\
& \equiv &
4\,
\Norm^{\,2} \, 
C_{f_j}
C_{f_l}\,
\mc E^{(i)}_{jr} \, 
\mc E^{(k)}_{lr} \,  
\bB^{(ijr, klr)}
\, ,
\eeq
\beq
\bbHC{ij}\, \bbHC{ijkl}^{\;(\bf c)}\, \RR
& \equiv &
\bbC{ij}\left( 1 - \bbS{i} - \bbS{j} \right)
\bbC{ijkl}
\left( 1 - \bbSC{kij} - \bbSC{lij} \right)
\RR
\\
& = &
\Norm^{\,2} \, 
\frac{P_{ij(r)}^{{\rm hc},\mu\nu}}{s_{ij}}  \; 
\frac{\bar P_{kl(r)}^{(ijr){\rm hc},\rho\sigma\!\!\!\!}}{\sk{kl}{ijr}} \; 
\bB^{(ijr, klr)}_{\mu \nu \rho \sigma}
\, .
\nnb
\eeq


\subsection{Improved limits of $\W{ijjk}$, $\W{ijkj}$, $\W{ijkl}$}
\label{app:LW}


\subsubsection*{Single-unresolved improved limits}

For the single-unresolved improved limits we have ($j \ne i$, $k \ne i$, 
$l \ne i,k$ and $r = r_{ijkl} \ne i,j,k,l$) 
\beq
\label{eq:bSiW}
\bbS{i} \, \W{ijkl}
& \equiv &
\bW{kl}
\,
\W{\!{\rm s},\,ij}^{(\alpha)}
\, ,
\\
\label{eq:bCijW}
\bbC{ij} \, \W{ijkl}
& \equiv &
\bW{kl} \,
\W{\!{\rm c},\,ij(r)}^{(\alpha)}
\, ,
\\
\label{eq:bSiCijW}
\bbS{i} \, \bbC{ij} \, \W{ijkl}
& \equiv &
\bW{kl}
\, .
\eeq


\subsection*{Uniform double-unresolved improved limits}

The double-soft improved limit is given by ($j \ne i$, $k \ne i$, 
$l \ne i,k$) 
\beq
\label{eq:bSikW}
\bbS{ik} \, \W{ijkl}
& \equiv &
\frac{\sigma_{ijkl}}
{
\sum_{b \neq i}\sum_{d \neq i,k} \sigma_{ibkd}
+ 
\sum_{b\neq k}\sum_{d\neq k,i} \sigma_{kbid}
}
\, .
\eeq
The soft-collinear improved limits $\bbSC{ikl}$ and $\bbSC{kij}$ 
as well as their double-soft versions $\bbS{ik}\,\bbSC{ikl}$ and $\bbS{ik}\,\bbSC{kij}$ 
read ($j \ne i$, $k \ne i$, $l \ne i,k$)  
\beq
\label{eq:bSCiklW}
\bbSC{ikl} \, \W{ijkl}
& \equiv & 
\frac{\sigma_{ij}^{(\alpha)}\frac{\sigma_{kl}}{w_{kr}}}
{
\sum_{b \neq i} \sigma_{ib}^{(\alpha)} 
\left( \frac{\sigma_{kl}}{w_{kr}} + \frac{\sigma_{lk}}{w_{lr}} \right)
+
\frac{\sigma_{kl}^{(\alpha)}}{w_{kr}}\sum_{d\neq i,k} \sigma_{id}
+
\frac{\sigma_{lk}^{(\alpha)}}{w_{lr}}\sum_{d\neq i,l} \sigma_{id}
}
\, ,
\quad
r = r_{ikl}
\, ,
\qquad
\\
\label{eq:bSCkijW}
\bbSC{kij} \, \W{ijkl}
& \equiv & 
\frac{\frac{\sigma_{ij}^{(\alpha)}}{w_{ir}}\sigma_{kl}}
{
\sum_{b \neq k} \sigma_{kb}^{(\alpha)} 
\left( \frac{\sigma_{ij}}{w_{ir}} + \frac{\sigma_{ji}}{w_{jr}} \right)
+
\frac{\sigma_{ij}^{(\alpha)}}{w_{ir}}\sum_{d\neq i,k} \sigma_{kd}
+
\frac{\sigma_{ji}^{(\alpha)}}{w_{jr}}\sum_{d\neq k,j} \sigma_{kd}
}
\, ,
\quad
r = r_{ijk}
\, ,
\qquad
\eeq
\vspace{-5mm}
\beq
\label{eq:bSikSCiklW}
\bbS{ik} \,\bbSC{ikl} \, \W{ijkl}
& \equiv &
\frac{\sigma_{ij}^{(\alpha)}\sigma_{kl}}
{
\sum_{b \neq i} \sigma_{ib}^{(\alpha)} 
\sigma_{kl}
+
\sigma_{kl}^{(\alpha)}\sum_{d\neq i,k} \sigma_{id}
}
\, ,
\quad
r = r_{ikl}
\, ,
\qquad
\\
\label{eq:bSikSCkijW}
\bbS{ik}\,\bbSC{kij} \, \W{ijkl}
& \equiv &
\frac{\sigma_{ij}^{(\alpha)}\sigma_{kl}}
{
\sum_{b \neq k} \sigma_{kb}^{(\alpha)} 
\sigma_{ij}
+
\sigma_{ij}^{(\alpha)}\sum_{d\neq i,k} \sigma_{kd}
}
\, ,
\quad
r = r_{ijk}
\, .
\qquad
\eeq
For the uniform double-unresolved limits involving $\bbC{ijk}$, 
we have ($j \ne i$, $k \ne i,j$ and $r = r_{ijk} \ne i,j,k$) 
\beq
\label{eq:bCijkW}
\bbC{ijk} \, \W{ijjk}
& \equiv & 
\frac{\hat\sigma_{ijjk(r)}}{\hat\sigma_{\{ijk\}(r)}}
\, ,
\qquad
\bbC{ijk} \, \W{ijkj}
\; \equiv \;
\frac{\hat\sigma_{ijkj(r)}}{\hat\sigma_{\{ijk\}(r)}}
\, ;
\qquad\qquad
\eeq
\beq
\label{eq:bSijCijkW}
\bbS{ij} \, \bbC{ijk} \, \W{ijjk}
& \equiv & 
\frac{\hat\sigma_{ijjk(r)}}
{
\hat\sigma_{ijjk(r)} \!+\! \hat\sigma_{ikjk(r)} 
\!+\!
\hat\sigma_{jiik(r)} \!+\! \hat\sigma_{jkik(r)}
}
\, ,
\\
\label{eq:bSikCijkW}
\bbS{ik} \, \bbC{ijk} \, \W{ijkj}
& \equiv & 
\frac{\hat\sigma_{ijkj(r)}}
{
\hat\sigma_{ijkj(r)} + \hat\sigma_{ikkj(r)}
+
\hat\sigma_{kiij(r)} + \hat\sigma_{kjij(r)}
}
\, ;
\qquad
\eeq
\vspace{-3mm}
\beq
\label{eq:bCijkSCijkW}
\bbC{ijk}\,\bbSC{ijk} \, \W{ijjk}
& \equiv &
\frac{\frac{\sigma_{ij}^{(\alpha)}}{w_{ir}^{\alpha}}
\frac{\sigma_{jk}}{w_{jr}}}
{
\frac{\sigma_{ij}^{(\alpha)} + \sigma_{ik}^{(\alpha)}}{w_{ir}^{\alpha}}
\left( 
\frac{\sigma_{jk}}{w_{jr}} 
+ 
\frac{\sigma_{kj}}{w_{kr}} 
\right) 
+
\frac{\sigma_{jk}^{(\alpha)}}{w_{jr}^{\alpha}}
\frac{\sigma_{ik}}{w_{ir}}
+ 
\frac{\sigma_{kj}^{(\alpha)}}{w_{kr}^{\alpha}}
\frac{\sigma_{ij}}{w_{ir}}
}
\, ,
\\
\bbC{ijk}\,\bbSC{ijk} \, \W{ijkj}
& \equiv &
\frac{\frac{\sigma_{ij}^{(\alpha)}}{w_{ir}^{\alpha}}
\frac{\sigma_{kj}}{w_{kr}}}
{
\frac{\sigma_{ij}^{(\alpha)} + \sigma_{ik}^{(\alpha)}}{w_{ir}^{\alpha}}
\left( 
\frac{\sigma_{jk}}{w_{jr}} 
+ 
\frac{\sigma_{kj}}{w_{kr}} 
\right) 
+
\frac{\sigma_{jk}^{(\alpha)}}{w_{jr}^{\alpha}}
\frac{\sigma_{ik}}{w_{ir}}
+ 
\frac{\sigma_{kj}^{(\alpha)}}{w_{kr}^{\alpha}}
\frac{\sigma_{ij}}{w_{ir}}
}
\, ,
\\
\label{eq:bCijkSCkijW}
\bbC{ijk}\,\bbSC{kij} \, \W{ijkj}
& \equiv &
\frac{\frac{\sigma_{ij}^{(\alpha)}}{w_{ir}^{\alpha}}
\frac{\sigma_{kj}}{w_{kr}}}
{
\frac{\sigma_{kj}^{(\alpha)} + \sigma_{ki}^{(\alpha)}}{w_{kr}^{\alpha}} 
\left( 
\frac{\sigma_{ji}}{w_{jr}} 
+ 
\frac{\sigma_{ij}}{w_{ir}} 
\right) 
+
\frac{\sigma_{ji}^{(\alpha)}}{w_{jr}^{\alpha}}
\frac{\sigma_{ki}}{w_{kr}}
+ 
\frac{\sigma_{ij}^{(\alpha)}}{w_{ir}^{\alpha}}
\frac{\sigma_{kj}}{w_{kr}}
}
\, ;
\eeq
\vspace{-3mm}
\beq
\label{eq:bSijCijkSCijkW}
\bbS{ij} \, \bbC{ijk} \, \bbSC{ijk} \, \W{ijjk}
& \equiv & 
\frac{\frac{\sigma_{ij}^{(\alpha)}}{w_{ir}^{\alpha}}
\frac{\sigma_{jk}}{w_{jr}}}
{
\frac{\sigma_{ij}^{(\alpha)} + \sigma_{ik}^{(\alpha)}}{w_{ir}^{\alpha}}
\frac{\sigma_{jk}}{w_{jr}} 
+
\frac{\sigma_{jk}^{(\alpha)}}{w_{jr}^{\alpha}}
\frac{\sigma_{ik}}{w_{ir}}
}
\, ,
\\
\label{eq:bSikCijkSCijkW}
\bbS{ik} \, \bbC{ijk} \, \bbSC{ijk} \, \W{ijkj}
& \equiv & 
\frac{\frac{\sigma_{ij}^{(\alpha)}}{w_{ir}^{\alpha}}
\frac{\sigma_{kj}}{w_{kr}}}
{
\frac{\sigma_{ij}^{(\alpha)} + \sigma_{ik}^{(\alpha)}}{w_{ir}^{\alpha}}
\frac{\sigma_{kj}}{w_{kr}} 
+
\frac{\sigma_{kj}^{(\alpha)}}{w_{kr}^{\alpha}}
\frac{\sigma_{ij}}{w_{ir}}
}
\, ,
\\
\label{eq:bSikCijkSCkijW}
\bbS{ik} \, \bbC{ijk} \, \bbSC{kij} \, \W{ijkj}
& \equiv & 
\frac{\frac{\sigma_{ij}^{(\alpha)}}{w_{ir}^{\alpha}}
\frac{\sigma_{kj}}{w_{kr}}}
{
\frac{\sigma_{kj}^{(\alpha)} + \sigma_{ki}^{(\alpha)}}{w_{kr}^{\alpha}} 
\frac{\sigma_{ij}}{w_{ir}} 
+
\frac{\sigma_{ij}^{(\alpha)}}{w_{ir}^{\alpha}}
\frac{\sigma_{kj}}{w_{kr}}
}
\, .
\eeq
Finally the limits involving $\bbC{ijkl}$ are given by 
($j \ne i$, $k \ne i,j$, $l \ne i,j,k$ and $r = r_{ijkl} \ne i,j,k,l$) 
\beq
\label{eq:bCijklW}
\bbC{ijkl}\, \W{ijkl}
& \equiv &
\frac{\frac{\sigma_{ijkl}}{w_{ir}w_{kr}}}
{
\frac{\sigma_{ijkl} \!+\! \sigma_{klij}}{w_{ir}w_{kr}}
\!+\! 
\frac{\sigma_{ijlk} \!+\! \sigma_{lkij}}{w_{ir}w_{lr}}
\!+\! 
\frac{\sigma_{jikl} \!+\! \sigma_{klji}}{w_{jr}w_{kr}}
\!+\! 
\frac{\sigma_{jilk} \!+\! \sigma_{lkji}}{w_{jr}w_{lr}}
}
\, ,
\qquad\qquad
\eeq
\vspace{-3mm}
\beq
\label{eq:bSikCijklW}
\hspace{-30mm}
\bbS{ik} \, \bbC{ijkl} \, \W{ijkl}
& \equiv &
\frac{\sigma_{ij}^{(\alpha)}\sigma_{kl}}
{
\sigma_{ij}^{(\alpha)}\sigma_{kl}
+
\sigma_{kl}^{(\alpha)}\sigma_{ij}
}
\, ,
\qquad
\eeq
\vspace{-3mm}
\beq
\label{eq:bSCiklCijklW}
\bbSC{ikl}\,\bbC{ijkl} \, \W{ijkl}
& \equiv &
\frac{\sigma_{ij}^{(\alpha)}\frac{\sigma_{kl}}{w_{kr}}}
{
\sigma_{ij}^{(\alpha)} \!\!
\left(\!
\frac{\sigma_{kl}}{w_{kr}} \!+\! \frac{\sigma_{lk}}{w_{lr}} 
\!\right)
\!+\!
\left(\!
\frac{\sigma_{kl}^{(\alpha)}}{w_{kr}} \!+\! \frac{\sigma_{lk}^{(\alpha)}}{w_{lr}} 
\!\right)
\!
\sigma_{ij}
}
\, ,
\qquad\qquad
\eeq
\vspace{-3mm}
\beq
\label{eq:bSCkijCijklW}
\bbSC{kij}\,\bbC{ijkl} \, \W{ijkl}
& \equiv &
\frac{\frac{\sigma_{ij}^{(\alpha)}}{w_{ir}}\sigma_{kl}}
{
\sigma_{kl}^{(\alpha)} \!\!
\left(\!
\frac{\sigma_{ij}}{w_{ir}} \!+\! \frac{\sigma_{ji}}{w_{jr}} 
\!\right)
\!+\!
\left(\!
\frac{\sigma_{ij}^{(\alpha)}}{w_{ir}} \!+\! \frac{\sigma_{ji}^{(\alpha)}}{w_{jr}} 
\!\right)
\!
\sigma_{kl}
}
\, .
\qquad\qquad
\eeq


\subsection*{Strongly-ordered double-unresolved improved limits}

The improved limit $\bbS{i}\,\bbS{ik}$ is given by ($j \ne i$, $k \ne i$, 
$l \ne i,k$) 
\beq 
\bbS{i} \, \bbS{ik} \, \W{ijkl}
& \equiv &
\bW{\!{\rm s},\,kl} \,
\W{\!{\rm s},\,ij}^{(\alpha)}
\, .
\eeq
For $\bbS{i}\,\bbSC{ikl}$ and $\bbS{i}\,\bbS{ik}\,\bbSC{ikl}$ we have 
($j \ne i$, $k \ne i$, $l \ne i,k$, and $r = r_{ikl} \ne i,k,l$)  
\beq
\label{eq:bSiSCiklW}
\bbS{i}\,\bbSC{ikl} \, \W{ijkl}
& \equiv & 
\bW{\!{\rm c},\,kl(r)} \,
\W{\!{\rm s},\,ij}^{(\alpha)}
\, ,
\eeq
\vspace{-5mm}
\beq
\label{eq:bSiSikSCiklW}
\bbS{i} \, \bbS{ik} \, \bbSC{ikl} \, \W{ijkl}
& \equiv &
\W{\!{\rm s},\,ij}^{(\alpha)}
\, .
\eeq
For the strongly-ordered double-unresolved limits involving 
$\bbS{i}\,\bbC{ijk}$, we have ($j \ne i$, $k \ne i,j$, 
$r = r_{ijk} \ne i,j,k$ and $\tau = jk,kj$) 
\beq
\bbS{i}\,\bbC{ijk} (1\!-\!\bbSC{ijk}) \W{ij\tau}
& \equiv &
\bW{\!{\rm c},\tau (r)} \,
\frac{\sigma_{ij}^{(\alpha)}}{\sigma_{ij}^{(\alpha)} \!+\! \sigma_{ik}^{(\alpha)}}
\, ,
\eeq
\beq 
\bbS{i} \, \bbS{ij} \, \bbC{ijk} (1\!-\!\bbSC{ijk})  \W{ijjk}
& \equiv &
\frac{\sigma_{ij}^{(\alpha)}}{\sigma_{ij}^{(\alpha)} \!+\! \sigma_{ik}^{(\alpha)}}
\, ,
\eeq
\vspace{-5mm}
\beq
\bbS{i} \, \bbS{ik} \, \bbC{ijk} (1\!-\!\bbSC{ijk})  \W{ijkj}
& \equiv &
\frac{\sigma_{ij}^{(\alpha)}}{\sigma_{ij}^{(\alpha)} \!+\! \sigma_{ik}^{(\alpha)}}
\, .
\eeq
For $\bbC{ij}\,\bbSC{kij}$ and $\bbS{i}\,\bbC{ij}\,\bbSC{kij}$ we have 
($j \ne i$, $k \ne i$, $l \ne i,k$, and 
$r = r_{ijkl} \ne i,j,k,l$
) 
\beq
\label{eq:bCijSCkijW}
\bbC{ij} \, \bbSC{kij} \, \W{ijkl}
& \equiv & 
\W{\!{\rm c},\,ij(r)}^{(\alpha)} \,
\bW{\!{\rm s},\,kl} 
\, ;
\\
\label{eq:bSiCijSCkijW}
\bbS{i} \, \bbC{ij} \, \bbSC{kij} \, \W{ijkl}
& \equiv & 
\bW{\!{\rm s},\,kl}
\, .
\eeq
The improved limits $\bbC{ij}\,\bbS{ij}\,\RR\,\W{ijjk}$ and 
$\bbS{i}\,\bbC{ij}\,\bbS{ij}\,\RR\,\W{ijjk}$ read ($j \ne i$, 
$k \ne i,j$ and $r = r_{ijk} \ne i,j,k$) 
\beq
\bbC{ij} \, \bbS{ij} \, \W{ijjk}
& \equiv &
\W{\!{\rm c},\,ij(r)}^{(\alpha)} \,
\bW{\!{\rm s},\,jk}
\, ;
\\
\bbS{i} \, \bbC{ij} \, \bbS{ij} \, \W{ijjk}
& \equiv &
\bW{\!{\rm s},\,jk}
\, .
\eeq
For the strongly-ordered double-unresolved limits involving 
$\bbC{ij}\,\bbC{ijk}$, we have ($j \ne i$, $k \ne i,j$, 
$r = r_{ijk} \ne i,j,k$, and $\tau = jk,kj$) 
\beq
\label{eq:bCijCijkW}
\bbC{ij} \, \bbC{ijk} \, \W{ij\tau}
& \equiv & 
\W{\!{\rm c},\,ij(r)}^{(\alpha)} \,
\bW{\!{\rm c},\,\tau(r)}
\, ;
\\
\label{eq:bSiCijCijkW}
\bbS{i} \, \bbC{ij} \, \bbC{ijk} \, \W{ij\tau}
& \equiv & 
\bW{\!{\rm c},\,\tau(r)}
\, ;
\eeq
\beq
\label{eq:bCijSijCijkW}
\bbC{ij} \, \bbS{ij} \, \bbC{ijk} \, \W{ijjk}
& \equiv &
\W{\!{\rm c},\,ij(r)}^{(\alpha)}
\, ;
\qquad\quad
\\
\label{eq:bSiCijSijCijkW}
\bbS{i} \, \bbC{ij} \, \bbS{ij} \, \bbC{ijk}\, \W{ijjk}
& \equiv & 
1
\, ;
\eeq
\vspace{-5mm}
\beq
\label{eq:bCijCijkSCkijW}
\bbC{ij} \, \bbC{ijk} \, \bbSC{kij} \, \W{ijkj}
& \equiv &
\W{\!{\rm c},\,ij(r)}^{(\alpha)}
\, ;
\\
\label{eq:bSiCijCijkSCkijW}
\bbS{i}\, \bbC{ij}\, \bbC{ijk}\, \bbSC{kij}\, \W{ijkj}
& \equiv & 
1
\, .
\eeq
Finally the limits involving $\bbC{ij}\,\bbC{ijkl}$ are given by 
($j \ne i$, $k \ne i,j$, $l \ne i,j,k$ and 
$r = r_{ijkl} \ne i,j,k,l$) 
\beq
\label{eq:bCijCijklW}
\bbC{ij}\, \bbC{ijkl}\, \W{ijkl}
& \equiv &
\W{\!{\rm c},\,ij(r)}^{(\alpha)} \,
\bW{\!{\rm c},\,kl(r)}
\, ;
\\
\label{eq:bSiCijCijklW}
\bbS{i}\, \bbC{ij}\, \bbC{ijkl}\, \W{ijkl}
& \equiv & 
\bW{\!{\rm c},\,kl(r)}
\, ;
\\
\label{eq:bCijSCkijCijklW}
\bbC{ij}\, \bbSC{kij}\, \bbC{ijkl}\, \W{ijkl}
& \equiv &
\W{\!{\rm c},\,ij(r)}^{(\alpha)}
\, ;
\\
\label{eq:bSiCijSCkijCijklW}
\bbS{i}\, \bbC{ij}\, \bbSC{kij}\, \bbC{ijkl}\, \W{ijkl}
& \equiv &
1
\, .
\eeq


\subsection{Improved limits of $\Z{ijk}$, $\Z{ijkl}$}
\label{app:LZ}


\subsubsection*{Single-unresolved improved limits}

For the single-unresolved improved limits in $K^{\one}_{\{ijk\}}$ we have ($j \ne i$, $k \ne i, j$)
\beq
\label{eq:bL1 Zijk}
\bbS{i} \, \Z{ijk}
& = &
\bZ{jk}
\Big( \Z{{\rm s},\,ij}^{(\alpha)} + \Z{{\rm s},\,ik}^{(\alpha)} \Big)
\, ,
\qquad
\bbHC{ij} \, \Z{ijk}
\; = \;
\bZ{jk}
\, ;
\eeq
while for $K^{\one}_{\{ijkl\}}$ we have ($j \ne i$, $k \ne i, j$)
\beq
\label{eq:bL1 Zijkl}
\bbS{i} \, \Z{ijkl}
& = &
\bZ{kl} \,
\Z{{\rm s},\,ij}^{(\alpha)}
\, ,
\qquad
\bbHC{ij} \, \Z{ijkl}
\; = \;
\bZ{kl}
\, .
\eeq


\subsection*{Uniform double-unresolved improved limits}

For $K^{\two}_{\{ijk\}}$ we have ($j \ne i$, $k \ne i, j$, and 
$r = r_{ijk} \ne i,j,k$)
\beq
\label{eq:bL2 Zijk}
\bbS{ik} \, \Z{ijk}
& = &
\frac{\sigma_{ikkj}+\sigma_{ijkj}+\sigma_{kiij}+\sigma_{kjij}}
{
\sum_{b \neq i}\sum_{d \neq i,k} \sigma_{ibkd}
+ 
\sum_{b\neq k}\sum_{d\neq k,i} \sigma_{kbid}
}
\, ,
\\
\bbSC{ijk} \, \Z{ijk}
& = &
\frac{
\left( \sigma_{ij}^{(\alpha)} + \sigma_{ik}^{(\alpha)} \right)
\left( \frac{\sigma_{jk}}{w_{jr}} + \frac{\sigma_{kj}}{w_{kr}} \right)
+
\frac{\sigma_{jk}^{(\alpha)}}{w_{jr}} 
\sigma_{ik}
+ 
\frac{\sigma_{kj}^{(\alpha)}}{w_{kr}}
\sigma_{ij}
}
{
\sum_{b \neq i} \sigma_{ib}^{(\alpha)} 
\left( \frac{\sigma_{jk}}{w_{jr}} + \frac{\sigma_{kj}}{w_{kr}} \right)
+
\frac{\sigma_{jk}^{(\alpha)}}{w_{jr}}\sum_{d\neq i,j} \sigma_{id}
+
\frac{\sigma_{kj}^{(\alpha)}}{w_{kr}}\sum_{d\neq i,k} \sigma_{id}
}
\, ,
\nnb\\
\bbS{ij} \, \bbSC{ijk} \, \Z{ijk}
& = &
\frac{
\left( \sigma_{ij}^{(\alpha)} + \sigma_{ik}^{(\alpha)} \right)\sigma_{jk}
+
\sigma_{jk}^{(\alpha)}\sigma_{ik}
}
{
\sum_{b \neq i} \sigma_{ib}^{(\alpha)} 
\sigma_{jk}
+
\sigma_{jk}^{(\alpha)}\sum_{d\neq i,j} \sigma_{id}
}
\, ,
\nnb\\
\bbHC{ijk} \, \Z{ijk}
& = &
1
\, ,
\nnb\\
\bbC{ijk} \, \bbSHC{ijk} \, \Z{ijk}
& = &
1
\, .
\nnb
\eeq
For
$K^{\two}_{\{ijkl\}}$ one has ($j \ne i$, $k \ne i, j$, $l \ne i,j,k$, and 
$r = r_{ikl} \ne i,k,l$)
\beq
\label{eq:bL2 Zijkl}
\bbS{ik} \, \Z{ijkl}
& = &
\frac{\sigma_{ijkl}+\sigma_{klij}}
{
\sum_{b \neq i}\sum_{d \neq i,k} \sigma_{ibkd}
+ 
\sum_{b\neq k}\sum_{d\neq k,i} \sigma_{kbid}
}
\, ,
\\
\bbSC{ikl} \, \Z{ijkl}
& = &
\frac{
\sigma_{ij}^{(\alpha)}
\left( \frac{\sigma_{kl}}{w_{kr}} + \frac{\sigma_{lk}}{w_{lr}} \right)
+
\left( 
\frac{\sigma_{kl}^{(\alpha)}}{w_{kr}} + \frac{\sigma_{lk}^{(\alpha)}}{w_{lr}} 
\right)
\sigma_{ij}
}
{
\sum_{b \neq i} \sigma_{ib}^{(\alpha)} 
\left( \frac{\sigma_{kl}}{w_{kr}} + \frac{\sigma_{lk}}{w_{lr}} \right)
+
\frac{\sigma_{kl}^{(\alpha)}}{w_{kr}}\sum_{d\neq i,k} \sigma_{id}
+
\frac{\sigma_{lk}^{(\alpha)}}{w_{lr}}\sum_{d\neq i,l} \sigma_{id}
}
\, ,
\nnb\\
\bbS{ik} \, \bbSC{ikl} \, \Z{ijkl}
& = &
\frac{
\sigma_{ij}^{(\alpha)}\sigma_{kl}
+
\sigma_{kl}^{(\alpha)}\sigma_{ij}
}
{
\sum_{b \neq i} \sigma_{ib}^{(\alpha)} 
\sigma_{kl}
+
\sigma_{kl}^{(\alpha)}\sum_{d\neq i,k} \sigma_{id}
}
\, ,
\nnb\\
\bbHC{ijkl} \, \Z{ijkl}
& = &
1
\, .
\nnb
\eeq


\subsection*{Strongly-ordered double-unresolved improved limits}

For $K^{\otwo}_{\{ijk\}}$ one has ($j \ne i$, $k \ne i,j$)
\beq
\label{eq:bL12 Zijk}
\bbS{i} \, \bbS{ij} \, \Z{ijk}
& = &
\bZ{{\rm s},\,jk} \, 
\Big( \Z{{\rm s},\,ij}^{(\alpha)} + \Z{{\rm s},\,ik}^{(\alpha)} \Big)
\, ,
\\
\bbS{i} \, \bbSHC{ijk} \, \Z{ijk}
& = &
\Z{{\rm s},\,ij}^{(\alpha)} + \Z{{\rm s},\,ik}^{(\alpha)} 
\, ,
\nnb\\
\bbS{i} \, \bbHC{ijk}^{\;(\bf s)} \, \Z{ijk}
& = &
1
\, ,
\nnb\\
\bbHC{ij} \, \bbS{ij} \, \Z{ijk}
& = &
\bZ{{\rm s},\,jk}
\, ,
\nnb\\
\bbHC{ij} \, \bbSC{kij} \, \Z{ijk}
& = &
\bZ{{\rm s},\,kj}
\, ,
\nnb\\
\bbHC{ij} \, \bbHC{ijk}^{\;(\bf c)} \,  \Z{ijk} 
& = &
1
\, .
\nnb
\eeq
For $K^{\otwo}_{\{ijkl\}}$ one has ($j \ne i$, $k \ne i,j$, $l \neq i,j,k$)
\beq
\label{eq:bL12 Zijkl}
\bbS{i} \, \bbS{ik} \, \Z{ijkl}
& = &
\bZ{{\rm s},\,kl} \, \Z{{\rm s},\,ij}^{(\alpha)}
\, ,
\\
\bbS{i} \, \bbSHC{ikl} \, \Z{ijkl}
& = &
\Z{{\rm s},\,ij}^{(\alpha)}
\, ,
\nnb\\
\bbHC{ij} \, \bbSC{kij} \, \Z{ijkl}
& = &
\bZ{{\rm s},\,kl}
\, ,
\nnb\\
\bbHC{ij} \, \bbHC{ijkl}^{\;(\bf c)} \, \Z{ijkl}
& = &
1
\, .
\nnb
\eeq


\section{Integration of azimuthal contributions}
\label{app:int Q}

The azimuthal parts of the collinear kernels $Q_{ij(r)}^{\mu\nu}$, 
$\tilde Q_{ij(r)}^{\mu\nu}$ and $Q_{ijk(r)}^{\mu\nu}$, defined in 
\appn{app:kernels}, contain $\kt_{a}^\mu \kt_{a}^\nu$, where 
$a=i$ for $P_{ij(r)}^{\mu\nu}$, $\tilde P_{ij(r)}^{\mu\nu}$ and $a=i,j,k$ 
for $P_{ijk(r)}^{\mu\nu}$.
In all counterterms, $Q_{ij(r)}^{\mu\nu}$ has to be integrated in the 
single-radiative phase space $d\Phi_{\rm rad}^{(ijr)}$, 
$d\Phi_{\rm rad}^{(irj)}$ or $d\Phi_{\rm rad}^{(jri)}$, while 
$\tilde Q_{ij(r)}^{\mu\nu}$ and $Q_{ijk(r)}^{\mu\nu}$ are always 
integrated in $d\Phi_{\rm rad}^{(ijr)}$ and $d\Phi_{\rm rad,2}^{(ijkr)}$, 
respectively. 
In all cases, when integrating $Q_{ij(r)}^{\mu\nu}$ and 
$\tilde Q_{ij(r)}^{\mu\nu}$ in their single-radiative phase space, or 
$Q_{ijk(r)}^{\mu\nu}$ in its double-radiative phase space, the integral 
of the tensor structure $\kt_{a}^\mu\kt_{a}^\nu$ must be a symmetric rank-2 
tensor constructed combining $g^{\mu\nu}$ and mapped momenta, see \cite{Catani:1996jh}.
Thus
\beq
\int d\Phi_{\rm rad}^{(\tau)} \,
f(\{k\}) \, \kt_{a}^\mu \kt_{a}^\nu 
=
A \, g^{\mu\nu} 
+ 
B \, \kb^{(\tau)\mu}\kb^{(\tau)\nu}
+ 
C 
\Big( 
\kb^{(\tau)\mu}\kb_{q}^{(\tau)\nu} + \kb_{q}^{(\tau)\mu}\kb^{(\tau)\nu} 
\Big)
+ 
D \, \kb_{q}^{(\tau)\mu}\kb_{q}^{(\tau)\nu}
\, ,
\eeq
where $\tau = ijr,irj,jri,ijkr$, $q=r$ if $\tau = ijr,irj,jri$, $q=r$ if $\tau = ijkr$, and 
\beq
\kb^{(ijr)} = \kk{j}{ijr}
\, ,
\qquad
\kb^{(irj)} = \kk{j}{irj}
\, ,
\qquad
\kb^{(jri)} = \kk{i}{jri}
\, ,
\qquad
\kb^{(ijkr)} = \kk{k}{ijkr}
\, .
\eeq
Since $\kt_{a}$ is orthogonal to $\kb^{(\tau)\mu}$ and 
$\kb_{q}^{(\tau)\mu}$, so must be also its integrals. 
This leads to the conditions $D=0$ and 
$A + C\,\kb^{(\tau)}\dt\kb_{q}^{(\tau)} = 0$. 
We have 
\beq
\int d\Phi_{\rm rad}^{(\tau)} \,
f(\{k\}) \, \kt_{a}^\mu \kt_{a}^\nu 
=
A \left[
g^{\mu\nu} 
- 
\frac{\kb^{(\tau)\mu}\kb_{q}^{(\tau)\nu} + \kb_{q}^{(\tau)\mu}\kb^{(\tau)\nu}}{\kb^{(\tau)}\dt\kb_{q}^{(\tau)}}
\right]
+
B \, \kb^{(\tau)\mu}\kb^{(\tau)\nu}
\, .
\eeq
In all counterterms this tensor is contracted with either 
\beq
\bR^{(\tau)}_{\mu\nu} \, ,
\qquad
\bB^{(\tau)}_{\mu\nu} \, ,
\qquad
\bB^{(\tau,\dots)}_{\mu\nu} \, ,
\qquad\mbox{or}\qquad
\Bigg[
\frac{\kk{c,\mu}{\tau}}{\sk{jc}{\tau}}
-
\frac{\kk{d,\mu}{\tau}}{\sk{jd}{\tau}}
\Bigg]
\Bigg[
\frac{\kk{c,\nu}{\tau}}{\sk{jc}{\tau}}
-
\frac{\kk{d,\nu}{\tau}}{\sk{jd}{\tau}}
\Bigg]
\, .
\eeq 
As a consequence, the terms proportional to $\kb^{(\tau)\mu}$ or to 
$\kb^{(\tau)\nu}$ vanish, and just $A\,g^{\mu\nu}$ contributes. 
On the other hand, since $\kb^{(\tau)}$ is on shell, $A$ can be obtained as follows:
\beq
g_{\mu\nu}
\int d\Phi_{\rm rad}^{(\tau)} \,
f(\{k\}) \, \kt_{a}^\mu \kt_{a}^\nu 
\; = \;
A \, (d - 2)
\quad\Longrightarrow\quad
A 
\; = \;
\frac{1}{d-2}
\int d\Phi_{\rm rad}^{(\tau)} \,
f(\{k\}) \, \kt_{a}^{2}
\, .
\eeq
Thus in all counterterms we can subtitute 
\beq
\int d\Phi_{\rm rad}^{(\tau)} \,
f(\{k\}) \, \kt_{a}^\mu \kt_{a}^\nu 
\; \to \;
A \, g^{\mu\nu}
\; = \;
\int d\Phi_{\rm rad}^{(\tau)} \,
f(\{k\}) \, 
\frac{g^{\mu\nu}}{d-2} \,
\kt_{a}^{2}
\, ,
\eeq
and the integrals of $Q_{ij(r)}^{\mu\nu}$, 
$\tilde Q_{ij(r)}^{\mu\nu}$ and $Q_{ijk(r)}^{\mu\nu}$ vanish in all counterterms: 
\beq
\int d\Phi_{\rm rad}^{(\tau)} \,
\frac{Q_{ij(r)}^{\mu\nu}}{s_{ij}} 
& = &
\int d\Phi_{\rm rad}^{(\tau)} \,
\frac{Q_{ij(r)}}{s_{ij}}
\bigg[
- \, g^{\mu\nu} + (d - 2) \, \frac{\kt_{i}^\mu \kt_{i}^\nu}{\kt_{i}^2} 
\bigg]
\; \to \;
0
\, ,
\qquad
\tau = ijr,irj,jri
\, ;
\nnb\\
\int d\Phi_{\rm rad}^{(\tau)} \,
\frac{\tilde Q_{ij(r)}^{\mu\nu}}{s_{ij}} 
& = &
\int d\Phi_{\rm rad}^{(\tau)} \,
\frac{\tilde Q_{ij(r)}}{s_{ij}}
\bigg[
- \, g^{\mu\nu} + (d - 2) \, \frac{\kt_{i}^\mu \kt_{i}^\nu}{\kt_{i}^2} 
\bigg]
\; \to \;
0
\, ,
\qquad
\tau = ijr
\, ;
\nnb\\
\int d\Phi_{\rm rad,2}^{(\tau)} \,
\frac{Q_{ijk(r)}^{\mu\nu}}{s_{ijk}^2} 
& = &
\sum_{a=i,j,k}
\int d\Phi_{\rm rad,2}^{(\tau)} \,
\frac{Q_{ijk(r)}^{(a)}}{s_{ijk}^2}
\bigg[
- \, g^{\mu\nu} + (d - 2) \, \frac{\kt_{a}^\mu \kt_{a}^\nu}{\kt_{a}^2} 
\bigg]
\; \to \;
0
\, ,
\qquad
\tau = ijkr
\, .
\qquad
\eeq


\section{Constituent integrals}
\label{app:master integrals}

In the following we report the constituent integrals relevant for the analytic integration of all 
counterterms at NNLO. Such integrals are schematically denoted as $J_t^\ell$, where $t$ 
indicates the type of integral, while $\ell$ is a set of labels whose different indices denote 
distinguished particles.

\noindent
The soft integrated kernel is 
\beq
J_{\rm s}^{ilm}
& \; \equiv \; &
\Norm \,
\int d\Phi_{\rm rad}^{(ilm)}\,
\mc E_{lm}^{(i)}
\; \equiv \;
\delta_{f_i g} \,J_s(\sk{lm}{ilm}) 
\, ,
\label{eq:Js^ilm}
\eeq
with 
\beq
\label{eq:Js}
J_{\rm s}(s) 
& = &
\frac{\as}{2\pi}
\left( \frac{s}{e^{\euler\!}\mu^2} \right)^{\!\!-\eps} 
\frac{\Gam(1 - \eps) \Gam(2 - \eps)}{\eps^2 \, \Gam(2 - 3 \eps)}
\nnb \\
& = &
\frac{\as}{2\pi}
\left(\! \frac{s}{\mu^2} \!\right)^{\!\!\!-\eps} 
\bigg[ \;\;
\frac1{\eps^2}
+
\frac2{\eps}
+
6
-
\frac{7}{12}\pi^2
+ 
\bigg( 18 - \frac{7}{6}\pi^2 - \frac{25}{3}\zeta_3 \bigg) \eps
\nnb\\ 
&& 
\hspace{18mm} 
+ \,
\bigg(
54 - \frac{7}{2}\pi^2 - \frac{50}{3}\zeta_3 - \frac{71}{1440}\pi^4 
\bigg) 
\eps^2
+
\mc O(\eps^3)
\bigg]
\, .
\eeq
The double-soft integrated kernels read 
\beq
\label{doublesoftint}
J_{\rm s \otimes s}^{ijcdef}
& \equiv &
\Norm^{\,2}
\int d\Phi_{\rm rad,2}^{(icd,jef)}\,
\mc E_{cd}^{(i)} \, \mc E_{ef}^{(j)}
\, \equiv \,
J_{\rm s \otimes s}^{(4)} \Big(\sk{cd}{icd,jef},\sk{ef}{icd,jef} \Big) \,
f_{ij}^{gg}
\, ,
\nnb \\
J_{\rm s \otimes s}^{ijcde}
& \equiv &
\Norm^{\,2}
\int d\Phi_{\rm rad,2}^{(icd,jed)}\,
\mc E_{cd}^{(i)} \, \mc E_{ed}^{(j)}
\, \equiv \, 
J_{\rm s \otimes s}^{(3)} \Big(\sk{cd}{icd,jed},\sk{ed}{icd,jed} \Big) \,
f_{ij}^{gg}
\, ,
\nnb \\
J_{\rm s \otimes s}^{ijcd}
& \equiv &
\Norm^{\,2}
\int d\Phi_{\rm rad,2}^{(ijcd)}\,
\mc E_{cd}^{(i)} \, \mc E_{cd}^{(j)}
\, \equiv \, 
J_{\rm s \otimes s}^{(2)} \Big(\sk{cd}{ijcd} \Big) \,
f_{ij}^{gg}
\, ,
\nnb\\
J_{\rm ss}^{ijcd}
& \equiv &
\Norm^{\,2}
\int d\Phi_{\rm rad,2}^{(ijcd)}\,
\mc E_{cd}^{(ij)}
\, \equiv \, 
2\,T_R\,
J_{\rm ss}^{(\rm q \bar q)}\Big(\sk{cd}{ijcd}\Big) \,
f_{ij}^{q \bar q}
-
2\,C_A\,
J_{\rm ss}^{(\rm gg)}\Big(\sk{cd}{ijcd}\Big) \,
f_{ij}^{gg}
\, ,
\eeq
with
\beq
\label{varJscrosss}
J_{\rm s \otimes s}^{(4)}(s, s')
& = &
\!\bigg(\! \frac{\as}{2\pi} \!\bigg)^{\!\!2}
\left(\! \frac{ss'}{\mu^4} \!\right)^{\!\!\!-\eps}
\bigg[ \;\;
\frac1{\eps^4}
+ 
\frac4{\eps^3}
+
\bigg( 16 - \frac{7}{6}\pi^2 \bigg) \frac1{\eps^2}
+
\bigg( 60 - \frac{14}{3}\pi^2 - \frac{50}{3}\zeta_3 \bigg) 
\frac1{\eps}
\\
&& 
\hspace{24mm}
+ \, 
216
-
\frac{56}{3}\pi^2 
-
\frac{200}{3}\zeta_3 
+
\frac{29}{120}\pi^4
\!
+
\mc O (\eps)
\bigg] 
\, ,
\nnb\\
J_{\rm s \otimes s}^{(3)}(s, s')
& = &
\!\bigg(\! \frac{\as}{2\pi} \!\bigg)^{\!\!2}
\left(\! \frac{ss'}{\mu^4} \!\right)^{\!\!\!-\eps}
\bigg[ \;\;
\frac1{\eps^4}
+ 
\frac4{\eps^3}
+
\!\left(\! 17 - \frac{4}{3}\pi^2 \!\right) \!\! \frac1{\eps^2}
+ 
\!\left(\! 70 - \frac{16}{3}\pi^2 - \frac{68}{3}\zeta_3 \!\right) \!\!
\frac1{\eps} 
\nnb \\
&&
\hspace{24mm}
+ \,  
284
-
\frac{68}{3}\pi^2 
-
\frac{272}{3}\zeta_3
+
\frac{13}{90}\pi^4
+
\mc O (\eps)
\bigg] 
\, ,
\nnb\\
J_{\rm s \otimes s}^{(2)}(s)
& = & 
\bigg(\! \frac{\as}{2\pi} \!\bigg)^{\!\!2} \!\!
\left(\! \frac{s}{\mu^2} \!\right)^{\!\!\!-2\eps} \! 
\bigg[ \;\;
\frac1{\eps^4}
+ 
\frac4{\eps^3}
+
\bigg( \! 18 - \frac{3}{2}\pi^2 \! \bigg) \frac1{\eps^2}
+
\bigg( \!76 - 6\pi^2 - \frac{74}{3}\zeta_3 \!\bigg) \frac1{\eps}
\nnb\\
&&
\hspace{24mm}
+ \,
312
-
27\pi^2 
-
\frac{308}{3}\zeta_3 
+
\frac{49}{120}\pi^4
\!
+
\mc O (\eps)
\bigg]
\, ,
\nnb \\
J_{\rm ss}^{({\rm q\bar q})}(s)
& = & 
\bigg(\! \frac{\as}{2\pi} \!\bigg)^{\!\!2} \!\!
\left(\! \frac{s}{\mu^2} \!\right)^{\!\!\!-2\eps} \!
\bigg[
\;
\frac{1}{6}  \, \frac1{\eps^3}
+ 
\frac{17}{18} \, \frac1{\eps^2}
+ 
\left( \frac{116}{27} - \frac{7}{36}\,\pi^2 \right) \frac1{\eps}
+ 
\frac{1474}{81} 
- 
\frac{131}{108}\,\pi^2 
- 
\frac{19}{9}\,\zeta_3
+
\mc O (\eps)
\bigg]
\, ,
\nnb\\
J_{\rm ss}^{({\rm gg})}(s)
& = &
\bigg(\! \frac{\as}{2\pi} \!\bigg)^{\!\!2} \!\!
\left(\! \frac{s}{\mu^2} \!\right)^{\!\!\!-2\eps} \!
\bigg[
\;\;
\frac12 \, \frac1{\eps^4}
+ 
\frac{35}{12} \, \frac1{\eps^3}
+
\left( \frac{487}{36} - \frac{2}{3}\,\pi^2 \right) \frac1{\eps^2}
+
\left( \frac{1562}{27} - \frac{269}{72}\,\pi^2 - \frac{77}{6}\,\zeta_3 \right) 
\frac1{\eps}
\nnb\\
&&
\hspace{24mm}
+ \,
\frac{19351}{81} 
-
\frac{3829}{216}\,\pi^2 
-
\frac{1025}{18}\,\zeta_3 
-
\frac{23}{240}\,\pi^4
+
\mc O (\eps)
\;
\bigg]
\, .
\nnb
\eeq
The soft real-virtual integrated kernels are 
\beq
\label{softRVintker}
\tilde J_{\rm s}^{icd}
& \equiv &
\Norm \!
\int \! d\Phi_{\rm rad}^{(icd)}\,
\tilde{\mc E}_{cd}^{(i)}
\, \equiv \,
\delta_{f_i g} \,
C_{\!A} \,
\tilde J_{\rm s} \Big(\sk{cd}{icd} \Big) 
\, ,
\nnb\\
J_{\!\!_{\Delta \rm s}}^{icd(e)}
& \equiv &
\Norm
\frac{2}{\eps^{2}}
\int d\Phi_{\rm rad}^{(icd)} \,
\mc E^{(i)}_{cd} \,
\bigg[ \bigg(\frac{s_{ed}}{\sk{ed}{icd}}\bigg)^{\!\!-\eps} - 1 \bigg] 
\, \equiv \,
f_{i}^{g} \,
J_{\!\!_{\Delta \rm s}}^{(3)}\Big(\sk{cd}{icd} \Big)
\, ,
\nnb\\
J_{\!\!_{\Delta \rm s}}^{icd}
& \equiv &
\Norm \,
\frac{1}{\eps^{2}}
\int d\Phi_{\rm rad}^{(icd)} \,
\mc E^{(i)}_{cd} \,
\bigg[ \bigg(\frac{s_{cd}}{\sk{cd}{icd}}\bigg)^{\!\!-\eps} - 1 \bigg]
\, \equiv \,
f_{i}^{g} \,
J_{\!\!_{\Delta \rm s}}^{(2)}\Big(\sk{cd}{icd} \Big)
\, ,
\nnb\\
\tilde J_{\rm s}^{\, icde}
& \equiv &
\Norm \,
\int d\Phi_{\rm rad}^{(icd)}\,
\tilde{\mc E}_{cde}^{(i)}
\, ,
\eeq
with 
\beq
\label{somesoftintrv}
\tilde J_{\rm s}(s) 
& = &
\frac{\as}{2\pi}
\left( \frac{s}{e^{\euler\!}\mu^2} \right)^{\!\!-2\eps} 
\frac{\Gamma^3(1+\eps)\Gamma^3(1-\eps)}
     {4\,\eps^4\,\Gamma(1+2\eps)\Gam(2 - 4 \eps)}
\\
& = &
\frac{\as}{2\pi}
\left( \frac{s}{\mu^2} \right)^{\!\!-2\eps} 
\bigg[ \;\;
\frac1{ 4 \eps^4}
+
\frac1{\eps^3}
+
\!\left(\! 4 - \frac{7}{24}\pi^2 \!\right)\!\!\frac1{\eps^2}
+
\!\left(\! 16 - \frac{7}{6}\pi^2 - \frac{14}{3}\zeta_3 \!\right) \!
\frac1{\eps}
\nnb \\
&& 
\hspace{20mm} 
+ \, 
64
-
\frac{14}{3}\pi^2 
-
\frac{56}{3}\zeta_3 
-
\frac{7}{480}\pi^4
+
\mc O(\eps)
\bigg]
\, ,
\nnb\\
J_{\!\!_{\Delta \rm s}}^{(3)}(s)
& = &
\frac{\as}{2\pi}
\left( \frac{s}{\mu^2} \right)^{\!\!-\eps} \!
\bigg[
\bigg( \! 2 - \frac{\pi^2}{3} \bigg) \frac1{\eps^2}
+
\bigg( \! 16 - \frac{2}{3}\pi^2 - 12\,\zeta_3 \!\bigg) \frac1{\eps}
+
92
-
\frac{7}{2}\pi^2 
-
24\,\zeta_3 
-
\frac{7}{18}\pi^4
+
\mc O (\eps)
\bigg]
\, ,
\nnb\\
J_{\!\!_{\Delta \rm s}}^{(2)}(s)
& = & 
\frac{\as}{2\pi} \!
\left( \frac{s}{\mu^2} \right)^{\!\!\!-\eps} \!
\bigg[
\bigg( \! 2 - \frac{\pi^2}{3} \bigg) \frac1{\eps^2}
+
\bigg( \! 14 - \frac{2}{3}\pi^2 - 10\,\zeta_3 \!\bigg) \frac1{\eps}
+
74
-
\frac{23}{6}\pi^2 
-
20\,\zeta_3 
-
\frac{7}{36}\pi^4
+
\mc O (\eps)
\bigg]
\, ,
\nnb
\eeq
\beq
\sum_{\substack{c \neq i, d\neq i,c \\ e \neq i,c,d}} \!\!\!\!
\tilde J_{\rm s}^{\, icde} \, 
B_{cde}
& = &
- \,
f_{i}^{g} \,
\frac{\as}{2\pi}
\sum_{\substack{c \neq i, d\neq i,c \\ e \neq i,c,d}} \!\!\!\!
B_{cde}
\bigg[
\frac{1}{2}
\ln\!\frac{\bar{s}_{ce}}{\bar{s}_{de}}
\ln^2\!\frac{\bar{s}_{cd}}{\mu^2}
+
\frac{1}{6}\ln^3\!\frac{\bar{s}_{ce}}{\bar{s}_{de}}
+
\Li_3\!\left(\!-\frac{\bar{s}_{ce}}{\bar{s}_{de}}\!\right)
+
\mc O(\eps)
\bigg]
\, .
\nnb
\eeq
The hard-collinear integrated kernels are given by
\beq
\label{Jhc^ijr}
J_{\rm hc}^{ijr}
& \equiv &
\Norm 
\int d\Phi_{\rm rad}^{(ijr)} \,
\frac{P_{ij(r)}^{\rm hc}}{s_{ij}} 
\nnb\\
& \equiv &
J_{\rm hc}^{\zg} \! \left(\sk{jr}{ijr}\right) \,
f_{ij}^{q \bar q}
+ 
J_{\rm hc}^{\og} \! \left(\sk{jr}{ijr}\right) \,
( f_{ij}^{g q} + f_{ij}^{g \bar q} )
+ 
J_{\rm hc}^{\tg} \! \left(\sk{jr}{ijr}\right) \,
f_{ij}^{gg}
\, ,
\eeq
where 
\beq
\label{eq:Jhc}
J_{\rm hc}^{\zg}(s)
& = & 
\frac{\as}{2\pi}
\left( \frac{s}{e^{\euler\!}\mu^2} \right)^{\!\!-\eps} 
\frac{\Gam(1-\eps)\Gam(2-\eps)}{\eps\,\Gam(2 - 3\eps)} \,
T_R \,
\frac{-2}{3-2\eps}
\nnb \\
& = &
\frac{\as}{2\pi}
\left( \frac{s}{\mu^2} \right)^{\! -\eps} \!
T_R 
\bigg[
-
\frac23\,\frac1{\eps}
-
\frac{16}{9}
-
\left(\! \frac{140}{27} - \frac{7}{18}\pi^2 \!\right) \eps
- 
\bigg( \frac{1252}{81} - \frac{28}{27}\pi^2 - \frac{50}{9}\zeta_3 \bigg) 
\eps^2
+
\mc O(\eps^3)
\bigg]
\, ,
\nnb\\
\label{eq:Jhc1g}
J_{\rm hc}^{\og}(s)
& = & 
\frac{\as}{2\pi}
\left( \frac{s}{e^{\euler\!}\mu^2} \right)^{\!\!-\eps} 
\frac{\Gam(1-\eps)\Gam(2-\eps)}{\eps\,\Gam(2-3\eps)} \,
C_F \,
\bigg(
-
\frac{1}{2}
\bigg)
\nnb \\
& = &
\frac{\as}{2\pi}
\left( \frac{s}{\mu^2} \right)^{\!\!-\eps} 
C_F 
\bigg[
-
\frac12\,\frac1{\eps}
-
1
-
\left(\! 3 - \frac{7}{24}\pi^2 \!\right) \eps
- 
\bigg( 9 - \frac{7}{12}\pi^2 - \frac{25}{6}\zeta_3 \bigg) 
\eps^2
+
\mc O(\eps^3)
\bigg]
\, ,
\nnb\\
\label{eq:Jhc2g}
J_{\rm hc}^{\tg}(s)
& = &
\frac{\as}{2\pi}
\left( \frac{s}{e^{\euler\!}\mu^2} \right)^{\!\!-\eps} 
\frac{\Gam(1-\eps)\Gam(2-\eps)}{\eps \,\Gam(2-3\eps)} \,
C_A \,
\bigg(
-
\frac{1}{3-2\eps}
\bigg)
\\
& = &
\frac{\as}{2\pi} \!
\left( \frac{s}{\mu^2} \right)^{\!\!-\eps} \!\!
C_A 
\bigg[
-
\frac{1}{3}\,\frac1{\eps}
-
\frac{8}{9}
-
\left(\! \frac{70}{27} - \frac{7}{36}\pi^2 \!\right) \eps
- 
\bigg( \frac{626}{81} - \frac{14}{27}\pi^2 - \frac{25}{9}\zeta_3 \bigg) 
\eps^2
+
\mc O(\eps^3)
\bigg] 
\, .
\nnb
\eeq
A useful combination of these constituent integrals is 
\beq
\label{Jhck}
J_{\rm hc}^{\,k}(s)
& = &
( f_{k}^{q} \!+\! f_{k}^{\bar q} ) \,
J_{\rm hc}^{\og}(s)
+
f_{k}^{g} \,
\bigg[ 
N_{f} \, J_{\rm hc}^{\zg}(s)
+
\frac12 \, J_{\rm hc}^{\tg}(s)
\bigg]
\nnb\\
& = &
\frac{\as}{2\pi} \!
\left( \frac{s}{\mu^2} \right)^{\!\!-\eps}
\bigg[
\frac{\gamma^{\rm hc}_{k}}{\eps}
+
\phi^{\rm hc}_{k}
+
\mc O(\eps)
\bigg]
\, .
\eeq
The hard double-collinear integrated kernels are given by
\beq
\label{Jhcc}
J_{\rm hcc}^{ijkr}
& \; \equiv \; &
\Norm^{\,2}
\int d\Phi_{\rm rad,2}^{(ijkr)} \,
\frac{P_{ijk(r)}^{\rm hc}}{s_{ijk}^2} 
\nnb\\ 
& \; \equiv \; &
J_{\rm hcc}^{\zg} \Big( \sk{kr}{ijkr} \Big) \,
( f_{ijk}^{q \bar q q'} + f_{ijk}^{q \bar q \bar q'} )
+ 
J_{\rm hcc}^{(\rm 0g, id)} \Big( \sk{kr}{ijkr} \Big) \,
( f_{ijk}^{q \bar q q} + f_{ijk}^{q \bar q \bar q} )
\nnb \\
&&
\hspace{-4mm}
+ \,
J_{\rm hcc}^{\og} \Big( \sk{kr}{ijkr} \Big) \,
f_{ijk}^{q \bar q g}
+ 
J_{\rm hcc}^{\tg} \Big( \sk{kr}{ijkr} \Big) \,
( f_{ijk}^{gg q} + f_{ijk}^{gg \bar q} )
+ 
J_{\rm hcc}^{(\rm 3g)} \Big( \sk{kr}{ijkr} \Big) \,
f_{ijk}^{ggg}
\, ,
\eeq
with 
\beq
\label{constinthcc}
J_{\rm hcc}^{\zg}(s)
& = &
\bigg( \frac{\as}{2\pi} \bigg)^{\!2} \!
\left( \frac{s}{\mu^2} \right)^{\!\!-2\eps} \!
C_F T_R \,
\bigg[
\frac{1}{6} \, \frac1{\eps^2}
+
\left( \frac{13}{36} + \frac{1}{9}\pi^2 \right) \frac1{\eps}
- \,
\frac{119}{216} 
+
\frac{17}{108}\,\pi^2 
+ 
\frac{14}{3}\,\zeta_3
+
{\cal O}(\eps)
\bigg]
\, ,
\nnb\\
J_{\rm hcc}^{(\rm 0g, id)}(s)
& = &
\bigg( \frac{\as}{2\pi} \bigg)^{\!2} 
\left( \frac{s}{\mu^2} \right)^{\!\!-2\eps} 
C_F \, \big( 2 C_F -  C_A \big) \,
\nnb\\
&&
\times \,
\bigg[
- 
\left( \frac{13}{8} - \frac{1}{4}\,\pi^2 + \zeta_3 \right) 
\frac1{\eps}
- \,
\frac{227}{16}
+
\pi^2 
+ 
\frac{17}{2}\,\zeta_3
-
\frac{11}{120}\,\pi^4
+
{\cal O}(\eps)
\bigg]
\, ,
\nnb\\
J_{\rm hcc}^{\og}(s)
& = &
\bigg( \frac{\as}{2\pi} \bigg)^{\!2} 
\left( \frac{s}{\mu^2} \right)^{\!\!-2\eps} 
\nnb\\
&&
\hspace{-2mm}
\times \,
\Bigg\{
C_F T_R \,
\bigg[
-
\frac{2}{3}  \, \frac1{\eps^3}
-
\frac{31}{9} \, \frac1{\eps^2}
-
\left( \frac{889}{54} - \pi^2 \right) \frac1{\eps}
- \,
\frac{23833}{324} 
+
\frac{31}{6}\,\pi^2 
+ 
\frac{160}{9}\,\zeta_3
+
{\cal O}(\eps)
\bigg]
\nnb\\
&&
\hspace{3mm}
+ \,
C_{\!A} T_R \,
\bigg[
-
\frac1{\eps^3}
- 
\frac{89}{18} \frac1{\eps^2}
-
\left( \frac{1211}{54} \! - \! \frac{3}{2}\pi^2 \right) \! \frac1{\eps}
-
\frac{2620}{27} 
+
\frac{89}{12}\pi^2 
+ 
\frac{80}{3}\zeta_3
+
{\cal O}(\eps)
\bigg]
\Bigg\}
\, ,
\nnb\\
J_{\rm hcc}^{\tg}(s)
& = &
\bigg(\! \frac{\as}{2\pi} \!\bigg)^{\!\!2} \!\!
\left(\! \frac{s}{\mu^2} \!\right)^{\!\!\!-2\eps} \!
\Bigg\{
C_F^2\,
\bigg[
\;\;
-
\frac{2}{\eps^3}
-
\frac{37}{4} \frac1{\eps^2}
-
\left( 
\frac{307}{8} - 3\,\pi^2 + 4\,\zeta_3 
\right) \frac1{\eps}
\nnb\\
&&
\hspace{30mm}
- \,
\frac{2361}{16} 
+
\frac{111}{8}\,\pi^2 
+
\frac{136}{3}\,\zeta_3
-
\frac{\pi^4}{3}
+
{\cal O}(\eps)
\;
\bigg]
\nnb\\
&&
\hspace{22mm}
+ \,
C_F C_A\,
\bigg[
\;\;
-
\frac{1}{2} \, \frac1{\eps^3}
- 
\frac{23}{12} \frac1{\eps^2}
-
\left( 
\frac{241}{36} - \frac{1}{18}\,\pi^2 - 4\,\zeta_3 
\right) 
\frac1{\eps}
\nnb\\
&&
\hspace{37mm}
- \,
\frac{4609}{216} 
+
\frac{53}{216}\,\pi^2 
-
\frac{47}{6}\,\zeta_3
+
\frac{7}{20}\,\pi^4
+
{\cal O}(\eps)
\;
\bigg]
\;
\Bigg\}
\, ,
\nnb\\
J_{\rm hcc}^{(\rm 3g)}(s)
& = &
\bigg(\! \frac{\as}{2\pi} \!\bigg)^{\!\!2} \!\!
\left(\! \frac{s}{\mu^2} \!\right)^{\!\!\!-2\eps} \!
C_A^2 \,
\bigg[
\;\;
-
\frac{5}{2} \, \frac{1}{\eps^3}
-
\frac{77}{6} \, \frac1{\eps^2}
-
\left( 
48 - \frac{11}{4}\,\pi^2 + 3\,\zeta_3 
\right) 
\frac1{\eps}
\nnb\\
&&
\hspace{30mm}
- \,
\frac{16943}{108} 
+
\frac{61}{4}\,\pi^2 
+
\frac{56}{3} \, \zeta_3
-
\frac{9}{40}\,\pi^4
+
{\cal O}(\eps)
\;
\bigg]
\, .
\eeq
For the hard-collinear times hard-collinear integrated kernels we have
\beq
\label{Jhchc}
J^{\,ijklr}_{\rm hc \otimes hc}
& \; \equiv \; &
\Norm^{\,2}\, 
\int d\Phi_{\rm rad,2}^{(ijr,klr)}\,
\frac{P_{ij(r)}^{{\rm hc}}(s_{ir}, s_{jr})}{s_{ij}} \,
\frac{P_{kl(r)}^{{\rm hc}}(s_{kr}, s_{lr})}{s_{kl}}
\nnb\\
& \equiv &
J_{\rm hc \otimes hc}^{\rm qqqq}\Big(\sk{jr}{ijr,klr}\sk{lr}{ijr,klr}\Big) \,
f_{ij}^{q \bar q}
f_{kl}^{q' \bar q'}
\nnb\\
&&
\hspace{-3mm}
+ \,
J_{\rm hc \otimes hc}^{\rm qqqg}\Big(\sk{jr}{ijr,klr}\sk{lr}{ijr,klr}\Big) \,
\Big[
f_{ij}^{q \bar q}
( f_{kl}^{g q'} \!\!+\! f_{kl}^{g \bar q'} )
\!+\!
( f_{ij}^{g q'} \!\!+\! f_{ij}^{g \bar q'} )
f_{kl}^{q \bar q}
\Big]
\nnb\\
&&
\hspace{-3mm}
+ \,
J_{\rm hc \otimes hc}^{\rm qqgg}\Big(\sk{jr}{ijr,klr}\sk{lr}{ijr,klr}\Big) \,
( f_{ij}^{q \bar q}f_{kl}^{gg} + f_{ij}^{gg}f_{kl}^{q \bar q} )
\nnb\\
&&
\hspace{-3mm}
+ \,
J_{\rm hc \otimes hc}^{\rm qgqg}\Big(\sk{jr}{ijr,klr}\sk{lr}{ijr,klr}\Big) \,
( f_{ij}^{g q} + f_{ij}^{g \bar q} )
( f_{kl}^{g q'} + f_{kl}^{g \bar q'} )
\nnb\\
&&
\hspace{-3mm}
+ \,
J_{\rm hc \otimes hc}^{\rm qggg}\Big(\sk{jr}{ijr,klr}\sk{lr}{ijr,klr}\Big) \,
\Big[
( f_{ij}^{g q} \!+\! f_{ij}^{g \bar q} )
f_{kl}^{gg}
\!+\!
f_{ij}^{gg}
( f_{kl}^{g q} \!+\! f_{kl}^{g \bar q} )
\Big]
\nnb\\
&&
\hspace{-3mm}
+ \,
J_{\rm hc \otimes hc}^{\rm gggg}\Big(\sk{jr}{ijr,klr}\sk{lr}{ijr,klr}\Big) \,
f_{ij}^{gg}f_{kl}^{gg}
\, ,
\eeq
with
\beq
\label{constinthchc}
J_{\rm hc \otimes hc}^{\rm qqqq}(ss')
& = &
\!\bigg(\! \frac{\as}{2\pi} \!\bigg)^{\!\!2}
\left(\! \frac{ss'}{\mu^4} \!\right)^{\!\!\!-\eps}
T_R^{2} 
\left[
\frac{4}{9}\,\frac1{\eps^{2}}
+
\frac{64}{27}\,\frac1{\eps}
+
\frac{284}{27}
-
\frac{16}{27}\,\pi^{2}
+
\mc O (\eps)
\right]
\, ,
\nnb\\
J_{\rm hc \otimes hc}^{\rm qqqg}(ss')
& = &
\!\bigg(\! \frac{\as}{2\pi} \!\bigg)^{\!\!2}
\left(\! \frac{ss'}{\mu^4} \!\right)^{\!\!\!-\eps}
T_R \, C_{F}
\left[
\frac{1}{3}\,\frac1{\eps^{2}}
+
\frac{14}{9}\,\frac1{\eps}
+
\frac{181}{27}
-
\frac{4}{9}\,\pi^{2}
+
\mc O (\eps)
\right]
\, ,
\nnb\\
J_{\rm hc \otimes hc}^{\rm qqgg}(ss')
& = &
\!\bigg(\! \frac{\as}{2\pi} \!\bigg)^{\!\!2}
\left(\! \frac{ss'}{\mu^4} \!\right)^{\!\!\!-\eps}
T_R \, C_{A}
\left[
\frac{2}{9}\,\frac1{\eps^{2}}
+
\frac{32}{27}\,\frac1{\eps}
+
\frac{142}{27}
-
\frac{8}{27}\,\pi^{2}
+
\mc O (\eps)
\right]
\, ,
\nnb\\
J_{\rm hc \otimes hc}^{\rm qgqg}(ss')
& = &
\!\bigg(\! \frac{\as}{2\pi} \!\bigg)^{\!\!2}
\left(\! \frac{ss'}{\mu^4} \!\right)^{\!\!\!-\eps}
C_{F}^{2} \,
\left[
\frac{1}{4}\,\frac1{\eps^{2}}
+
\frac1{\eps}
+
\frac{17}{4}
-
\frac{1}{3}\,\pi^{2}
+
\mc O (\eps)
\right]
\, ,
\nnb\\
J_{\rm hc \otimes hc}^{\rm qggg}(ss')
& = &
\!\bigg(\! \frac{\as}{2\pi} \!\bigg)^{\!\!2}
\left(\! \frac{ss'}{\mu^4} \!\right)^{\!\!\!-\eps}
C_{A} \, C_{F}
\left[
\frac{1}{6}\,\frac1{\eps^{2}}
+
\frac{7}{9}\,\frac1{\eps}
+
\frac{181}{54}
-
\frac{2}{9}\,\pi^{2}
+
\mc O (\eps)
\right]
\, ,
\nnb\\
J_{\rm hc \otimes hc}^{\rm gggg}(ss')
& = &
\!\bigg(\! \frac{\as}{2\pi} \!\bigg)^{\!\!2}
\left(\! \frac{ss'}{\mu^4} \!\right)^{\!\!\!-\eps}
C_{A}^{2} \,
\left[
\frac{1}{9}\,\frac1{\eps^{2}}
+
\frac{16}{27}\,\frac1{\eps}
+
\frac{71}{27}
-
\frac{4}{27}\,\pi^{2}
+
\mc O (\eps)
\right]
\, .
\eeq
The soft-times-hard-collinear integrated kernels read 
\beq
\label{Jsofttimescoll}
J_{\rm s \otimes hc}^{\,jkricd}
& \equiv &
\Norm^{\,2}
\int \! d\Phi_{\rm rad,2}^{(jkr,icd)} \
\frac{P^{{\rm hc}}_{jk(r)}}{s_{jk}} \,
\mc E^{(i)}_{cd} 
\nnb\\
& \equiv &
f_{i}^{g} \,
\Big[
J_{\rm s \otimes hc}^{\,4\rm(1g)} \! \left(\sk{kr}{\mu}\!,\sk{cd}{\mu}\right) \!
f_{jk}^{q \bar q}
+ 
J_{\rm s \otimes hc}^{\,4\rm(2g)} \! \left(\sk{kr}{\mu}\!,\sk{cd}{\mu}\right) \!
( f_{jk}^{g q} \!+\! f_{jk}^{g \bar q} )
+
J_{\rm s \otimes hc}^{\,4\rm(3g)} \! \left(\sk{kr}{\mu}\!,\sk{cd}{\mu}\right) \!
f_{jk}^{gg}
\Big]_{\mu=jkr,icd}
\, ,
\nnb\\
J_{\rm s \otimes hc}^{\,jkricr} \,
& \equiv &
\Norm^{\,2}
\int \! d\Phi_{\rm rad,2}^{(jkr,icr)} \,
\frac{P^{{\rm hc}}_{jk(r)}}{s_{jk}} \,
\mc E^{(i)}_{cr} 
\nnb\\
& \equiv &
f_{i}^{g} \,
\Big[
J_{\rm s \otimes hc}^{\,3\rm(1g)} \! \left(\sk{kr}{\mu}\!,\sk{cr}{\mu}\right) \!
f_{jk}^{q \bar q}
+ 
J_{\rm s \otimes hc}^{\,3\rm(2g)} \! \left(\sk{kr}{\mu}\!,\sk{cr}{\mu}\right) \!
( f_{jk}^{g q} \!+\! f_{jk}^{g \bar q} )
+
J_{\rm s \otimes hc}^{\,3\rm(3g)} \! \left(\sk{kr}{\mu}\!,\sk{cr}{\mu}\right) \!
f_{jk}^{gg}
\Big]_{\mu=jkr,icr}
\, ,
\nnb\\
J_{\rm s \otimes hc}^{\,krjic} \,
& \equiv &
\Norm^{\,2}
\int \! d\Phi_{\rm rad,2}^{(krj,icj)} \,
\frac{P^{{\rm hc}}_{jk(r)}}{s_{jk}} \,
\mc E^{(i)}_{jc} 
\nnb\\
& \equiv &
f_{i}^{g} \,
\Big[
J_{\rm s \otimes hc}^{\,3\rm(1g)} \! \left(\sk{jr}{\mu}\!,\sk{jc}{\mu}\right) \!
f_{jk}^{q \bar q}
+ 
J_{\rm s \otimes hc}^{\,3\rm(2g)} \! \left(\sk{jr}{\mu}\!,\sk{jc}{\mu}\right) \!
( f_{jk}^{g q} \!+\! f_{jk}^{g \bar q} )
+
J_{\rm s \otimes hc}^{\,3\rm(3g)} \! \left(\sk{jr}{\mu}\!,\sk{jc}{\mu}\right) \!
f_{jk}^{gg}
\Big]_{\mu=krj,icj}
\, ,
\nnb\\
J_{\rm s \otimes hc}^{\,krjir} \,
& \; \equiv \; &
\Norm^{\,2}
\int \! d\Phi_{\rm rad,2}^{(\mu)} \,
\frac{P^{{\rm hc}}_{jk(r)}}{s_{jk}} \,
\mc E^{(i)}_{jr}
\nnb\\
& \equiv &
f_{i}^{g} \,
\Big[ \;
J_{\rm s \otimes hc}^{gqq} \! \left(\sk{jr}{\mu}\right) \!
f_{jk}^{q \bar q}
+ 
J_{\rm s \otimes hc}^{ggq} \! \left(\sk{jr}{\mu}\right) \!
f_{j}^{g} ( f_{k}^{q} \!+\! f_{k}^{\bar q} )
\nnb\\
&&
\hspace{4mm}
+ \,
J_{\rm s \otimes hc}^{gqg} \! \left(\sk{jr}{\mu}\right) \!
( f_{j}^{q} \!+\! f_{j}^{\bar q} ) f_{k}^{g}
+
J_{\rm s \otimes hc}^{ggg} \! \left(\sk{jr}{\mu}\right) \!
f_{jk}^{gg}
\Big]_{\mu= \{ krj,irj; \; krj,ijr  \}}
\, ,
\eeq
with 
\beq
\label{constintshc}
J_{\rm s \otimes hc}^{\,4 \rm(1g)}(s,\! s')
& = &
\!\bigg(\! \frac{\as}{2\pi} \!\bigg)^{\!\!\!2} \!\!
\left(\! \frac{ss'}{\mu^4} \!\right)^{\!\!\!-\eps} \!\!\!
T_R \!
\left[
-
\frac{2}{3} \frac1{\eps^{3}}
-
\frac{28}{9} \frac1{\eps^{2}}
-
\left( \frac{344}{27} \!-\! \frac{7}{9} \pi^{2} \!\right) \! \frac1{\eps}
-
\frac{3928}{81}
+
\frac{98}{27} \pi^{2}
+
\frac{100}{9} \zeta(3)
+
\mc O (\eps) \!
\right]
\! ,
\nnb\\
J_{\rm s \otimes hc}^{\,4 \rm(2g)}(s,\! s')
& = &
\!\bigg(\! \frac{\as}{2\pi} \!\bigg)^{\!\!\!2} \!\!
\left(\! \frac{ss'}{\mu^4} \!\right)^{\!\!\!-\eps} \!\!\!
C_F \!
\left[
-
\frac{1}{2} \frac1{\eps^{3}}
-
\frac2{\eps^{2}}
-
\left( 8 - \frac{7}{12} \pi^{2} \right) \! \frac1{\eps}
-
30
+
\frac{7}{3} \pi^{2}
+
\frac{25}{3} \zeta(3)
+
\mc O (\eps) \!
\right]
\, ,
\nnb\\
J_{\rm s \otimes hc}^{\,4 \rm(3g)}(s,\! s')
& = &
\!\bigg(\! \frac{\as}{2\pi} \!\bigg)^{\!\!\!2} \!\!
\left(\! \frac{ss'}{\mu^4} \!\right)^{\!\!\!-\eps} \!\!\!
C_A \!
\left[
-
\frac{1}{3} \frac1{\eps^{3}}
-
\frac{14}{9} \frac1{\eps^{2}}
-
\left( \frac{172}{27} \!-\! \frac{7}{18} \pi^{2} \!\right) \! \frac1{\eps}
-
\frac{1964}{81}
+
\frac{49}{27} \pi^{2}
+
\frac{50}{9} \zeta(3)
+
\mc O (\eps) \!
\right]
\!,
\nnb
\eeq
\beq
\label{eq:J(3,ng)}
J_{\rm s \otimes hc}^{\,3 \rm(1g)}(s,\! s')
& = &
\!\bigg(\! \frac{\as}{2\pi} \!\bigg)^{\!\!\!2} \!\!
\left(\! \frac{ss'}{\mu^4} \!\right)^{\!\!\!-\eps} \!\!\!
T_R \!
\left[
-
\frac{2}{3} \frac1{\eps^{3}}
-
\frac{28}{9} \frac1{\eps^{2}}
-
\!\left( \frac{362}{27} \!-\! \frac{8}{9} \pi^{2} \!\right) \! \frac1{\eps}
-
\frac{4504}{81}
+
\frac{112}{27} \pi^{2}
+
\frac{136}{9} \zeta(3)
+
\mc O (\eps) \!
\right]
\! ,
\nnb\\
J_{\rm s \otimes hc}^{\,3 \rm(2g)}(s,\! s')
& = &
\!\bigg(\! \frac{\as}{2\pi} \!\bigg)^{\!\!\!2} \!\!
\left(\! \frac{ss'}{\mu^4} \!\right)^{\!\!\!-\eps} \!\!\!
C_F \!
\left[
-
\frac{1}{2} \frac1{\eps^{3}}
-
\frac2{\eps^{2}}
-
\left( \frac{17}{2} - \frac{2}{3} \pi^{2} \right) \! \frac1{\eps}
-
35
+
\frac{8}{3} \pi^{2}
+
\frac{34}{3} \zeta(3)
+
\mc O (\eps) \!
\right]
\, ,
\nnb\\
J_{\rm s \otimes hc}^{\,3 \rm(3g)}(s,\! s')
& = &
\!\bigg(\! \frac{\as}{2\pi} \!\bigg)^{\!\!\!2} \!\!
\left(\! \frac{ss'}{\mu^4} \!\right)^{\!\!\!-\eps} \!\!\!
C_A \!
\left[
-
\frac{1}{3} \frac1{\eps^{3}}
-
\frac{14}{9} \frac1{\eps^{2}}
-
\left( \frac{181}{27} \!-\! \frac{4}{9} \pi^{2} \!\right) \! \frac1{\eps}
-
\frac{2252}{81}
+
\frac{56}{27} \pi^{2}
+
\frac{68}{9} \zeta(3)
+
\mc O (\eps) \!
\right]
,
\nnb
\eeq

\beq
J_{\rm s \otimes hc}^{gqq}(s)
& = & 
\!\bigg(\! \frac{\as}{2\pi} \!\bigg)^{\!\!\!2} \!\!
\left(\! \frac{s}{\mu^2} \!\right)^{\!\!\!-2\eps} \!\!
T_R \!
\left[
-
\frac{2}{3} \frac1{\eps^{3}}
-
\frac{28}{9} \frac1{\eps^{2}}
-
\left( \frac{344}{27} \!-\! \frac{17}{18} \pi^{2} \!\right) \! \frac1{\eps}
-
\frac{4225}{81}
+
\frac{128}{27} \pi^{2}
+
\frac{139}{9} \zeta(3)
+
\mc O (\eps) \!
\right]
\! ,
\nnb\\
J_{\rm s \otimes hc}^{gqg}(s)
& = & 
\!\bigg(\! \frac{\as}{2\pi} \!\bigg)^{\!\!\!2} \!\!
\left(\! \frac{s}{\mu^2} \!\right)^{\!\!\!-2\eps} \!\!
C_F \!
\left[
-
\frac12\frac1{\eps^{3}}
-
\frac2{\eps^{2}}
-
\left( 9 \!-\! \frac{5}{6} \pi^{2} \!\right) \! \frac1{\eps}
-
38
+
\frac{19}{6} \pi^{2}
+
\frac{101}{6} \zeta(3)
+
\mc O (\eps) \!
\right]
\, ,
\nnb\\
J_{\rm s \otimes hc}^{ggq}(s)
& = & 
\!\bigg(\! \frac{\as}{2\pi} \!\bigg)^{\!\!\!2} \!\!
\left(\! \frac{s}{\mu^2} \!\right)^{\!\!\!-2\eps} \!\!
C_F \!
\left[
-
\frac12\frac1{\eps^{3}}
-
\frac2{\eps^{2}}
-
\left( 8 \!-\! \frac{2}{3} \pi^{2} \!\right) \! \frac1{\eps}
-
32
+
\frac{17}{6} \pi^{2}
+
\frac{59}{6} \zeta(3)
+
\mc O (\eps) \!
\right]
\, ,
\\
J_{\rm s \otimes hc}^{ggg}(s)
& = & 
\!\bigg(\! \frac{\as}{2\pi} \!\bigg)^{\!\!\!2} \!\!
\left(\! \frac{s}{\mu^2} \!\right)^{\!\!\!-2\eps} \!\!
C_A \!
\left[
-
\frac{1}{3} \frac1{\eps^{3}}
-
\frac{14}{9} \frac1{\eps^{2}}
-
\left( \frac{199}{27} \!-\! \frac{5}{9} \pi^{2} \!\right) \! \frac1{\eps}
-
\frac{2477}{81}
+
\frac{119}{54} \pi^{2}
+
\frac{101}{9} \zeta(3)
+
\mc O (\eps) \!
\right]
.
\nnb
\eeq
Finally the hard-collinear real-virtual integrated kernels read 
\beq
\label{Jhc^ijr tilde}
\tilde J_{\rm hc}^{ijr}
& \equiv &
\Norm 
\int d\Phi_{\rm rad}^{(ijr)} \,
\frac{\tilde P_{ij(r)}^{\rm hc}}{s_{ij}} 
\, \equiv \,
\tilde J_{\rm hc}^{\zg} \!\! \left( \sk{jr}{ijr} \right)\! 
f_{ij}^{q \bar q}
+ 
\tilde J_{\rm hc}^{\og} \!\! \left(\sk{jr}{ijr}\right) \!
( f_{ij}^{g q} \!+\! f_{ij}^{g \bar q} )
+ 
\tilde J_{\rm hc}^{\tg} \!\! \left(\sk{jr}{ijr}\right) \!
f_{ij}^{gg}
\, ,
\nnb\\
J_{\!\!_{\Delta {\rm hc}}}^{ijr}
& \equiv &
\Norm \,
\frac{2}{\eps^2} 
\int d\Phi_{\rm rad}^{(ijr)} \,
\frac{P_{ij(r)}^{\rm hc}}{s_{ij}} 
\bigg[ \bigg(\frac{s_{cr}}{\sk{cr}{ijr}}\!\bigg)^{\!\!\!-\eps} - 1 \bigg]
\nnb\\
& \equiv &
J_{\!\!_{\Delta {\rm hc}}}^{\zg}\!\left(\sk{jr}{ijr}\right)
f_{ij}^{q \bar q}
+ 
J_{\!\!_{\Delta {\rm hc}}}^{\og}\!\left(\sk{jr}{ijr}\right)
( f_{ij}^{g q} \!+\! f_{ij}^{g \bar q} )
+
J_{\!\!_{\Delta {\rm hc}}}^{\tg}\!\left(\sk{jr}{ijr}\right)
f_{ij}^{gg}
\, ,
\nnb\\
J_{\!\!_{\Delta {\rm hc}}}^{ijrc}
& \equiv &
\Norm \,
\frac{2}{\eps^2} \!
\int \! d\Phi_{\rm rad}^{(ijr)}
\bigg\{
\frac{P_{\!ij(r)}^{\rm hc}\!}{s_{ij}} 
\bigg[ 1 \!-\! \bigg(\frac{\sk{jc}{ijr}}{s_{[ij]r}}\bigg)^{\!\!\!-\eps} \bigg]
+
2 
C_{\!f_j} 
\mc E^{(i)}_{jr} \!
\bigg[ 1 \!-\! \bigg(\!\frac{s_{jr}}{s_{[ij]r}}\!\bigg)^{\!\!\!-\eps} \bigg]
+
2 
C_{\!f_i} 
\mc E^{(j)}_{ir} \!
\bigg[ 1 \!-\! \bigg(\!\frac{s_{ir}}{s_{[ij]r}}\!\bigg)^{\!\!\!-\eps} \bigg]
\bigg\}
\nnb\\
& \equiv &
\left[ 
J_{\!\!_{\Delta {\rm hc,A}}}^{\zg}\!\!\left(\!\sk{jr}{ijr}\!\right)
\!+\!
J_{\!\!_{\Delta {\rm hc,B}}}^{\zg}\!\!\left(\!\sk{jc}{ijr}\!\right)
\right]
\!
f_{ij}^{q \bar q}
+ 
\left[ 
J_{\!\!_{\Delta {\rm hc,A}}}^{\og}\!\!\left(\!\sk{jr}{ijr}\!\right)
\!+\!
J_{\!\!_{\Delta {\rm hc,B}}}^{\og}\!\!\left(\!\sk{jc}{ijr}\!\right)
\right]
( f_{ij}^{g q} \!+\! f_{ij}^{g \bar q} )
\nnb\\
&&
\hspace{-3mm}
+ \,
\left[ 
J_{\!\!_{\Delta {\rm hc,A}}}^{\tg}\!\!\left(\!\sk{jr}{ijr}\!\right)
\!+\!
J_{\!\!_{\Delta {\rm hc,B}}}^{\tg}\!\!\left(\!\sk{jc}{ijr}\!\right)
\right]
f_{ij}^{gg}
\, ,
\nnb\\
J_{\!\!_{\Delta \rm hc}}^{jri,c}
& \equiv &
\Norm \,
\frac{\rho^{\scriptscriptstyle (C)}_{ij}}{\eps^2} 
\int d\Phi_{\rm rad}^{(jri)} \,
\frac{P_{ij(r)}^{\rm hc}}{s_{ij}} 
\bigg[ 
\bigg(\frac{\sk{ic}{jri}}{\sk{ir}{jri}}\bigg)^{\!\!\!-\eps} 
\!-\!
\bigg(\frac{s_{ir}\sk{ic}{jri}}{\sk{ir}{jri}s_{ic}}\bigg)^{\!\!\!-\eps}
\bigg]
\nnb\\
& \equiv &
\left[ 
J_{\!\!_{\Delta {\rm hc,A}}}^{\rm qq}\!\!\left(\!\sk{ir}{jri}\!\right)
\!+\!
J_{\!\!_{\Delta {\rm hc,B}}}^{\rm qq}\!\!\left(\!\sk{ic}{jri}\!\right)
\right]
f_{ij}^{q \bar q}
+ 
\left[ 
J_{\!\!_{\Delta {\rm hc,A}}}^{\rm qg}\!\!\left(\!\sk{ir}{jri}\!\right)
\!+\!
J_{\!\!_{\Delta {\rm hc,B}}}^{\rm qg}\!\!\left(\!\sk{ic}{jri}\!\right)
\right]
( f_{i}^{q} \!+\! f_{i}^{\bar q} )f_{j}^{g}
\nnb\\
&&
\hspace{-3mm}
+ \,
\left[ 
J_{\!\!_{\Delta {\rm hc,A}}}^{\rm gq}\!\!\left(\!\sk{ir}{jri}\!\right)
\!+\!
J_{\!\!_{\Delta {\rm hc,B}}}^{\rm gq}\!\!\left(\!\sk{ic}{jri}\!\right)
\right]
f_{i}^{g}( f_{j}^{q} \!+\! f_{j}^{\bar q} )
+
\left[ 
J_{\!\!_{\Delta {\rm hc,A}}}^{\rm gg}\!\!\left(\!\sk{ir}{jri}\!\right)
\!+\!
J_{\!\!_{\Delta {\rm hc,B}}}^{\rm gg}\!\!\left(\!\sk{ic}{jri}\!\right)
\right]
f_{ij}^{gg}
\, ,
\nnb\\
\tilde J_{\!\!_{\Delta \rm hc}}^{jri,c}
& \equiv &
\frac{\Norm}{\eps^2} 
\int \! d\Phi_{\rm rad}^{(jri)} \,
\frac{T_R}{s_{ij}} 
\bigg( 1 - \frac{2}{1 \!-\! \eps}\,\frac{s_{ir}s_{jr}}{s_{[ij]r}^{2}} \bigg) 
\bigg[ \!
\bigg(\!\frac{\sk{ic}{jri}}{\mu^2}\!\bigg)^{\!\!\!-\eps} \!\!
\!-\! 
\bigg(\!\frac{\sk{ic}{jri}}{s_{ic}}\!\bigg)^{\!\!\!-\eps}
\bigg]
\, \equiv \,
\tilde J_{\!\!_{\Delta \rm hc}}^{c}\!\!\left(\sk{ir}{jri}\!,\sk{ic}{jri}\right)
\, ,
\eeq
where 
\beq
\label{tildehcint}
\tilde J_{\rm hc}^{\zg}(s)
& = &
\frac{\as}{2\pi} \!
\left( \frac{s}{\mu^2} \!\right)^{\!\!\!-2\eps} \!\!
T_R \,
\bigg\{ \;
N_f T_R \,
\bigg[
\frac{4}{9} \frac1{\eps^2}
+
\frac{64}{27} \frac1{\eps}
+
\frac{284}{27}
-
\frac{2}{3}\pi^2 
+
\mc O(\eps)
\bigg]
\nnb\\
&&
\hspace{22mm}
+ \,
C_F 
\bigg[
\frac{2}{3}  \frac1{\eps^3}
+
\frac{31}{9} \frac1{\eps^2}
+
\bigg( \frac{431}{27} - \pi^2 \bigg) \frac1{\eps}
+
\frac{5506}{81}
-
\frac{31}{6}\pi^2 
-
\frac{124}{9}\zeta_3 
+
\mc O(\eps)
\bigg]
\nnb\\
&&
\hspace{22mm}
+ \,
C_A 
\bigg[\!
-
\frac1{3\eps^3}
-
\frac{31}{18} \frac1{\eps^2}
-
\bigg( \frac{211}{27} \!-\! \frac{1}{2}\pi^2 \!\bigg) \frac1{\eps}
-
\frac{5281}{162}
\!+\!
\frac{31}{12}\pi^2 
\!+\!
\frac{62}{9}\zeta_3 
+
\mc O(\eps)
\bigg]
\bigg\}
, 
\nnb\\[2mm]
\tilde J_{\rm hc}^{\og}(s)
& = & 
\frac{\as}{2\pi} 
\!\left(\! \frac{s}{\mu^2} \!\right)^{\!\!\!-2\eps} \!\!\!
C_F \,
\bigg\{ \;
C_F \,
\bigg[
-
\bigg( \frac{5}{4} - \frac{\pi^2}{3} \bigg) \frac1{\eps^2}
-
\bigg( \frac{15}{2} - \frac{2}{3}\pi^2 - 10\zeta_3 \bigg) \frac1{\eps}
\nnb\\ 
&& 
\hspace{39mm} 
- \, 
\frac{141}{4}
+
\frac{109}{24}\pi^2 
+
20\zeta_3 
-
\frac{7}{45}\pi^4
+
\mc O(\eps)
\bigg]
\nnb\\
&&
\hspace{21mm}
+ \,
C_A 
\bigg[
\frac1{4\eps^3}
\!+\!
\frac1{2\eps^2}
\!+\!
\bigg(\! 1 \!-\! \frac{\pi^2}{24} \!-\! 4 \zeta_3 \!\bigg) \frac1{\eps}
+
\frac{9}{4}
+
\frac{7}{12}\pi^2 
-
\frac{67}{6}\zeta_3 
-
\frac{11}{90}\pi^4
+
\mc O(\eps)
\bigg]
\bigg\}
\, ,
\nnb\\
\tilde J_{\rm hc}^{\tg}(s)
& = &
\frac{\as}{2\pi} 
\left( \frac{s}{\mu^2} \right)^{\!\!-2\eps} \!\!
C_A \,
\bigg\{ \;
N_f T_R \,
\bigg[
\frac{1}{3} \frac1{\eps}
+
\frac{25}{9}
+
\mc O(\eps)
\bigg]
\nnb
\\
&&
\hspace{27mm}
+ \,
C_A \,
\bigg[ \;
\frac{1}{6} \frac1{\eps^3}
-
\bigg( \frac{28}{9} - \frac{2}{3}\pi^2 \bigg) \frac1{\eps^2}
-
\bigg( \frac{61}{3} - \frac{7}{4}\pi^2 - 12\,\zeta_3 \bigg) \frac1{\eps}
\nnb\\
&&
\hspace{36mm}
- \,
\frac{15805}{162}
+
\frac{38}{3}\pi^2 
+
\frac{221}{9}\zeta_3 
-
\frac{5}{9}\pi^4
+
\mc O(\eps)
\bigg]
\bigg\}
\, ;
\eeq
\beq
\label{somedeltaint}
J_{\!_{\Delta {\rm hc}}}^{\zg}(s)
& = & 
\frac{\as}{2\pi}
\left( \frac{s}{\mu^2} \right)^{\! -\eps} \!
T_R 
\bigg[
-
\left( \frac{4}{3} - \frac{2}{9}\pi^2 \right)\frac1{\eps}
-
\frac{104}{9}
+
\frac{16}{27}\pi^2
+
8\zeta_3
+
\mc O(\eps)
\bigg]
\, ,
\nnb\\
J_{\!_{\Delta {\rm hc}}}^{\og}(s)
& = & 
\frac{\as}{2\pi} \!
\left( \frac{s}{\mu^2} \right)^{\!\! -\eps} \!\!
C_F 
\bigg[
- \bigg( 1 - \frac{\pi^2}{6} \bigg) \frac1{\eps}
- 8 + \frac{\pi^2}{3} + 6\zeta_3
+
\mc O(\eps)
\bigg]
\, ,
\nnb\\
J_{\!_{\Delta {\rm hc}}}^{\tg}(s)
& = & 
\frac{\as}{2\pi} \!
\left( \frac{s}{\mu^2} \right)^{\!\!\! -\eps} \!\!\!
C_A 
\bigg[
- \bigg( \frac{2}{3} - \frac{\pi^2}{9} \bigg) \frac1{\eps}
- \frac{52}{9} + \frac{8}{27}\pi^2 + 4\zeta_3
+
\mc O(\eps)
\bigg]
\, ;
\eeq
\beq
\label{someotherdeltaint}
J_{\!_{\Delta {\rm hc,A}}}^{\zg}(s)
& = & 
\frac{\as}{2\pi} \,
\!\left( \frac{s}{\mu^2} \right)^{\!\! -\eps} \!\!\!
T_R \,
\bigg[
- 
\frac{4}{3} \frac1{\eps^{3}}
-
\frac{32}{9} \frac1{\eps^{2}}
-
\bigg( \frac{280}{27} \!-\! \frac{7}{9}\pi^2 \!\bigg) \frac1{\eps}
-
\frac{2504}{81} 
+
\frac{56}{27}\pi^2 
+ 
\frac{100}{9}\zeta_3
+
\mc O(\eps)
\bigg]
\, ,
\nnb\\
J_{\!\!_{\Delta {\rm hc,B}}}^{\zg}(s)
& = & 
\frac{\as}{2\pi} \,
\!\left( \frac{s}{\mu^2} \right)^{\!\! -\eps} \!\!\!
T_R \,
\bigg[ \,
\frac{4}{3}
\frac1{\eps^{3}}
+
\frac{32}{9}
\frac1{\eps^{2}}
+
\bigg( \frac{244}{27} - \frac{5}{9}\pi^2 \bigg) 
\frac1{\eps}
+
\frac{1784}{81} 
- 
\frac{40}{27}\pi^2 
- 
\frac{52}{9}\zeta_3
+
\mc O(\eps)
\bigg]
\, ,
\nnb\\
J_{\!\!_{\Delta {\rm hc,A}}}^{\og}(s)
& = & 
\frac{\as}{2\pi} \,
\!\left( \frac{s}{\mu^2} \right)^{\!\! -\eps} \!\!\!
C_F \,
\bigg[
- \frac{1}{\eps^{3}}
- \bigg( 6 - \frac{2}{3}\pi^2 \bigg) \frac1{\eps^{2}}
- \bigg( 30 - \frac{23}{12}\pi^2 - 16\zeta_3 \bigg) \frac1{\eps}
\nnb\\
&&
\hspace{26mm}
- \, 126 + \frac{49}{6}\pi^2 + \frac{121}{3}\zeta_3 + \frac{\pi^4}{9} 
+
\mc O(\eps)
\bigg]
\, ,
\\
J_{\!\!_{\Delta {\rm hc,B}}}^{\og}(s)
& = & 
\frac{\as}{2\pi} \,
\!\left( \frac{s}{\mu^2} \right)^{\!\! -\eps} \!\!\!
C_F \,
\bigg[
\frac{1}{\eps^{3}}
+ \frac{2}{\eps^{2}}
+ \bigg( 5 - \frac{5}{12}\pi^2 \bigg) \frac1{\eps}
+ 12 - \frac{5}{6}\pi^2 - \frac{13}{3}\zeta_3 
+
\mc O(\eps)
\bigg]
\, ,
\nnb\\
J_{\!\!_{\Delta {\rm hc,A}}}^{\tg}(s)
& = &
\frac{\as}{2\pi} \,
\!\left( \frac{s}{\mu^2} \right)^{\!\! -\eps} \!\!\!
C_A \,
\bigg[
- \frac{2}{3} \frac1{\eps^{3}}
- \bigg( \frac{88}{9} - \frac{4}{3}\pi^2 \bigg)
\frac1{\eps^{2}}
- \bigg( \frac{1436}{27} - \frac{55}{18}\pi^2 - 32 \zeta_3 \bigg)
\frac1{\eps}
\nnb\\
&&
\hspace{26mm}
- \,\frac{18748}{81} + \frac{406}{27}\pi^2 + \frac{626}{9}\zeta_3 
+ \frac{2}{9}\pi^4 
+
\mc O(\eps)
\bigg]
\, ,
\nnb\\
J_{\!\!_{\Delta {\rm hc,B}}}^{\tg}(s)
& = &
\frac{\as}{2\pi} \,
\!\left( \frac{s}{\mu^2} \right)^{\!\! -\eps} \!\!\!
C_A \,
\bigg[
\frac{2}{3}\frac1{\eps^{3}}
+ \frac{16}{9} \frac1{\eps^{2}}
+ \bigg( \frac{122}{27} - \frac{5}{18}\pi^2 \bigg) \frac1{\eps}
+ \frac{892}{81} - \frac{20}{27}\pi^2 - \frac{26}{9}\zeta_3
+
\mc O(\eps)
\bigg]
\, ;
\nnb
\eeq
\beq
J_{\!\!_{\Delta \rm hc,A}}^{\rm qq}(s)
& = & 
\frac{\as}{2\pi} \,
\left( \frac{s}{\mu^2} \right)^{\!\!\! -\eps} \!\!
T_R \,
\bigg[
\frac{2}{3}
\frac1{\eps^{3}}
+
\frac{16}{9}
\frac1{\eps^{2}}
+
\bigg( \frac{122}{27} - \frac{4}{9}\pi^2 \bigg)
\frac1{\eps}
+
\frac{1108}{81} 
- 
\frac{44}{27}\pi^2 
- 
\frac{47}{9}\zeta_3
+
\mc O(\eps)
\bigg]
\, ,
\nnb\\
J_{\!\!_{\Delta \rm hc,B}}^{\rm qq}(s)
& = & 
\frac{\as}{2\pi} \,
\left( \frac{s}{\mu^2} \right)^{\!\!\! -\eps} \!\!
T_R
\bigg[
-
\frac{2}{3}
\frac1{\eps^{3}}
-
\frac{16}{9}
\frac1{\eps^{2}}
-
\bigg( \frac{140}{27} \!-\! \frac{7}{18}\pi^2 \!\bigg)
\frac1{\eps}
-
\frac{1252}{81} 
+
\frac{28}{27}\pi^2 
+
\frac{50}{9}\zeta_3
+
\mc O(\eps)
\bigg]
\, ,
\nnb\\
J_{\!\!_{\Delta \rm hc,A}}^{\rm qg}(s)
& = & 
\frac{\as}{2\pi} \,
\left( \frac{s}{\mu^2} \!\right)^{\!\!\! -\eps} \!\!\!\!
(2C_F\!-\!C_A)
\bigg[
\frac{1}{2} \frac1{\eps^{3}}
+ \frac1{\eps^{2}}
+ \bigg( \frac{7}{2} - \frac{11}{24}\pi^2 \bigg) \frac1{\eps}
+ \frac{23}{2} - \frac{2}{3}\pi^2 - \frac{32}{3}\zeta_3
+
\mc O(\eps)
\bigg]
\, ,
\nnb\\
J_{\!\!_{\Delta \rm hc,B}}^{\rm qg}(s)
& = & 
\frac{\as}{2\pi} \,
\left( \frac{s}{\mu^2} \right)^{\!\!\! -\eps} \!\!\!\!
(2C_F\!-\!C_A)
\bigg[
- \frac1{2\eps^{3}}
- \frac1{\eps^{2}}
- \bigg( 3 - \frac{7}{24}\pi^2 \bigg) \frac1{\eps}
- 9 + \frac{7}{12}\pi^2 + \frac{25}{6}\zeta_3
+
\mc O(\eps)
\bigg]
\, ,
\nnb\\
J_{\!\!_{\Delta \rm hc,A}}^{\rm gq}(s)
& = & 
\frac{\as}{2\pi} \,
\left( \frac{s}{\mu^2} \!\right)^{\!\!\! -\eps} \!\!
C_A \,
\bigg[
\frac1{2\eps^{3}}
+ \frac1{\eps^{2}}
+ \bigg( \frac{5}{2} - \frac{7}{24}\pi^2 \bigg) \frac1{\eps}
+ \frac{13}{2} - \frac{5}{6}\pi^2 - \frac{5}{3}\zeta_3
+
\mc O(\eps)
\bigg]
\, ,
\label{yetsomeotherdeltaint}
\\
J_{\!\!_{\Delta \rm hc,B}}^{\rm gq}(s)
& = & 
\frac{\as}{2\pi} \,
\left( \frac{s}{\mu^2} \right)^{\!\!\! -\eps} \!\!
C_A \,
\bigg[
- \frac1{2\eps^{3}}
- \frac1{\eps^{2}}
- \bigg( 3 - \frac{7}{24}\pi^2 \bigg) \frac1{\eps}
- 9 + \frac{7}{12}\pi^2 + \frac{25}{6}\zeta_3
+
\mc O(\eps)
\bigg]
\, ,
\nnb\\
J_{\!\!_{\Delta \rm hc,A}}^{\rm gg}(s)
& = &
\frac{\as}{2\pi} \,
\left( \frac{s}{\mu^2} \!\right)^{\!\!\! -\eps} \!\!
C_A \,
\bigg[
\frac1{3\eps^{3}}
+ \frac{8}{9} \frac1{\eps^{2}}
+ \bigg( \frac{88}{27} - \frac{11}{36}\pi^2 \bigg) \frac1{\eps}
+ \frac{716}{81} - \frac{17}{54}\pi^2 - \frac{64}{9}\zeta_3
+
\mc O(\eps)
\bigg]
\, ,
\nnb\\
J_{\!\!_{\Delta \rm hc,B}}^{\rm gg}(s)
& = &
\frac{\as}{2\pi} \,
\left( \frac{s}{\mu^2} \right)^{\!\!\! -\eps} \!\!
C_A \,
\bigg[
- \frac1{3\eps^{3}}
- \frac{8}{9} \frac1{\eps^{2}}
- \bigg( \frac{70}{27} - \frac{7}{36}\pi^2 \bigg)
\frac1{\eps}
- \frac{626}{81} + \frac{14}{27}\pi^2 + \frac{25}{9}\zeta_3
+
\mc O(\eps)
\bigg]
\, .
\nnb
\eeq


\bibliographystyle{JHEP}
\bibliography{subt}


\end{document}